\documentclass[a4paper,11pt,twoside,openright,fleqn]{report}

\usepackage{cite}
\usepackage{amssymb}
\usepackage{axodraw}

\newcommand{\spa}[2]{\langle#1#2\rangle}
\newcommand{\spb}[2]{[#1#2]}
\newcommand{\spab}[3]{\langle#1#2#3]}
\newcommand{\spba}[3]{[#1#2#3\rangle}
\newcommand{\spaa}[4]{\langle#1#2#3#4\rangle}
\newcommand{\spbb}[4]{[#1#2#3#4]}

\newcommand{\spxa}[1]{|#1\rangle}
\newcommand{\spax}[1]{\langle #1|}
\newcommand{\spxb}[1]{|#1]}
\newcommand{\spbx}[1]{[#1|}
\newcommand{\spxxa}[2]{|#1#2\rangle}
\newcommand{\spaxx}[2]{\langle#1#2|}
\newcommand{\spxxb}[2]{|#1#2]}

\newcommand{\spxxxa}[3]{|#1#2#3\rangle}
\newcommand{\spxxxb}[3]{|#1#2#3]}

\newcommand{\spbxxx}[3]{[#1#2#3|}

\newcommand{\hcyc}{\langle\cdots\rangle}

\newcommand{\wh}{\widehat}
\newcommand{\whP}{\widehat P}
\newcommand{\nn}{\nonumber\\}
\newcommand{\sla}{\slash\!\!\!}
\newcommand{\sign}{\mathrm{sign}}

\newcommand{\Tr}{\mathrm{Tr}}
\newcommand{\N}{\mathcal{N}}
\newcommand{\A}{\mathcal{A}}
\newcommand{\M}{\mathcal{M}}
\renewcommand{\H}{\mathcal{H}}
\newcommand{\Li}{\mathrm{Li}_{2}}

\newcommand{\mhv}{\mathrm{MHV}}

\begin{document}

\begin{titlepage}
  \begin{center}
    \ \\ \vspace{60mm}
    \huge\textbf{Unitarity and On-Shell Recursion Methods for
    Scattering Amplitudes}\\
    \vspace{10mm}
    \large by\\
    \vspace{6mm}
    \Large Kasper Risager\\
    \vspace{68mm}
    \large
    Submitted for the degree of Ph.D.~in Physics from the \\
    Niels Bohr Institute, Faculty of Science,\\
    University of Copenhagen.\\
    \vspace{8mm}
    November 2007\\
(April 2008)
    \normalsize
  \end{center}
\end{titlepage}
\begin{abstract}
\setcounter{page}{2}
This thesis describes some of the recent (and some less recent)
developments in calculational techniques for scattering amplitudes in
quantum field theory. The focus is on on-shell recursion relations in
complex momenta and on the use of unitarity methods for loop
calculations. In particular, on-shell recursion is related to the MHV
rules for computing tree-level gauge amplitudes and used to extend the
MHV rules to graviton scattering. Combinations of unitarity cut
techniques and recursion are used to argue for the ``No-Triangle
Hypothesis'' in $\N=8$ supergravity which is related to its UV
behaviour. Finally, combinations of unitarity and recursion are used to
demonstrate the full calculation of a one-loop amplitude involving a
Higgs particle and four gluons in the limit of large top mass. The
present version is edited to incorporate some of the comments and
suggestions of the evaluation committee, but has not been updated for
developments in the meantime.
\end{abstract}

\setcounter{page}{1}
\tableofcontents

\chapter{Introduction}
\label{cha:intro}

\section{Scattering Amplitudes in Theory and Experiment}

Ever since Physics arose as an empirical science, one of its main
objectives has been the identification and understanding of the
fundamental principles and constituents of Nature. Over time,
fundamental principles have changed or been extended, and
constituents regarded as fundamental have been found to contain even
more fundamental parts. For the last 100 years, this hunt for the
fundamental has been tied to the principles of quantum mechanics which
governs Nature at the fundamental level.

Particle physics as we know it today studies objects which are so
small and intangible that they can rarely be held in one place for
study, either because they are massless and move at the speed of light,
or because they simply decay too fast. At the same time, the measurement
apparatus cannot possibly be small enough to explore a fundamental
particle without being a particle itself. Thus, the only viable method
of exploration is to study the interactions between pairs of particles
that collide by identifying the consequences of this collision: What
comes out where, and how fast.

The physical quantity that describes this is the scattering cross
section. It specifies the area within which one particle must hit
the other for a specific process to take place, or rather, the area
where the particles do collide times the quantum mechanical
probability (density) that the given process will occur if the
particles collide. Such quantum mechanical probability densities
always arise as the absolute value squared of a quantum mechanical
amplitude, which is in this case called a scattering amplitude.

A scattering amplitude can receive contributions from different
processes which are indistinguishable in the quantum sense, that is,
processes which look the same to the measurement apparatus. To form a
scattering cross section, all these processes have to be summed before
squaring. Thus, if we wish to say something meaningful about a theory
with certain particles but not all of those we know to exist, we
should not start squaring the amplitude as we would be missing
interference terms from the other processes. This point is made to argue
that the scattering amplitude is the furthest one can calculate in a
theory without taking into account other physical processes or
experimental conditions. In that sense, the scattering amplitude is
the most proper way to describe a result in any reductionist approach
to particle physics.

Scattering amplitudes are normally calculated by perturbation theory
as power series in coupling constants. The lowest order results are
easily accessible with today's methods, but already the next order is
an analytical and numerical challenge. This is unfortunate because, as
we shall see, more precise (approximate or exact) results are in high
demand.

\subsection{Background Processes at the LHC}

For the next many years, particle physics will be dominated by the
experimental programme at CERN, where the Large Hadron Collider (LHC)
is being built. This machine will collide (for the most part) protons
on protons with a center-of-mass energy of 14 TeV. This is hoped to
produce an array of novel particles, such as supersymmetric partners
of all the known particles,
and expected to produce decisive statistical evidence for the
existence of the Higgs boson.

These interesting results will, however, be hiding behind massive
amounts of particle reactions that we already know and love from
earlier investigations of QCD and electroweak interactions. A good
knowledge of the production rates of these well-known events is
essential because they contaminate the interesting events and need to
be subtracted. Knowing only these background processes to the lowest
order in the coupling constants is likely to introduce large
theoretical uncertainties, some of them prohibitively large.

\begin{table}[htb]
\begin{center}
\begin{tabular}{|l|l|}
\hline
&\\
process&relevant for\\
($V\in\{Z,W,\gamma\}$)&\\
\hline
&\\
1. $pp\to V\,V$\,jet &  $t\bar{t}H$, new physics\\
2. $pp\to t\bar{t}\,b\bar{b}$ &  $t\bar{t}H$\\
3. $pp\to t\bar{t}+2$\,jets  &  $t\bar{t}H$\\
4. $pp\to V\,V\,b\bar{b}$ &  VBF$\to H\to VV$,
$t\bar{t}H$, new physics\\
5. $pp\to V\,V+2$\,jets &  VBF$\to H\to VV$\\
6. $pp\to V+3$\,jets &  various new physics signatures\\
7. $pp\to V\,V\,V$ &  SUSY trilepton\\
&\\
\hline
\end{tabular}
\caption{The LHC ``priority" wishlist,
  extracted from \cite{Buttar06Leshouches}. \label{tab:wishlist}}
\end{center}
\end{table}

Theorists and phenomenologists working on the LHC have issued a
prioritized wishlist of next-to-leading order calculations they want
done before the LHC turns on \cite{Buttar06Leshouches} (scheduled for
Spring 2008 at the time of writing) included as table
\ref{tab:wishlist}. At the time of writing, only the first and the
last seem to be close to solved \cite{Campbell07WWjet, Dittmaier07,
Lazopoulos07zzz}. In principle, it is known how to calculate all
these, but the conventional methods seem to have hit a roadblock; the
calculational complexity of the required calculations is rising faster
than the available resources, be they human or electronic. Decisive
progress in this area must come from both improved methods of
calculation and a strengthened effort.

\subsection{Approaching Quantum Gravity?}

Apart from the prosaic undertaking of extracting useful results from a
multi-billion Euro experiment, scattering amplitudes also play a role
in the theoretical understanding of quantum field theories. Amplitudes
express and reveal both the internal symmetries of a theory and its
internal inconsistencies.

One inconsistency plays a special role in the problems defining a
quantum theory of gravity, namely that of non-renormalizability. In
contrast to other quantum field theories where infinities that pop up
in intermediate results can be arranged to vanish in physical
quantities, gravity theories normally require more and more new terms
in their defining equations to counter the infinities which necessarily
appear.  

However, it so happens that the maximally supersymmetric ($\N=8$)
quantum field theory of gravity does not seem to have such
infinities. Calculation of the necessary scattering amplitudes in this
theory (and others) of gravity are unfortunately extremely cumbersome
and evidence is only accumulating slowly that this may indeed be the
case.

This understanding is driven by new methods of calculating
amplitudes. Thus, it may just be that such new methods can give a
glimpse of something which, without being a true description of all
known particles and interactions, can be called a consistent quantum
theory of gravity.

\section{Scattering Amplitudes in the Complex Plane}

Scattering amplitudes are usually calculated perturbatively using
Feynman rules. Feynman rules are derived directly from an action
principle, are understood by all particle physicists alike, and have
well-studied mathematical properties. A calculation done with Feynman
rules is rarely called in question, save for the pointing out of simple
algebraic errors. This makes Feynman rules the preferred choice of
almost everyone in the community, and with good reason.

Unfortunately, there are some issues with Feynman rules that make them
unfit in several situations. The first is the issue of complexity:
Since the rules work by summing over contributions from all graphs of
certain kinds, the complexity of the calculation is factorial in the
number of participating particles. When going beyond leading order in
coupling constants, there are integrals whose complexities are also
factorial in the number of particles. This leads to the roadblock
mentioned earlier. The second issue is that of unphysical
singularities: The results of the calculation are most often quoted in
a form where some singularities have to cancel between terms. When
evaluating such expressions numerically, roundoff error may introduce
apparent effects which are not really there.

A contributing factor to these problems is that Feynman rules throw
out information in exchange for more mechanical calculations. The
prime example of this is non-abelian gauge theories, where gauge
invariance enforces a particular structure of the three and four-point
interactions. In the actual calculations, however, no reference is
made to the gauge invariance, which might as well not have been
there. In the end, the constrained form of the Feynman rules ensure
that the amplitude is gauge invariant, but this has been obscured by
the intermediate calculation. This is just one of several pieces of
information which may be used to circumvent Feynman rules and obtain
amplitudes faster, in more compact forms, and with fewer unphysical
singularities.

\subsection{The Unitary and Analytic $S$-Matrix}

The focus of this thesis will be the use of methods of complex
analysis and methods derived from such considerations. Although most of
the methods described here have been developed since 2004, the idea
itself is by no means new. In some sense, it can be traced back to the
so-called $S$-matrix programme of the 60's.

The $S$-matrix is the evolution operator and is conventionally
decomposed into a unit piece (which describes no scattering) and into
the rest (which describes the actual scattering). 
\begin{equation}
\label{eq:tmatrixdefine}
S=1+iT.
\end{equation}
The scattering operator $iT$ sandwiched between a state of particles
coming in from the infinite past and a state of particles coming out
at infinite future, gives the scattering amplitude (also known as an
$S$-matrix element if the unit matrix is included). Requiring that the
$S$-matrix be unitary imposes a relation for the $T$ operator commonly
known as the optical theorem,
\begin{equation}
2\mathrm{Im}T=TT^\dagger.
\end{equation}
If we insert a complete set of states between $T$ and $T^\dagger$ on the right
side, we get the formal diagrammatic expression
\begin{equation}
\label{eq:optical}
\raisebox{35pt}{\Large 2Im~}
\begin{picture}(80,80)(0,0)
\Line(40,40)(80,40)
\Line(40,40)(60,5)
\Line(40,40)(20,5)
\Line(40,40)(0,40)
\Line(40,40)(20,75)
\Line(40,40)(60,75)
\GCirc(40,40){10}{.5}
\end{picture}
\raisebox{35pt}{\Large ~$\displaystyle{=\sum\int}$~}
\begin{picture}(120,80)(0,0)
\Line(40,40)(20,5)
\Line(40,40)(0,40)
\Line(40,40)(20,75)
\Line(40,40)(58,58)
\Line(40,40)(58,22)
\GCirc(40,40){8}{.5}
\Line(80,40)(100,75)
\Line(80,40)(120,40)
\Line(80,40)(100,5)
\Line(80,40)(62,22)
\Line(80,40)(62,58)
\GCirc(80,40){8}{.5}
\DashLine(60,0)(60,80)4
\end{picture}
\end{equation}
where the sum is over all possible insertions of internal states and
the integral is over their on-shell momenta. In perturbation theory,
the left side could correspond to a one-loop amplitude and the right
could correspond to a product of two tree amplitudes. 

If we also view the $S$-matrix as an analytic function of the
kinematical variables, we can gain more information. In general, a
one-loop amplitude as a complex function of one of the kinematic
variables will have a branch cut at real positive values of the
variable. The optical theorem can then be twisted to give the
discontinuity across the branch cut in terms of the insertions of all
possible states between two tree amplitudes, and the hope would be to
reconstruct the one-loop amplitude from its branch cut
discontinuities. The really interesting thing would then be if this
information could somehow be used to bootstrap amplitudes to all
orders in perturbation theory.

This was the goal of the $S$-matrix programme: to produce a theory for the
strong interactions (and, in the long run, all of particle physics)
from unitarity and analyticity constraints on the $S$-matrix. That
programme failed, partly because a host of other assumptions turned
out to be needed for the full reconstruction of the $S$-matrix, partly
because QCD, a Lagrangian quantum field theory, became established as
the correct theory of the strong interactions. A review of the methods
and motivations of the $S$-matrix programme can be found in
\cite{Eden}. 

In the 90's, it was realized that some of the tools of
$S$-matrix theory were more powerful than public opinion in the
community had assumed. The power, however, came from the input of
facts known from Feynman rules and dimensional regularization, two
concepts which were absent in the original $S$-matrix theory. Bern,
Dixon, Dunbar, Kosower and others used the known structure of certain
one-loop amplitudes to write them in a form where the discontinuities
across branch cuts could be expressed easily so that a (not even
complete) computation of the right side of (\ref{eq:optical}) would
almost completely determine the amplitude
\cite{Bern94oneloopn4mhv,Bern95oneloopn1mhv,Bern95massiveloop
,Bern96sdym}. We return to a detailed describtion of these methods in
chapter \ref{cha:loop}.

Although this so-called ``unitarity cut technique'' produced results
which were unthinkable to compute with Feynman graph techniques, and
attempted to use all known information as efficiently as possible, it
did lack in generality only being useful for some particular cases
(some of which were nevertheless relevant to experiment). A central
point in these methods is that only on-shell tree amplitudes are used
as input for the method, and this hints that we may, after all, be
able to bootstrap our way through the perturbative expansion.

This is not as far fetched as it may sound, because there are
already results that indicate that this is feasible in principle. The
Feynman Tree Theorem \cite{Feynman00ftt,Klauder72magic} gives exactly
such a construction. It works by using retarded propagators $\Delta_R$
in all propagators in a loop, which set the amplitude to zero because
of time ordering. By using that the retarded propagator is the Feynman
propagator plus a delta function
\begin{equation}
\Delta_R(P)=\Delta_F(P)-2\pi i\delta(P^2-m^2)\theta(-P_0)
\end{equation}
the expression can be rewritten as a sum of ordinary Feynman diagrams
where some number of propagators are Feynman and others are set
on-shell by the delta function. This construction shows directly that
the all-orders perturbative $S$-matrix can be written as some
complicated convolution of the tree-level $S$-matrix. The Feynman Tree
Theorem is discussed in more detail in \cite{Brandhuber05bstproof}.

\subsection{Generalizing from Loops to Trees}

As we saw above, the use of complex momentum space methods is not a
new invention, and at one-loop level, it has been used for some time
as a tool for the most efficient evaluation of scattering
amplitudes. Most tree-level calculations were still being conducted
in real Minkowski space using Feynman diagrams (or related methods) at
the turn of the century. The change in affairs would, however, come
from a completely different direction.

It had been known since the mid-80's that a certain class of tree
amplitudes were considerably simpler than a Feynman diagram
calculation would suggest when using the spinor helicity formalism, to
which we will turn in chapter \ref{cha:prelim}. These were the
so-called Maximally Helicity Violating (MHV) amplitudes where (when
all gluons are considered outgoing) two gluons have negative helicity
and any number have positive helicity. Building on some critical
earlier insight by Nair\cite{Nair88}, Witten
\cite{Witten03twistorstring} was able, in 2003, to describe gluon
tree-amplitudes from the topological B-model string theory in twistor
space, a type of theory not commonly associated to the nuts-and-bolts
of scattering calculations. Soon after it was realized that in the
same sense that MHV amplitdes are lines in twistor space,
tree-amplitudes with more negative helicity gluons were collections of
intersecting lines in twistor space. From this, it was only a small
leap to conjecture that there existed a formalism for calculating
gluon amplitudes whose vertices were MHV amplitudes.

That this was indeed the case was shown by Cachazo, Svr\v cek and
Witten in the beginning of 2004 \cite{Cachazo04csw}. The construction
is commonly known as either MHV rules or CSW rules. This thesis will
use the first, knowing that it may cause a bit of confusion (Feynman
rules generate Feynman amplitudes; MHV rules \emph{do not} just generate
MHV amplitudes). A description of MHV rules and some extensions can be
found in chapter \ref{cha:tree}. A good deal of calculations followed
which verified the MHV rules \cite{Kosower04nmhv, Zhu04paritycheck,
  Georgiou04mhv, Wu04paritycheck}. Brandhuber, Spence and
Travaglini were then able to use them for one-loop calculations
\cite{Brandhuber04bst} which was slightly surprising at the time.

All of these events were directly related to the twistor picture of
gauge amplitudes, which in itself did not involve complex momenta, but
to some extent it did involve momenta in $2+2$ dimensions. In that
metric signature, Britto, Cachazo and Feng \cite{Britto04quad}
realized that the unitarity cut technique described above can be
extended to insert four on-shell internal states rather than just two,
basically because the on-shell demand is easier to satisfy with two
time directions. When there are four on-shell constraints in a loop
integral it becomes trivial because the four coordinates of loop
momentum are fixed. This so-called ``Quadruple Cut Technique'' allowed
\emph{e.g.}~the conversion of the calculation of one-loop amplitudes
in maximally supersymmetric ($\N=4$) Yang--Mills to a purely algebraic
one in terms of products of tree amplitudes. It was also realized that
the constraint of being in 2+2 dimensions could effectively be removed
by assuming to be in 4 complex dimensions. We return to this
technology in chapter \ref{cha:loop}.

Since there are relations between tree and one-loop amplitudes in
$\N=4$ super-Yang--Mills \cite{Giele91ir, Catani98ir, Kunszt94ir,
Bern04sevengluon}, this allowed a quartic recursion relation for
on-shell tree-level Yang--Mills amplitudes \cite{Roiban04looptotree},
but with complex momenta. Britto, Cachazo and Feng \cite{Britto04bcf}
were then able to rewrite this as a quadratic recursion relation, but
they also showed together with Witten \cite{Britto05bcfw} that it
resulted from taking the external momenta seriously as complex and
using simple complex analysis. Moreover, they showed that these
recursion relations could be iterated such that only complex
three-point amplitudes contributed; in other words, the four-point
vertex in Yang--Mills can be thought of purely as an artifact of gauge
invariance necessary for Feynman rule calculations. These results will
be presented in chapter \ref{cha:tree}.

The remarkable developments described here have presented a completely
new view on the calculation of scattering amplitudes, and tie together
concepts across loop order in ways which are still not properly
understood. The consequences of treating scattering amplitude
calculations as complex analysis is, at its essence, the topic of this
thesis.

\subsection{Full Amplitudes from Trees}

The concept of on-shell recursion relations derived from
complex-momentum methods have numerous applications. Most obvious are
concrete tree-level calculations which come out in a form which is
more compact than the form from Feynman rules. Another tree-level
application is to show how the MHV rules mentioned above can be
viewed from recursion as a particular reorganization of the Feynman
rule calculation. The connection between MHV rules and on-shell
recursion will be explored in detail in chapter
\ref{cha:mhv}. 

Another use of on-shell recursion is for the calculation of one-loop
amplitudes. Parts of one-loop amplitudes can be computed using the
unitarity cut technique, and exactly the parts that are missed in that
technique are rational ones which have similarities to tree
amplitudes. This means that they are also computable by recursion
relations \cite{Bern05looprecursion1,Bern05looprecursion2,
  Bern05lastofthefinite} although in a slightly more involved
form. Such a determination of a one-loop amplitude from unitarity cuts
and recursion is presented in chapter \ref{cha:higgs}.

Numerous developments of unitarity combined with recursion have
accumulated over the last years and have improved one-loop
calculation. Interestingly, some results extend to a higher number of
loops, although they are primarily unitarity based. Founded on earlier
insight by Anastasiou, Bern, Dixon and Kosower
\cite{Anastasiou03twoloopbds}, Bern, Dixon and Smirnov
\cite{Bern05bds} have been able to propose a closed form result for
the all-loops $\N=4$ four and five-point amplitudes. Similarly,
there have been calculations up to three loops in $\N=8$
supergravity which lends credibility to the notion that it is
perturbatively UV finite, and hence a (comparably) well-defined theory
of gravity outside string theory. The input into the calculations are
tree amplitudes and the results are consequences of their
structure. These multi-loop results continue to attract attention from
outside the circle of specialists.

\section{About This Thesis}

\subsection{Omissions}

The purpose of this thesis is to review some of the central complex
momentum methods developed in recent years and to convey the results
of the authors own works in this area. As the reader will notice, the
review part already takes up a large part of the thesis, but still the
topics that are relevant and interesting in this context are so
numerous and complex that many of these things have had to be
omitted. 

Twistors are omitted completely, a choice which could be percieved as
radical when the field has such close connections to twistor string
theory. In the approach taken here, however, this connection is more
historical since complex momentum methods has managed to outrun the
twistor methods in many cases, in particular those cases chosen
here. Twistor string theory has been well reviewed
by Cachazo and Svr\v cek \cite{Cachazo05twistorreview} to which the
reader is referred. Gauge theories in twistor spaces have been
developed by \emph{e.g.}~\cite{Boels07twistormhv}. 

Another area which is not covered is that dealing with MHV rules as
derived from ordinary Yang--Mills in light-cone gauge. In this
approach \cite{Mansfield05lightconemhv, Gorsky06mhv}, the light-cone
Lagrangian is rewritten in a form where the MHV rules appear
naturally, and this permits extensions to loop calculations. Some
important works in this area are \cite{Ettle06mhv,
Brandhuber07bstregularized}. The viewpoint taken on MHV rules here is
that MHV rules are a special case of recursion relations, a viewpoint
which is admittedly too narrow.

There have also been refinements of many of the unitarity and
recursions mentioned here which would lead us too far if they were to
be explained in detail. These are mentioned in the relevant
connections. Within more traditional calculational approaches, there
have also been recent advances which will not be covered.

\subsection{Outline}

The thesis is roughly divided in two parts, having to do with
tree-level and one-loop level, and is not in historical order. It
starts out by explaining some basic results in Yang--Mills theory,
some notation, and the concept of gravity as a quantum field theory in
chapter \ref{cha:prelim}. Armed with those methods, we can take a look at
the two important tree-level developments, namely the MHV rules and
on-shell recursion in chapter \ref{cha:tree}. Based primarily on the
articles \cite{Risager05csw} by the author and \cite{Risager05grcsw}
together with Bjerrum-Bohr, Dunbar, Ita and Perkins, chapter
\ref{cha:mhv} describes how MHV rules can appear as a special case of
recursion relations, and how that can be used to derive MHV rules for
a gravitational theory. This highlights some of the similarities
between gravity and Yang--Mills that are obscure in their Lagrangian
formulation. 

The purpose of chapter \ref{cha:loop} is to introduce the loop level
methods that will guide us through the rest of the thesis, as well as
some related ones, and the remaining three chapters present some more
concrete one-loop calculations which serve different purposes. The
calculations of chapter \ref{cha:swi} serve to explain how amplitudes
in $\N=4$ super-Yang--Mills are computed using a variety of
techniques. The part of the calculation which regards amplitudes with
other external field content than gluons is based on
\cite{Risager05swi} together with Bidder and Perkins. Chapter
\ref{cha:sugra} contains calculations which elucidate the one-loop
structure of maximal ($\N=8$) supergravity, which is necessary for the
understanding of the all-loop structure. As such, it serves to support
the notion that $\N=8$ supergravity is UV finite, and is based on
\cite{Risager06notri} together with Bjerrum-Bohr, Dunbar, Ita and
Perkins.  The thesis ends on a more prosaic note with the calculation
in chapter \ref{cha:higgs} of the amplitude for a Higgs going to four
gluons of particular helicities in the limit where the top mass is
large. It is based on work together with Badger and Glover
\cite{Badger07phimhv}, but the treatment is slightly different from
the article.

\section{Conclusion and Outlook}

Rather than ending the thesis with a conclusion, we will sum up the
main points that the thesis will support, and give an outlook for the
future. 

\subsection{A Competitor in the Calculation Race}

The combined methods of unitarity and recursion have provided some
extremely effective means of calculating one-loop scattering
amplitudes. More specifically, they have allowed for the calculation
of hitherto unknown six-parton QCD amplitudes, and hold the promise of
making the progression to seven-parton QCD amplitudes smoother. As
demonstrated in chapter \ref{cha:higgs}, they have also allowed for
calculations with fewer external particles to be sufficiently
simplified that compact algebraic expressions for amplitudes can be
generated without the use of computer algebra. Thus, it has opened up
a new frontier for the specialists, and hopefully provided tools for
the many smaller groups of physicists with simpler calculations in
mind. New twists on these methods seem to keep coming in.

This is not to say that these methods are the only game in town. A
purely Lagrangian approach is still favoured by the broad community,
and there have been several refinements of conventional technology,
both semi-numerical and algebraic, which are reaching a comparable
level. One can only hope that a combined effort of all the methods on
the market can have the results in place when they are needed in the
near future.

\subsection{Understanding Quantum Field Theories}

The use of complex momentum methods have shown us that perturbative
quantum field theory is still far from understood. The investigation
of scattering amplitudes as complex functions gives several
puzzles. At tree level, the simplest version of on-shell recursion
described in chapter \ref{cha:tree} is known to work for both
Yang--Mills theories and gravity, but more complicated versions, such
as those used in chapter \ref{cha:mhv}, are not proven but seem
extremely likely to hold. Superficially, the existence of on-shell
recursion seems related to the UV behaviour of tree amplitudes, a
relation which is not clear either.

Equally, if not more, interesting are the relations \emph{between}
loop orders. Using unitarity, statements about loop amplitudes seem to
be reformulable as statements about tree amplitudes. In chapter
\ref{cha:sugra}, we will see how the existence of on-shell recursion
for tree gravity amplitudes comes close to proving that the one-loop
structure of $\N=8$ supergravity is the same as $\N=4$
super-Yang--Mills. Putting such relations on a firm footing would
contribute greatly to both our fundamental understanding of
perturbative quantum field theory, as well as our ability to do
calculations.

Together with the understanding of the full perturbative expansion of
$\N=4$ Yang--Mills and the prospects for a UV finite theory of
gravity, this makes the topic of this thesis most exciting, an
excitement that I hope that you, dear reader, will share.

Enjoy!

\chapter{Preliminaries}
\label{cha:prelim}

\section{Colour Ordering}
\label{sec:colourordering}

One of the more annoying aspects of doing gauge theory calculations is
the proliferation of indices, both for colour and for space-time. A
strategy to avoid this is to specify quantum numbers (such as
colour) as early as possible in the calculation. This can provide
expressions with less index mess, compensated by more expressions to
compute. The two chief methods for doing this are colour ordering as
described in this section and spinor helicity notation described in
the next. The contents of sections \ref{sec:colourordering} to
\ref{sec:ymtree} are also reviewed in \cite{Dixon96efficiently}.

The strategy inherent in colour ordering is to identify the possible
colour structures we can obtain when we have finished an on-shell
calculation. If we consider adjoint particles only, the colour
structures are products of structure constants, $f^{abc}$, which can
be written in terms of traces of $N_C\times N_C$ fundamental colour
matrices,
\begin{equation}
\label{eq:strucdef}
  f^{abc}=-\frac i{\sqrt2}\Tr[t^at^bt^c-t^at^ct^b],
\end{equation}
where we normalise the matrices as
\begin{equation}
  [t^a,t^b]=i\sqrt2 f^{abc}t^c,\quad \Tr[t^at^b]=\delta^{ab}.
\end{equation}
By using contraction identities such as (for $SU(N_C)$)
\begin{equation}
\label{eq:colourcompleteness1}
  (t^a)_i^{\bar j}(t^a)_k^{\bar l}=\delta_i^{\bar l}\delta_k^{\bar j}
-\frac1{N_C}\delta_i^{\bar j}\delta_k^{\bar l},
\end{equation}
we can write all colour structures in terms of traces of fundamental
colour matrices. Note that when we are contracting structure constants
only, the last term of (\ref{eq:colourcompleteness1}) will always drop
out. Thus, we can write the completeness relation as
\begin{equation}
  \label{eq:colourcompleteness2}
  \Tr[t^a\mathsf{M}]\Tr[t^a\mathsf{N}]=\Tr[\mathsf{MN}]
\end{equation}
where \textsf{M} and \textsf{N} are arbitrary matrix strings. 

At the end of the calculation, (the tree-level part of) an amplitude can
be written as a linear combination of the distinct trace structures,
\begin{equation}
  A_n=\sum_{\{\sigma_i\}\in S_n(1,\ldots,n)/Z_n}
\Tr[t^{a_{\sigma_1}}t^{a_{\sigma_2}}\cdots t^{a_{\sigma_n}}]
\A_n(\sigma_1,\sigma_2,\ldots,\sigma_n), 
\end{equation}
the simplest examples being
\begin{equation}
A_3=\Tr[t^{a_1}t^{a_2}t^{a_3}]\A_3(1,2,3)
+\Tr[t^{a_1}t^{a_3}t^{a_2}]\A_3(1,3,2)
\end{equation}
and
\begin{eqnarray}
A_4
&=&\Tr[t^{a_1}t^{a_2}t^{a_3}t^{a_4}]\A_4(1,2,3,4)
+\Tr[t^{a_1}t^{a_2}t^{a_4}t^{a_3}]\A_4(1,2,4,3)\nn
&&+\Tr[t^{a_1}t^{a_3}t^{a_2}t^{a_4}]\A_4(1,3,2,4)
+\Tr[t^{a_1}t^{a_3}t^{a_4}t^{a_2}]\A_4(1,3,4,2)\nn
&&+\Tr[t^{a_1}t^{a_4}t^{a_2}t^{a_3}]\A_4(1,4,2,3)
+\Tr[t^{a_1}t^{a_4}t^{a_3}t^{a_2}]\A_4(1,4,3,2).
\end{eqnarray}
The coefficients of the trace structures are called \emph{colour
  ordered} amplitudes, because they correspond to a particular
ordering of the colour matrices. Instead of calculating the amplitude,
using (\ref{eq:strucdef}) to (\ref{eq:colourcompleteness2}), and
  deducing the colour ordered amplitudes we can calculate them
  directly by using colour ordered Feynman rules. A diagram then
  contributes to a colour ordered amplitude if the cyclic ordering of
  the external particles corresponds to that of the wanted colour
  ordered amplitude. 

If we are dealing with particles in the fundamental representation,
the strings of matrices will not be traced, because the colour indices
of the fundamental particles are free. It is still possible to define
a colour ordering, \emph{e.g.},
\begin{equation}
A_4(q_1^i,\bar q_{2\bar j},3^{a_3},4^{a_4})
=[t^{a_3}t^{a_4}]^i_{\bar j}\A_4(q_1,3,4,\bar q_2)
+[t^{a_4}t^{a_3}]^i_{\bar j}\A_4(q_1,4,3,\bar q_2).
\end{equation}
If we work in an $SU(N_C)$ gauge theory, we now have to take the last
term of (\ref{eq:colourcompleteness1}) into account. This corresponds
to subtracting at each vertex the contribution of a $U(1)$ gauge
boson.

At loop level, we will also encounter multi-trace terms which we need
to compute. At one-loop level, however, it can be shown that all
two-trace terms are permutations of the (``planar'') single-trace
terms \cite{Bern94oneloopn4mhv}.

\section{Spinor Helicity Notation}
\label{sec:spinor}

After trading the colour information for more diagrams, we can now
trade the use of polarization vectors for yet more diagrams. This
consists of writing the amplitudes completely (or rather, to the
extent possible) in terms of Weyl spinors of massless particles. We
will find that in this notation we can write down explicit
polarization vectors corresponding to positive and negative helicity
gluons. 

We start by deriving some basic facts about Weyl spinors. The Weyl
spinors corresponding to a massless momentum can be found by solving
the massless Dirac equation
\begin{equation}
\label{eq:diracequation}
  p_\mu\gamma^\mu u(p)=\sla pu(p)=0
\end{equation}
where $u(p)$ is a four component vector and $\gamma^\mu$ is a vector
of $4\times 4$ matrices obeying
\begin{equation}
  \{\gamma^\mu,\gamma^\nu\}=2g^{\mu\nu}I.
\end{equation}
We use ``Peskin and Schroeder'' conventions \cite{PeskinSchroeder} where
\begin{equation}
  \gamma^\mu=\left(
    \begin{array}{c c}
      0&\sigma^\mu\\
      \bar\sigma^\mu&0
    \end{array}\right)
\end{equation}
Switching now to Weyl--van der Waerden notation the Dirac spinor can
be written as two two-component spinors
\begin{equation}
  u(p)=\left(\begin{array}{c}u_\alpha(p)\\\tilde u^{\dot\alpha}
(p)\end{array}\right)
\end{equation}
whose indices are raised and lowered with $\epsilon_{\alpha\beta}$, 
$\epsilon_{\dot\alpha\dot\beta}$, \emph{etc.}, using
southwest-northeast contraction.
We define antisymmetric inner products between spinors,
\begin{equation}
  \spa \lambda\eta=\epsilon_{\alpha\beta}\lambda^\alpha\eta^\beta
=\epsilon^{\alpha\beta}\lambda_\alpha\eta_\beta=\lambda^\alpha
\eta_\alpha,
\end{equation}
\begin{equation}
  \spb {\tilde\lambda}{\tilde\eta}=\epsilon_{\dot\beta\dot\alpha}
\tilde\lambda^{\dot\alpha}\tilde\eta^{\dot\beta}=
\epsilon^{\dot\beta\dot\alpha}
\tilde\lambda_{\dot\alpha}\tilde\eta_{\dot\beta}=
\tilde\lambda_{\dot\alpha}\tilde\eta^{\dot\alpha},
\end{equation}
thus we are working with the ``QCD'' sign convention rather than the
``string'' sign convention. In this notation (\ref{eq:diracequation})
becomes 
\begin{equation}
  p\cdot\sigma_{\beta\dot\alpha}\tilde u^{\dot\alpha}(p)
=p\cdot\sigma^{\dot\beta\alpha}u_\alpha(p)=0,
\end{equation}
which can further be elucidated by noting that
$p\cdot\sigma^{\dot\alpha\alpha}$ can be decomposed into a product of
two two-component spinors because $\det(p\cdot\sigma)=p^2=0$:
\begin{equation}
  p\cdot\sigma^{\dot\alpha\alpha}=\tilde\lambda^{\dot\alpha}(p)
\lambda^\alpha(p),
\end{equation}
which is solved by
\begin{equation}
  \lambda^1=\sqrt{|p_0+p_3|},\quad \lambda^2=\sign(p_0)\frac{p_1+ip_2}{\sqrt{|
p_0+p_3|}},
\end{equation}
\begin{equation}
  \lambda_1=\sign(p_0)\frac{p_1+ip_2}{\sqrt{|p_0+p_3|}},\quad \lambda_2
=-\sqrt{|p_0+p_3|}.
\end{equation}
\begin{equation}
  \tilde\lambda^{\dot\alpha}=\sign(p_0)(\lambda^\alpha)^*
\end{equation}
From the above properties of the two-component spinors it follows
immediately that the Dirac spinors satisfying the Dirac equation are
spanned by
\begin{equation}
  u^+(p)=\left(
    \begin{array}{c}
      \lambda_\alpha(p)\\0
    \end{array}\right)\equiv \spxa p,\qquad
u^-(p)=\left(
  \begin{array}{c}
    0\\\tilde\lambda^{\dot\alpha}(p)
  \end{array}\right)\equiv \spxb p,
\end{equation}
whose conjugate spinors are
\begin{equation}
  \bar u^-(p)=\left(
    \begin{array}{c}
      \lambda^\alpha(p)\\0
    \end{array}\right)^T\equiv \spax p,\qquad
\bar u^+(p)=\left(
  \begin{array}{c}
    0\\\tilde\lambda_{\dot\alpha}(p)
  \end{array}\right)^T\equiv \spbx p.
\end{equation}
The definitions of $\spxa p$, $\spxb p$ \emph{etc.}, ensure that $\spa
pq =\spa{\lambda(p)}{\lambda(q)}$ and $\spb
pq =\spb{\tilde\lambda(p)}{\tilde\lambda(q)}$ thereby helping to simplify the
algebra significantly. In terms of momenta, the antisymmetric inner
products are 
\begin{equation}
  \spa pq=-\sign(p_0)\sqrt{\left|\frac{q_0+q_3}{p_0+p_3}\right|}(p_1+ip_2)
+\sign(q_0)\sqrt{\left|\frac{p_0+p_3}{q_0+q_3}\right|}(q_1+iq_2),
\end{equation}
\begin{equation}
  \spb pq=\sign(q_0)
\sqrt{\left|\frac{q_0+q_3}{p_0+p_3}\right|}(p_1-ip_2)
-\sign(p_0)\sqrt{\left|\frac{p_0+p_3}{q_0+q_3}\right|}(q_1-iq_2),
\end{equation}
which obey
\begin{equation}
  \spa pq\spb qp=(p+q)^2=2p\cdot q,\qquad \spb qp=\sign(p_0q_0)
\spa pq^*.
\end{equation}

We will often encounter these spinors strung together using gamma
matrices. The following identities can be shown by using standard
Dirac algebra:
\begin{equation}
  \spab p{|\gamma^\mu|}q=\spba q{|\gamma^\mu|}p,
\end{equation}
\begin{equation}
  \spab q{|\gamma^\mu|}q=2q^\mu,
\end{equation}
\begin{equation}
  \spab p{|\gamma^\mu|}q\spab r{|\gamma_\mu|}s=-2\spa pr\spb qs\qquad
\textrm{(Fierz rearrangement)},
\end{equation}
\begin{equation}
  \spaa p{|\gamma^\mu}{\gamma^\nu|}q=-\spaa q{|\gamma^\nu}{\gamma^\mu|}p,
\end{equation}
\begin{equation}
  \spa pq\spax r+\spa qr\spax p+\spa rp\spax q=0\qquad
\textrm{(Schouten Identity)}.
\end{equation}
The last two have equivalent forms with $\spax\cdot\to\spbx\cdot$. We
will suppress Feynman slashes in general unless possibilities of
misunderstandings arise. In fact, we will quite often write momenta as
their slashed versions rather than their vector versions, as the whole
point of spinor helicity notation is to get rid of explicit
vectors. When we write massless momenta in matrix form we should
really write
\begin{equation}
  p=\spxa p\spbx p+\spxb p \spax p,
\end{equation}
but since only one of them avoids being projected out in spinor strings
such as $\spab q{|p|}r$ or $\spaa q{|p|}{r|}s$, we will allow
ourselves the sloppiness of writing
\begin{equation}
  p=\spxa p\spbx p\quad\textrm{or}\quad p=\spxb p\spax p.
\end{equation}

We now know how to treat expressions with Weyl spinors and
momenta. The last objects which may occur in a colour ordered
scattering amplitude are polarization vectors. We use the two explicit
realizations
\begin{equation}
  \epsilon_\mu^+(p)=\frac{\spab q{|\gamma_\mu|}p}{\sqrt 2\spa
  qp},\qquad
\epsilon_\mu^-(p)=\frac{\spab p{|\gamma_\mu|}q}{\sqrt 2\spb pq}.
\end{equation}
In both of these, $q$ is some massless reference momentum. Using the
rules above it is quite straightforward to verify the standard
properties of polarization vectors,
\begin{equation}
  \epsilon^\pm(p)\cdot p=0,\qquad \epsilon^+_\mu(p)=
(\epsilon_\mu^-(p))^*,
\end{equation}
\begin{equation}
  \epsilon^+(p)\cdot\epsilon^-(p)=-1,\qquad
\epsilon^+(p)\cdot\epsilon^-(p)=0.
\end{equation}
We can argue that $q$ is an arbitrary (apart from $p\cdot q\neq 0$)
gauge vector by considering the difference beween two different
choices of $q$:
\begin{eqnarray} 
&&\epsilon_\mu^+(p;q^\prime)-\epsilon_\mu^+(p;q)\\
&=&\frac{\spab {q^\prime}{|\gamma_\mu|}p}{\sqrt2\spa{q^\prime}p}
-\frac{\spab q{|\gamma_\mu|}p}{\sqrt2\spa qp}\\
&=&\frac{-\spaa {q^\prime}{|\gamma_\mu}{\gamma_\nu|}q
+\spaa q{|\gamma_\mu}{\gamma_\nu|}{q^\prime}}{\sqrt2\spa{q^\prime}p
\spa qp}p^\nu\\
&=&\frac{\langle q|\{\gamma_\mu,\gamma_\nu\}|q^\prime\rangle}
{\sqrt2\spa{q^\prime}p\spa qp}p^\nu\\
&=&\frac{\sqrt2\spa q{q^\prime}}{\spa{q^\prime}p\spa qp}p_\mu.
\end{eqnarray}
Thus, the two choices give the same results due to the Ward
identity. Finally, we note that the polarization sum is that of a
lightlike axial gauge with gauge vector $q$:
\begin{equation}
  \epsilon_\mu^+\epsilon_\nu^-+\epsilon_\mu^-\epsilon_\nu^+
=-g_{\mu\nu}+\frac{p_\mu q_\nu+q_\mu p_\nu}{p\cdot q}.
\end{equation}
This particular gauge is ghost-free, which will extremely helpful in 
later chapters when we deal with loops. 

Recently, a Mathematica package for manipulations with spinors in this
context called S@M has been published \cite{Mastrolia07satm}, but
using slightly different conventions.

\section{Tree-Level Structure of Yang--Mills Theory}
\label{sec:ymtree}

This section collects some facts of Yang--Mills theory that we will
need in the future, and is not intended as a review of the subject. We
will return to the one-loop structure of Yang--Mills amplitudes in
section \ref{sec:loopstructure}.

\subsection{MHV Amplitudes}

It has already been mentioned that Yang--Mills scattering amplitudes
take on very simple forms in spinor helicity notation. The most
striking example of this was conjectured by Parke and Taylor
\cite{Parke86mhv} and
proven by Berends and Giele \cite{Berends87mhv}, and states that
\begin{eqnarray}
\label{eq:parketaylor}
\A(1^+,\ldots,n^+)&=&0,\nn
\A(1^+,\ldots,i^-,\ldots,n^+)&=&0,\nn
\A(1^+,\ldots,i^-,\ldots,j^-,\ldots,n^+)
&=&i\frac{\spa ij^4}{\spa 12\spa 23\cdots\spa n1},
\end{eqnarray}
where $\ldots$ denotes any number of positive helicity gluons. The
last of these is called the ``Maximally Helicity Violating''
amplitude, or MHV amplitude for short. There is a simiar expression
for the amplitude with two positive and the rest negative helicity
obtained by complex conjugation,
\begin{equation}
\A(1^-,\ldots,i^+,\ldots,j^+,\ldots,n^-)
=(-1)^ni\frac{\spb ij^4}{\spb 12\spb 23\cdots\spb n1},
\end{equation}
called the ``googly MHV amplitude''\footnote{The word ``googly'' is a
  cricket reference often used in twistor theory. In cricket it
  denotes a ball bowled with the opposite spin of the normal, thrown
  by a right-handed bowler.}. These formulas determine all gluon
  amplitudes up to and including five points, since all those
  amplitudes are either zero, MHV or googly-MHV.

It should be noted that the ``Maximally'' in MHV becomes a misnomer
at loop level since the two first equations of (\ref{eq:parketaylor}) cease
to hold in non-supersym\-metric theories.

\subsection{Supersymmetry}
\label{sec:swi}

Tree-level amplitudes in Yang--Mills theory have an apparent
supersymmetry \cite{Parke85susy,Kunszt85susy} because, say, a gluino
and any adjoint fermion have the same Feynman rules. Also, at the
level of individual colour ordered amplitudes fundamental and adjoint
fermions are really indistinguishable since their colour ordered
Feynman rules are the same. The difference only shows when we compute
amplitudes with non-adjacent fermions where the fundamental fermion
amplitude must vanish. The actual supersymmetry only reveals itself at
loop level where the number of particle species becomes important.

The tree-level supersymmetry can be exploited to constrain tree
amplitudes by ``Supersymmetric Ward Identities'', SWI's
\cite{Grisaru76swi1,Grisaru77swi2}, which relate
amplitudes of different external particle content but with the same
momenta and the same sum of helicities. Following
\cite{Dixon96efficiently}, SWI's are obtained by writing the amplitude
as a string of operators $\Phi_i$ cretaing helicity eigenstates by
acting on the vacuum
\begin{equation}
\langle 0|\Phi_1\Phi_2\cdots\Phi_n|0\rangle.
\end{equation}
We now introduce the spinorial supercharge $Q_\alpha$ and contract it as
$Q(q,\theta)=\theta q^\alpha Q_\alpha$ where $\theta$ is a Grassmann
parameter and $q^\alpha$ is the Dirac spinor of some lightlike
momentum. The inclusion of $\theta$ makes $Q(q,\theta)$ a bosonic
operator. Since $Q_\alpha$ annihilates the vacuum, we can write
\begin{equation}
0=\langle 0 |\,[Q(q,\theta),\Phi_1\Phi_2\cdots\Phi_n]\,|0\rangle
=\sum_{i=1}^n\langle 0|\Phi_1\cdots[Q(q,\theta),\Phi_i]\cdots
\Phi_n|0\rangle,
\end{equation}
where the last expression must represent a sum over amplitudes where
$Q$ has exchanged one particle with another. If one computes all the
commutators, one will get
\begin{eqnarray}
[Q(q,\theta),g^+(k)]&=&\theta\spb kq f^+,\nn\
[Q(q,\theta),f^+(k)]&=&\theta\spa kq g^+,\nn\
[Q(q,\theta),f^-(k)]&=&\theta\spb qk g^-,\nn\
[Q(q,\theta),g^-(k)]&=&\theta\spa qk f^-.
\end{eqnarray}
This permits us to prove the two first formulas of
(\ref{eq:parketaylor}). The second can be proven by acting on $\langle
0|g^-_1f^+_2g^+_3\cdots g^+_n|0\rangle$,
\begin{equation}
0=\spa q1\A(f^-_1,f^+_2,\ldots)-\spa q2\A(g^-_1,g^+_2,\ldots),
\end{equation}
and choosing $q=1$ to eliminate the first term. Notice that the
amplitudes with two $f^+$'s are zero because of the requirement of
conservation of fermion helicity.

We can also extend the formula for MHV gluon amplitudes to amplitudes
with a fermion pair by acting on an amplitude with two negative
helicity gluons at 1 and $i$ and a positive helicity fermion at
$j$ and the rest positive helicity gluons. This gives the SWI
\begin{equation}
0=\spa q1 \A(f^-_1,g^-_i,f^+_j)+\spa qi\A(g^-_1,f^-_i,f^+_j)
-\spa qj\A(g^-_1,g^-_i,g^+_j),
\end{equation}
and choosing $q=1$ reduces this to
\begin{equation}
\label{eq:fermionmhv}
\A(g^-_1,f^-_i,f^+_j)=\frac{\spa 1j}{\spa 1i}\A(g^-_1,g^-_i,g^+_j)
=i\frac{\spa 1j\spa 1i^3}{\spa 12\spa 23\cdots\spa n1}.
\end{equation}
This can be explained as moving a half unit of negative helicity from
$i$ to $j$, thereby exchanging a factor of $\spa 1i$ with a factor of
$\spa 1j$.

SWI's can be extended to describe other supersymmetries. For example,
$\N=2$ supersymmetry would contain a scalar and thus allow
(\ref{eq:fermionmhv}) to be extended to include those. The $\N=4$
supersymmetry algebra will be of special importance to us since that
theory enters at both tree and loop level in several places below. It
has a gluon $g^\pm$, four fermions $f_a^\pm$, six real scalars
$s_{ab}=-s_{ba}$, and the algebra
\begin{eqnarray}
\label{eq:n4algebra}
[Q_a(q,\theta),g^+(k)]&=&\theta\spb kq f_a^+,\nn\
[Q_a(q,\theta),f_b^+(k)]&=&\theta\delta_{ab}\spa kq g^++\theta
\spb kqs_{ab},\nn\
[Q_a(q,\theta),s_{bc}(k)]&=&\theta\delta_{ab}\spa kq f_c^+
-\theta\delta_{ac}\spa kqf_b^+
+\theta\spb qk\epsilon_{abcd}f_d^-,\nn\
[Q_a(q,\theta),f_b^-(k)]&=&\theta\delta_{ab}\spb qk g^-+\frac12\theta
\spa qk\epsilon_{abcd}s_{cd}\nn\
[Q_a(q,\theta),g^-(k)]&=&\theta\spa qk f_ a^-.
\end{eqnarray}
Notice that we might as well have used $\tilde
s_{ab}=\frac12\epsilon_{abcd}s_{cd}$. Using these, we can deduce SWI's
for theories with several different fermions and scalars, as well as
SWI's for $\N=4$ amplitudes to any loop order. In particular, we can
deduce the exact form of all tree-level MHV amplitudes in $\N=4$ SYM
by a procedure like (\ref{eq:fermionmhv}).

\subsection{Limit Behaviour}
\label{sec:treelimit}

We will also briefly touch on the behaviour of Yang--Mills tree
amplitudes in different limits. In a limit where a multi-particle
kinematic invariant (and thus a propagator) goes on-shell, the residue
factorizes into two on-shell amplitudes, summed over possible internal
states. In the pure-glue case we have
\begin{eqnarray}
  \A(\ldots)&\to& \sum_{h=\pm}
\frac{\A(\ldots,P^h)\A
(-P^{-h},\ldots)}{P^2}
\end{eqnarray}
as $P^2\to0$. This factorization is present in most theories described
by Feynman rules and not just Yang--Mills.

When two-particle kinematic invariants go on-shell the situation is
different in massless Yang--Mills theory. Because of kinematics we
should really be looking at the limit where two colour adjacent
momenta become collinear. In this limit, the amplitude factorizes into
the amplitude with one fewer external state times a so-called splitting
amplitude which depends on the particles going collinear and the
internal state which is going on-shell. Again, considering only gluons
we have, when $a\to zP$ and $b\to(1-z)P$,
\begin{eqnarray}
\A(\ldots,a^{h_a},b^{h_b},\ldots)&\to&\sum_{h_i=\pm}
\A(\ldots,P^{-h_i},\ldots)\mathrm{Split}_{h_i}(z,a^{h_a},b^{h_b}).
\end{eqnarray}
The splitting amplitudes for gluons are
\begin{eqnarray}
\label{eq:treesplit}
\mathrm{Split}^{\mathrm{tree}}_-(a^-,b^-)&=&0,\nn
\mathrm{Split}^{\mathrm{tree}}_-(a^+,b^+)&=&
\frac1{\sqrt{z(1-z)}\spa ab},\nn
\mathrm{Split}^{\mathrm{tree}}_+(a^+,b^-)&=&
\frac{(1-z)^2}{\sqrt{z(1-z)}\spa ab},\nn
\mathrm{Split}^{\mathrm{tree}}_-(a^+,b^-)&=&
-\frac{z^2}{\sqrt{z(1-z)}\spb ab},
\end{eqnarray}
from which the remaining can be deduced by pariaty. Similarly, one can deduce
splitting functions involving fermions and scalars from the
corresponding MHV amplitudes. The one-loop limits and simgularities
will be considered in section \ref{sec:looplimit}.

\section{Gravity as a Quantum Field Theory}
\label{sec:gravity}

It is well known that General Relativity has serious problems when
implemented as a quantum field theory. It has a dimensionful coupling
which means that it is non-renormalizable. At each loop order one has
to introduce new counterterms to cancel divergences, and since nothing
prevents finite parts of those counterterms, we automatically
introduce infinitely many coupling constants into the theory.

Because of this, the ``conventional wisdom'' is that a QFT version of
GR is non-sensical. The situation, however, is not as bad as one may
think. Gravity can be treated as an effective field theory
\cite{Weinberg78effective}, where the theory is only renormalized to
the loop order needed. This is legal because we know that an
all-orders renormalized calculation can still be Taylor expanded
around zero coupling. As long as the coupling constant is small, we
can expect tree-level computations to be close to the actual result,
and even at loop level the counterterms contribute to local
interactions but not to \emph{e.g.}~long range corrections to the
gravitational potential \cite{Donoghue93newtoncorrection,
Donoghue94grcorrection, BjerrumBohr02correction}. One could also
remark that Yang--Mills theory in more than four dimensions is also
non-renormalizable, a fact which neither does nor should deter people
from working with the theory.

Another problem plaguing quantum gravity theories is that of
algebra. When the Einstein--Hilbert action,
\begin{equation}
S_{EH}=\frac1{2\kappa^2}\int d^4x \sqrt{-g}R,
\end{equation}
is written out in terms of a perturbation around flat
\begin{equation}
g_{\mu\nu}=\eta_{\mu\nu}+\kappa h_{\mu\nu}\quad
\textrm{or curved}\quad g_{\mu\nu}=g^0_{\mu\nu}+\kappa h_{\mu\nu}
\end{equation}
space, it contains vertices to all orders, and even the lowest order
terms suffer from severe congestion of indices. The three-point
Feynman vertex takes up about half a page in condensed notation, the
four point requires one or two. This means that only amplitudes where
gravitons enter in the simplest manner can be calculated
analytically. The only computational point we will make here is that
of polarization tensors: Gravitons are spin 2 particles which have two
on-shell helicity states with traceless polarization tensors
$\epsilon_{\mu\nu}^\pm$. It so happens that one can use products of
gluon polarization,
\begin{equation}
\epsilon_{\mu\nu}^\pm=\epsilon_\mu^\pm \epsilon_\nu^\pm.
\end{equation}
With the choices of polarization vectors described in section
\ref{sec:spinor}, we may even choose different reference momenta for
$\epsilon^\pm_\mu$ and $\epsilon^\pm_\nu$.

\subsection{The KLT Relations}

What one \emph{can} do to compute tree-level amplitudes in quantum
gravity is to use a result from string theory by Kawai, Lewellen and
Tye \cite{Kawai85klt}, the KLT relations. These are obtained by
rewriting the string computation for closed string scattering in
terms of products of open string scatterings, and by taking the
$\alpha^{'}\to 0$ limit, they make a statement about a relation
between graviton and gluon scattering amplitudes. The relation is
valid on-shell at tree-level in any number of dimensions. The first
three are (setting irrelevant factors to 1)
\begin{eqnarray}
\M(1,2,3)&=&\A(1,2,3)\widetilde\A(1,2,3)\nn
\M(1,2,3,4)&=&s_{12}\A(1,2,3,4)\widetilde\A(1,2,4,3)\nn
\M(1,2,3,4,5)&=&s_{12}s_{34}\A(1,2,3,4,5)\widetilde\A(1,4,3,5,2)\nn
&&\qquad+s_{13}s_{24}\A(1,3,2,4,5)\widetilde\A(1,4,2,5,3).
\label{eq:klt}
\end{eqnarray}
The $\M$'s are graviton amplitudes and the $\A$ and $\widetilde\A$ are
colour ordered gauge amplitudes. In fact, there are many forms of
these relations since an (irrelevant) relabeling on the gravity side
gives another combination of orderings on the gauge side. The simplest
non-trivial example is the four graviton amplitude in four dimensions
which is
\begin{eqnarray}
\M(1^-,2^-,3^+,4^+)&=&s_{12}\A(1^-,2^-,3^+,4^+)\A(1^-,
2^-,4^+,3^+)\nn
&=&\spa 34\spb 43\frac{\spa 12^3}{\spa 23\spa 34\spa 41}
\frac{\spa 12^3}{\spa 24\spa 43\spa 31}\nn
&=&\frac{\spb 43\spa 12^7}{\spa 12\spa 23\spa 34\spa 41\spa 24\spa 13}.
\end{eqnarray}

The KLT relations also imply that graviton states decompose into a
tensor product of two gluon states. Since the proof of the KLT
relations does not rely on the particular string theory in question,
we may imagine that there are more states than gravitons on the
gravity side and gluons on the gauge side and that the gravity states
are tensor products of two gauge states. The two gauge theories need
not be the same. If we imagine the gravity side to be a heterotic
string theory, the one gauge theory corresponds to the right moving
sector and the other corresponds to the left moving sector
\cite{Bern99heteroticklt}.

\subsection{The $\N=8\,/\,\N=4$ Relation}

At this point we can take a closer look at one of the main themes of
this thesis, namely the relation between the maximally supersymmetric
gravity and gauge theories in four dimensions. At tree level, the KLT
relations give us a correspondence between the two theories both at
the level of the states of the two theories (something which is
independent of the loop order) and between the actual amplitudes. 

The spectrum (and Lagrangian) of $\N=8$ supergravity
\cite{Cremmer78n81,Cremmer79n82} can be derived by dimensionally
reducing 11 dimensional $\N=1$ supergravity \cite{Cremmer7811d} and
contains a graviton, eight Rarita--Schwinger gravitini, 28 vectors, 56
Majorana fermions and 70 real scalars, a total of 256 states when
helicities are taken into account. The $\N=4$ multiplet has a gluon,
four (Majorana or Weyl) fermions and six real scalars, a total of 16
states. By the KLT relation, any state in $\N=8$ supergravity can be
seen as a tensor product of two $\N=4$ states such that the sum of the
two $\N=4$ state helicities add up to the $\N=8$ state
helicity. Supergravity can have the same construction as we saw for
supersymmetric gauge theories above where the states are related by
supersymmetry charges. In that sense, four of the $\N=8$ charges map
to the charges of one of the $\N=4$ theories and the remaining four to
the other.

This, however, does not account for all the symmetry on the gauge
side. The symmetry of Yang--Mills theory at tree level is the same as
$\N=4$ SYM at loop level, namely the superconformal symmetry. Since
the spectrum, the supersymmetry and the amplitudes are directly
related to tree-level $\N=8$ supergravity, one might guess that there
is an additional symmetry in gravity theories which mirrors the
additional conformal symmetry in Yang--Mills. We will return to the
relations between $\N=8$ and $\N=4$ at loop level in chapter
\ref{cha:sugra}.

\chapter{Tree Level Methods}
\label{cha:tree}

\section{Earlier Methods}
\label{sec:earlytree}

Calculation of tree-level scattering amplitudes is normally done by
applying the Feynman rules. This procedure is completely safe, in the
sense that it is completely mechanical and results are trusted. The
expressions generated in this way are known not to be the most
compact, but they also avoid unphysical singularities which may cause
problems for numerical evaluation. For calculations by hand, however,
they become impractical beyond five-particle processes because the
number of diagrams becomes too big. Taking as an example the $n+2$
point gluon amplitude, the number of diagrams increases factorially, as
shown in the table below \cite{Mangano90review,Kleiss88multigluon}.

\renewcommand{\arraystretch}{1.8}
    \begin{table}[tbh]
	\begin{center}
       	\begin{tabular}{|c||r|r|r|r|r|r|r|} \hline
       $ n $  &     2   &  3  &  4  &  5  &    6   &  7     &    8  
\\  \hline                                                       
$ \# $ of diagrams    
	&   4	& 25 &  220 & 2485 & 34300 & 559405 & 10525900 
\\ \hline                       
	\end{tabular} 
	\end{center}
	\caption{The number of Feynman diagrams contributing to the scattering
process $gg\to$ n $g$. Extracted from \cite{Mangano90review}.}
\end{table} 
\renewcommand{\arraystretch}{1}

The problem of the number of diagrams can, to a large extent, be
solved by implementing Feynman rules (or an equivalent recursive
formulation) on a computer, where the running
time for generating an amplitude and evaluating it at a specific
kinematical point is not prohibitively long even for a number of
points of the order of 15. There are several programs that can do such
calculations \cite{Maltoni02madevent, Pukhov99comphep, Krauss01amegic,
  Mangano02alpgen, Kanaki00helac}.

Problems may still arise if we want to generate a \emph{large}
number of events with many final states, since reevaluating the
amplitude many times is time consuming. On the other hand, deriving
the amplitude analytically and subsequently inserting numbers is also
problematic since the analytic expressions from Feynman rules are
extremely large. What we would ideally like to have is some means of
computing the amplitude in a form which is compact and well suited for
insertions of numerical values.

One method for this, based on Feynman rules, is Berends--Giele
recursion \cite{Berends87mhv}. It uses colour ordering and the spinor
helicity formalism and in addition keeps as little off-shell
information as possible. In this way it takes advantage of gauge
invariance and other on-shell conditions to simplify intermediate
results instead of waiting till the end. Berends--Giele recursion for
gluon amplitudes works by introducing the off-shell $n$-point gluon
current $J^\mu_n$ which is the (gauge dependent) off-shell amplitude
for $n$-gluon scattering with $n-1$ legs taken on-shell by multiplying
in polarization vectors and using $k^2=0$. By looking back into the
way $J^\mu_n$ was calculated, we see that the off-shell leg had to be
attached to either a three or a four vertex. Whatever was at the other
legs of that vertex would sum up exactly to some $J^\mu_m$ where
$m<n$. This allows us to write down a cubic recursion relation for
$J^\mu_n$, which can be terminated by setting the momentum of the
off-shell leg on-shell and multiplying by the polarization. We can
either try to solve the recursion relation exactly or implement it on
a computer. A good description of Berends--Giele recursion can be
found in \cite{Dixon96efficiently}.

Even though this sort of recursion can speed up the calculation of
amplitudes, it still suffers from very long expressions, although less
severely than the traditional application of Feynman rules. 2004 saw the
invention of two methods which, in their separate ways, have
contributed to both more compact results and theoretical insight into
the structure of gauge theory amplitudes. We now turn to them in
historical order.

\section{MHV Rules}
\label{sec:mhvrules}

\subsection{Basic Construction}
\label{sec:mhvbasic}

As described in the introduction, the analysis of Yang--Mills
amplitudes in twistor space led to an intuitive picture of MHV
vertices connected by propagators. The MHV vertex, however, was not a
well defined concept since it is only well defined on-shell where
the momenta are all light-like. To make the MHV picture concrete, it
was necessary to construct an off-shell continuation of the MHV
amplitude. This was achieved by Cachazo, Svr\v cek, and Witten
\cite{Cachazo04csw} in March of 2004. The rules presented below were
not derived as such, but rather proposed and subsequently
motivated. For multi-gluon amplitudes, the rules state:
\begin{enumerate}
\item Draw all possible graphs (``MHV diagrams'')
  \begin{itemize}
  \item where external lines represent external gluons and have their
  helicity marked,
  \item where internal lines have opposite helicities ($\pm$) at each end,
    \item where each vertex has exactly two edges of negative
    helicity---and at least one of positive helicity---attached,
    \item and which is planar and respects the colour ordering of the
    external gluons.
  \end{itemize}
\item For each graph, assign
  \begin{itemize}
  \item to each external line the outgoing momentum of the
  respective gluon,
\item off-shell momenta $P_i$ to the internal lines by assuming
  conservation of momentum at each vertex,
\item holomorphic spinors to internal lines as
  \begin{equation}
    \label{eq:flatted}
    \spxa{P_i^\flat}=\spxxb{P_i}\eta
  \end{equation}
  where the same (almost arbitrary) anti-holomorphic spinor
  $\spxb\eta$ must be used for all internal lines of all graphs.
  \end{itemize}
\item For each graph, multiply,
  \begin{itemize}
  \item for each internal line of momentum $P_i$,
    \begin{equation}
      \frac1{P_i^2},
    \end{equation}
  \item for each vertex, the MHV formula using $\spxa{P_i^\flat}$ for
  the holomorphic spinors of the internal lines. 
  \end{itemize}
\item Add the contributions of all graphs.
\end{enumerate}

The combinatorics of these rules are similar to that of Feynman rules,
but with a somewhat gentler sprawl of diagrams. It is simple to show
that an MHV diagram with $p$ negative helicity external gluons will
have $p-1$ MHV vertices and $p-2$ propagators; significantly fewer
than for Feynman rules. For fixed $p$, the number of diagrams grows
polynomially in the number of external gluons, thus presenting a
significant simplification over Feynman rules, but if we need all
helicity configurations (and thus all $p$ up to half the number of
gluons) \emph{e.g.}~for computing an unpolarized cross section, the
asymptotic behaviour of the number of MHV diagrams is again factorial,
thus not providing a significant advantage over Feynman rules for many
external gluons.

For few gluons, however, the simplification is remarkable. By taking
advantage of parity, a six gluon amplitude can be computed with the
use of only 32 diagrams (depending on how you count) three of which
are MHV and thus trivial.

To gain some familiarity with the rules, let us consider the simplest
non-trivial case of $A(1^-,2^-,3^-,4^+,5^+)$. It is of course slightly
trivial because it is a googly-MHV amplitude, but it will be a welcome
first check to verify that the rules obey parity. Having three
negative helicity gluons, this amplitude is called ``Next-to-MHV'', or
NMHV, and the diagrams must have two vertices and one
propagator. There are four permitted diagrams,
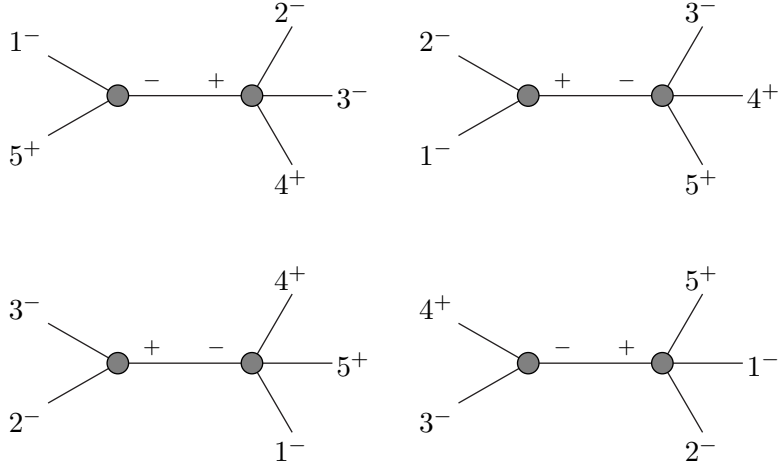
\begin{center}
\begin{picture}(150,100)(0,0)
\Line(50,50)(100,50)
\Text(63,53)[b]{\scriptsize $-$}
\Text(87,53)[b]{\scriptsize $+$}
\Line(50,50)(24,35)
\Line(50,50)(24,65)
\Text(22,33)[rt]{$5^+$}
\Text(22,67)[rb]{$1^-$}
\GCirc(50,50){4}{.5}
\Line(100,50)(115,76)
\Line(100,50)(130,50)
\Line(100,50)(115,24)
\Text(115,78)[b]{$2^-$}
\Text(132,50)[l]{$3^-$}
\Text(115,22)[t]{$4^+$}
\GCirc(100,50){4}{.5}
\end{picture}
\begin{picture}(150,100)(0,0)
\Line(50,50)(100,50)
\Text(63,53)[b]{\scriptsize $+$}
\Text(87,53)[b]{\scriptsize $-$}
\Line(50,50)(24,35)
\Line(50,50)(24,65)
\Text(22,33)[rt]{$1^-$}
\Text(22,67)[rb]{$2^-$}
\GCirc(50,50){4}{.5}
\Line(100,50)(115,76)
\Line(100,50)(130,50)
\Line(100,50)(115,24)
\Text(115,78)[b]{$3^-$}
\Text(132,50)[l]{$4^+$}
\Text(115,22)[t]{$5^+$}
\GCirc(100,50){4}{.5}
\end{picture}
\\
\begin{picture}(150,100)(0,0)
\Line(50,50)(100,50)
\Text(63,53)[b]{\scriptsize $+$}
\Text(87,53)[b]{\scriptsize $-$}
\Line(50,50)(24,35)
\Line(50,50)(24,65)
\Text(22,33)[rt]{$2^-$}
\Text(22,67)[rb]{$3^-$}
\GCirc(50,50){4}{.5}
\Line(100,50)(115,76)
\Line(100,50)(130,50)
\Line(100,50)(115,24)
\Text(115,78)[b]{$4^+$}
\Text(132,50)[l]{$5^+$}
\Text(115,22)[t]{$1^-$}
\GCirc(100,50){4}{.5}
\end{picture}
\begin{picture}(150,100)(0,0)
\Line(50,50)(100,50)
\Text(63,53)[b]{\scriptsize $-$}
\Text(87,53)[b]{\scriptsize $+$}
\Line(50,50)(24,35)
\Line(50,50)(24,65)
\Text(22,33)[rt]{$3^-$}
\Text(22,67)[rb]{$4^+$}
\GCirc(50,50){4}{.5}
\Line(100,50)(115,76)
\Line(100,50)(130,50)
\Line(100,50)(115,24)
\Text(115,78)[b]{$5^+$}
\Text(132,50)[l]{$1^-$}
\Text(115,22)[t]{$2^-$}
\GCirc(100,50){4}{.5}
\end{picture}
\end{center}
Following the rules above, we get the contribution of the first
diagram, 
\begin{eqnarray}
&&\frac{\spa 1{(-P_{51}^\flat)}^3}{\spa{(-P_{51}^\flat)}5\spa 51}
\frac1{P_{51}^2}\frac{\spa23^3}{\spa 34\spa 4{P_{51}^\flat}
\spa{P_{51}^\flat}2}\\
&=&\frac{\spab 1{P_{51}}\eta^3}{\spab 5{P_{51}}\eta\spa 51}
\frac1{\spb 15\spa 51}\frac{\spa 23^3}{\spa 34\spab 4{P_{51}}\eta
\spab 2{P_{51}}\eta}\\
&=&\frac{\spb 5\eta^3\spa 23^3}{\spb 1\eta\spb 51\spa 34\spab 4{(5+1)}
\eta\spab 2{(5+1)}\eta}.
\end{eqnarray}
At this point we can make a choice of $\spxb\eta$ if we like. The
choice $\spxb\eta=\spxb5$ will immediately set the diagram to zero,
where as the choices $\spxb\eta=\spxb1$, $\spxxa{(5+1)}4$,
$\spxxa{(5+1)}2$ will render it undefined. The latter three correspond
to unphysical poles which cancel among the diagrams. If we continue
with $\spxb\eta=\spxb5$, the three other diagrams contribute
\begin{equation}
  -\frac{\spa 34^2\spb 45^3}{\spb 12\spb 51\spa 45\spb 52
\spab 5{P_{12}}5},
\end{equation}
\begin{equation}
  -\frac{\spa14^2\spb 45^3}{\spb 23\spb 51\spa 45\spb 52
\spa 15\spb 53}
\end{equation}
and,
\begin{equation}
  -\frac{\spa 12^2\spb45^3}{\spb 34\spb 51\spa 15\spb 53\spab 5{P_{34}}5},
\end{equation}
which adds up to the expected result,
\begin{equation}
  -\frac{\spb 45^3}{\spb 12\spb 23\spb 34\spb 51}.
\end{equation}
Though the result of this calculation was already known to be fairly
simple, it is remarkable that we can compute a five point amplitude
using only three contributing diagrams of such limited complexity.

A comment on the choice of the off-shell extension (\ref{eq:flatted})
is in order here. Because internal lines must
have opposite helicities at each end, every diagram is invariant under
scaling of internal spinors. In other words, the requirement
(\ref{eq:flatted}) should really read
\begin{equation}
  \spxa{P_i^\flat}\sim \spxxb{P_i}\eta.
\end{equation}
As noted by Kosower \cite{Kosower04nmhv} such a
relation can be achieved by defining
\begin{equation}
  P_i^{\flat\mu}=P_i^\mu+\alpha \eta^\mu
\end{equation}
and choosing $\alpha$ such that $P_i^\flat$ is lightlike. This results
in 
\begin{eqnarray}
  &&P_i^{\flat\mu}=P_i^\mu-\frac{P_i^2}{2P_i\cdot \eta}\eta^\mu\\
&&\spxa{P_i^\flat}\spbx{P_i^\flat}=\frac{\spxxb
  {P_i}\eta\spaxx\eta{P_i}}
{\spab \eta{P_i}\eta}
\end{eqnarray}
In the Yang--Mills MHV rules, only the holomorphic spinor is used, so
in a sense, the above rewriting adds information which is not used in
practise.

\subsection{Gravity: A Puzzling Failure}
\label{sec:reformgr}

The success of MHV rules for Yang--Mills theory immediately raised the
question of whether MHV rules existed for gravity. Even though such
rules were not directly implied by the KLT relationship, the
possibility was not excluded by them, nor by any other known
principle. The main obstacle, however, was that graviton MHV
amplitudes were not holomorphic (\emph{cf.}~(\ref{eq:klt})) and thus the
extension $\spxa{P_i^\flat}\sim\spxxb{P_i}\eta$ was insufficient. As
an example, take the graviton equivalent of the diagram treated in
detail above. MHV rules would give that this diagram was
\begin{equation}
M(5^+,1^-,-P_{51}^{\flat-})\frac1{P_{51}^2}M(2^-,3^-,4^+,
P_{51}^{\flat+}),
\end{equation}
where the graviton amplitudes can be obtained from the KLT
relations. For that we would need $P_{23}^2$ or,
equivalently, $(P_4+P_{51}^\flat)^2$, but these are not obviously the
same because we are now ``inside the MHV vertex'' where we must use
$P_{51}^\flat$ rather than $P_{51}$. For this to be consistent, we
must have
\begin{eqnarray}
  \spab 4{P_{51}^\flat}4&=&\spab 4{P_{51}}4+P_{51}^2\\
\frac{\spab 4{P_{51}}\eta\spab\eta{P_{51}}4}{\spab \eta{P_{51}}\eta}
&=&\spab 4{P_{51}}4+P_{51}^2\\
-P_{51}^2\frac{\spa 4\eta\spb\eta 4}{\spab \eta{P_{51}}\eta}
&=&P_{51}^2\\
\spab\eta{P_{451}}\eta&=&0\\
\spab\eta{(2+3)}\eta&=&0.
\end{eqnarray}
The last requirement states that $\eta^\mu$ must be a lightlike
momentum orthogonal to the time-like momentum $P_{23}$. By going to
the rest frame of $P_{23}$ it can be seen clearly that no such
(non-zero) $\eta^\mu$ exists. Thus, any gravitational attempt at MHV
rules along these lines seems impossible. 

Although it seems that the problem at hand is rooted in the
non-holomorphic character of graviton MHV amplitudes, it is in fact
much deeper. The amplitude $M(2^-,3^-,4^+,P_{51}^{\flat+})$ is
ill-defined because we have taken $P_{51}$ away from its real value
without compensating in other places; the amplitude isn't just
off-shell, it disobeys conservation of momentum. Had this been an
amplitude proper, the implicit delta function in the momenta would
have set it to zero before we even started. And this argument relates
as much to Yang--Mills as to gravity.

In this light, the fact that Yang--Mills MHV rules \emph{do} work
becomes even more puzzling. It gives the impression of being an
inherently ill-defined procedure which magically becomes meaningful
for theories with holomorphic MHV amplitudes. The missing ingredient
to understand why the situation is significantly brighter than this,
will appear in section \ref{sec:recursion}, and the puzzle will be
resolved in the next chapter.

\subsection{Fermions, Higgses, and Vector Bosons}
\label{sec:extmhv}

Meanwhile, we can turn to some of the additional successes of MHV
rules. As described above, the success of the MHV rules seemed to rely
somewhat on the holomorphic nature of MHV amplitudes, so similar
theories with holomorphic MHV amplitudes ought to have associated MHV
rules. 

The primary example of this is Yang--Mills theory with coloured
(adjoint or fundamental) massless scalars and fermions
\cite{Georgiou04mhv, Wu04paritycheck, Wu04fermions,
Georgiou04fermionsscalars}. As described in section \ref{sec:swi},
replacing gluons with fermions or scalars in an MHV amplitude only
changes the numerator, and there is nothing preventing us from
extending the MHV rules with fermionic and scalar internal
lines. Particles in the fundamental representation are handled by
requiring them to be adjacent at all times. If, for instance, we
wanted to compute the NMHV amplitude $A_5(q_1^-,2^-,3^-,4^+,\bar q_5^+)$,
the diagrams would be similar to those of our previous example, just
with some of the gluons being replaced by fermions denoted by dashed
lines,
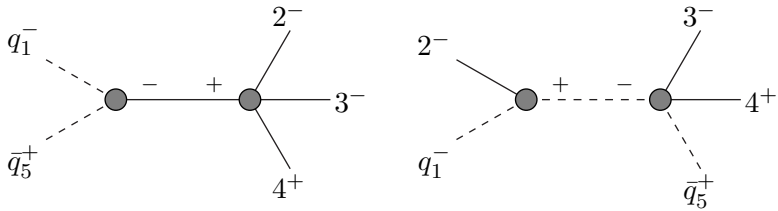
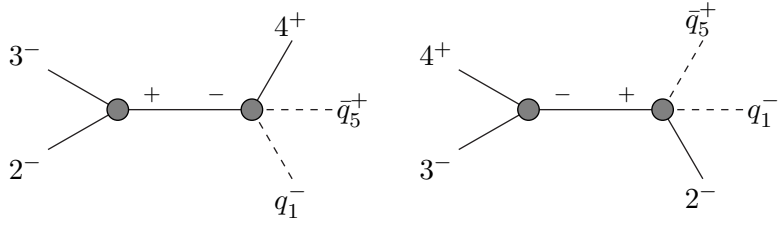
\begin{center}
\begin{picture}(150,100)(0,0)
\Line(50,50)(100,50)
\Text(63,53)[b]{\scriptsize $-$}
\Text(87,53)[b]{\scriptsize $+$}
\DashLine(50,50)(24,35)3
\DashLine(50,50)(24,65)3
\Text(22,33)[rt]{$\bar q_5^+$}
\Text(22,67)[rb]{$q_1^-$}
\GCirc(50,50){4}{.5}
\Line(100,50)(115,76)
\Line(100,50)(130,50)
\Line(100,50)(115,24)
\Text(115,78)[b]{$2^-$}
\Text(132,50)[l]{$3^-$}
\Text(115,22)[t]{$4^+$}
\GCirc(100,50){4}{.5}
\end{picture}
\begin{picture}(150,100)(0,0)
\DashLine(50,50)(100,50)3
\Text(63,53)[b]{\scriptsize $+$}
\Text(87,53)[b]{\scriptsize $-$}
\DashLine(50,50)(24,35)3
\Line(50,50)(24,65)
\Text(22,33)[rt]{$q_1^-$}
\Text(22,67)[rb]{$2^-$}
\GCirc(50,50){4}{.5}
\Line(100,50)(115,76)
\Line(100,50)(130,50)
\DashLine(100,50)(115,24)3
\Text(115,78)[b]{$3^-$}
\Text(132,50)[l]{$4^+$}
\Text(115,22)[t]{$\bar q_5^+$}
\GCirc(100,50){4}{.5}
\end{picture}
\\
\begin{picture}(150,100)(0,0)
\Line(50,50)(100,50)
\Text(63,53)[b]{\scriptsize $+$}
\Text(87,53)[b]{\scriptsize $-$}
\Line(50,50)(24,35)
\Line(50,50)(24,65)
\Text(22,33)[rt]{$2^-$}
\Text(22,67)[rb]{$3^-$}
\GCirc(50,50){4}{.5}
\Line(100,50)(115,76)
\DashLine(100,50)(130,50)3
\DashLine(100,50)(115,24)3
\Text(115,78)[b]{$4^+$}
\Text(132,50)[l]{$\bar q_5^+$}
\Text(115,22)[t]{$q_1^-$}
\GCirc(100,50){4}{.5}
\end{picture}
\begin{picture}(150,100)(0,0)
\Line(50,50)(100,50)
\Text(63,53)[b]{\scriptsize $-$}
\Text(87,53)[b]{\scriptsize $+$}
\Line(50,50)(24,35)
\Line(50,50)(24,65)
\Text(22,33)[rt]{$3^-$}
\Text(22,67)[rb]{$4^+$}
\GCirc(50,50){4}{.5}
\DashLine(100,50)(115,76)3
\DashLine(100,50)(130,50)3
\Line(100,50)(115,24)
\Text(115,78)[b]{$\bar q_5^+$}
\Text(132,50)[l]{$q_1^-$}
\Text(115,22)[t]{$2^-$}
\GCirc(100,50){4}{.5}
\end{picture}
\end{center}
The contribution of the second diagram would be
\begin{eqnarray}
\quad&&\hspace{-10mm}\frac{\spa 12^3\spa {(-P_{12}^\flat)}2}{\spa 12\spa 2{(-P_{12}^\flat)}
 \spa {(-P_{12}^\flat)}1}\frac1{P_{12}^2}\frac{\spa 3{P_{12}^\flat}^3\spa
35}{\spa{P_{12}^\flat}3\spa 34\spa 45\spa 5{P_{12}^\flat}}\\
&=&-\frac{\spa 12^2}{\spa 1{P_{12}^\flat}}\frac1{P_{12}^2}
\frac{\spa 3{P_{12}^\flat}^2\spa 35}{\spa 34\spa 45\spa 5{P_{12}^\flat}},
\end{eqnarray}
which differs from the corresponding gluon diagram by the factor
\begin{equation}
  \frac{\spa 2{P_{12}^\flat}\spa 35}{\spa 21\spa 3{P_{12}^\flat}}
=\frac{\spb 1\eta\spa 35}{\spab 3{P_{12}}\eta}.
\end{equation}
If we now try to compute $A_5(s_1,2^-,3^-,4^+,s_5)$ we will again have
four diagrams similar to those above and for the one treated we would
obtain the factor
\begin{equation}
  \bigg(\frac{\spb 1\eta\spa 35}{\spab 3{P_{12}}\eta}\bigg)^2
\end{equation}
relative to the gluon result. 

Another case where MHV rules apply is one to which we will turn in
chapter \ref{cha:higgs}, namely that of a massive, uncoloured particle $H$
(such as a Higgs) which couples to a gauge field through the term
$CHF_{\mu\nu}F^{\mu\nu}$ in the Lagrangian. 

The last case is that of the coupling of gauge fields to electroweak
vector boson currents. The formalism is somewhat involved, and rather
than trying to explain it here, we refer the reader to the original
article on the matter \cite{Bern04ewcsw}.

\section{On-Shell Recursion Relations}
\label{sec:recursion}

As was discussed in the introduction, a decisive turn came with the
realization that spinors could be taken away from 3+1 dimensions, in
which $\lambda_a$ and $\widetilde \lambda_{\dot a}$ are related by
complex conjugation, to complexified Minkowski space $\mathbb C^4$
where they are complex and independent. This allows for the use of
standard methods of complex analysis, in particular Cauchy's
Theorem. Although on-shell recursion relations were first discovered
as the combined consequence of $\N=4$ IR consistency conditions and
the quadruple cut method \cite{Britto04bcf}, we will take the complex
analysis viewpoint introduced by Britto, Cachazo, Feng, and Witten
(BCFW) \cite{Britto05bcfw}. We will, however, also touch on the
connection with IR consistency conditions in later chapters.

\subsection{Using Cauchy's Theorem on Yang--Mills Amplitudes}
\label{sec:usingcauchy}

BCFW on-shell recursion works by deforming a tree-level amplitude $\A$
by a complex parameter $z$ to $\widehat\A(z)$ by choosing two external
particles $i$ and $j$ and deforming their spinors
\begin{equation}
\label{eq:deformation}
\spxb{\widehat i}=\spxb i+z\spxb j,\qquad
\spxa{\widehat j}=\spxa j-z\spxa i.
\end{equation}
This deformations obeys conservation of momentum since
\begin{equation}
\widehat p_i+\widehat p_j=\spxa i\spbx{\widehat i}+
\spxa{\widehat j}\spbx j=\spxa i\spbx i +z\spxa i\spbx j
+\spxa j\spbx j-z\spxa i\spbx j=p_i+p_j,
\end{equation}
so $\widehat \A(z)$ is a well defined amplitude in $\mathbf C^4$.

For now, we assume that $\widehat\A(z)\to 0$ as $z\to \infty$. This
allows us to use Cauchy's Theorem to write
\begin{equation}
\label{eq:cauchy}
0=\frac1{2\pi i}\oint_{\mathcal C_\infty}  \frac{dz}z\widehat\A(z)
=\widehat\A(0)+\sum_{\mathrm{poles~}p}\frac{\mathrm{Res}_p\widehat\A(z)}
{z_p}
\end{equation}
The poles of tree amplitudes are well known; they correspond to the
Feynman propagators of internal momenta going on-shell, and the
``residue'' is given by
\begin{equation}
\label{eq:ymsing}
\lim_{P_{k,m}^2\to 0}\Big[P_{k,m}^2 \A\Big]=
\sum_{h=\pm 1}\A(k,\ldots,m,-P_{k,m}^{-h})\A(P_{k,m}^h,m+1,\ldots,k-1).
\end{equation}
If $i\in\{k,\ldots,m\}$ and $j\notin\{k,\ldots,m\}$ (or vice
versa) there will be a corresponding pole in $z$ because 
\begin{equation}
\widehat P_{k,m}(z)^2=(P_{k,m}+z\spxa i\spbx j)^2=P_{k,m}^2
+z\spab i{P_{k,m}}j,
\end{equation}
thus
\begin{equation}
z_{k,m}=-\frac{P_{k,m}^2}{\spab i{P_{k,m}}j}.
\end{equation}
The residue of this pole is
\begin{eqnarray}
&&\lim_{z\to z_{k,m}}\Big[z\A\Big]\nn
&=&\frac1{\spab i{P_{k,m}}j}\lim_{P_{k,m}^2\to 0}
\Big[P_{k,m}^2\A\Big]\nn
&=&\sum_{h=\pm 1}\frac{\A(k,\ldots,m,-\widehat P_{k,m}^{-h}(z_{k,m}))
\A(\widehat P_{k,m}^h(z_{k,m}),m+1,\ldots,k-1)}
{\spab i{P_{k,m}}j}.
\end{eqnarray}
Combining this with (\ref{eq:cauchy}) leads to
\begin{eqnarray}
\label{eq:recursion}
\A&=&\sum_{\scriptsize
\begin{array}{c}
k\in\{j+1,\ldots,i\}\\
m\in\{i,\ldots,j-1\}\\
k\neq m,m+2\\
h=\pm 1
\end{array}}\hspace{-7mm}
\frac{\A(k,\ldots,m,-\widehat P_{k,m}^{-h}(z_{k,m}))
\A(\widehat P_{k,m}^h(z_{k,m}),m+1,\ldots,k-1)}
{P_{k,m}^2}.\nn
&&
\end{eqnarray}
In human language, the conditions under the sum state that we must sum
over all internal momenta affected by the deformation. We can draw
this as a diagram consisting of two deformed amplitudes and an
internal propagator, together with the sum,
\begin{center}
\raisebox{50pt}{$\displaystyle{\sum_{\mathrm{all~such~diagrams}}}$}
\begin{picture}(150,100)(0,0)
\Line(50,50)(100,50)
\Text(63,53)[b]{\scriptsize $\pm$}
\Text(87,53)[b]{\scriptsize $\mp$}
\Text(75,47)[tt]{$\widehat P_{k,m}$}
\Line(50,50)(35,24)
\Line(50,50)(24,35)
\Line(50,50)(20,50)
\Line(50,50)(24,65)
\Line(50,50)(35,76)
\Text(22,67)[rb]{$\widehat i$}
\DashCArc(50,50)(22,120,240){2}
\BCirc(50,50){7}
\Line(100,50)(115,76)
\Line(100,50)(126,65)
\Line(100,50)(130,50)
\Line(100,50)(126,35)
\Line(100,50)(115,24)
\Text(128,33)[tl]{$\widehat j$}
\DashCArc(100,50)(22,300,60){2}
\BCirc(100,50){7}
\end{picture}
\end{center}
$\widehat
P_{k,m}$ is on-shell and has the form
\begin{equation}
\widehat P_{k,m}=P_{k,m}-\frac{P_{k,m}^2}{\spab i{P_{k,m}}j}\spxa
i\spbx j = \frac{\spxxb {P_{k,m}}j\spaxx i{P_{k,m}}}
{\spab i{P_{k,m}}j}.
\end{equation}
Very often this
recursion relation is used with $i$ and $j$ adjacent,
\emph{e.g.~}$(i,j)=(n,1)$ in an $n$-point amplitude. This simplifies
the above relation to
\begin{eqnarray}
\A_n&=&\sum_{m=2}^{n-2}\sum_{h=\pm 1}
\frac{\A(m+1,\ldots,n,\widehat P_{1,m}^{-h}(z_m))
\A(-\widehat P_{1,m}^h(z_m),1,\ldots,m)}
{P_{1,m}^2}
\end{eqnarray}
where
\begin{equation}
z_m=\frac{P_{1,m}^2}{\spab n{P_{1,m}}1},\qquad
\widehat P_{1,m}=\frac{\spxxb {P_{1,m}}n\spaxx 1{P_{1,m}}}
{\spab 1{P_{1,m}}n}=\frac{\spxxb {P_{n,m}}n\spaxx 1{P_{2,m}}}
{\spab 1{P_{1,m}}n}.
\end{equation}

Before we can go and apply this decomposition on poles to the
calculation of real gluon amplitudes, there are two subtleties to
consider, having to do with three-point amplitudes and the behaviour
of $\widehat \A(z)$ as $z\to\infty$. 

\subsection{Three-Point Amplitudes (and an Example)}

In real Minkowski space, three-point gauge amplitudes are zero. This
basically stems from the fact that in order for the three momenta to
be light-like, they must also be orthogonal,
\begin{equation}
0=p_3^2=(-p_1-p_2)^2=\spa 12\spb 21,\quad
\Rightarrow \quad\spa 12=\spb 12^*=0,
\quad\mathit{etc.}
\end{equation}
\begin{equation}
\Rightarrow p_1\sim p_2\sim p_3.
\end{equation}
When we move away from real Minkowski space, however, the restriction
that $\lambda_a=\widetilde\lambda_{\dot a}$ is lifted and the on-shell
constraints have two solutions,
\begin{equation}
  \spxa 1\sim\spxa 2\sim\spxa 3\quad\mathrm{or}\quad
  \spxb 1\sim\spxb 2\sim\spxb 3.
\end{equation}
These each allow one of the two helicity configurations to be non-zero
because the expressions
\begin{equation}
  \frac{\spb 12^3}{\spb 23\spb 31}\quad\mathrm{and}\quad
  \frac{\spa 12^3}{\spa 23\spa 31},
\end{equation}
respectively, are well-defined and contribute to
(\ref{eq:recursion}). 

We can illustrate the method and the considerations about three-point
amplitudes with the calculation of the googly-MHV five-point amplitude
$\A(1^-,2^+,3^-,4^+,5^-)$. We define the deformation by taking
\begin{equation}
  \spxb{\widehat 3}=\spxb 3+z\spxb 4,\qquad
\spxa{\widehat 4}=\spxa 4-z\spxa 3.
\end{equation}
The internal momenta affected by this deformation are $P_{1,3}$ and
$P_{2,3}$ each corresponding to a pole in $z$ at
\begin{equation}
  z_{1,3}=-\frac{P_{1,3}^2}{\spab 3{P_{1,3}}4}=\frac{\spa 45}{\spa 35}
,\qquad z_{2,3}=-\frac{P_{2,3}^2}{\spab 3{P_{2,3}}4}=
-\frac{\spb 23}{\spb 24}.
\end{equation}
Summing also over internal helicities gives four diagrams
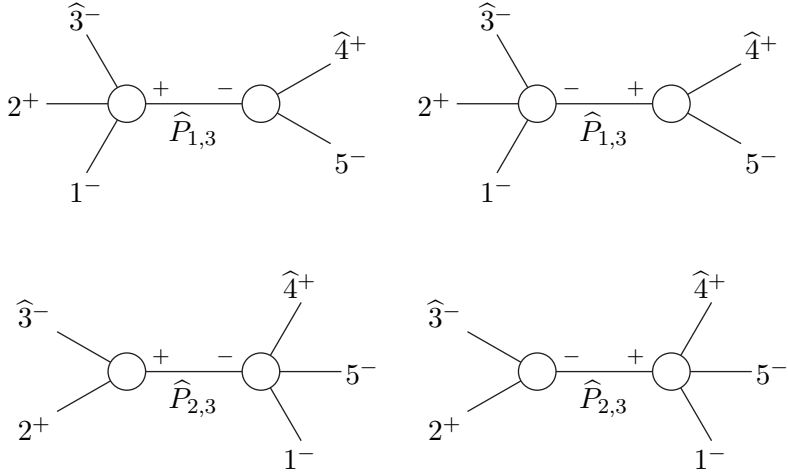
\begin{center}
\begin{picture}(150,100)(0,0)
\Line(50,50)(100,50)
\Text(63,53)[b]{\scriptsize $+$}
\Text(87,53)[b]{\scriptsize $-$}
\Text(75,47)[t]{$\widehat P_{1,3}$}
\Line(50,50)(35,24)
\Line(50,50)(20,50)
\Line(50,50)(35,76)
\Text(35,22)[t]{$1^-$}
\Text(18,50)[r]{$2^+$}
\Text(35,78)[b]{$\widehat 3^-$}
\BCirc(50,50){7}
\Line(100,50)(126,65)
\Line(100,50)(126,35)
\Text(128,67)[lb]{$\widehat 4^+$}
\Text(128,33)[lt]{$5^-$}
\BCirc(100,50){7}
\end{picture}
\begin{picture}(150,100)(0,0)
\Line(50,50)(100,50)
\Text(63,53)[b]{\scriptsize $-$}
\Text(87,53)[b]{\scriptsize $+$}
\Text(75,47)[t]{$\widehat P_{1,3}$}
\Line(50,50)(35,24)
\Line(50,50)(20,50)
\Line(50,50)(35,76)
\Text(35,22)[t]{$1^-$}
\Text(18,50)[r]{$2^+$}
\Text(35,78)[b]{$\widehat 3^-$}
\BCirc(50,50){7}
\Line(100,50)(126,65)
\Line(100,50)(126,35)
\Text(128,67)[lb]{$\widehat 4^+$}
\Text(128,33)[lt]{$5^-$}
\BCirc(100,50){7}
\end{picture}
\\
\begin{picture}(150,100)(0,0)
\Line(50,50)(100,50)
\Text(63,53)[b]{\scriptsize $+$}
\Text(87,53)[b]{\scriptsize $-$}
\Text(75,47)[t]{$\widehat P_{2,3}$}
\Line(50,50)(24,35)
\Line(50,50)(24,65)
\Text(22,33)[rt]{$2^+$}
\Text(22,67)[rb]{$\widehat 3^-$}
\BCirc(50,50){7}
\Line(100,50)(115,76)
\Line(100,50)(130,50)
\Line(100,50)(115,24)
\Text(115,78)[b]{$\widehat 4^+$}
\Text(132,50)[l]{$5^-$}
\Text(115,22)[t]{$1^-$}
\BCirc(100,50){7}
\end{picture}
\begin{picture}(150,100)(0,0)
\Line(50,50)(100,50)
\Text(63,53)[b]{\scriptsize $-$}
\Text(87,53)[b]{\scriptsize $+$}
\Text(75,47)[t]{$\widehat P_{2,3}$}
\Line(50,50)(24,35)
\Line(50,50)(24,65)
\Text(22,33)[rt]{$2^+$}
\Text(22,67)[rb]{$\widehat 3^-$}
\BCirc(50,50){7}
\Line(100,50)(115,76)
\Line(100,50)(130,50)
\Line(100,50)(115,24)
\Text(115,78)[b]{$\widehat 4^+$}
\Text(132,50)[l]{$5^-$}
\Text(115,22)[t]{$1^-$}
\BCirc(100,50){7}
\end{picture}
\end{center}
with the contributions
\begin{eqnarray}
  &&\frac{\A(1^-,2^+,\widehat 3^-(z_{1,3}),-\widehat P_{1,3}^+(z_{1,3}))
    \A(\widehat P_{1,3}^-(z_{1,3}),\widehat 4^+(z_{1,3}),5^-)}
  {P_{1,3}^2}\nn
  &&\frac{\A(1^-,2^+,\widehat 3^-(z_{1,3}),-\widehat P_{1,3}^-(z_{1,3}))
    \A(\widehat P_{1,3}^+(z_{1,3}),\widehat 4^+(z_{1,3}),5^-)}
  {P_{1,3}^2}\nn
  &&\frac{\A(2^+,\widehat 3^-(z_{2,3}),-\widehat P_{2,3}^+(z_{2,3}))
    \A(\widehat P_{2,3}^-(z_{2,3}),\widehat 4^+(z_{2,3}),5^-,1^-)}
  {P_{2,3}^2}\nn
  &&\frac{\A(2^+,\widehat 3^-(z_{2,3}),-\widehat P_{2,3}^-(z_{2,3}))
    \A(\widehat P_{2,3}^+(z_{2,3}),\widehat 4^+(z_{2,3}),5^-,1^-)}
  {P_{2,3}^2}.
\end{eqnarray}
Two of these contributions can be discarded immediately. Number 2 and
3 contain an amplitude of the form $\A(---+)$ which is zero. In
contribution number 1, $z_{1,3}$ is found by requiring that
\begin{equation}
\widehat P_{1,3}^2=\spa{\widehat 4}5\spb 54=0,\quad
\mathrm{or}\quad \spxa{\widehat 4}\sim\spxa 5,
\end{equation}
so the three-point MHV amplitude in that contribution is zero, even in
complex momenta. The corresponding condition in the last contribution
is $\spxb{\widehat 3}\sim \spxb 2$ which kills the googly-MHV
three-point but not the MHV three-point we need. Thus it gives the
only contribution, for which we need the solutions
\begin{equation}
  \spxa{\widehat 4}=\spxa 4+\frac{\spb 23}{\spb 24}\spxa 3
=\frac{\spxxb {P_{3,4}}2}{\spb 42},\qquad
\widehat P_{23}=\frac{\spxxb {P_{2,3}}4\spaxx 3{P_{2,3}}}
{\spab 3{P_{2,3}}4}=\frac{\spxxb {P_{2,4}}4\spbx 2}
{\spb 24}.
\end{equation}
This allows us to compute
\begin{eqnarray}
&&\frac{\spa 3{(-\widehat P_{2,3})}^3}{\spa{(-\widehat P_{2,3})}2
\spa 23}\frac1{P_{2,3}^2}\frac{\spa 51^3}{\spa 1{\widehat P_{2,3}}
\spa{\widehat P_{2,3}}{\widehat 4}\spa {\widehat 4}5}\nn
&=&\frac{\spa 51^3\spa 3{\widehat P_{2,3}}^3}{\spa 23^2\spb 32
\spa {\widehat 4}5
\spa 2{\widehat P_{2,3}}\spa 1{\widehat P_{2,3}}\spa {\widehat 4}{
\widehat P_{2,3}}}\nn
&=&\frac{\spa 51^3\spab 3{P_{2,3}}4^3\spb 24^2}{\spa 23^2\spb 32
\spba 2{P_{3,4}}5
\spab 2{P_{2,3}}4\spab 1{P_{2,3}}4\spbb 2{P_{3,4}}{P_{2,3}}4}\nn
&=&\frac{\spa 51^3\spa 32^3\spb 24^5}{\spa 23^2\spb 32
\spb 21\spa 15
\spa 23\spb 34\spa 15\spb 54\spb 24\spa 51\spb 15}\nn
&=&-\frac{\spb 24^4}{\spb 12\spb 23\spb 34\spb 45\spb 51},
\end{eqnarray}
which is, of course, the known expression for a five-point googly-MHV
amplitude.

\subsection{Asymptotic Behaviour of the Deformed Amplitude}

Until now we postponed the issue of whether $\lim_{z\to\infty}\widehat
\A(z)=0$. For certain choices of the spinors to deform we can
immediately argue that this must be the case by looking at the
calculation from Feynman diagrams and tracking the dependence on
$z$. Every propagator in a Feynman diagram affected by the deformation
must asymptotically
contribute $z^{-1}$, and every three vertex affected by the
deformation must contribute $z^1$. The ``worst case scenario'' for the
asymptotic $z$ dependence will be when the deformation only affects
three-vertices. Before multiplying by the external polarizations, such
a diagram will go as $z^1$.

If we make the deformation (\ref{eq:deformation}) the dependences of
$\epsilon_i^\mu$ and $\epsilon_j^\mu$ on $z$ will depend on their
helicities. For $i$ the reaction to the deformation is
\begin{equation}
  \widehat \epsilon_i^{\mu+}=\frac{\spab q{\gamma^\mu}{\widehat i}}
{\sqrt 2\spa iq}
\sim z^1,\qquad
\widehat \epsilon_i^{\mu-}=\frac{\spab i{\gamma^\mu}q}
{\sqrt 2\spb{\widehat i}q}\sim z^{-1},
\end{equation}
and for $j$ the asymptotic behaviour is the opposite. This leads us to
conclude that the worst behaviour of $\widehat\A(z)$ as $z\to\infty$
for helicities $(h_i,h_j)$ is
\begin{equation}
  (-,+)\sim z^{-1},\quad (+,+),(-,-)\sim z^1,\quad (+,-)\sim z^3.
\end{equation}
This tells us that using BCFW on-shell recursion relations is always
justified when we deform the antiholomorphic spinor of a negative
helicity particle and the holomorphic spinor of a positive helicity
particle. 

In practice, however, the behaviour may be more benign. If we consider
the MHV amplitude $\A(1^-,2^-,\ldots,n^+)$, the choice
$(i,j)=(1^-,2^-)$ goes as $z^{-1}$ and so does \emph{e.g.}~the choice
$(3^+,4^+)$ while $(2^-,4^+)$ and $(3^+,5^+)$ go as $z^{-2}$. This
shows us that we can often expect an actual behaviour which is a few
powers of $z$ better than the worst case. This is another expression
of the fact that Feynman rules are not taking advantage of all the
available symmetries of the theory.

\subsection{Some Consequences of BCFW on-shell recursion}

Two points are worth making at this point. First, since we have
rigorously proven that we can always perform on-shell recursion with
the choice $(-,+)$ it follows that any tree amplitude can, by
successive recursion, be reduced to three-amplitudes. In this
perspective, the four gluon vertex can be completely neglected for
practical calculations, and its presence in the Lagrangian is only
required for gauge invariance. 

Second, the existence of on-shell recursion shows that other
formalisms that claim to reproduce Yang--Mills amplitudes are correct
if they have the right singularity structure (\ref{eq:ymsing}) and
Lorentz covariance. This was proven together with the proposal
of the MHV rules \cite{Cachazo04csw}, thus, BCFW recursion provides an
indirect proof that the MHV rules are correct.

Both of these points are elaborated on in the original article
\cite{Britto05bcfw}.

\subsection{Extension to Other Theories and Deformations}

The deformation (\ref{eq:deformation}) chosen by BCFW is by no means
unique. For instance, one may choose three gluons $i,j,k$ and deform
them to \cite{Bern05looprecursion1}
\begin{equation}
\spxb{\widehat i}=\spxb i+z\spxb j+\alpha z\spxb k,\quad
\spxa{\widehat j}=\spxa j-z\spxa i,\quad
\spxa{\widehat k}=\spxa k-\alpha z\spxa i,
\end{equation}
which also obeys momentum conservation for any $\alpha$. In general
one will have to redo the derivations of section
\ref{sec:usingcauchy}, but most features of the relations
persist. Using extended deformations may be helpful for improved
$z\to\infty $ behaviour and for constructing relations to the MHV
rules as done in the next chapter.

The methods of deriving on-shell recursion relations are clearly very
generic, relying mostly on the fact that the theory at hand is an
ordinary quantum field theory. Gauge invariance or colour ordering may
be practical tools for obtaining simpler expressions, but are not
required as such for doing on-shell recursion in this way. One
extension is to allow other states than gluons to participate, such as
charged scalars and fermions (massless, massive, $D$-dimensional)
 \cite{Luo05fermionbcfw,Luo05partonbcfw,Badger05massivebcfw,
Badger05massivespinbcfw,Quigley05ddimbcfw}, another is to take a
completely different theory, such as QED \cite{Ozeren05qedbcfw},
gravity \cite{Bedford05gravbcfw,Cachazo05gravbcfw}, or more exotic
theories \cite{Benincasa07highspin}. The challenge in most of these
cases is to prove the appropriate $z\to\infty$ behaviour, which is
sometimes done and sometimes left to faith. Especially in gravity,
where the ``worst case scenario'' analysis gives very bad asymptotic
behaviour, the on-shell recursion relations were used for quite some
time on this faith until a proof appeared \cite{Benincasa07gravrecur}.

The applications to loop-level calculations will be reviewed in
section \ref{sec:looprecursion}.

\chapter{MHV Rules from Recursion}
\label{cha:mhv}

In the present chapter we will connect the two subjects of the
previous chapter and show how MHV rule constructions follow from
on-shell recursion. The original articles are \cite{Risager05csw} for the
Yang--Mills case and \cite{Risager05grcsw} for the gravity case. 

\section{MHV Rules in Light of Recursion Relations}
\label{sec:light}

The MHV rules as described in section \ref{sec:mhvrules} distance
themselves from recursion relations in one obvious way: recursion
relations have one unshifted propagator, while the MHV rules have
several. This immediately tells us that a simple derivation of the MHV
rules from recursion is not obtainable, but also that the NMHV case
can serve as an initial test case. With this in mind, we can note four
features of MHV rules that will guide us:
\begin{enumerate}
\item An internal momentum is shifted on-shell in a manner close to
  that of recursion relations. Setting
  \begin{equation}
    P^\flat=P-\frac{P^2}{2P\cdot \eta}\eta,
  \end{equation}
is practically the same as shifting $P$ as
\begin{equation}
  \whP=P-z\eta,
\end{equation}
setting $\whP^2=0$, and solving for $z$.
\item The asymmetry between holomorphic and anti-holomorphic spinors
  is deeply rooted. The attempt to construct MHV rules for gravity by
  choosing $\spxa \eta=(\spxb \eta)^*$ failed because it produced
  non-Lorentz invariant results, thereby providing a hint that $\eta$
  is a complex momentum whose holomorphic and anti-holomorphic spinors
  are unrelated. Admittedly, there is a bit of hindsight in this
  argument, since the existence of MHV rules for gravity was not
  guaranteed.
\item If $\eta$ is a complex momentum, we might as well continue all
  momenta. If we do that, all the anti-holomorphic spinors of the
  external momenta are unspecified by the MHV rules, just as $\spxa
  \eta$. In fact, $\spxa\eta$ may even differ between different
  internal momenta.
\item The MHV rules contain MHV three-point vertices and no googly
  three-point vertices. The easiest way of ensuring this would be to
  shift only anti-holomorphic spinors. This fits well the fact that
  these do not show up in the MHV rules anyway.
\end{enumerate}
This leads us to conclude that if MHV rules can be proven from
recursion relations, the proof must involve a non-minimal shift of
anti-holomorphic spinors, and all these shifts must be proportional to
the (otherwise unspecified) $\spxb\eta$.

\section{The NMHV Case}
\label{sec:nmhv}

We first apply the above considerations to the calculation of the amplitude
$\A_n(m_1^-,\ldots,m_2^-,\ldots,m_3^-,\ldots)$ where the $\ldots$ have
helicity $+$. If we choose to shift only the three negative helicity
gluons, the shift is unique up to a constant, 
\begin{equation}
  \spxb {\wh m_1}=\spxb {m_1}+z\spxb \eta\spa {m_2}{m_3},
\end{equation}
\begin{equation}
  \spxb {\wh m_2}=\spxb {m_2}+z\spxb \eta\spa {m_3}{m_1},
\end{equation}
\begin{equation}
  \spxb {\wh m_3}=\spxb {m_3}+z\spxb \eta\spa {m_1}{m_2},
\end{equation}
where conservation of momentum follows from the Schouten identity
\begin{equation}
  \spxa {m_1}\spa {m_2}{m_3}+\spxa {m_2}\spa {m_3}{m_1}
+\spxa {m_3}\spa {m_1}{m_2}=0.
\end{equation}
We will postpone the proof of the legality of this shift to section
\ref{sec:mhvlargez}.  With the above shift, the $z$ dependent internal
momenta are those that have at least one of $m_{1,2,3}$ at each end,
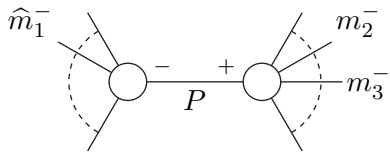
\emph{e.g.}
\begin{center}
\begin{picture}(150,100)(0,0)
\Line(50,50)(100,50)
\Text(63,53)[b]{\scriptsize $-$}
\Text(87,53)[b]{\scriptsize $+$}
\Text(75,47)[t]{$P$}
\Line(50,50)(35,24)
\Line(50,50)(24,65)
\Line(50,50)(35,76)
\Text(22,67)[rb]{$\widehat m_1^-$}
\DashCArc(50,50)(22,120,240){2}
\BCirc(50,50){7}
\Line(100,50)(115,76)
\Line(100,50)(126,65)
\Line(100,50)(130,50)
\Line(100,50)(115,24)
\Text(128,67)[lb]{$m_2^-$}
\Text(132,50)[l]{$m_3^-$}
\DashCArc(100,50)(22,300,60){2}
\BCirc(100,50){7}
\end{picture}
\end{center}
The opposite helicity assignment for the inner line does not
contribute, so we have split the amplitude into two (shifted) MHV
amplitudes. The solution for $z$ is found by setting
\begin{equation}
  \whP^2=(P-z\spxa {m_1}\spa{m_2}{m_3}\spbx\eta)^2=P^2-z\spa{m_2}{m_3}
  \spab {m_1}P\eta=0
\end{equation}
\begin{equation}
  \whP=\frac{\spxxb P\eta\spaxx {m_1}P}{\spab {m_1}P\eta}
\qquad\spxa{\whP}\sim \spxxb P\eta
\end{equation}
It should now be clear that we have reproduced the NMHV version of the
MHV rules for Yang--Mills: We must sum over all diagrams that split
the amplitude into two MHV vertices. For the external momenta we must
use the shifted spinors (but because the MHV amplitude only involves
holomorphic spinors, it makes no difference) and for the internal we
must use $\spxa{\whP}\sim\spxxb P\eta$.

We have, however, obtained more than just the Yang--Mills MHV rules.
We have obtained the full expression for the shifted momenta, allowing
us to construct MHV rules for theories with other MHV amplitudes than
Yang--Mills; most notably gravity. We will return to the MHV rules for
gravity in section \ref{sec:grmhvrules} after treating the general
case with more than three negative helicity particles.

\section{The NNMHV and General Cases}
\label{sec:generalcase}

From the above discussion, it should be clear that setting up a
recursion by shifting the anti-holomorphic spinors of all negative
helicity gluons by the same reference spinor $\spxb\eta$ allows a
description where a) the left and right amplitudes both have fewer
negative helicity gluons than the full amplitude, b) there are no
three-point googly amplitudes, and c) the spinor to be used for the
internal line is the same as in the MHV rules. The remaining problem
is that there are several propagators in each MHV diagram while the
recursion relations can only provide one.

The solution will be to perform similar shifts one after the other and
to reach a reperesentation which can subsequently be proven to be
equivalent to the MHV rules. Starting with an NNMHV amplitude, we can
choose a set $c_i^{(1)}$ and shift
\begin{equation}
  \spxb{\wh m_i}= \spxb{m_i}+zc^{(1)}_i\spxb\eta.
\end{equation}
For conservation of momentum, the $c$'s must be chosen such that
\begin{equation}
  \sum_{i=1}^4 c^{(1)}_i\spax{m_i}=0.
\end{equation}
If we label the possible internal lines by $a$, we can write the
result of the recursion as
\begin{equation}
\label{eq:firststep}
  \A_n(m_{1\ldots4}^-)=\sum_a\frac{\A(\wh m^-,\wh P_a^\pm)
\A(\wh m^-,-\wh P_a^\mp)}{P_a^2},
\end{equation}
where $\spxa{\wh P_a}=\spxxb{P_a}\eta$, and there is an implicit
summation over helicities where relevant. The two sub amplitudes are
MHV and NMHV, respectively. Such contributions can be depicted as
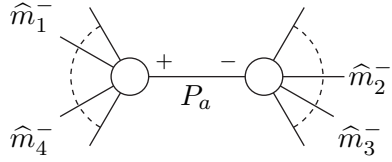
\begin{center}
\begin{picture}(150,100)(0,0)
\Line(50,50)(100,50)
\Text(63,53)[b]{\scriptsize $+$}
\Text(87,53)[b]{\scriptsize $-$}
\Text(75,47)[t]{$P_a$}
\Line(50,50)(35,24)
\Line(50,50)(24,35)
\Line(50,50)(24,65)
\Line(50,50)(35,76)
\Text(22,33)[rt]{$\wh m_4^-$}
\Text(22,67)[rb]{$\widehat m_1^-$}
\DashCArc(50,50)(22,120,240){2}
\BCirc(50,50){7}
\Line(100,50)(115,76)
\Line(100,50)(130,50)
\Line(100,50)(126,35)
\Line(100,50)(115,24)
\Text(132,50)[l]{$\widehat m_2^-$}
\Text(128,33)[lt]{$\widehat m_3^-$}
\DashCArc(100,50)(22,300,60){2}
\BCirc(100,50){7}
\end{picture}
\end{center}
To uncover a second propagator, we can set up a similar recursion for each
term in the above sum. However, we have the constraint on those
recursions that we should not uncover the same propagator $1/P_a^2$
again, or we would not have progressed. Thus, for each $a$ we choose a
set $c^{(2)}_{a,i}$ and set up the recursion as
\begin{equation}
  \spxb{\wh{\wh m}_i}=\spbx{\wh m_i}+zc^{(2)}_{a,i}\spbx\eta
\end{equation}
such that $\wh P^2_a=P_a^2$. This gives one additional constraint on
the $c_{a,i}^{(2)}$ apart from conservation of momentum. Performing
the recursion for all terms in the sum (\ref{eq:firststep}), we arrive at
\begin{equation}
\label{eq:secondstep}
  \A_n(m_{1\ldots 4}^-)=\sum_a\sum_{b\neq a}\frac{\A(\wh{\wh m}^-,
\wh{\wh P}_a^\pm)\A(\wh{\wh m}^-,-\wh{\wh P}_a^\mp,\wh{\wh P}_b^\pm)
\A(\wh{\wh m}^-,-\wh{\wh P}_ b^\mp)}{P_a^2\wh P_b^2}
\end{equation}
where the terms in the sum can be drawn as (placement of $m$'s and
$\pm$'s can differ)
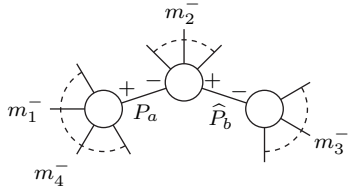
\begin{center}
\begin{picture}(140,90)(0,0)
\Line(40,40)(70,50)
\Line(70,50)(100,40)
\Text(49,45)[b]{\scriptsize $+$}
\Text(59,48)[b]{\scriptsize $-$}
\Text(81,48)[b]{\scriptsize $+$}
\Text(91,44)[b]{\scriptsize $-$}
\Text(56,42)[t]{\scriptsize $P_a$}
\Text(84,42)[t]{\scriptsize $\wh P_b$}
\Line(40,40)(50,23)
\Line(40,40)(20,40)
\Line(40,40)(30,23)
\Line(40,40)(30,57)
\Text(28,21)[tr]{\scriptsize $m_4^-$}
\Text(18,40)[r]{\scriptsize $m_1^-$}
\BCirc(40,40){7}
\DashCArc(40,40)(16,120,300)2
\Line(70,50)(56,64)
\Line(70,50)(70,70)
\Line(70,50)(84,64)
\Text(70,72)[b]{\scriptsize $m_2^-$}
\BCirc(70,50){7}
\DashCArc(70,50)(16,45,135)2
\Line(100,40)(117,50)
\Line(100,40)(117,30)
\Line(100,40)(100,20)
\Text(119,28)[l]{\scriptsize $m_3^-$}
\BCirc(100,40){7}
\DashCArc(100,40)(16,270,30)2
\end{picture}
\end{center}
Since our recursion cannot split MHV amplitudes, it is clear that all
three amplitudes in the numerator must be MHV and thus that the
helicities of the internal lines are fixed by the external
helicities. Moreover, if we write
the twice shifted $\spxb m$'s as
\begin{equation}
  \spxb{\wh{\wh m}_i}=\spxb{m_i}+r_i\spxb\eta
\end{equation}
and remind ourselves that the coefficients were constructed to obey
momentum conservation and $\wh{\wh P}_a^2=\wh{\wh P}_ b^2=0$ we can
fix the $\spxb{\wh{\wh m}_i}$ uniquely as a function of $a$ and
$b$. This solution is independent of which of these was uncovered
first, so we can collect terms in (\ref{eq:secondstep}) two by two and
write it as
\begin{eqnarray}
   \A_n(m_{1\ldots 4}^-)&=&\sum_{\{a,b\}}\A(\wh{\wh m}^-,
\wh{\wh P}_a^\pm)\A(\wh{\wh m}^-,-\wh{\wh P}_a^\mp,\wh{\wh P}_b^\pm)
\A(\wh{\wh m}^-,-\wh{\wh P}_ b^\mp)\\
&&\qquad\qquad\times\bigg(\frac1{P_a^2\wh P_{b(a)}^2}
+\frac1{P_b^2\wh P_{a(b)}^2}\bigg)
\end{eqnarray}
where $\wh P_{b(a)}$ means $P_b$ shifted using $c_i^{(1)}$ as to put
$P_a$ on-shell. The last factor, however, is exactly what we would get
if we took
\begin{equation}
  \frac1{P_a^2P_b^2}
\end{equation}
and did recursion on it using $c_i^{(1)}$. In terms of diagrams, what
we are arguing is that
\begin{center}
\begin{picture}(140,90)(0,0)
\Line(40,40)(70,50)
\Line(70,50)(100,40)
\Text(49,45)[b]{\scriptsize $+$}
\Text(59,48)[b]{\scriptsize $-$}
\Text(81,48)[b]{\scriptsize $+$}
\Text(91,44)[b]{\scriptsize $-$}
\Text(56,42)[t]{\scriptsize $P_a$}
\Text(84,42)[t]{\scriptsize $\wh P_{b(a)}$}
\Line(40,40)(50,23)
\Line(40,40)(20,40)
\Line(40,40)(30,23)
\Line(40,40)(30,57)
\Text(28,21)[tr]{\scriptsize $\wh{\wh m}_4^-$}
\Text(18,40)[r]{\scriptsize $\wh{\wh m}_1^-$}
\BCirc(40,40){7}
\DashCArc(40,40)(16,120,300)2
\Line(70,50)(56,64)
\Line(70,50)(70,70)
\Line(70,50)(84,64)
\Text(70,72)[b]{\scriptsize $\wh{\wh m}_2^-$}
\BCirc(70,50){7}
\DashCArc(70,50)(16,45,135)2
\Line(100,40)(117,50)
\Line(100,40)(117,30)
\Line(100,40)(100,20)
\Text(119,28)[l]{\scriptsize $\wh{\wh m}_3^-$}
\BCirc(100,40){7}
\DashCArc(100,40)(16,270,30)2
\end{picture}
\raisebox{40pt}{\Large +}
\begin{picture}(140,90)(0,0)
\Line(40,40)(70,50)
\Line(70,50)(100,40)
\Text(49,45)[b]{\scriptsize $+$}
\Text(59,48)[b]{\scriptsize $-$}
\Text(81,48)[b]{\scriptsize $+$}
\Text(91,44)[b]{\scriptsize $-$}
\Text(56,42)[t]{\scriptsize $\wh P_{a(b)}$}
\Text(84,42)[t]{\scriptsize $P_b$}
\Line(40,40)(50,23)
\Line(40,40)(20,40)
\Line(40,40)(30,23)
\Line(40,40)(30,57)
\Text(28,21)[tr]{\scriptsize $\wh{\wh m}_4^-$}
\Text(18,40)[r]{\scriptsize $\wh{\wh m}_1^-$}
\BCirc(40,40){7}
\DashCArc(40,40)(16,120,300)2
\Line(70,50)(56,64)
\Line(70,50)(70,70)
\Line(70,50)(84,64)
\Text(70,72)[b]{\scriptsize $\wh{\wh m}_2^-$}
\BCirc(70,50){7}
\DashCArc(70,50)(16,45,135)2
\Line(100,40)(117,50)
\Line(100,40)(117,30)
\Line(100,40)(100,20)
\Text(119,28)[l]{\scriptsize $\wh{\wh m}_3^-$}
\BCirc(100,40){7}
\DashCArc(100,40)(16,270,30)2
\end{picture}
\\
\raisebox{40pt}{\Large =}
\begin{picture}(140,90)(0,0)
\Line(40,40)(70,50)
\Line(70,50)(100,40)
\Text(49,45)[b]{\scriptsize $+$}
\Text(59,48)[b]{\scriptsize $-$}
\Text(81,48)[b]{\scriptsize $+$}
\Text(91,44)[b]{\scriptsize $-$}
\Text(56,42)[t]{\scriptsize $P_a$}
\Text(84,42)[t]{\scriptsize $P_b$}
\Line(40,40)(50,23)
\Line(40,40)(20,40)
\Line(40,40)(30,23)
\Line(40,40)(30,57)
\Text(28,21)[tr]{\scriptsize $\wh{\wh m}_4^-$}
\Text(18,40)[r]{\scriptsize $\wh{\wh m}_1^-$}
\BCirc(40,40){7}
\DashCArc(40,40)(16,120,300)2
\Line(70,50)(56,64)
\Line(70,50)(70,70)
\Line(70,50)(84,64)
\Text(70,72)[b]{\scriptsize $\wh{\wh m}_2^-$}
\BCirc(70,50){7}
\DashCArc(70,50)(16,45,135)2
\Line(100,40)(117,50)
\Line(100,40)(117,30)
\Line(100,40)(100,20)
\Text(119,28)[l]{\scriptsize $\wh{\wh m}_3^-$}
\BCirc(100,40){7}
\DashCArc(100,40)(16,270,30)2
\end{picture}
\end{center}
Thus, we can write the amplitude as
\begin{equation}
   \A_n(m_{1\ldots 4}^-)=\sum_{\{a,b\}}\frac{\A(\wh{\wh m}^-,
\wh{\wh P}_a^\pm)\A(\wh{\wh m}^-,-\wh{\wh P}_a^\mp,\wh{\wh P}_b^\pm)
\A(\wh{\wh m}^-,-\wh{\wh P}_ b^\mp)}{P_a^2P_b^2}.
\end{equation}
This is exactly the MHV rules for NNMHV amplitudes. 

It should now be clear how to proceed with the general case. To
uncover the first propagator in an N$^p$MHV amplitude, $a$, use the
shift
\begin{equation}
  \spxb{\wh m_i}=\spxb{m_i}+zc_i^{(1)}\spxb\eta.
\end{equation}
For each $a$, uncover the next propagator $b$ by using the shift
\begin{equation}
  \spxb{\wh{\wh m}_i}=\spxb{\wh m_i}+zc_{i,a}^{(2)}\spxb\eta
\end{equation}
where $c_{i,a}^{(2)}$ is chosen such that $\wh{\wh P}_a^2=0$. The
third propagator $c$ can now be uncovered by using the shift
\begin{equation}
  \spxb{\wh{\wh{\wh m}}_i}=\spxb{\wh{\wh m}_i}+c_ {i,a,b}^{(3)}\spxb
\eta
\end{equation}
where $c_ {i,a,b}^{(3)}$ is chosen such that $\wh{\wh{\wh P}}_a^2=\wh{
\wh{\wh P}}_b^2=0$. We can evidently continue in this way until we
have uncovered $p$ propagators, at which point we have shifted the
$\spxb{m_i}$ to $\spxb{m_i}+r_i\spxb\eta$ and required that the
shifted momenta of the uncovered propagators are on-shell. The latter,
together with conservation of momentum, fixes $r_i$ independently of
the order in which the propagators were uncovered. Writing the amplitude
formally as
\begin{equation}
  \A_n=\sum_{\scriptsize \begin{array}{c}\mathrm{ordered}
\\\mathrm{sets~of~p}\\\mathrm{propagators}\end{array}}
\frac{\A^\mhv \A^\mhv\cdots \A^\mhv}{P_a^2\wh P_b^2 
\wh{\wh P}_c^2\cdots},
\end{equation}
this allows the formal rewriting
\begin{equation}
  \A_n=\sum_{\scriptsize \begin{array}{c}\mathrm{unordered}
\\\mathrm{sets~of~p}\\\mathrm{propagators}\end{array}}
\A^\mhv \A^\mhv\cdots \A^\mhv
\sum_{\scriptsize\mathrm{orderings}}\frac1{P_a^2\wh P_b^2 
\wh{\wh P}_c^2\cdots}.
\end{equation}
The last sum is exactly what one would get by considering
\begin{equation}
  \frac1{P_a^2P_b^2P_c^2\cdots}
\end{equation}
and applying recursion, first using $c_i^{(1)}$, then, depending on
which propagator was uncovered, applying recursion using
$c_{i,a}^{(2)}$ and then (depending on $a$ and $b$) applying recursion
using $c_{i,a,b}^{(3)}$, \emph{etc.} In other words, 
\begin{equation}
  \A_n=\sum_{\scriptsize \begin{array}{c}\mathrm{unordered}
\\\mathrm{sets~of~p}\\\mathrm{propagators}\end{array}}
\frac{\A^\mhv \A^\mhv\cdots \A^\mhv}{P_a^2P_b^2P_c^2\cdots},
\end{equation}

We can now state what we have derived: To compute an amplitude, 
\begin{enumerate}
\item Draw all permitted MHV diagrams,
\item For each diagram, shift the negative helicity particles
  according to
  \begin{equation}
    \spxb{m_i^\flat}=\spxb{m_i}+r_i\spxb\eta
  \end{equation}
  and solve for $r_i$ under the condition that all shifted internal momenta
  $P_a^\flat$ are on-shell, and under the condition of conservation of
  momentum, 
  \begin{equation}
    \sum_ir_i\spax{m_i}=0.
  \end{equation}
\item For each MHV vertex, multiply by the amplitude as a function of
  the shifted momenta.
\item For each internal line with unshifted momentum $P_a$, multiply
  by $1/P_a^2$.
\end{enumerate}
For pure Yang--Mills, step 2 is irrelevant because it only provides
information on the anti-holomorphic spinors. For internal momenta,
$\spxa{P^\flat}\sim\spxxb P\eta$ is independent of the actual values
of the $r_i$. In the gravity application of these rules, to which we
now turn, this step becomes crucial.

\section{MHV Rules for Gravity}
\label{sec:grmhvrules}

Postponing again the question of the required large $z$ behaviour, we
can apply the above version of the MHV rules to gravity amplitudes. We
will do so through the example of the amplitude
$M_6(1^-,2^-,3^-,4^-,5^+,6^+)$, which is N$^2$MHV, but also
googly MHV, a fact which will not concern us here. This amplitude has
a total of 90 MHV diagrams that are permutations of six basic ones,
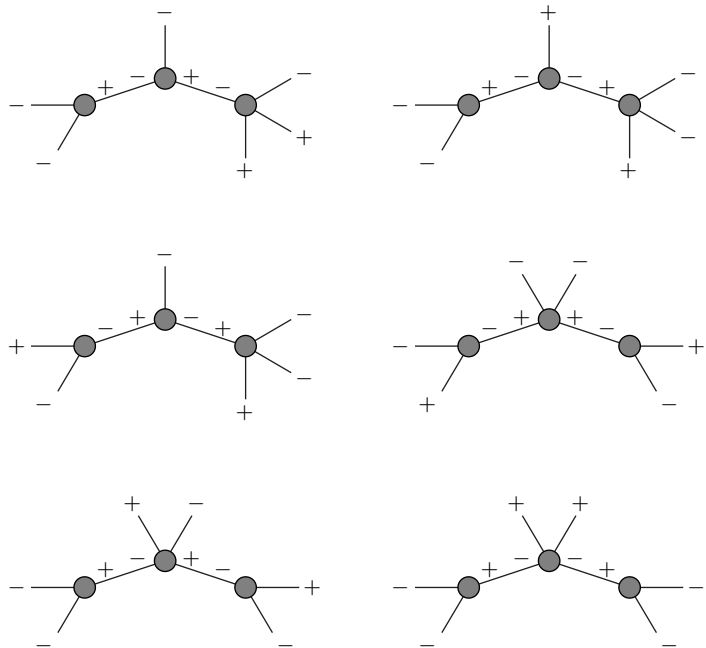
\begin{center}
\begin{picture}(140,90)(0,0)
\Line(40,40)(70,50)
\Line(70,50)(100,40)
\Text(48,44)[b]{\scriptsize $+$}
\Text(60,48)[b]{\scriptsize $-$}
\Text(80,48)[b]{\scriptsize $+$}
\Text(92,44)[b]{\scriptsize $-$}
\Line(40,40)(20,40)
\Line(40,40)(30,23)
\Text(28,21)[tr]{\scriptsize $-$}
\Text(18,40)[r]{\scriptsize $-$}
\GCirc(40,40){4}{.5}
\Line(70,50)(70,70)
\Text(70,72)[b]{\scriptsize $-$}
\GCirc(70,50){4}{.5}
\Line(100,40)(117,50)
\Line(100,40)(117,30)
\Line(100,40)(100,20)
\Text(119,52)[l]{\scriptsize $-$}
\Text(119,28)[l]{\scriptsize $+$}
\Text(100,18)[t]{\scriptsize $+$}
\GCirc(100,40){4}{.5}
\end{picture}
\begin{picture}(140,90)(0,0)
\Line(40,40)(70,50)
\Line(70,50)(100,40)
\Text(48,44)[b]{\scriptsize $+$}
\Text(60,48)[b]{\scriptsize $-$}
\Text(80,48)[b]{\scriptsize $-$}
\Text(92,44)[b]{\scriptsize $+$}
\Line(40,40)(20,40)
\Line(40,40)(30,23)
\Text(28,21)[tr]{\scriptsize $-$}
\Text(18,40)[r]{\scriptsize $-$}
\GCirc(40,40){4}{.5}
\Line(70,50)(70,70)
\Text(70,72)[b]{\scriptsize $+$}
\GCirc(70,50){4}{.5}
\Line(100,40)(117,50)
\Line(100,40)(117,30)
\Line(100,40)(100,20)
\Text(119,52)[l]{\scriptsize $-$}
\Text(119,28)[l]{\scriptsize $-$}
\Text(100,18)[t]{\scriptsize $+$}
\GCirc(100,40){4}{.5}
\end{picture}
\\
\begin{picture}(140,90)(0,0)
\Line(40,40)(70,50)
\Line(70,50)(100,40)
\Text(48,44)[b]{\scriptsize $-$}
\Text(60,48)[b]{\scriptsize $+$}
\Text(80,48)[b]{\scriptsize $-$}
\Text(92,44)[b]{\scriptsize $+$}
\Line(40,40)(20,40)
\Line(40,40)(30,23)
\Text(28,21)[tr]{\scriptsize $-$}
\Text(18,40)[r]{\scriptsize $+$}
\GCirc(40,40){4}{.5}
\Line(70,50)(70,70)
\Text(70,72)[b]{\scriptsize $-$}
\GCirc(70,50){4}{.5}
\Line(100,40)(117,50)
\Line(100,40)(117,30)
\Line(100,40)(100,20)
\Text(119,52)[l]{\scriptsize $-$}
\Text(119,28)[l]{\scriptsize $-$}
\Text(100,18)[t]{\scriptsize $+$}
\GCirc(100,40){4}{.5}
\end{picture}
\begin{picture}(140,90)(0,0)
\Line(40,40)(70,50)
\Line(70,50)(100,40)
\Text(48,44)[b]{\scriptsize $-$}
\Text(60,48)[b]{\scriptsize $+$}
\Text(80,48)[b]{\scriptsize $+$}
\Text(92,44)[b]{\scriptsize $-$}
\Line(40,40)(20,40)
\Line(40,40)(30,23)
\Text(28,21)[tr]{\scriptsize $+$}
\Text(18,40)[r]{\scriptsize $-$}
\GCirc(40,40){4}{.5}
\Line(70,50)(60,67)
\Line(70,50)(80,67)
\Text(58,69)[b]{\scriptsize $-$}
\Text(82,69)[b]{\scriptsize $-$}
\GCirc(70,50){4}{.5}
\Line(100,40)(120,40)
\Line(100,40)(110,23)
\Text(122,40)[l]{\scriptsize $+$}
\Text(112,21)[tl]{\scriptsize $-$}
\GCirc(100,40){4}{.5}
\end{picture}
\\
\begin{picture}(140,90)(0,0)
\Line(40,40)(70,50)
\Line(70,50)(100,40)
\Text(48,44)[b]{\scriptsize $+$}
\Text(60,48)[b]{\scriptsize $-$}
\Text(80,48)[b]{\scriptsize $+$}
\Text(92,44)[b]{\scriptsize $-$}
\Line(40,40)(20,40)
\Line(40,40)(30,23)
\Text(28,21)[tr]{\scriptsize $-$}
\Text(18,40)[r]{\scriptsize $-$}
\GCirc(40,40){4}{.5}
\Line(70,50)(60,67)
\Line(70,50)(80,67)
\Text(58,69)[b]{\scriptsize $+$}
\Text(82,69)[b]{\scriptsize $-$}
\GCirc(70,50){4}{.5}
\Line(100,40)(120,40)
\Line(100,40)(110,23)
\Text(122,40)[l]{\scriptsize $+$}
\Text(112,21)[tl]{\scriptsize $-$}
\GCirc(100,40){4}{.5}
\end{picture}
\begin{picture}(140,90)(0,0)
\Line(40,40)(70,50)
\Line(70,50)(100,40)
\Text(48,44)[b]{\scriptsize $+$}
\Text(60,48)[b]{\scriptsize $-$}
\Text(80,48)[b]{\scriptsize $-$}
\Text(92,44)[b]{\scriptsize $+$}
\Line(40,40)(20,40)
\Line(40,40)(30,23)
\Text(28,21)[tr]{\scriptsize $-$}
\Text(18,40)[r]{\scriptsize $-$}
\GCirc(40,40){4}{.5}
\Line(70,50)(60,67)
\Line(70,50)(80,67)
\Text(58,69)[b]{\scriptsize $+$}
\Text(82,69)[b]{\scriptsize $+$}
\GCirc(70,50){4}{.5}
\Line(100,40)(120,40)
\Line(100,40)(110,23)
\Text(122,40)[l]{\scriptsize $-$}
\Text(112,21)[tl]{\scriptsize $-$}
\GCirc(100,40){4}{.5}
\end{picture}
\end{center}
To compute the contribution of a representative of the first basic
diagram, 
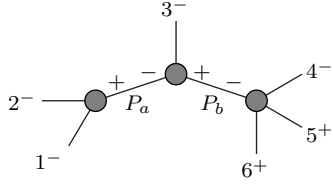
\begin{center}
\begin{picture}(140,90)(0,0)
\Line(40,40)(70,50)
\Line(70,50)(100,40)
\Text(48,44)[b]{\scriptsize $+$}
\Text(60,48)[b]{\scriptsize $-$}
\Text(56,42)[t]{\scriptsize $P_a$}
\Text(80,48)[b]{\scriptsize $+$}
\Text(92,44)[b]{\scriptsize $-$}
\Text(84,42)[t]{\scriptsize $P_b$}
\Line(40,40)(20,40)
\Line(40,40)(30,23)
\Text(28,21)[tr]{\scriptsize $1^-$}
\Text(18,40)[r]{\scriptsize $2^-$}
\GCirc(40,40){4}{.5}
\Line(70,50)(70,70)
\Text(70,72)[b]{\scriptsize $3^-$}
\GCirc(70,50){4}{.5}
\Line(100,40)(117,50)
\Line(100,40)(117,30)
\Line(100,40)(100,20)
\Text(119,52)[l]{\scriptsize $4^-$}
\Text(119,28)[l]{\scriptsize $5^+$}
\Text(100,18)[t]{\scriptsize $6^+$}
\GCirc(100,40){4}{.5}
\end{picture}
\end{center}
we first write down the conditions for the on-shell'ness of the shifted
internal momenta,
\begin{eqnarray}
  (P_a^\flat)^2&=&(P_a+r_1\spxa 1\spbx\eta+r_2\spxa 2\spbx\eta)^2
=P_a^2+r_1\spab 1{P_a}\eta+r_2\spab 2{P_a}\eta\nn
&=&\spa 12(\spb 21+r_1\spb 2\eta+r_2\spb\eta 1)=0,\\
(P_b^\flat)^2&=&(P_b+r_4\spxa 4\spbx\eta)^2=P_b^2+r_4\spab 4{(5+6)}
\eta=0.
\end{eqnarray}
Two additional conditions can be obtained by hitting the conservation
of momentum equation with two spinors (here we choose 3 and 4):
\begin{eqnarray}
  r_1\spa 13+r_2\spa 23+r_4\spa 43&=&0\\
r_1\spa 14+r_2\spa 24+r_3\spa 34&=&0.
\end{eqnarray}
The solution is
\begin{eqnarray}
  r_1&=&-\frac{\spab 4{(5+6)}1}{\spab 4{(5+6)}\eta},\qquad
\spxb{1^\flat}=\frac{\spxxa{(5+6)}4\spb\eta 1}{\spba \eta{(5+6)}4},\\
r_2&=&-\frac{\spab 4{(5+6)}2}{\spab 4{(5+6)}\eta},\qquad
\spxb{2^\flat}=\frac{\spxxa{(5+6)}4\spb\eta 2}{\spba \eta{(5+6)}4},\\
r_3&=&-\frac{\spab 4{(5+6)}3}{\spab 4{(5+6)}\eta},\qquad
\spxb{3^\flat}=\frac{\spxxa{(5+6)}4\spb\eta 3}{\spba \eta{(5+6)}4},\\
r_4&=&-\frac{(4+5+6)^2}{\spab 4{(5+6)}\eta},\qquad
\spxb{4^\flat}=-\frac{\spxxxb{(5+6)}{(4+5+6)}\eta}
{\spab 4{(5+6)}\eta},
\end{eqnarray}
such that
\begin{eqnarray}
  P_a^\flat&=&\frac{\spxxb{(1+2)}\eta
\spaxx 4{(5+6)}}{\spab4{(5+6)}\eta},\\
P_b^\flat&=&\frac{\spxxb{(4+5+6)}\eta\spaxx 4{(5+6)}}
{\spab 4{(5+6)}\eta}.
\end{eqnarray}
The contribution of this diagram is thus
\begin{eqnarray}
  &&M_3^\mhv(1^{\flat -},2^{\flat -},-P_a^{\flat +})
\frac1{P_a^2}\nn
&&\qquad\times
M_3^\mhv(P_a^{\flat -},3^{\flat -},P_b^{\flat +})
\frac1{P_b^2}M_4^\mhv(-P_b^{\flat -},4^{\flat -},5^+,6^+)\\
&=&\frac{\spa 12^6}{\spa 2{P_a^\flat}^2\spa{P_a^\flat}1^2}
\frac1{P_a^2}\frac{\spa{P_a^\flat}3^6}{\spa 3{P_b^\flat}^2
\spa{P_b^\flat}{P_a^\flat}^2}\frac1{P_b^2}
\frac{\spb 56 \spa{P_b^\flat}4^6}{\spa 56\spa 45
\spa 6{P_b^\flat}\spa 46\spa 5{P_b^\flat}}\\
&=&\frac{\spa 12^2}{\spb 1\eta^2\spb\eta2^2}\frac1{\spa 12\spb 21}
\frac{\spba\eta{(1+2+3)}3^6}{\spab 3{(4+5+6)}\eta^2
\spbb\eta{(4+5+6)}{(1+2)}\eta^2}\nn
&&\qquad\times\frac1{(4+5+6)^2}
\frac{\spb 56\spba\eta{(5+6)}4^6}{\spa 45\spa 56\spa 46
\spab 6{(4+5)}\eta\spab 5{(4+6)}\eta}\\
&=&\frac{\spa 12\spb 56\spab 3{P_b}\eta^2\spab 4{P_b}\eta^6}
{\spb 1\eta^2\spb 2\eta^2
\spb 3\eta^2\spb 21\spa 45\spa 56\spa 46\spab 6{P_b}\eta
\spab 5{P_b}\eta P_b^2}.
\end{eqnarray}
Though the procedure to obtain one out of six contributions to the
amplitude is in itself quite complicated, it is nothing in comparison
to the Lagrangian approach. But again, the on-shell recursion
relations for gravity will produce results faster.

Rather than computational speed, the interesting thing here is that it
is possible to formulate MHV rules for a gravity theory. This
highlights one of the mysterious connections between Yang--Mills theory
and gravity which is not properly understood. In this case, the MHV rules
for YM drop out of a description in twistor space, and such a
description does not seem likely for gravity, although there have been
attempts which reproduce the correct spectrum
\cite{AbouZeid06gravitytwistor} but not the correct dynamical theory
\cite{Nair07chiralgrav}. Thus, an explanation outside of on-shell
recursion relations is still an outstanding problem.

\section{Large $z$ Properties of the MHV Rule Shift}
\label{sec:mhvlargez}

The question of the $z\to\infty$ behaviour of deformed amplitudes
should now be addressed. To see what behaviour we need, consider the
very last step of the successive recursions where, for an N$^n$MHV
amplitude, we have uncovered $n-1$ propagators. To uncover the last
propagator, the last shift must have the amplitude go as
$z^{-1}$. When we get to this point, the antiholomorphic spinors have
already been shifted many times,
\begin{equation}
\spxb {m_i}\to \spxb {m_i}+(s_i+t_iz)\spxb\eta
\end{equation}
where $s_i$ are the consequences of all the previous shifts and the
$t_i$ impose conservation of momentum and the vanishing of the $z$
dependence of all propagator terms we have already uncovered. Had we
not imposed this, every propagator already uncovered would have
contributed $z^{-1}$ to the large $z$ limit. In other words, had the
$t_i$ been unconstrained (apart from momentum conservation) the large
$z$ behaviour of the final shift would be $z^{-n}$. The dependence on
the $s_i$ is subleading in $z$ and mainly a consequence of the
$c_{i,\ldots}^{(\cdot)}$'s chosen along the way, so their presence
should not disturb this conclusion for generic external momenta. This
leads us to conclude that the condition for the MHV rules to hold as
described here is that under the shift
\begin{equation}
\spxb {m_i}\to \spxb m_i+\tilde t_iz\spxb\eta
\end{equation}
where $\tilde t_i$ impose momentum conservation but are otherwise
free, an N$^n$MHV amplitude must go as $z^{-n}$ as $z\to\infty$.

For gluon amplitudes in Yang--Mills this can be proven directly by
looking at the worst possible Feynman diagram. In such a diagram, the
$n+2$ external negative helicity gluons must be connected by some
number $v$ of three-vertices and $v-1$ propagators. Since
three-vertices go as $z^1$ and propagators go as $z^{-1}$, these
internal parts of the diagram go as $z^1$. The dependence of the
polarizations is $z^{-1}$ each, because we are shifting the
antiholomorphic spinor of a negative helicity gluon, so the
polarizations together go as $z^{-(n+2)}$. In total the whole
amplitude goes as $z^{-(n+1)}$ which is one power better than
required. This forms the last ingredient of the on-shell recursive
proof of the MHV rules for gluons.

For gravitons it is a different story. If we do the same analysis as
above, we find that the $v$ vertices contribute $z^{2v}$, the propagators
contribute $z^{-v+1}$ and the external polarizations contribute
$z^{-2(n+2)}$. This gives in total $z^{v-2n-3}$. The maximal number of
vertices that may have $z$ dependence in an $m$-point amplitude is
$m-2$, so the worst possible Feynman diagram has the behaviour
$z^{m-2n-5}$ where we require $z^{-n}$. An analysis using the above
result for Yang--Mills together with the KLT relations gives the same
estimate. 

Luckily, these types of arguments are known to be way off for
gravity. If we do the same analysis for the BCFW shift
(\ref{eq:deformation}) of an $m$-point amplitude we would obtain
$z^{m-5}$, where the real result is known to be $z^{-2}$
\cite{Benincasa07gravrecur, Arkani-Hamed08}. Cancellations of this
order of magnitude in the shifts used here should in general ensure
that the MHV rules for gravity are justified. In the original article
on gravity MHV rules \cite{Risager05grcsw}, the NMHV six and seven
point amplitudes were found to go as $z^{-5}$ which is two powers
better than needed, and four and five powers, respectively, better
than the worst expectation. If we allow ourselves some speculation,
this hints that the actual $z\to\infty$ behaviour is $z^{-n-4}$,
implying that $m-n+1$ powers are removed by obscure symmetries in the
gravity theory.

\chapter{Loop-Level Methods}
\label{cha:loop}

Until now we have mostly concerned ourselves with tree-level
amplitudes. For most practical applications though, amplitudes at one
or more loops are required for interesting results. This chapter will
review four general tools out of strict historical order. For loop-level
calculations it is often necessary (or just useful) to use
combinations of them. 

\section{Structure of One-Loop Amplitudes}
\label{sec:loopstructure}

First, however, we will start out by reviewing what is known in
general about one-loop amplitudes, mostly in Yang--Mills theory, but
also more generally.

\subsection{Integral Reduction}
\label{sec:pv}

When doing one-loop calculations with Feynman rules, one will
encounter integrals with arbitrary numbers of propagators and tensors
depending on the loop momentum of arbitrary rank. In Yang--Mills
theory, an $n$-point calculation will give up to $n$ loop propagators
and up to rank $n$ tensors, while \emph{e.g.}~a gravity calculation
will give up to $n$ loop propagators and rank $2n$ tensors in the loop
momentum. Such integrals are impossible to do directly, so one will
normally do some sort of integral reduction.

The most important of these is Passarino--Veltman reduction
\cite{Passarino78pv}, which reduces the powers of loop momentum
appearing in the numerator by using the knowledge that the result must
be writable as a Lorentz tensor, together with the knowledge of which
Lorentz tensors it can depend on. As a simple example, take the
integral
\begin{equation}
\int d^DL\frac{L^\mu L^\nu}{(L^2+i\epsilon)((L+P)^2+i\epsilon)}.
\end{equation}
Since this integral can only depend on $g_{\mu\nu}$ and $P^\mu$, it
must be writable as
\begin{equation}
A(P)g^{\mu\nu}+B(P)P^\mu P^\nu.
\end{equation}
By contracting with $g_{\mu\nu}$ and $P_\mu$ on both sides of this
equation and using similar results from simpler integrals, one can
find expressions for $A$ and $B$ in terms of integrals fewer powers of
loop momentum in the numerator and the same number of, or fewer, loop
propagators. Iterating this procedure can remove all dependence on the
loop momentum in the numerator and leave us with scalar integrals
only. 

Depending on the dimension of spacetime, the scalar integrals may be
reduced further. Take for instance the massless hexagon integral in
four dimensions (with implicit $i\epsilon$'s)
\begin{equation}
\int d^4L\frac1{L^2(L+K_1)^2(L+K_2)^2(L+K_3)^2(L+K_4)^2(L+K_5)^2}.
\end{equation}
Because the external momenta are ``twice'' linearly dependent in four
dimensions, we can always adjust constants $a_0$ through $a_5$ such
that 
\begin{equation}
  a_0L^2+\sum_{i=1}^5 a_i(L+K_i)^2=1,
\end{equation}
which allows us to split the hexagon integral into six
pentagons. Notice that this depends on the dimension in which the
external momenta live; dimensionally regulating the loop momentum does
not change this. 

By using these types of integral reduction, one can arrive at
expressions for the wanted amplitudes which contain a reduced set of
integrals that are possible to compute, or more often, look up.

\subsection{Massless Integral Basis in Four Dimensions}
\label{sec:integralbasis}

For four dimensional theories with massless states running in the
loop, there is an additional identity that allows pentagons to be
written in terms of lower point integrals \cite{Bern93pentagon}. This
leaves a basis for all one-loop amplitudes consisting of scalar box
integrals, triangle integrals, bubble integrals, and
1\footnote{Remember that massless tadpoles vanish in dimensional
regularization}. The integrals are defined as (implicit $i\epsilon$'s)
\begin{eqnarray}
I_4(P_1,P_2,P_3)&=&-i\mu^{2\epsilon}\int
\frac{d^{4-2\epsilon}L}{(2\pi)^{4-2\epsilon}}\nn
&&\qquad\times\frac1{L^2(L+P_1)^2
(L+P_1+P_2)^2(L+P_1+P_2+P_3)^2},\nn
I_3(P_1,P_2)&=&i\mu^{2\epsilon}
\int \frac{d^{4-2\epsilon}L}
{(2\pi)^{4-2\epsilon}}\frac1{L^2(L+P_1)^2
(L+P_1+P_2)^2},\nn
I_2(P)&=&-i\mu^{2\epsilon}
\int \frac{d^{4-2\epsilon}L}{(2\pi)^{4-2\epsilon}}\frac1{L^2(L+P)^2},
\end{eqnarray}
These integrals (and 1) have coefficients which are rational
functions of the external spinors and momenta. In addition, box and
triangle integrals come out as expressions of mass dimension
$-2\epsilon$, divided by their respective Gram determinants which may
or may not be rational. This prompts us to define the box and triangle
\emph{integral functions} $F_4$ and $F_3$ where the Gram determinant
is taken into the coefficient instead, where we will often find that
it is cancelled. 

Box and triangle integrals are classified according to which of their
external momenta are massless. For triangles there are zero-mass
through three-mass ($0m$, $1m$, $2m$, $3m$) while for boxes there
zero-, one-, three-, and four-mass together with the so-called
two-mass-easy ($2me$) and two-mass-hard ($2mh$) configurations where
the massless corners are opposite and adjacent, respectively. The
integral functions and Gram determinants are given in appendix
\ref{app:integrals}. 

\subsection{Supersymmetric Decomposition}
\label{sec:susydecom}

When calculating gauge theory amplitudes, the answer depends on the
number of massless fermions and scalars in the gauge theory because
they can run in the loop. There are particular choices for these
numbers that give dramatically simpler answers, namely those
corresponding to supersymmetric theories. It is standard to use the
``basis'' for the particle content consisting of an $\N=4$ multiplet
(1 gluon, 4 fermions, 3 complex scalars), an $\N=1$ chiral multiplet
(1 fermion, 1 complex scalar), and the so-called $\N=0$ multiplet
containing just a complex scalar. A theory with $n_f$ adjoint fermions
and $n_s$ adjoint scalars can thus be written as
\begin{equation}
(\N=4)-\Big[4-n_f\Big](\N=1)+\Big[1-n_f+n_s\Big](\N=0).
\end{equation}
If the fermions or scalars are in the fundamental representation,
$n_{f,s}$ should be substituted with $n_{f,s}/N_C$ where $N_C$ is the
number of colours.

In background field gauge, it can be shown explicitly that there are
diagram-by-diagram cancellations that remove four powers of loop
momentum in the numerator for $\N=4$ and two powers for $\N=1$. When
those results are Passarino--Veltman reduced, they can only give rise
to boxes in the $\N=4$ case and only to boxes, triangles and bubbles
(that is, no additional rational terms) in $\N=1$. For $\N=0$ there
are no such cancellations, but the calculation is made simpler by the
reduced number of Lorentz indices for a scalar running in the loop as
compared to a gluon.

Supersymmetric decompositions can be used in other theories as well,
however, without a formal handle on the cancellation of numerator loop
momentum. Since the expression of the supersymmetry at tree level is
the same as in gauge theory (\emph{e.g.}~SWI's, \emph{cf.}~section
\ref{sec:swi}) and tree level results are recycled in most loop
calculations, it is generally believed that the same regularity holds
for other theories, in particular gravity. For arbitrary $\N$ it is
believed that an even $\N$ multiplet will have $\N$ cancellations of
loop momenta in the numerator while an odd $\N$ multiplet will have
$\N+1$ \cite{Bern07gravitycancellations}.

\subsection{Limits and Singularities}
\label{sec:looplimit}

To aid in both checking and construction expressions for one-loop
amplitudes, it becomes important to look at these expressions in
certain limits. The tree-level limiting behaviour was studied in
section \ref{sec:treelimit} and this is extended here to one-loop
level. 

The multiparticle poles of Yang--Mills amplitudes can occur in the
same way as at tree-level when a non-loop propagator goes on shell and
splits the diagram in two, the loop being on one side or the other. We
may also have a singularity coming from other parts of the calculation
which has a more messy interpretation, though. This gives the
multiparticle limit \cite{Bern95factorization}
\begin{eqnarray}
\label{eq:ymmultiparticle}
  \A^{\mathrm{1-loop}}(\ldots)&\to& \sum_{h=\pm}\bigg(
\frac{\A^{\mathrm{tree}}(\ldots,P^h)\A^{\mathrm{1-loop}}
(-P^{-h},\ldots)}{P^2}\nn
&&\qquad +\frac{\A^{\mathrm{1-loop}}(\ldots,P^h)\A^{\mathrm{tree}}
(-P^{-h},\ldots)}{P^2}\nn
&&\qquad +\mathcal F\frac{\A^{\mathrm{tree}}(\ldots,P^h)\A^{\mathrm{tree}}
(-P^{-h},\ldots)}{P^2}\bigg)
\end{eqnarray}
as $P^2\to 0$, where $\mathcal F$ is a function depending (in a
cut-containing and non-factorized way) on all incoming momenta, but
not on the helicities. It also depends on the particle content of the
theory. 

Another type of singularity that Yang--Mills theory has, is
collinear singularities. Again, we get something similar to the
tree-level case, only now we have to introduce 1-loop splitting
amplitudes which depend on the particle content of the theory. We have
\begin{eqnarray}
\label{eq:ymcollinear}
\A^{\mathrm{1-loop}}(a^{h_a},b^{h_b},\ldots)&\to&
\sum_{h=\pm}\bigg[\mathrm{Split}^{\mathrm{tree}}_{h}(z,a^{h_a},b^{h_b})
\A^{\mathrm{1-loop}}(P^{-h},\ldots)\nn
&&\qquad +\mathrm{Split}^{\mathrm{1-loop}}_{h}(z,a^{h_a},b^{h_b})
\A^{\mathrm{tree}}(P^{-h},\ldots)\bigg]
\end{eqnarray}
as $a\to zP$, $b\to (1-z)P$. The one-loop splitting functions for both
gluons and fermions are given in \cite{Bern94oneloopn4mhv} (and to
all orders in $\epsilon$ in \cite{Bern99nnlo}). The pure-glue splitting
functions are
\begin{eqnarray}
\label{eq:fullsplit}
\mathrm{Split}_+^{\mathrm{1-loop}}(a^+,b^+)
&=&-\frac1{48\pi^2}\bigg(1-\frac{n_f}N
+\frac{n_s}N\bigg)\sqrt{z(1-z)}\frac{\spb ab}{\spa ab^2}\nn
\mathrm{Split}_-^{\mathrm{1-loop}}(a^+,b^+)
&=&c_\Gamma\mathrm{Split}_-^{\mathrm{tree}}(a^+,b^+)
\bigg[ U+\frac13\bigg(1-\frac{n_f}N+\frac{n_s}N\bigg)z(1-z)\bigg]\nn
\mathrm{Split}_+^{\mathrm{1-loop}}(a^\pm,b^\mp)
&=&c_\Gamma\mathrm{Split}_+^{\mathrm{tree}}(a^\pm,b^\mp)U
\end{eqnarray}
where 
\begin{equation}
U=-\frac1{\epsilon^2}\bigg(\frac{\mu^2}{z(1-z)(-s_{ab})}\bigg)^\epsilon
+2\log z\log (1-z)-\frac{\pi^2}6
\end{equation}
and $c_\Gamma$ is defined in (\ref{eq:cGamma}).

The general form of infrared divergences in Yang--Mills is also known
from general arguments. This imposes the constraint that
\cite{Giele91ir, Catani98ir, Kunszt94ir}
\begin{equation}
\A^1=-\frac{c_\Gamma}{\epsilon^2}A^0\sum_{i=1}^4\bigg(\frac{
\mu^2}{-s_{i,i+1}}\bigg)^\epsilon+\mathcal O(\epsilon^0).
\end{equation}
This imposes some quite stringent constraints on coeffiecients of box
and triangle functions because there needs to be complete cancellation of
\emph{e.g.}~terms of the kind
\begin{equation}
\frac1{\epsilon^2}\bigg(\frac{\mu^2}{-t_{\cdots}}\bigg)^\epsilon
\end{equation}
where $t_{\cdots}$ is any multiparticle kinematic invariant
\cite{Bern04sevengluon}. In one-loop gravity, a similar relation holds
\cite{Dunbar95softgravity},
\begin{equation}
\M^{\mathrm{1-loop}}=\frac{i}{(4\pi)^2}
\bigg[\frac{\sum_{i<j}s_{ij}\log[-s_{ij}]}{2\epsilon}\bigg]
\M^{\mathrm{tree}}+\mathcal O(\epsilon^0)
\end{equation}
with similar consequences \cite{BjerrumBohr05notri} to be explored
further in chapter \ref{cha:sugra}.

\section{Unitarity Cuts}
\label{sec:unitarity}

As mentioned in the introduction, the concepts of unitarity and
analyticity of the $S$-matrix came to play a large role again in the
90's. That was primarily because the insights of the previous section
were sufficient to fill in the missing parts of the puzzle, the parts
which (helped by experimental evidence) sent $S$-matrix theory out in
the cold.

\subsection{Ordinary $D=4$ Unitarity Cuts}

Instead of trying to directly reconstruct the one-loop amplitude from
its cut discontinuities---a procedure which is often ambiguous---it is
possible to use all the insights of the previous section to come up
with an ansatz for the result and match cuts on both sides of the
equation to arrive at an answer. This was first used by Bern, Dixon,
Dunbar and Kosower to compute the one-loop MHV amplitudes in $\N=4$
Yang--Mills \cite{Bern94oneloopn4mhv} as follows: It is known that the
$\N=4$ amplitude only contains boxes, so we can write
\begin{equation}
\label{eq:ansatz}
\A^{\mathrm{1-loop}}_{\mathrm{MHV}}(i^-,j^-)
=\sum_{a\in \mathrm{boxes}}c_aI_a.
\end{equation}
On both sides we can pick out the discontinuity in $P^2_{m_1,m_2}$
by replacing the loop propagators between $m_1$ and $m_1+1$ and
between $m_2-1$ and $m_2$ by delta functions in the square of the
loop momenta. On the left side of (\ref{eq:ansatz}) we get an integral
over the product of two tree amplitudes, possibly summed over
internal helicities in the loop,
\begin{eqnarray}
\lefteqn{\Delta_{m_1,m_2}\A^{1-loop}}\nn
&=&\sum_{h,h^{'}}\int d^Dl_1d^Dl_2\delta(l_1^2)\delta(l_2^2)
\delta^{(D)}(l_1-l_2-P_{m_1,m_2})\nn
&&\times\A(m_2+1,\ldots,m_1-1,l_1^h,(-l_2)^{-h^{'}})
\A(m_1,\ldots,m_2,l_2^{h^{'}},(-l_1)^{-h}).
\end{eqnarray}
Pictorically, this becomes
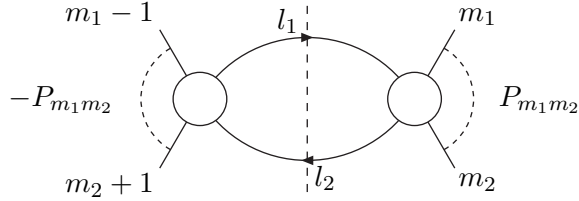
\begin{center}
\begin{picture}(180,100)(0,0)
\ArrowArcn(90,26.9)(46.2,150,30)
\ArrowArcn(90,73.1)(46.2,330,210)
\Text(87,75)[br]{$l_1$}
\Text(93,25)[tl]{$l_2$}
\DashLine(90,15)(90,85){3}
\Line(50,50)(35,24)
\Line(50,50)(35,76)
\DashCArc(50,50)(22,120,240){2}
\BCirc(50,50){10}
\Text(18,50)[r]{$-P_{m_1m_2}$}
\Text(33,22)[tr]{$m_2+1$}
\Text(33,78)[br]{$m_1-1$}
\Line(130,50)(145,76)
\Line(130,50)(145,24)
\DashCArc(130,50)(22,300,60){2}
\BCirc(130,50){10}
\Text(162,50)[l]{$P_{m_1m_2}$}
\Text(147,78)[bl]{$m_1$}
\Text(147,22)[tl]{$m_2$}
\end{picture}
\end{center}
where the dashed line denotes the cutting of the two internal lines. 
In Lorentz invariant gauges such as Feynman gauge, we would have to
include a ghost loop, but by using the spinor-helicity formalism, which
is implicitly in a light-like axial gauge as mentioned at the end of
section \ref{sec:spinor}, we can avoid them.

For the two tree amplitudes to be non-zero, they must both be
MHV. Then, if $i$ and $j$ are both on the left [right] side we have
$(h,h^{'})=(1,-1)\,[(-1,1)]$ while if they are on opposite sides
$h=h^{'}$ can take any value in the $\N=4$ multiplet and we need to do
an actual summation. These two distinct cases can be drawn as
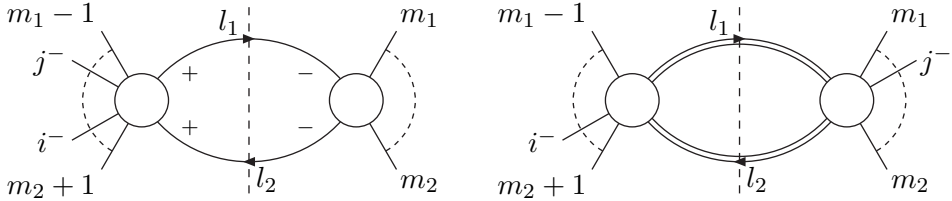
\begin{center}
\begin{picture}(180,100)(0,0)
\ArrowArcn(90,26.9)(46.2,150,30)
\ArrowArcn(90,73.1)(46.2,330,210)
\Text(65,63)[tl]{\scriptsize $+$}
\Text(115,63)[tr]{\scriptsize $-$}
\Text(115,37)[br]{\scriptsize $-$}
\Text(65,37)[bl]{\scriptsize $+$}
\Text(87,75)[br]{$l_1$}
\Text(93,25)[tl]{$l_2$}
\DashLine(90,15)(90,85){3}
\Line(50,50)(35,24)
\Line(50,50)(24,35)
\Line(50,50)(24,65)
\Line(50,50)(35,76)
\DashCArc(50,50)(22,120,240){2}
\BCirc(50,50){10}
\Text(33,22)[tr]{$m_2+1$}
\Text(22,35)[r]{$i^-$}
\Text(22,65)[r]{$j^-$}
\Text(33,78)[br]{$m_1-1$}
\Line(130,50)(145,76)
\Line(130,50)(145,24)
\DashCArc(130,50)(22,300,60){2}
\BCirc(130,50){10}
\Text(147,78)[bl]{$m_1$}
\Text(147,22)[tl]{$m_2$}
\end{picture}
\begin{picture}(180,100)(0,0)
\ArrowArcn(90,26.9)(46.2,150,30)
\ArrowArcn(90,73.1)(46.2,330,210)
\CArc(90,26.9)(44.2,30,150)
\CArc(90,73.1)(44.2,210,330)
\Text(87,75)[br]{$l_1$}
\Text(93,25)[tl]{$l_2$}
\DashLine(90,15)(90,85){3}
\Line(50,50)(35,24)
\Line(50,50)(24,35)
\Line(50,50)(35,76)
\DashCArc(50,50)(22,120,240){2}
\BCirc(50,50){10}
\Text(33,22)[tr]{$m_2+1$}
\Text(22,35)[r]{$i^-$}
\Text(33,78)[br]{$m_1-1$}
\Line(130,50)(145,76)
\Line(130,50)(156,65)
\Line(130,50)(145,24)
\DashCArc(130,50)(22,300,60){2}
\BCirc(130,50){10}
\Text(147,78)[bl]{$m_1$}
\Text(158,65)[l]{$j^-$}
\Text(147,22)[tl]{$m_2$}
\end{picture}
\end{center}
The summation over the $\N=4$ multiplet, however, gives exactly the
same result as when both $i$ and $j$ are on the same side, namely that
the integrand can be written as
\begin{equation}
\A^{\mathrm{tree}}(i^-,j^-)\frac{\spa{(m_1-1)}{m_1}\spa {l_1}{l_2}}
{\spa {(m_1-1)}{l_1}\spa {l_1}{m_1}}\frac{\spa{m_2}{(m_2+1)}
\spa{l_2}{l_1}}{\spa {m_2}{l_2}\spa{l_2}{(m_2+1)}}.
\end{equation}
This can again be rewritten as
\begin{eqnarray}
\label{eq:integrandasboxes}
&&\frac12\A^{\mathrm{tree}}\bigg[
\frac{P^2_{m_1,m_2}P^2_{m_1-1,m_2+1}-P^2_{m_1-1,m_2}P^2_{m_1,m_2+1}}
{(l_1+(m_1-1))^2(l_2-(m_2+1))^2}\nn
&&\qquad+\frac{P^2_{m_1,m_2-1}P^2_{m_1-1,m_2}-P^2_{m_1-1,m_2-1}
P^2_{m_1,m_2}}{(l_1+(m_1-1))^2(l_2+m_2)^2}\nn
&&\qquad+\frac{P^2_{m_1+1,m_2}P^2_{m_1,m_2+1}-P^2_{m_1,m_2}
P^2_{m_1+1,m_2+1}}{(l_1-m_1)^2(l_2-(m_2+1))^2}\nn
&&\qquad +\frac{P^2_{m_1+1,m_2-1}P^2_{m_1,m_2}-P^2_{m_1,m_2-1}
P^2_{m_1+1,m_2}}{(l_1-m_1)^2(l_2+m_2)^2}\bigg],
\end{eqnarray}
plus terms that vanish when we integrate.  In this way all the
dependence on the loop momenta has been confined to two propagators in
each of the four terms. That, however, is exactly what one would get by
cutting four particular 2-mass easy boxes. And even better, the
coefficients (apart from $\A^{\mathrm{tree}}$) are the respective Gram
determinants, so this is a sum of cuts of four box functions ($F$'s)
with the same coefficient. In equations,
\begin{eqnarray}
\Delta_{m_1,m_2}\A^{\mathrm{1-loop}}&=&
\A^{\mathrm{tree}}\Delta_{m_1,m_2}\Big(
F_4(m_1-1,P_{m_1,m_2},m_2+1)\nn
&&\qquad+F_4(m_1-1,P_{m_1,m_2-1},m_2)\nn
&&\qquad+F_4(m_1,P_{m_1+1,m_2},m_2+1)\nn
&&\qquad +F_4(m_1,P_{m_1+1,m_2-1},m_2)\Big).
\end{eqnarray}
This suggests that the one-loop amplitude can be written roughly as
the tree amplitude times the sum of all 2-mass easy box functions. The
'roughly' comes in through careful considerations of which choices of
$m_1$ and $m_2$ are allowed and independent. The result can now be
checked using other means such as collinear limits and it does indeed
hold up.

Staying inside Yang--Mills theory there are two extensions we can make
on this \cite{Bern95oneloopn1mhv}. Firstly, we can consider the $\N=1$
or the $\N=0$ theory where triangle and bubble integrals are
introduced on the right side of (\ref{eq:ansatz}). This complicates
the calculation by giving integrands whose numerators depend on the
cut loop momenta while there may be one less propagator. This then
requires integral reduction \emph{a la} Passarino--Veltman (section
\ref{sec:pv}) with some simplifying conditions on the loop
momenta. For the $\N=0$ theory there is the additional complication
that the rational terms are not picked up by four-dimensional
unitarity cuts. As we will see later, it can still be quite useful to
know the cut-containing parts of the amplitude. Secondly, we can go
beyond MHV amplitudes where the integrand does not immediately have
the nice decomposition into boxes as \emph{e.g.}~in
(\ref{eq:integrandasboxes}). Still, some $\N=4$ six-point results can
be obtained.

Inspired by the recent revival of unitarity arguments, some
improvements have been made on the technology which allows for quite
efficient extraction of integral coefficients
\cite{Britto05sqcd1,Britto06sqcd2}. 

\subsection{Generalized Unitarity}
\label{sec:genuni}

The process of computing a unitarity cut where two propagators are put
on-shell has a clear interpretation as the discontinuity across a cut
in the amplitude as a complex function of the kinematic variables. In
the way it is applied, however, this interpretation of the cut is not
used. Although the interpretation of the cutting of more than two loop
propagators may not be as clear, it seems just as reasonable to use it
to constrain an amplitude.

The problem is that many integrals cut on more than two propagators
turn out to be zero because there are no solutions in Minkowski space
for the cut momenta. A triple cut is only non-zero when applied to
three-mass triangles and three- and four-mass boxes, while a quadruple
cut is only non-zero for four-mass boxes. Higher cuts are zero because
they over-constrain the four-dimensional loop momentum. There are some
situations where generalized unitarity does come in handy in Minkowski
space; the NMHV $\N=4$ amplitude has been calculated to all $n$ in
this way \cite{Bern04sevengluon,Bern04nmhv}.

Since the comments here apply only to amplitudes in real Minkowski
space, they will be altered completely when we move to complex momenta
below.

\subsection{Unitarity in $D=4-2\epsilon$}
\label{sec:ddimunitarity}

If we want to know a full one-loop amplitude we must do the
dimensional regularization properly. This involves taking the loop
momentum to be $4-2\epsilon$-dimensional rather than
four-dimensional. This poses a very deep problem when using the spinor
helicity formalism as it is tied to exactly four dimensions. One
solution could be to use spinor helicity notation for all non-loop
quantities and ordinary Lorentz notation for loop quantities. This
comes at the expense of the compact form of tree amplitudes.

When computing gluon scattering amplitudes, the easiest way to proceed
is to use the fact that only the $\N=0$ component can have rational
terms, and thus we only need the less compact amplitudes for scalar
particles in the loop \cite{Bern95massiveloop,Bern96sdym}. In this way
issues of summing over polarization states are also avoided. The cut
momentum of the scalar is written as
\begin{equation}
L_{(4-2\epsilon)}=L_{(4)}+\mu e_{(-2\epsilon)},
\end{equation}
where $L_{(4)}$ is a four-dimensional momentum of mass $\mu$ and
$e_{(-2\epsilon)}$ is a $-2\epsilon$-dimensional unit vector. The
integration will then split into an integral over a massive scalar and
an integral over the mass. Using the unitarity method, this can be
used to write the amplitude in terms of loop integrals whose rational
parts as $\epsilon\to 0$ are known. Recent versions of such methods
are developed in \cite{Anastasiou06duni1,Britto06duni2,Anastasiou06duni3}.

\section{MHV Rules for Loops: The BST Prescription}
\label{sec:bst}

When the MHV rules were conceived, the perspective for their use at
loop level was rather dim. It was argued that the loops would
necessarily contain states of conformal supergravity, a theory which
is non-unitary. Not everyone was deterred by this argument, at least
not Brandhuber, Spence and Travaglini (BST) who decided to recalculate
the one-loop $N=4$ MHV amplitude using MHV rules \cite{Brandhuber04bst}. 

Their prescription consists of drawing all diagrams with two MHV
vertices and two internal lines to form a loop,
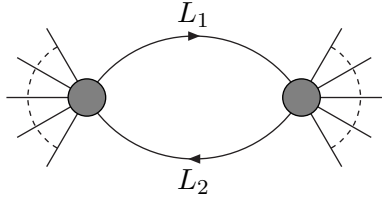
\begin{center}
\begin{picture}(180,100)(0,0)
\ArrowArcn(90,26.9)(46.2,150,30)
\ArrowArcn(90,73.1)(46.2,330,210)
\Text(90,77)[b]{$L_1$}
\Text(90,23)[t]{$L_2$}
\Line(50,50)(35,24)
\Line(50,50)(24,35)
\Line(50,50)(20,50)
\Line(50,50)(24,65)
\Line(50,50)(35,76)
\DashCArc(50,50)(22,120,240){2}
\GCirc(50,50){7}{.5}
\Line(130,50)(145,76)
\Line(130,50)(156,65)
\Line(130,50)(160,50)
\Line(130,50)(156,35)
\Line(130,50)(145,24)
\DashCArc(130,50)(22,300,60){2}
\GCirc(130,50){7}{.5}
\end{picture}
\end{center}
The
momenta $L_1,L_2$ of these internal lines are off-shell, so on-shell
continuations of them are required for use in the MHV
vertices. This is performed by taking the integration measure
\begin{equation}
\label{eq:mhvmeasure}
  d^4 L_1 d^4L_2 \frac{\delta^{(4)}(L_1-L_2+P)}{(L_1^2+i\epsilon)
(L_2^2+i\epsilon)}
\end{equation}
and changing variables to $l_{1,2}$ and $z_{1,2}$ such that $l_{1,2}$
are lightlike and
\begin{equation}
L_{1,2}=l_{1,2}+z_{1,2}\eta
\end{equation}
where $\eta$ is an arbitrary lightlike vector. As shown in
\cite{Brandhuber05bstproof}, the combination
$z_1+z_2$ can be integrated out by standard (albeit involved) methods
of complex analysis, while the combination $z_1-z_2$ can be replaced
by $P_z^2=(P+z_1\eta-z_2\eta)^2$. This yields the measure
\begin{equation}
\label{eq:bstmeasure}
2\pi id^4l_1d^4l_2\delta^{(+)}(l_1^2)\delta^{(+)}(l_2^2)
\frac{dP_z^2\theta(P_z^2)}
{P_z^2-P^2-i\epsilon}\delta^{(4)}(l_1-l_2+P_z).
\end{equation}
where the $^{(+)}$ on the delta functions impose that $l_1$ and $l_2$
must have positive energy.

For $\N=4$ matter running in the loop, BST found that the product of
the two MHV vertices (summed over the multiplet) became---in a way
similar to the unitarity example above---a sum over terms of the kind
\begin{equation}
\label{eq:bstintegrand}
\A^{\mathrm{tree}}\frac{(P_z+m_1)^2(P_z-m_2)^2-P_z^2(P_z+m_1-m_2)^2}
{2(l_1+m_1)^2(l_2+m_2)^2}.
\end{equation}
At this point the loop momenta are described as 4-momenta rater than
spinors, and that permits a dimensional regularization by taking the
number of dimensions to be $4-2\epsilon$ as usual. 

With sufficient hard work, the integration of (\ref{eq:bstintegrand})
with the measure (\ref{eq:bstmeasure}) can be performed and terms
(four of each) which have the same numerator in
(\ref{eq:bstintegrand}) combine to form two-mass easy box
functions. The proof relies in part on an identity between nine dilog
functions and in this way gives a different form of the two-mass easy
function derived earlier by Duplan\v ci\'c and Ni\v zi\'c
\cite{Duplancic00} but not widely known.

There are generalizations to the cases with $\N=1$
\cite{Bedford04bstchiral,Quigley04bstchiral} and scalar matter
\cite{Bedford04bstscalar} in the loop which require triangle and
bubble functions. In the scalar case there are additional rational
pieces which are not predicted by this version of the loop MHV rules
because they contain cuts almost by construction. These rational
terms, however, appear to be generated in the light-cone reformulation
of the MHV rules with a particular non-dimensional regularization
\cite{Brandhuber07bstregularized}.

In spite of the successful application to several one-loop calculations
of MHV (and similar \cite{Badger06allplus,Badger07phimhv}) amplitudes,
the BST prescription appears to be unpractical for higher numbers of
negative helicity gluons than two. Although the formal correctness of
the prescription has been proven using the Feynman Tree Theorem
\cite{Brandhuber05bstproof} for any number of negative helicity
gluons, there have been no concrete calculations, mainly because the
method has been superceded by others mentioned in this chapter.

The prescription also inherits a formal problem from the tree-level
MHV rules, namely the one mentioned in section \ref{sec:reformgr} that
the vertices do not obey momentum conservation. From that perspective,
it may be seen as odd that the right results appear, as any (momentum
conserving) rewriting of the vertices will change the results to
something wrong. Resolution of this problem comes either from
formulating the loop-level MHV rules in a light-cone form or by noting
that any cutting of internal lines (as in the Feynman Tree Theorem
mentioned in the introduction) reduces the amplitude to integrals over
on-shell tree amplitudes where the conservation-of-momentum problem
was resolved by viewing the rules as coming from recursion relations
in complexified Minkowski space (chapter \ref{cha:mhv}).

\section{Quadruple Cuts}
\label{sec:quadcuts}

The unitarity method of section \ref{sec:unitarity} employed cuts on
two internal states, partly because the cut conditions are guaranteed
to have solutions for which the internal momenta are real in Minkowski
space. Clearly, there is nothing wrong with applying more cuts in
principle, but the cut conditions often turn out not to allow any real
solutions, thereby setting the cut to zero and rendering its
information useless. 

While working with twistor methods, first described in 2+2 dimensions
where the spinors are real and independent, Britto, Cachazo and Feng
realized \cite{Britto04quad} that working in that signature (or,
more generally, in complexified Minkowski space) would allow
simultaneous solutions to the constraints of four cuts. Since the loop
momentum is four dimensional this would fix it completely, reducing
the problem of finding the coefficient of a box function to a purely
algebraic one. This is particularly useful for calculating $\N=4$
amplitudes which only contain box integrals. 

\subsection{Basic idea in $\N=4$}

Consider first a one-loop amplitude and choose four internal lines to
take on-shell. There is, of course, a one-to-one correspondence
between the boxes and the sets of four internal lines, so there is no
double counting issue. The chosen box with coefficient $c$ is
\begin{eqnarray}
  &&c\int \frac{d^4L}{(2\pi)^4}\frac1{[L^2+i\epsilon]
[(L+K_1)^2+i\epsilon][(L+K_1+K_2)^2+i\epsilon]
[(L-K_4)^2+i\epsilon]}\nn
&&
\end{eqnarray}
which is cut to
\begin{equation}
c\sum_{\mathrm{soln.}}\int d^4L\,\delta^{(+)}[L^2]\,\delta^{(+)}[(L+K_1)^2]\,
\delta^{(+)}[(L+K_2)^2]\,\delta^{(+)}[(L-K_4)^2].
\end{equation}
The constraints have two solutions which give rise to the same
Jacobian for the integral. The Jacobian is inversely proportional to
the Gram determinant but will otherwise not concern us. The result of
the cutting is now
\begin{equation}
2c\,\mathrm{Jac}(K_1,K_2,K_4).
\end{equation}
We can do the same considerations for each Feynman diagram in the
calculation of the one-loop amplitude. Adding up contributions from
all diagrams that cut our four chosen internal lines will give
\begin{eqnarray}
&&\int
  d^4L\sum_{\mathrm{soln.,states}}\A(\ldots,-L,L+K_1)
\A(\ldots,-L-K_1,L+K_1+K_2)\nn
&&\qquad\times\A(\ldots,-L-K_1-K_2,L-K_4)\A(\ldots,-L+K_4,L)\nn
&&\qquad\times\delta^{(+)}[L^2]\,\delta^{(+)}[(L+K_1)^2]\,
\delta^{(+)}[(L+K_2)^2]\,\delta^{(+)}[(L-K_4)^2],
\end{eqnarray}
where the summation over states refers to the possibility that
different particles may have those momenta, and that we must sum over
all possibilities. In the same way as above, this gives
\begin{equation}
\bigg(\sum_{\mathrm{soln.,states}}\A_A\A_B\A_C\A_D\bigg)
\mathrm{Jac}(K_1,K_2,K_4),
\end{equation}
or,
\begin{equation}
c=\frac12\sum_{\mathrm{soln.,states}}\A_A\A_B\A_C\A_D.
\end{equation}
In other words, the coefficient is given by products of four on-shell
tree amplitudes in four dimensions. The quadruple cut can be depicted
as a box with four corner tree amplitudes connected by the (implicitly
cut) loop propagators,
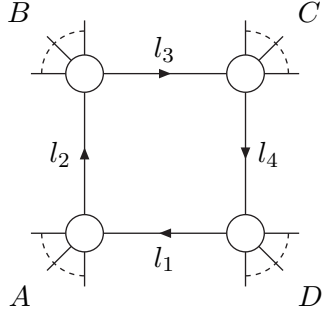
\begin{center}
\begin{picture}(140,140)(0,0)
\ArrowLine(40,40)(40,100)
\ArrowLine(40,100)(100,100)
\ArrowLine(100,100)(100,40)
\ArrowLine(100,40)(40,40)
\Text(70,35)[t]{$l_1$}
\Text(35,70)[r]{$l_2$}
\Text(70,105)[b]{$l_3$}
\Text(105,70)[l]{$l_4$}
\Line(40,40)(40,20)
\Line(40,40)(26,26)
\Line(40,40)(20,40)
\Text(20,20)[tr]{$A$}
\DashCArc(40,40)(16,180,270){2}
\BCirc(40,40){7}
\Line(40,100)(20,100)
\Line(40,100)(26,114)
\Line(40,100)(40,120)
\Text(20,120)[br]{$B$}
\DashCArc(40,100)(16,90,180){2}
\BCirc(40,100){7}
\Line(100,100)(100,120)
\Line(100,100)(114,114)
\Line(100,100)(120,100)
\Text(120,120)[bl]{$C$}
\DashCArc(100,100)(16,0,90){2}
\BCirc(100,100){7}
\Line(100,40)(120,40)
\Line(100,40)(114,26)
\Line(100,40)(100,20)
\Text(120,20)[tl]{$D$}
\DashCArc(100,40)(16,270,0){2}
\BCirc(100,40){7}
\end{picture}
\end{center}
In concrete examples the internal momenta have helicity labels if there
is only one helicity configuration which gives non-vanishing
contributions.

\subsection{A Simple Example}

As a first example, we may try to calculate the one-loop $(--++)$
amplitude. This has only one quadruple cut, namely
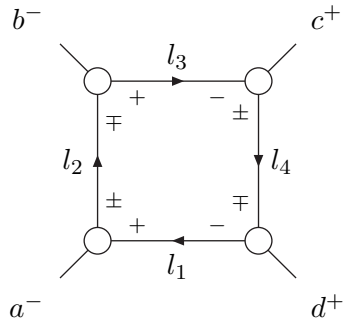
\begin{center}
\begin{picture}(140,140)(0,0)
\ArrowLine(40,40)(40,100)
\ArrowLine(40,100)(100,100)
\ArrowLine(100,100)(100,40)
\ArrowLine(100,40)(40,40)
\Text(70,35)[t]{$l_1$}
\Text(35,70)[r]{$l_2$}
\Text(70,105)[b]{$l_3$}
\Text(105,70)[l]{$l_4$}
\Text(85,43)[b]{\scriptsize $-$}
\Text(55,43)[b]{\scriptsize $+$}
\Text(43,55)[l]{\scriptsize $\pm$}
\Text(43,85)[l]{\scriptsize $\mp$}
\Text(55,97)[t]{\scriptsize $+$}
\Text(85,97)[t]{\scriptsize $-$}
\Text(97,88)[r]{\scriptsize $\pm$}
\Text(97,55)[r]{\scriptsize $\mp$}
\Line(40,40)(26,26)
\Text(20,20)[tr]{$a^-$}
\BCirc(40,40){5}
\Line(40,100)(26,114)
\Text(20,120)[br]{$b^-$}
\BCirc(40,100){5}
\Line(100,100)(114,114)
\Text(120,120)[bl]{$c^+$}
\BCirc(100,100){5}
\Line(100,40)(114,26)
\Text(120,20)[tl]{$d^+$}
\BCirc(100,40){5}
\end{picture}
\end{center}
There are two solutions for the internal momenta $l_1$ through
$l_4$. The first is
\begin{eqnarray}
\label{eq:0msol}
&&l_1=\spxa d\frac{\spa ba}{\spa bd}\spbx a,\qquad
l_2=\spxa b\frac{\spa da}{\spa bd}\spbx a,\nn
&&l_3=\spxa b\frac{\spa dc}{\spa db}\spbx c,\qquad
l_4=\spxa d\frac{\spa bc}{\spa db}\spbx c,
\end{eqnarray}
with the consequence that MHV three-point amplitudes at corners $B$
and $D$ are zero while googly-MHV three-point amplitudes at corners
$A$ and $C$ are zero. Thus, the only non-vanishing assignment of
internal states is
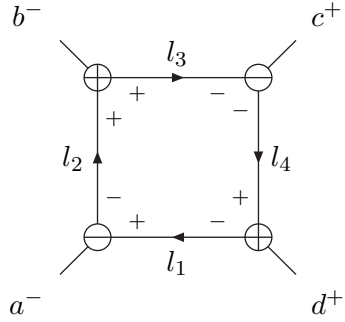
\begin{center}
\begin{picture}(140,140)(0,0)
\ArrowLine(40,40)(40,100)
\ArrowLine(40,100)(100,100)
\ArrowLine(100,100)(100,40)
\ArrowLine(100,40)(40,40)
\Text(70,35)[t]{$l_1$}
\Text(35,70)[r]{$l_2$}
\Text(70,105)[b]{$l_3$}
\Text(105,70)[l]{$l_4$}
\Text(85,43)[b]{\scriptsize $-$}
\Text(55,43)[b]{\scriptsize $+$}
\Text(43,55)[l]{\scriptsize $-$}
\Text(43,85)[l]{\scriptsize $+$}
\Text(55,97)[t]{\scriptsize $+$}
\Text(85,97)[t]{\scriptsize $-$}
\Text(97,88)[r]{\scriptsize $-$}
\Text(97,55)[r]{\scriptsize $+$}
\Line(40,40)(26,26)
\Text(20,20)[tr]{$a^-$}
\BCirc(40,40){5}
\Line(35,40)(45,40)
\Line(40,100)(26,114)
\Text(20,120)[br]{$b^-$}
\BCirc(40,100){5}
\Line(35,100)(45,100)
\Line(40,95)(40,105)
\Line(100,100)(114,114)
\Text(120,120)[bl]{$c^+$}
\BCirc(100,100){5}
\Line(95,100)(105,100)
\Line(100,40)(114,26)
\Text(120,20)[tl]{$d^+$}
\BCirc(100,40){5}
\Line(95,40)(105,40)
\Line(100,35)(100,45)
\end{picture}
\end{center}
giving the contribution\footnote{Remember that $\spxa{(-k)}=\spxa k$ and
  $\spxb {(-k)}=-\spxb k$.}
\begin{eqnarray}
&&\frac12\frac{\spa a{l_2}^3}{\spa {l_2}{(-l_1)}\spa{(-l_1)}a}
\frac{\spb {l_3}{(-l_2)}^3}{\spb{(-l_2)}b\spb b{l_3}}
\frac{\spa {l_4}{(-l_3)}^3}{\spa {(-l_3)}c\spa c{l_4}}
\frac{\spb {(-l_4)}d^3}{\spb d{l_1}\spb {l_1}{(-l_4)}}\nn
&=& \frac12\frac{\spa a{l_2}^3}{\spa a{l_1}\spa{l_1}{l_2}}
\frac{\spb {l_2}{l_3}^3}{\spb{l_2}b\spb b{l_3}}
\frac{\spa {l_3}{l_4}^3}{\spa {l_3}c\spa c{l_4}}
\frac{\spb {l_4}d^3}{\spb {l_4}{l_1}\spb {l_1}d}\nn
&=&\frac12\frac{\spa da^2\spa dc^2\spa bc^2}{\spa bd^6}
\frac{\spa ab^3}{\spa a{l_1}\spa {l_1}b}
\frac{\spb ac^3}{\spb ab\spb bc}
\frac{\spa bd^3}{\spa bc\spa bd}
\frac{\spb cd^3}{\spb c{l_1}\spb {l_1}d}\nn
&=&\frac12\frac{\spa da^2\spa dc^2\spa bc^2}{\spa bd^4\spa ab^2}
\frac{\spa ab^3}{\spa ad\spa db}
\frac{\spb ac^3}{\spb ab\spb bc}
\frac{\spa bd^3}{\spa bc\spa bd}
\frac{\spb cd^3}{\spb ca\spb ad}\nn
&=&-\frac12st\A^{\mathrm{tree}},
\end{eqnarray}
where $s$ and $t$ are the normal Mandelstam variables. The other
solution for the internal momenta comes about by 'flipping'
$\spxa\cdot\leftrightarrow\spxb\cdot$ in (\ref{eq:0msol}) and
appropriately reversing the conclusions about the helicity assignments
to get 
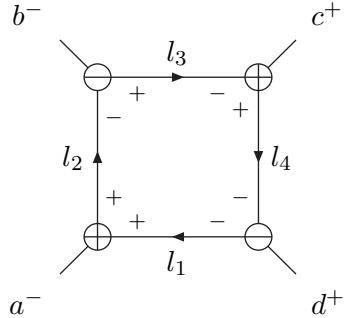
\begin{center}
\begin{picture}(140,140)(0,0)
\ArrowLine(40,40)(40,100)
\ArrowLine(40,100)(100,100)
\ArrowLine(100,100)(100,40)
\ArrowLine(100,40)(40,40)
\Text(70,35)[t]{$l_1$}
\Text(35,70)[r]{$l_2$}
\Text(70,105)[b]{$l_3$}
\Text(105,70)[l]{$l_4$}
\Text(85,43)[b]{\scriptsize $-$}
\Text(55,43)[b]{\scriptsize $+$}
\Text(43,55)[l]{\scriptsize $+$}
\Text(43,85)[l]{\scriptsize $-$}
\Text(55,97)[t]{\scriptsize $+$}
\Text(85,97)[t]{\scriptsize $-$}
\Text(97,88)[r]{\scriptsize $+$}
\Text(97,55)[r]{\scriptsize $-$}
\Line(40,40)(26,26)
\Text(20,20)[tr]{$a^-$}
\BCirc(40,40){5}
\Line(35,40)(45,40)
\Line(40,35)(40,45)
\Line(40,100)(26,114)
\Text(20,120)[br]{$b^-$}
\BCirc(40,100){5}
\Line(35,100)(45,100)
\Line(100,100)(114,114)
\Text(120,120)[bl]{$c^+$}
\BCirc(100,100){5}
\Line(95,100)(105,100)
\Line(100,95)(100,105)
\Line(100,40)(114,26)
\Text(120,20)[tl]{$d^+$}
\BCirc(100,40){5}
\Line(95,40)(105,40)
\end{picture}
\end{center}
Notice how the symbols
\begin{picture}(10,10)
\BCirc(5,5){5}
\Line(0,5)(10,5)
\end{picture}
and 
\begin{picture}(10,10)
\BCirc(5,5){5}
\Line(0,5)(10,5)
\Line(5,0)(5,10)
\end{picture}
are used for three-point MHV and googly-MHV corner amplitudes,
respectively.  Since $\A^{\mathrm{tree}}$ is invariant under flipping
we get the same result as before. The one-loop amplitude then becomes
\begin{equation}
\A^{\mathrm{1-loop}}=-st\A^{\mathrm{tree}}I_4(a,b,c)
=2\A^{\mathrm{tree}}F_4(a,b,c).
\end{equation}
In practice, a calculation will be a good deal more involved as it may
include more complicated corner amplitudes, especially non-MHV
amplitudes, and summing over the full $\N=4$ multiplet possibly
running in the loop. Such an example (pieces of which were calculated
in the original article on quadruple cuts \cite{Britto04quad}) is
presented in the next chapter.

To aid our calculations in the remainder of this thesis, the solutions
of the internal constraints have been assembled in Appendix
\ref{app:quad}.

\subsection{Box Coefficients in Other Theories} 

The quadruple cut method extends smoothly to other theories, but
without being ``complete'' in the sense that the entire amplitude can
be found from that method. What it does calculate is the contribution
to the amplitude from box functions, since the quadruple cut singles
out the boxes. When calculating a general Yang--Mills one-loop
amplitude, the contributions to boxes from the $\N=1$ chiral multiplet
running in the loop can be calculated just by adjusting the internal
states. As is done in chapter \ref{cha:higgs}, the method may also be
used where some (massive) external states do not run in the
loop. Another extension is to gravity theories such as $\N=8$
supergravity, where it is possible in general to use the KLT relations
between tree amplitudes to construct KLT-like relations between the
box coefficients. The use of quadruple cuts in gravity will be
considered in more detail in chapter \ref{cha:sugra}.

\subsection{Generalized Unitarity in Complexified Minkowski Space}

As briefly mentioned in section \ref{sec:genuni}, taking quadruple
cuts was not a genuine new idea in itself, rather it was the use of
internal momenta in complexified Minkowski space. Thus, the method
also allows for wider use of the triple cut. This leads to different
methods for obtaining the cut-containing parts of amplitudes
\cite{Forde07coefficients,BjerrumBohr07threemassrecursion, Ossola06,
Ellis07}.

Also, both triple and quadruple cuts can be done with massive
particles or, equivalently (\emph{cf.}~section
\ref{sec:ddimunitarity}) with massless particles exactly in
$4-2\epsilon$ dimensions to yield also the rational parts of
amplitudes \cite{Brandhuber05massivequad,Mastrolia06triplecuts}

The idea of generalized unitarity can be extended to higher loop
orders, where cutting a large portion of the propagators can reveal
information as to the structure of both the momentum-dependent
numerator and any other coefficients, and can likewise provide useful
checks of higher loop unitarity cut calculation where a more
conventional generalized unitarity approach is taken. An amusing case
\cite{Buchbinder05twoloop}, which unfortunately does not seem to
generalize, is $\N=4$ two-loop integral coefficients up to and
including six external gluons where completely cut integrals such as
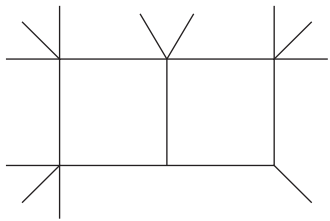
\begin{center}
\begin{picture}(120,80)
\EBox(20,20)(100,60)
\Line(60,20)(60,60)
\Line(20,20)(20,0)
\Line(20,20)(6,6)
\Line(20,20)(0,20)
\Line(20,60)(0,60)
\Line(20,60)(6,74)
\Line(20,60)(20,80)
\Line(60,60)(50,77)
\Line(60,60)(70,77)
\Line(100,60)(100,80)
\Line(100,60)(114,74)
\Line(100,60)(120,60)
\Line(100,20)(114,6)
\end{picture}
\end{center}
have an additional propagator-like singularity coming from the
Jacobian of the right box. This singularity can again be cut providing
a total of eight constraints which fix the integration completely and
gives the coefficient as a product of tree amplitudes. This, however,
requires that there is a box which is two-mass hard or one-mass;
otherwise the Jacobian is not propagator-like.

Recently, Cachazo and Skinner has shed some additional light on the
use of cutting procedures at higher loops \cite{Cachazo08cutting}.

\section{Recursion at Loop Level}
\label{sec:looprecursion}

Up till now, this chapter has been concerned with deriving one-loop
amplitudes from tree amplitudes. In that perspective it should come as
no surprise that recursion can also play a role here. Indeed it does
play a role in determining coefficients of the integral functions
\cite{Bern05coefficientrecursion,BjerrumBohr07threemassrecursion},
although it does not seem to be as strong a tool as at tree level.

The problem of determining integral coefficients is, however, not our
main objective in this section. Rather, it is the calculation of the
rational parts of non-supersymmetric one-loop amplitudes directly in
four dimensions without appealing to $D$-dimensional (generalized)
unitarity. The methods described here were developed by Bern, Dixon
and Kosower, first for amplitudes with no cut-containing parts
\cite{Bern05looprecursion1,Bern05lastofthefinite} and later for
amplitudes with cut-containing parts
\cite{Bern05looprecursion2}. Since this section is more intended as
background material for chapter \ref{cha:higgs} than a review of
advanced topics, the treatment will be adapted to that case and other
complications will only be mentioned briefly.

As in the tree case, on-shell recursion starts out by deforming two
external momenta with a complex parameter $z$ and writing the $z=0$
result as an integral along a closed contour enclosing the origin,
\begin{equation}
\A^{\mathrm{1-loop}}=\frac1{2\pi i}\oint_{\mathrm{around~0}}
\frac{dz}z\widehat\A^{\mathrm{1-loop}}(z).
\end{equation}
If $\widehat\A^{\mathrm{1-loop}}(z)\to 0$ as $z\to\infty$, we can
write this formally as a sum over integrations around all
singularities of $\widehat\A^{\mathrm{1-loop}}(z)$, which may be both
poles and branch cuts,
\begin{eqnarray}
\A^{\mathrm{1-loop}}&=&-\frac1{2\pi i}\sum_{\mathrm{singularities}~i}
\oint_{\mathrm{around}~i}
\frac{dz}z\widehat\A^{\mathrm{1-loop}}(z)\nn
&=&-\sum_{\mathrm{poles}~i}\frac{\mathrm{Res}_i\widehat
\A^{\mathrm{1-loop}}(z)}{z_i}-\int_B\frac{dz}z\mathrm{Disc}_B
\A^{\mathrm{1-loop}}(z)
\end{eqnarray}
where $B$ is a branch cut and $\mathrm{Disc}_B$ is the discontinuity
across it. In the first term, the residues are given by the
appropriate multiparticle or collinear factorization and depend on
lower-point one-loop amplitudes or splitting functions. Since we
assume that the cut-containing parts of the amplitude can be
calculated by other methods described in this chapter, it makes sense
to split all one-loop amplitudes and one-loop splitting functions into
their cut-containing\footnote{Rational terms proportional to $\pi^2$
are included with the cut-containing part.} and rational parts. Then
it can be argued that cut-containing parts can be deduced from the
cut-containing parts of lower-point one-loop amplitudes and splitting
functions, and that rational parts can be deduced from the rational
parts of them.

This removes the complication of cuts in the amplitude, but
introduces a new one. When we make the split of an amplitude into
cut-containing and rational parts
\begin{equation}
\label{eq:cutratsplit}
\A^{\mathrm{1-loop}}=C+R,
\end{equation}
we cannot trust that $C$ and $R$ individually only have
physical (\emph{i.e.}~multiparticle and collinear) singularities. In
fact, the cut containing parts will often contain terms like
\begin{equation}
\frac{\log(s/t)}{(s-t)^n},
\end{equation}
which do not correspond to physical singularities of
$\A^{\mathrm{1-loop}}$. To handle this, we add and subtract a rational
\emph{cut-completion} term $CR$,
\begin{equation}
\A^{\mathrm{1-loop}}=(C+CR)+(R-CR)
\end{equation}
which has (minus) the same unphysical singularities as $C$. The
singularities of $R-CR$ are poles (single or double) in kinematic
invariants, so we can do the usual on-shell recursion trick (again,
assuming $R(z)-CR(z)\to 0$ as $z\to\infty$),
\begin{equation}
R-CR=-\sum_{\mathrm{poles~}i}\frac{\mathrm{Res}_i(R(z)-CR(z))}
{z_i},
\end{equation}
where the poles are those that arise from known singularities of
$\A^{\mathrm{1-loop}}$. The terms coming from $R$ and $CR$ are known as the
direct recursive $DR$ and overlap $O$ terms, respectively. The latter can be
computed directly from $CR(z)$ while the former is more intricate.

If the one-loop splitting functions had only single poles, we could
interpret them---in the same way as tree splitting functions---as
recursion diagrams of the type
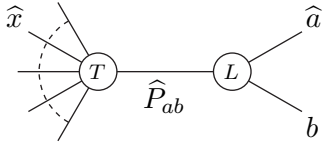
\begin{center}
\begin{picture}(150,100)(0,0)
\Line(50,50)(100,50)
\Text(75,47)[t]{$\widehat P_{ab}$}
\Line(50,50)(35,24)
\Line(50,50)(24,35)
\Line(50,50)(20,50)
\Line(50,50)(24,65)
\Line(50,50)(35,76)
\Text(35,22)[t]{$$}
\Text(22,33)[rt]{$$}
\Text(18,50)[r]{$$}
\Text(22,67)[rb]{$\widehat x$}
\Text(35,78)[b]{$$}
\DashCArc(50,50)(22,120,240){2}
\BCirc(50,50){7}
\Text(50,50)[]{\scriptsize $T$}
\Line(100,50)(126,65)
\Line(100,50)(126,35)
\Text(128,67)[lb]{$\widehat a$}
\Text(128,33)[lt]{$b$}
\BCirc(100,50){7}
\Text(100,50)[]{\scriptsize $L$}
\end{picture}
\end{center}
Drawing it like that, we can now use the rational parts of the
splitting functions (\ref{eq:fullsplit}) for a gluon it the loop,
\begin{eqnarray}
\label{eq:rationalsplit}
\mathrm{Split}^{\mathrm{1-loop,R}}_\pm(x,a^+,b^-)&=&0,\nn
\mathrm{Split}^{\mathrm{1-loop,R}}_+(x,a^+,b^+)&=&
-\frac{c_\Gamma}3\sqrt{x(1-x)}\frac{\spb ab}{\spa ab^2},\nn
\mathrm{Split}^{\mathrm{1-loop,R}}_-(x,a^+,b^+)&=&
\frac{c_\Gamma}3\frac{\sqrt{x(1-x)}}{\spa ab}
\end{eqnarray}
to say something about the contribution of such a diagram. Firstly, if
$a$ and $b$ have opposite helicities, there is no contribution. This
is also the case if all three helicities sitting on the three-point
``vertex'' and $a$ had its antiholomorphic spinor shifted, since the
requirement of $\hat a$ and $b$ being collinear is $\spb{\hat a}b=0$;
conversely, if $a$ had its holomorphic spinor shifted the all-minus
splitting function would give zero.

If the holomorphic spinor of $a$ is shifted and we have to deal with
an all-plus splitting amplitude, we immediately see that it has a
double pole in $z$. This poses a serious problem because finding the
residue requires us to seperate the pure double pole from any single
pole which may be hiding underneath it. However, Bern, Dixon and
Kosower found a way to circumvent this by introducing a slightly
different three-point vertex which seems to do the job in
certain cases. Lastly, the complex factorization of the last splitting
amplitude above is quite murky, and it should generally be avoided
\cite{Bern05looprecursion1}. 

Keeping these matters in mind, we can obtain the direct recursive terms
as 
\begin{eqnarray}
DR&=&\sum_{i,h=\pm}\bigg(\frac{\A^{\mathrm{tree}}(\ldots,-P_i^h)
R(P_i^{-h},\ldots)}{P_i^2}+\frac{R(\ldots,-P_i^h)\A^{\mathrm{tree}}
(P_i^{-h},\ldots)}{P_i^2}\nn
&&\qquad+\mathcal F_R\frac{\A^{\mathrm{tree}}(\ldots,-P_i^h)
\A^{\mathrm{tree}}(P_i^{-h},\ldots)}{P_i^2}\bigg)
\end{eqnarray}
where $i$ runs over all channels affected by the shift, $R$ are the
rational parts of lower-point amplitudes or properly defined
three-point loop vertices and $\mathcal F_R$ is the rational part of
the factorization function. Summing up, the rational part of the
amplitude we are after is
\begin{equation}
R=DR+CR-O.
\end{equation}

A last complication is that of the $z\to\infty$ behaviour of the
amplitudes. It is not always possible to find a shift which is both
zero as $z\to\infty$ and avoids problematic splitting
functions. However, it has been shown \cite{Berger06generalbootstrap}
that all problems can be circumvented with imaginative combinations of
shifts. This means that any one-loop amplitude in massless QCD is in
principle obtainable by the combination of unitarity methods for the
cut-constructible parts and recursion for the rational part. A
computer program under the name of \emph{BlackHat} which performs such
calculations is under development \cite{Berger08blackhat}. In chapter
\ref{cha:higgs} we will turn to a complete calculation of a particular
one-loop amplitude in QCD with an effective coupling to the Higgs
particle.

\chapter{One-Loop Amplitudes in $\N=4$ Super-Yang--Mills}
\label{cha:swi}

In this chapter we take a closer look at the calculation of one-loop
amplitudes in $\N=4$ Yang--Mills. As described in section
\ref{sec:quadcuts}, these amplitudes can be written as
linear combinations of scalar Feynman integrals, where the coefficients
are essentially products of four tree amplitudes. Here we will
calculate some concrete one-loop amplitudes as a demonstration of the
technique. The first calculation, of the one-loop MHV amplitude, is
primarily included for demonstrational purposes. The part of the
calculation of the one-loop NMHV amplitude where scalars and fermions
in external states are taken into account is based on
\cite{Risager05swi}.

\section{MHV Constructibility}
\label{sec:mhvconstructibility}

Although the quadruple cut method allows for the efficient numerical
evaluation of one-loop amplitudes from tree amplitudes, there are many
analytic simplifications to exploit. The most important of these is
the combination with the simple all-$n$ expression for MHV tree
amplitudes. For certain configurations of external helicities, and for
some box coefficients, the
required tree amplitudes are all either MHV or googly-MHV, leading to
easily obtainable all-$n$ expressions. A box coefficient with only MHV
and googly-MHV corners is called \emph{MHV constructible}.

The simplest class of MHV constructible box coefficients are those of
the MHV one-loop amplitude. They were first computed by the unitarity
method of section \ref{sec:unitarity} by Bern, Dixon and Kosower
\cite{Bern94oneloopn4mhv}, but we will review the calculation from the
quadruple cut perspective here for the sake of familiarity with the
method.

As in any quadruple cut calculation, we must start by determining
which box functions contribute. This can often be reduced to a matter
of counting pluses and minuses: The MHV amplitude has two external
minuses. Inside the cut box there are four internal lines with one
minus each. Thus, the four tree-level corner amplitudes must have a
total of six minuses. The only way these can be distributed without
giving zero amplitudes is to have two MHV corners and two
googly-MHV three-point corners, and the latter must be at opposite
corners. In other words, we can only have 2-mass easy boxes where the
massless corners are googly. Of course, this argument only holds for
gluons in the internal and external states, but the conclusions hold
in general because all non-zero amplitudes are related to the
gluons-only amplitude with the same sum of helicities.
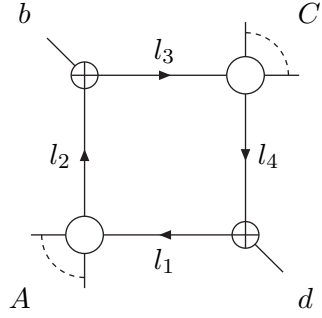
\begin{center}
\begin{picture}(140,140)(0,0)
\ArrowLine(40,40)(40,100)
\ArrowLine(40,100)(100,100)
\ArrowLine(100,100)(100,40)
\ArrowLine(100,40)(40,40)
\Text(70,35)[t]{$l_1$}
\Text(35,70)[r]{$l_2$}
\Text(70,105)[b]{$l_3$}
\Text(105,70)[l]{$l_4$}
\Line(40,40)(40,20)
\Line(40,40)(20,40)
\Text(20,20)[tr]{$A$}
\DashCArc(40,40)(16,180,270){2}
\BCirc(40,40){7}
\Line(40,100)(26,114)
\Text(20,120)[br]{$b$}
\BCirc(40,100){5}
\Line(35,100)(45,100)
\Line(40,95)(40,105)
\Line(100,100)(100,120)
\Line(100,100)(120,100)
\Text(120,120)[bl]{$C$}
\DashCArc(100,100)(16,0,90){2}
\BCirc(100,100){7}
\Line(100,40)(114,26)
\Text(120,20)[tl]{$d$}
\BCirc(100,40){5}
\Line(95,40)(105,40)
\Line(100,35)(100,45)
\end{picture}
\end{center}

Rather than now classifying the positions of the two negative helicity
external legs, we can start out by using the simple form of the MHV
amplitude to write down part of the coefficients, namely the part that
depends on the denominator of (\ref{eq:parketaylor}),
\begin{eqnarray}
&&\frac12\frac1{\spa{(-l_1)}{d+1}\langle d+1\cdots b-1\rangle
\spa{b-1}{l_2}\spa{l_2}{(-l_1)}}\frac1{\spb{(-l_2)}b\spb b{l_3}
\spb{l_3}{(-l_2)}}\nn
&&\times\frac1{\spa{(-l_3)}{b+1}\langle b+1\cdots d-1\rangle
\spa {d-1}{l_4}\spa{l_4}{(-l_3)}}\frac1{\spb{(-l_4)}d\spb d {l_1}
\spb{l_1}{(-l_4)}}\nn
&=&\frac1{2\langle\cdots\rangle}\frac{\spa d{d+1}}{\spb d{l_1}
\spa{l_1}{d+1}}\frac{\spa {b-1}b}{\spa {b-1}{l_2}\spb {l_2}b}
\frac{\spa b{b+1}}{\spb b{l_3}\spa{l_3}{b+1}}
\frac{\spa {d-1}d}{\spa {d-1}{l_4}\spb {l_4}d}\nn
&&\times \frac1{\spa{l_1}{l_2}\spb{l_2}{l_3}\spa{l_3}{l_4}\spb
{l_4}{l_1}}.
\end{eqnarray}
Using the solutions of the internal constraints
(\ref{eq:2meplussoln}), we can write this as
\begin{eqnarray}
&&\frac1{2\langle\cdots\rangle}\frac{\spa db}{\spba dAb}\frac{\spa bd
}{\spab dAb}\frac{\spa bd}{\spba bCd}\frac{\spa db}{\spab bCd}
\frac{\spa bd^4}{\spa db\spaa dACd\spa bd\spaa bCAb}\nn
&=&\frac{\spab bCd\spab dAb}{2\langle\cdots\rangle}
\bigg(\frac{\spa bd}{\spab bCd\spab dAb}\bigg)^4.\label{eq:loopmhvdenom}
\end{eqnarray}
Notice that
\begin{eqnarray}
&&\spab bCd\spab dAb\nn
&=&-\langle bC(A+b+C)Ab]\nn
&=&-A^2\spab bCb-C^2\spab bAb-\spab bAb\spab bCb\nn
&=&A^2(C^2-(b+C)^2)+C^2(A^2-(A+b)^2)\nn
&&\qquad-(A^2-(A+b)^2)(C^2-(C+b)^2)\nn
&=&-(A+b)^2(C+b)^2+A^2C^2\label{eq:2megram}
\end{eqnarray}
is twice the Gram determinant. Thus, the denominator consists of the
Gram determinant, the conventional $\langle\cdots\rangle$ spinor
product, and a remaining term to the fourth power.

The term coming from the numerator depends on the position of the two
external negative helicity legs. These can be placed in four distinct
ways, but we will content ourselves with the calculation of two of
them. The first case is when the two negative helicity legs are on the
same MHV corner, say corner $A$, 
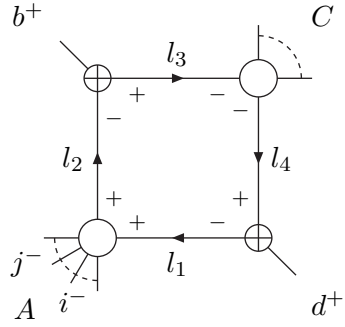
\begin{center}
\begin{picture}(140,140)(0,0)
\ArrowLine(40,40)(40,100)
\ArrowLine(40,100)(100,100)
\ArrowLine(100,100)(100,40)
\ArrowLine(100,40)(40,40)
\Text(70,35)[t]{$l_1$}
\Text(35,70)[r]{$l_2$}
\Text(70,105)[b]{$l_3$}
\Text(105,70)[l]{$l_4$}
\Text(85,43)[b]{\scriptsize $-$}
\Text(55,43)[b]{\scriptsize $+$}
\Text(43,55)[l]{\scriptsize $+$}
\Text(43,85)[l]{\scriptsize $-$}
\Text(55,97)[t]{\scriptsize $+$}
\Text(85,97)[t]{\scriptsize $-$}
\Text(97,88)[r]{\scriptsize $-$}
\Text(97,55)[r]{\scriptsize $+$}
\Line(40,40)(40,20)
\Line(40,40)(30,23)
\Line(40,40)(23,30)
\Line(40,40)(20,40)
\Text(17,17)[tr]{$A$}
\Text(31,21)[t]{$i^-$}
\Text(20,31)[r]{$j^-$}
\DashCArc(40,40)(16,180,270){2}
\BCirc(40,40){7}
\Line(40,100)(26,114)
\Text(20,120)[br]{$b^+$}
\BCirc(40,100){5}
\Line(35,100)(45,100)
\Line(40,95)(40,105)
\Line(100,100)(100,120)
\Line(100,100)(120,100)
\Text(120,120)[bl]{$C$}
\DashCArc(100,100)(16,0,90){2}
\BCirc(100,100){7}
\Line(100,40)(114,26)
\Text(120,20)[tl]{$d^+$}
\BCirc(100,40){5}
\Line(95,40)(105,40)
\Line(100,35)(100,45)
\end{picture}
\end{center}
If we call them $i$ and $j$,
respectively, the numerator becomes
\begin{eqnarray}
&&\spa ij^4\spb b{l_3}^4\spa {(-l_3)}{l_4}^4\spb {(-l_4)}d^4\nn
&=&\spa ij^4\bigg(\frac{\spba bCd\spa bd\spab bCd}
{\spa bd^2}\bigg)^4\nn
&=&\spa ij^4\bigg(\frac{\spab bCd\spab dAb}{\spa bd}\bigg)^4.
\end{eqnarray}
We see that the paranthesis cancels with the one in
(\ref{eq:loopmhvdenom}) and leaves only the Gram determinant times the
tree amplitude. 

We now turn to the case where $i$ and $j$ are on corners $A$ and $C$,
respectively. In this case, the whole $\N=4$ multiplet can run in the
loop which we draw as an extra internal line, 
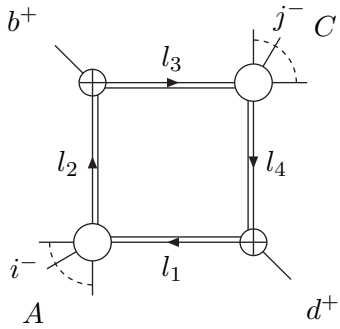
\begin{center}
\begin{picture}(140,140)(0,0)
\ArrowLine(40,40)(40,100)
\ArrowLine(40,100)(100,100)
\ArrowLine(100,100)(100,40)
\ArrowLine(100,40)(40,40)
\Text(70,35)[t]{$l_1$}
\Text(35,70)[r]{$l_2$}
\Text(70,105)[b]{$l_3$}
\Text(105,70)[l]{$l_4$}
\EBox(42,42)(98,98)
\Line(40,40)(40,20)
\Line(40,40)(23,30)
\Line(40,40)(20,40)
\Text(22,17)[tr]{$A$}
\Text(20,31)[r]{$i^-$}
\DashCArc(40,40)(16,180,270){2}
\BCirc(40,40){7}
\Line(40,100)(26,114)
\Text(20,120)[br]{$b^+$}
\BCirc(40,100){5}
\Line(35,100)(45,100)
\Line(40,95)(40,105)
\Line(100,100)(100,120)
\Line(100,100)(110,117)
\Line(100,100)(120,100)
\Text(123,118)[bl]{$C$}
\Text(114,120)[b]{$j^-$}
\DashCArc(100,100)(16,0,90){2}
\BCirc(100,100){7}
\Line(100,40)(114,26)
\Text(120,20)[tl]{$d^+$}
\BCirc(100,40){5}
\Line(95,40)(105,40)
\Line(100,35)(100,45)
\end{picture}
\end{center}
The numerator contribution from the gluons can easily be seen to
be
\begin{equation}
\spa i{l_2}^4\spb {(-l_2)}b^4\spa j{l_4}^4\spb{(-l_4)}d^4
+\spa {(-l_1)}i^4\spb b{l_3}^4\spa {(-l_3)}j^4\spb d{l_1}^4.
\end{equation}
The contribution of the fermions can be computed by noting that the
amplitude $A^{\mhv}(g^-,f^-,f^+,\ldots)$ is simply related to $A^\mhv
(g^-,g^-,\ldots)$, namely by the exchange of one of the numerator
factors with another. This yields
\begin{eqnarray}
&&-4\Big(\spa i{l_2}^3\spa i{(-l_1)}\,\spb {(-l_2)}b^3\spb {l_3}b\,
\spa j{l_4}^3\spa j{(-l_3)}\,\spb {(-l_4)}d^3\spb {l_1}d\nn
&&+\spa {l_2}i\spa {(-l_1)}i^3\,\spb b{(-l_2)}\spb b{l_3}^3\,
\spa {l_4}j\spa {(-l_3)}j^3\,\spb d{(-l_4)}\spb d{l_1}^3\Big),
\end{eqnarray}
where the prefactor takes into account that there are four species and
that they must have fermionic sign. The scalar contribution becomes
\begin{eqnarray}
&&6\,\spa i{l_2}^2\spa i{(-l_1)}^2\,\spb {(-l_2)}b^2\spb {l_3}b^2\,
\spa j{l_4}^2\spa j{(-l_3)}^2\,\spb {(-l_4)}d^2\spb {l_1}d^2.
\end{eqnarray}
When we add these three expressions together, we see a common
situation in $\N=4$ SYM: the particle multiplicities make sure that
the result is simplified, in this case by making it writable as a
fourth power of some term:
\begin{eqnarray}
&&\Big(\spa i{l_2}\spb {(-l_2)}b\spa j{l_4}\spb{(-l_4)}d
-\spa {(-l_1)}i\spb b{l_3}\spa {(-l_3)}j\spb d{l_1}\Big)^4\nn
&=&\Big(\spa i{l_3}\spb {l_3}b\spa j{l_1}\spb {l_1}d-
\spa i{l_1}\spb {l_1}d\spa j{l_3}\spb {l_3}b\Big)^4\nn
&=&\spa ij^4\Big(\spa {l_3}{l_1}\spb {l_3}b\spb{l_1}d\Big)^4\nn
&=&\spa ij^4\bigg(\frac{\spba bCd\spa bd\spab bAd}{\spa bd^2}
\bigg)^4\nn
&=&\spa ij^4\bigg(\frac{\spab bCd\spab dAb}{\spa bd}\bigg)^4.
\end{eqnarray}
In words, the particle states of $\N=4$ running in the loop conspire
to make the numerator factor the same as before. It is not difficult
to imagine (and true) that all four positionings of the negative
helicity legs $i$ and $j$ give the same numerator factor.

We can now write down the complete one-loop MHV amplitude in $\N=4$
SYM. It is
\begin{equation}
A^{\mathrm{1-loop}}_\mhv=c_\Gamma A^{\mathrm{tree}}_\mhv
\sum_{i=1}^n\sum_{j=i+2}^{\min(n,i-2+n)} F_4(i,P_{i+1,j-1},j).
\end{equation}

This was the result for gluons in the external states, but in fact, it
holds for any MHV choice of external particles. Remember from section
\ref{sec:swi} that SWI's fix all tree-level MHV amplitudes up to a
constant; since the one-loop result should also respect the full
$\N=4$ supersymmetry, the same is the case here. Only the common
factor is corrected from its tree-level value to its one-loop value.

\section{NMHV Amplitudes}

The next natural step would be to compute all NMHV amplitudes. The
purely gluonic case was worked out by the unitarity methods of section
\ref{sec:unitarity} by Bern, Dixon and Kosower \cite{Bern04nmhv} prior
to the publication of the quadruple cut method. With this method, its
inventors computed the MHV constructible parts of the NMHV amplitude
\cite{Britto04quad}. The case with other external particles was worked
out by Bidder, Perkins and the author \cite{Risager05swi}, but since
the latter calculation naturally incorporates the pure gluon case,
both cases will be reviewed here.

For most (but not all) of the box coefficients of the NMHV amplitude,
we can again exploit MHV constructibility. Doing the same analysis of
the ``number of minuses'' as above, we see that there are a total of
seven minuses, which can be distributed among the four corner
amplitudes in two ways: Either there is one googly-MHV three point
corner and three MHV corners, or there are two googly-MHV three point
corners, one MHV corner, and one NMHV corner. The latter solution is
not MHV constructible and will be postponed until we have considered
the former which is. The diagram for the calculation and the solution
for the internal spinors are\\
\parbox{55mm}{
\begin{picture}(140,140)(0,0)
\ArrowLine(40,40)(40,100)
\ArrowLine(40,100)(100,100)
\ArrowLine(100,100)(100,40)
\ArrowLine(100,40)(40,40)
\Text(70,35)[t]{$l_1$}
\Text(35,70)[r]{$l_2$}
\Text(70,105)[b]{$l_3$}
\Text(105,70)[l]{$l_4$}
\Line(40,40)(40,20)
\Line(40,40)(20,40)
\Text(20,20)[tr]{$A$}
\Text(40,20)[tl]{$A_1$}
\Text(20,35)[br]{$A_{-1}$}
\DashCArc(40,40)(16,180,270){2}
\BCirc(40,40){7}
\Line(40,100)(20,100)
\Line(40,100)(40,120)
\Text(20,120)[br]{$B$}
\Text(20,105)[tr]{$B_1$}
\Text(40,120)[bl]{$B_{-1}$}
\DashCArc(40,100)(16,90,180){2}
\BCirc(40,100){7}
\Line(100,100)(100,120)
\Line(100,100)(120,100)
\Text(120,121)[bl]{$C$}
\Text(105,120)[br]{$C_1$}
\Text(123,105)[tl]{$C_{-1}$}
\DashCArc(100,100)(16,0,90){2}
\BCirc(100,100){7}
\Line(100,40)(114,26)
\Text(120,20)[tl]{$d$}
\BCirc(100,40){5}
\Line(95,40)(105,40)
\Line(100,35)(100,45)
\end{picture}}
\parbox{80mm}{
\begin{eqnarray}
&&l_1=\frac{\spxa d\langle dCBA|}{\spaa dACd}\nn
&&l_2=\frac{|BCd\rangle\spaxx dA}{\spaa dACd}\nn
&&l_3=\frac{|BAd\rangle\spaxx dC}{\spaa dACd}\nn
&&l_4=\frac{\spxa d\langle dABC|}{\spaa dACd}.
\end{eqnarray}}

\subsection{3 Mass Boxes, Gluons Only}

As in the MHV case, we start by getting the denominator factors out of
the way since they are independent of the external helicities. To fix
the ambiguity in scale of the spinors, we assume the convention
\begin{equation}
\spxa{l_1}=\spxa d,\quad
\spxa{l_2}=\spxxxa BCd,\quad
\spxa{l_3}=\spxxxa BAd,\quad
\spxa{l_4}=\spxa d.
\end{equation}
The numerators now become
\begin{eqnarray}
&&\frac12\frac1{\spa{(-l_1)}{A_1}\langle A_1\cdots A_{-1}\rangle
\spa{A_{-1}}{l_2}\spa{l_2}{(-l_1)}}\nn
&&\times\frac1{\spa{(-l_2)}{B_1}\langle B_1\cdots B_{-1}\rangle
\spa{B_{-1}}{l_3}\spa{l_3}{(-l_2)}}\nn
&&\times\frac1{\spa{(-l_3)}{C_1}\langle C_1\cdots C_{-1}\rangle
\spa{C_{-1}}{l_4}\spa{l_4}{(-l_3)}}\nn
&&\times\frac1{\spb {(-l_4)}d\spb d{l_1}\spb{l_1}{(-l_4)}}\\
&=&\frac1{2\hcyc}\frac{\spa d{A_1}\spa{A_{-1}}{B_1}\spa{B_{-1}}
{C_1}\spa{C_{-1}}d}{\spa{A_1}{l_1}\spb{l_1}{l_4}\spa{l_4}{C_{-1}}
\spb d{l_4}\spa{l_4}{l_3}\spa{l_2}{l_1}\spb{l_1}d}\nn
&&\times\frac1{\spa{A_{-1}}{l_2}
\spa{l_2}{B_1}\spa{l_3}{l_2}\spa{B_{-1}}{l_3}\spa{l_3}{C_1}}\\
&=&\frac1{2\hcyc}\frac{-\spaa dACd^2}{\langle dCBACBAd\rangle}\nn
&&\times
\frac{-\spa{A_{-1}}{B_1}\spa{B_{-1}}{C_1}}{\spba dC{l_3}\spab
{l_2}Ad\spa{A_{-1}}{l_2}
\spa{l_2}{B_1}\spa{l_3}{l_2}\spa{B_{-1}}{l_3}\spa{l_3}{C_1}}\\
&=&\frac1{2\hcyc}\frac{\spaa dACd}{\langle dCBAd]}
\frac{\spa{A_{-1}}{B_1}\spa{B_{-1}}{C_1}}{[dCBAd\rangle
\langle dCBAd]\langle dCBBAd\rangle}\nn
&&\times\frac1{\spaa{A_{-1}}BCd\spaa dCB{B_1}
\spaa{B_{-1}}BAd\spaa dAB{C_1}}\\
&=&\frac{\langle dCBAd]}2\frac1{\langle dCBAd]^4}\nn
&&\times\frac{-\spa{A_{-1}}{B_1}\spa{B_{-1}}{C_1}}
{\hcyc P_B^2\spaa dCB{A_{-1}}\spaa dCB{B_1}
\spaa dAB{B_{-1}}\spaa dAB{C_1}}.
\end{eqnarray}
By manipulations of the same kind as (\ref{eq:2megram}) it can be
shown that $\langle dCBAd]/2$ is the Gram determinant. We see the same
  kind of structure as for MHV: the Gram determinant, something to the
  power 4 (which we hope disappears) and a factor involving $\hcyc$
  which looks natural as a box coefficient.

We can now turn to the numerators. The easiest way to proceed is to
consider the gluon case first and then use NMHV SWI's and the known
facts about the MHV amplitude for other-than-gluons to relate them to
the gluon amplitudes. There are seven distinct non-zero configurations
of the three negative helicity gluons and we will show the
calculations for three of them.

The first case is shown in the box diagram below
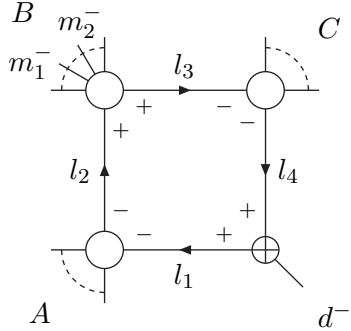
\begin{center}
\begin{picture}(140,140)(0,0)
\ArrowLine(40,40)(40,100)
\ArrowLine(40,100)(100,100)
\ArrowLine(100,100)(100,40)
\ArrowLine(100,40)(40,40)
\Text(70,35)[t]{$l_1$}
\Text(35,70)[r]{$l_2$}
\Text(70,105)[b]{$l_3$}
\Text(105,70)[l]{$l_4$}
\Text(85,43)[b]{\scriptsize $+$}
\Text(55,43)[b]{\scriptsize $-$}
\Text(43,55)[l]{\scriptsize $-$}
\Text(43,85)[l]{\scriptsize $+$}
\Text(55,97)[t]{\scriptsize $+$}
\Text(85,97)[t]{\scriptsize $-$}
\Text(97,88)[r]{\scriptsize $-$}
\Text(97,55)[r]{\scriptsize $+$}
\Line(40,40)(40,20)
\Line(40,40)(20,40)
\Text(20,20)[tr]{$A$}
\DashCArc(40,40)(16,180,270){2}
\BCirc(40,40){7}
\Line(40,100)(20,100)
\Line(40,100)(23,110)
\Line(40,100)(30,117)
\Line(40,100)(40,120)
\Text(14,126)[br]{$B$}
\Text(21,111)[r]{$m_1^-$}
\Text(31,120)[b]{$m_2^-$}
\DashCArc(40,100)(16,90,180){2}
\BCirc(40,100){7}
\Line(100,100)(100,120)
\Line(100,100)(120,100)
\Text(120,120)[bl]{$C$}
\DashCArc(100,100)(16,0,90){2}
\BCirc(100,100){7}
\Line(100,40)(114,26)
\Text(120,20)[tl]{$d^-$}
\BCirc(100,40){5}
\Line(95,40)(105,40)
\Line(100,35)(100,45)
\end{picture}
\end{center}
The numerator is given by
\begin{eqnarray}
&&\Big(\spa{(-l_1)}{l_2}\spa{m_1}{m_2}\spa{(-l_3)}{l_4}\spb{(-l_4)}
{l_1}\Big)^4\nn
&=&\Big(-\spa{m_1}{m_2}\spa{l_3}{l_4}\spb{l_4}{l_1}\spa{l_1}{l_2}
\Big)^4\nn
&=&(\spa {m_1}{m_2}\spa {l_3}d\spba dA{l_2})^4\nn
&=&(\spa{m_1}{m_2}\spaa dABd)^4\langle dCBAd]^4.
\end{eqnarray}
We should be happy to see the factor $\langle dCBAd]^4$ appearing as
  it cancels the corresponding (unphysical) factor in the
  denominator. The second placement of the negative helicity gluons we
  will explore is
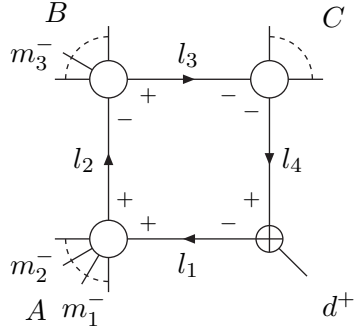
\begin{center}
\begin{picture}(140,140)(0,0)
\ArrowLine(40,40)(40,100)
\ArrowLine(40,100)(100,100)
\ArrowLine(100,100)(100,40)
\ArrowLine(100,40)(40,40)
\Text(70,35)[t]{$l_1$}
\Text(35,70)[r]{$l_2$}
\Text(70,105)[b]{$l_3$}
\Text(105,70)[l]{$l_4$}
\Text(85,43)[b]{\scriptsize $-$}
\Text(55,43)[b]{\scriptsize $+$}
\Text(43,55)[l]{\scriptsize $+$}
\Text(43,85)[l]{\scriptsize $-$}
\Text(55,97)[t]{\scriptsize $+$}
\Text(85,97)[t]{\scriptsize $-$}
\Text(97,88)[r]{\scriptsize $-$}
\Text(97,55)[r]{\scriptsize $+$}
\Line(40,40)(40,20)
\Line(40,40)(30,23)
\Line(40,40)(23,30)
\Line(40,40)(20,40)
\Text(17,17)[tr]{$A$}
\Text(31,20)[t]{$m_1^-$}
\Text(20,30)[r]{$m_2^-$}
\DashCArc(40,40)(16,180,270){2}
\BCirc(40,40){7}
\Line(40,100)(20,100)
\Line(40,100)(23,110)
\Line(40,100)(40,120)
\Text(25,122)[br]{$B$}
\Text(20,110)[r]{$m_3^-$}
\DashCArc(40,100)(16,90,180){2}
\BCirc(40,100){7}
\Line(100,100)(100,120)
\Line(100,100)(120,100)
\Text(120,120)[bl]{$C$}
\DashCArc(100,100)(16,0,90){2}
\BCirc(100,100){7}
\Line(100,40)(114,26)
\Text(120,20)[tl]{$d^+$}
\BCirc(100,40){5}
\Line(95,40)(105,40)
\Line(100,35)(100,45)
\end{picture}
\end{center}
which has the numerator
\begin{eqnarray}
&&\Big( \spa{m_1}{m_2}\spa{(-l_2)}{m_3}\spa{(-l_4)}{l_3}
\spb{(-l_4)}d\Big)^4\nn
&=&\Big(\spa{m_1}{m_2}\spa{l_2}{m_3}\spab {l_3}{l_4}d\Big)^4\nn
&=&(-\spa{m_1}{m_2}\spa{l_2}{m_3}\spab{l_3}Cd)^4\nn
&=&(\spa{m_1}{m_2}\spaa dCBd)^4\langle dABCd]^4\nn
&=&(\spa{m_1}{m_2}\spaa dCBd)^4\langle dCBAd]^4.
\end{eqnarray}
As before, we see the wanted factor to the power four. The last case
is the one where the three negative helicity gluons are situated on
the three massive corners,
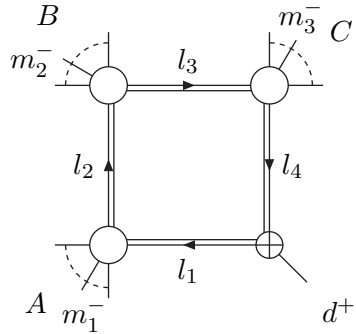
\begin{center}
\begin{picture}(140,140)(0,0)
\ArrowLine(40,40)(40,100)
\ArrowLine(40,100)(100,100)
\ArrowLine(100,100)(100,40)
\ArrowLine(100,40)(40,40)
\Text(70,35)[t]{$l_1$}
\Text(35,70)[r]{$l_2$}
\Text(70,105)[b]{$l_3$}
\Text(105,70)[l]{$l_4$}
\EBox(42,42)(98,98)
\Line(40,40)(40,20)
\Line(40,40)(30,23)
\Line(40,40)(20,40)
\Text(17,22)[tr]{$A$}
\Text(31,20)[t]{$m_1^-$}
\DashCArc(40,40)(16,180,270){2}
\BCirc(40,40){7}
\Line(40,100)(20,100)
\Line(40,100)(23,110)
\Line(40,100)(40,120)
\Text(22,123)[br]{$B$}
\Text(20,110)[r]{$m_2^-$}
\DashCArc(40,100)(16,90,180){2}
\BCirc(40,100){7}
\Line(100,100)(100,120)
\Line(100,100)(110,117)
\Line(100,100)(120,100)
\Text(123,117)[bl]{$C$}
\Text(112,120)[b]{$m_3^-$}
\DashCArc(100,100)(16,0,90){2}
\BCirc(100,100){7}
\Line(100,40)(114,26)
\Text(120,20)[tl]{$d^+$}
\BCirc(100,40){5}
\Line(95,40)(105,40)
\Line(100,35)(100,45)
\end{picture}
\end{center}
This configuration is rather special, as the whole $\N=4$ multiplet
can run in the loop and we have to sum over all the particle
species. The two gluon contributions can be constructed as for the
cases above,
\begin{eqnarray}
&&\Big(\spa{m_1}{l_2}\spa{m_2}{l_3}\spa{m_3}{l_4}\spb d{(-l_4)}
\Big)^4+\Big(\spa{m_1}{(-l_1)}\spa{m_2}{(-l_2)}\spa{m_3}{(-l_3)}
\spb d{l_1}\Big)^4\nn
&=&\bigg(\spaa{m_1}BCd\spaa{m_2}BAd\frac{\spa{m_3}d\langle dABCd]}
{\spaa dACd}\bigg)^4\nn
&&\qquad+\bigg(-\frac{\spa{m_1}d\langle dCBAd]}{\spaa dACd}
\spaa{m_2}BCd\spaa{m_3}BAd\bigg)^4\nn
&=&\Big[\Big(\spaa{m_1}BCd\spaa{m_2}BAd\spa{m_3}d\Big)^4
+\Big(-\spa{m_1}d\spaa{m_2}BCd\spaa{m_3}BAd\Big)^4\Big]\nn
&&\qquad\times\frac{\langle dCBAd]^4}
{\spaa dACd^4}.
\end{eqnarray}
As in the one-loop MHV calculation above, the fermion and scalar
contributions follow from the gluon ones by taking products of the two
parantheses above with powers summing to four and by adjusting the
prefactors. This exactly turns the final result into a complete power
of four:
\begin{eqnarray}
&&\Big(\spa{m_1}{l_2}\spa{m_2}{l_3}\spa{m_3}{l_4}\spb d{(-l_4)}
+\spa{m_1}{(-l_1)}\spa{m_2}{(-l_2)}\spa{m_3}{(-l_3)}
\spb d{l_1}\Big)^4\nn
&=&\Big(\spaa{m_1}BCd\spaa{m_2}BAd\spa{m_3}d
-\spa{m_1}d\spaa{m_2}BCd\spaa{m_3}BAd\Big)^4
\frac{\langle dCBAd]^4}{\spaa dACd^4}\nn
&=&\Big[\spa{m_3}{m_1}\spaa dBCd\spaa{m_2}BAd\nn
&&+\spa{m_1}d\Big(\spaa{m_3}BCd\spaa{m_2}BAd-\spaa{m_2}BCd
\spaa{m_3}BAd\Big)\Big]^4\frac{\langle dCBAd]^4}{\spaa dACd^4}\nn
&=&\Big(-\spa{m_3}{m_1}\spaa{m_2}BAd+K_B^2\spa{m_1}d\spa{m_2}{m_3}
\Big)^4\langle dCBAd]^4\nn
&=&\Big(\spa{m_1}{m_2}\spaa{m_3}BAd+\spa{m_3}{m_2}\spaa{m_1}BCd
\Big)^4\langle dCBAd]^4.
\end{eqnarray}
Again, we see the same picture emerging as before. 

When doing all negative helicity gluon
configurations, we finally obtain the result
\begin{equation}
\label{eq:threemasscoef}
c^{3m}(m_{1,2,3}^-)=\frac{-\spa{A_{-1}}{B_1}\spa{B_{-1}}{C_1}
\mathcal H_k^4}{\hcyc P_B^2\spaa dCB{A_{-1}}\spaa dCB{B_1}
\spaa dAB{B_{-1}}\spaa dAB{C_1}}
\end{equation}
where $k$ denotes the configuration and
\begin{eqnarray}
\mathcal H_{BBd}&=&\spa{m_1}{m_2}\spaa dACd,\nn
\mathcal H_{ABd}&=&\spa{m_1}d\spaa{m_2}BCd,\nn
\mathcal H_{BCd}&=&\spa{m_2}d\spaa{m_1}BAd,\nn
\mathcal H_{ACd}&=&\spa{m_1}d\spa{m_2}dK_B^2,\nn
\mathcal H_{AAB}&=&\spa{m_1}{m_2}\spaa{m_3}BCd,\nn
\mathcal H_{BCC}&=&\spa{m_2}{m_3}\spaa{m_1}BAd,\nn
\mathcal H_{ABB}&=&\spa{m_2}{m_3}\spaa{m_1}BCd,\nn
\mathcal H_{BBC}&=&\spa{m_1}{m_2}\spaa{m_3}BAd,\nn
\mathcal H_{AAC}&=&\spa{m_1}{m_2}\spa{m_3}dK_B^2,\nn
\mathcal H_{ACC}&=&\spa{m_2}{m_3}\spa{m_1}dK_B^2,\nn
\mathcal H_{ABC}&=&\spa{m_1}{m_2}\spaa{m_3}BAd+\spa{m_3}{m_2}
\spaa{m_1}BCd.
\end{eqnarray}

\subsection{NMHV SWIs}
\label{sec:nmhvswis}

Before proceeding to the equivalent calculations with other particle
content than gluons, we should consider some of the constraints on
those amplitudes coming from supersymmetry. In section \ref{sec:swi}
we looked at the effective supersymmetry of tree amplitudes,
supersymmetry algebras and the supersymmetric Ward identities that
follow. Since we are now working with a genuinely $\N=4$
supersymmetric theory we can use these considerations for loop
amplitudes also. Even better, since the box functions are all
independent, $\N=4$ SWIs for the full amplitude imply that they must
also be satisfied coefficient by coefficient. Here we will derive SWIs
for amplitudes $\A$ where all $g^+$ states are implicit.

To derive NMHV SWIs relating pure glue amplitudes to amplitudes with a
pair of fermions, we start from $\A(g_1^-,g_i^-,g_j^-,f_{a,k}^+)$ and
apply $Q_a(q,\theta)$ to find
\begin{eqnarray}
\label{eq:fermionnmhvswi}
0&=&\spa 1q\A(f_{a,1}^-,g_i^-,g_j^-,f_{a,k}^+)
+\spa iq\A(g_1^-,f_{a,i}^-,g_j^-,f_{a,k}^+)\nn
&&\quad+\spa jq\A(g_1^-,g_i^-,f_{a,j}^-,f_{a,k}^+)
-\spa kq\A(g_1^-,g_i^-,g_j^-,g_k^+).
\end{eqnarray}
This relation holds irrespective of the number of supersymmetries
since we are only using one. In a notation closer in spirit to what we
will use here, the SWI is
\begin{eqnarray}
0&=&\spa 1q\A(1^{-1/2},i^-,j^-,k^{+1/2})
+\spa iq\A(1^-,i^{-1/2},j^-,k^{+1/2})\nn
&&\quad+\spa jq\A(1^-,i^-,j^{-1/2},k^{+1/2})
-\spa kq\A(1^-,i^-,j^-,k^+).
\end{eqnarray}
We can extend this to scalars in external states by starting from the
amplitude $\A(g_1^-,g_i^-,f_{b,j}^-,s_{ab,k})$ and applying
again $Q_a(q,\theta)$. Using (\ref{eq:n4algebra}) we get
\begin{eqnarray}
\label{eq:scalarnmhvswi}
0&=&\spa 1q\A(f_{a,1}^-,g_i^-,f_{b,j}^-,s_{ab,k})
+\spa iq\A(g_1^-,f_{a,i}^-,f_{b,j}^-,s_{ab,k})\nn
&&\quad+\spa jq\A(g_1^-,g_i^-,\tilde s_{ab,j},s_{ab,k})
-\spa kq\A(g_1^-,g_i^-,f_{b,j}^-,f_{b,k}^-),
\end{eqnarray}
which relates amplitudes with both gluons, fermions and scalars. For
this SWI we used that an amplitude involving $\A(\ldots,s_{ab},f_a^+)$
vanishes, something which is not immediately obvious. This follows
roughly by noting that $s_{ab}\sim Q_a f_b^+\sim Q_b f_a^+$ which can be
interpreted in the way that $s_{ab}$ contains a $f_a^+$, and this sets
the amplitude to zero by fermion conservation of helicity. These
arguments can be made more exact by considering a superfield
formulation of the theory such as Nair's \cite{Nair88} which plays a
major role in some twistor studies.

To obtain SWIs involving amplitudes with no negative helicity gluons
we can start from $\A(g_1^-,f_{c,i}^-,f_{b,j}^-,s_{ab,k},f_{c,l}^+)$
and apply $Q_a(q,\theta)$ to get
\begin{eqnarray}
0&=&\spa 1q\A(f_{a,1}^-,f_{c,i}^-,f_{b,j}^-,s_{ab,k},f_{c,l}^+)
+\spa iq\A(g_1^-,\tilde s_{ac,i},f_{b,j}^-,s_{ab,k},f_{c,l}^+)\nn
&&\quad+\spa jq\A(g_1^-,f_{c,i}^-,\tilde s_{ab,j},s_{ab,k},f_{c,l}^+)
-\spa kq\A(g_1^-,f_{c,i}^-,f_{b,j}^-,f_{b,k}^+,f_{c,l}^+).
\end{eqnarray}
It is possible to deduce quite a few of these NMHV SWI's and we shall
not attempt to derive all of them here. The general procedure for
generating them is this: Start with an amplitude which has three and a
half units of negative helicity (N$^{3/2}$MHV) and where the SUSY
indices $b$, $c$ and $d$ are each maximally used once on $f_a^-$ and
$\tilde s_{ab}$ states, and where the indices $a$, $b$, $c$ and $d$
are used maximally once on $f_a^+$ and $s_{ab}$ states (and $a$ must
be used once). Then, assuming the four SUSY indices to be different,
act with $Q_a(q,\theta)$ to get the SWI. Resulting amplitudes
involving $a$ twice in  $f_a^+$ and $s_{ab}$ states are zero for the
reason discussed above.

As should be clear from the discussion of SWI's, they are not
concerned with colour ordering, or numbering in general.

\subsection{3 Mass Boxes with Two Fermions}

We now want to calculate the three mass box function coefficients when
we have two negative helicity gluons and a pair of fermions. We first
review some general arguments and then give the full results without
detailed calculation.

The easiest way of classifying the configurations is to view it as a
process of moving a half unit of negative helicity from one of the
$m_i$'s to a leg $q$ situated on one of the four corners. Take as an
example the configuration $q,m_1\in A$, $m_2\in B$ and $m_3=d$ and
move a half unit of negative helicity from $m_1$ on $A$ to $q$ which
is also on $A$. Since this only involves corner $A$ the ratio of the
coefficients must be the ratio of the $A$ corner amplitudes,\\
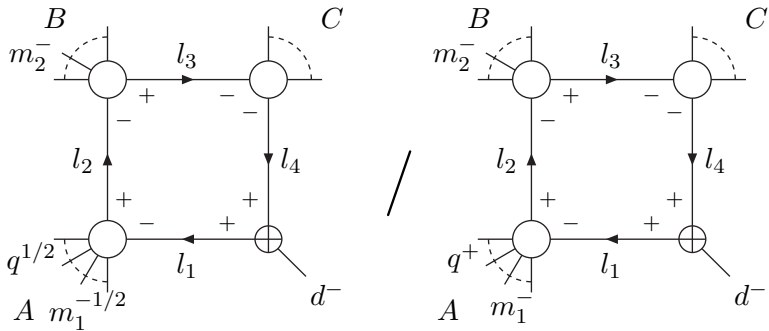
\begin{picture}(140,140)(0,0)
\ArrowLine(40,40)(40,100)
\ArrowLine(40,100)(100,100)
\ArrowLine(100,100)(100,40)
\ArrowLine(100,40)(40,40)
\Text(70,35)[t]{$l_1$}
\Text(35,70)[r]{$l_2$}
\Text(70,105)[b]{$l_3$}
\Text(105,70)[l]{$l_4$}
\Text(85,43)[b]{\scriptsize $+$}
\Text(55,43)[b]{\scriptsize $-$}
\Text(43,55)[l]{\scriptsize $+$}
\Text(43,85)[l]{\scriptsize $-$}
\Text(55,97)[t]{\scriptsize $+$}
\Text(85,97)[t]{\scriptsize $-$}
\Text(97,88)[r]{\scriptsize $-$}
\Text(97,55)[r]{\scriptsize $+$}
\Line(40,40)(40,20)
\Line(40,40)(30,23)
\Line(40,40)(23,30)
\Line(40,40)(20,40)
\Text(13,17)[tr]{$A$}
\Text(33,21)[t]{$m_1^{-1/2}$}
\Text(21,33)[r]{$q^{1/2}$}
\DashCArc(40,40)(16,180,270){2}
\BCirc(40,40){7}
\Line(40,100)(20,100)
\Line(40,100)(23,110)
\Line(40,100)(40,120)
\Text(25,120)[br]{$B$}
\Text(20,110)[r]{$m_2^-$}
\DashCArc(40,100)(16,90,180){2}
\BCirc(40,100){7}
\Line(100,100)(100,120)
\Line(100,100)(120,100)
\Text(120,120)[bl]{$C$}
\DashCArc(100,100)(16,0,90){2}
\BCirc(100,100){7}
\Line(100,40)(114,26)
\Text(116,24)[tl]{$d^-$}
\BCirc(100,40){5}
\Line(95,40)(105,40)
\Line(100,35)(100,45)
\end{picture}
\raisebox{55pt}{\Huge /}
\begin{picture}(140,140)(0,0)
\ArrowLine(40,40)(40,100)
\ArrowLine(40,100)(100,100)
\ArrowLine(100,100)(100,40)
\ArrowLine(100,40)(40,40)
\Text(70,35)[t]{$l_1$}
\Text(35,70)[r]{$l_2$}
\Text(70,105)[b]{$l_3$}
\Text(105,70)[l]{$l_4$}
\Text(85,43)[b]{\scriptsize $+$}
\Text(55,43)[b]{\scriptsize $-$}
\Text(43,55)[l]{\scriptsize $+$}
\Text(43,85)[l]{\scriptsize $-$}
\Text(55,97)[t]{\scriptsize $+$}
\Text(85,97)[t]{\scriptsize $-$}
\Text(97,88)[r]{\scriptsize $-$}
\Text(97,55)[r]{\scriptsize $+$}
\Line(40,40)(40,20)
\Line(40,40)(30,23)
\Line(40,40)(23,30)
\Line(40,40)(20,40)
\Text(13,17)[tr]{$A$}
\Text(33,21)[t]{$m_1^-$}
\Text(21,33)[r]{$q^+$}
\DashCArc(40,40)(16,180,270){2}
\BCirc(40,40){7}
\Line(40,100)(20,100)
\Line(40,100)(23,110)
\Line(40,100)(40,120)
\Text(25,120)[br]{$B$}
\Text(20,110)[r]{$m_2^-$}
\DashCArc(40,100)(16,90,180){2}
\BCirc(40,100){7}
\Line(100,100)(100,120)
\Line(100,100)(120,100)
\Text(120,120)[bl]{$C$}
\DashCArc(100,100)(16,0,90){2}
\BCirc(100,100){7}
\Line(100,40)(114,26)
\Text(116,24)[tl]{$d^-$}
\BCirc(100,40){5}
\Line(95,40)(105,40)
\Line(100,35)(100,45)
\end{picture}
\begin{eqnarray}
&=&\frac{c^{3m}(\{q^{1/2},m_1^{-1/2}\},\{m_2\},\{\},d^-)}
{c^{3m}(\{m_1^-\},\{m_2\},\{\},d^-)}\nn
&=&\frac{\spa{l_1}q}{\spa{l_1}{m_1}}\nn
&=&\frac{\spa qd}{\spa{m_1}d}\nn
&=&\frac{\H_{ABd}(q,m_2)}{\H_{ABd}(m_1,m_2)}.
\end{eqnarray}
If the $-1/2$ helicity came from $d$ we could track the
effect on each corner and derive the relative factor in a similar
way. However, we could also use the SWI
\begin{eqnarray}
&&\spa q\eta c^{3m}(\{m_1^-\},\{m_2^-\},\{\},d^-)\nn
&=&\spa{m_1}\eta c^{3m}(\{q^{1/2},m_1^{-1/2}\},\{m_2^-\},\{\},d^-)\nn
&&+\spa{m_2}\eta c^{3m}(\{q^{1/2},m_1^-\},\{m_2^{-1/2}\},\{\},d^-)\nn
&&+\spa d\eta c^{3m}(\{q^{1/2},m_1^-\},\{m_2^-\},\{\},d^{-1/2}),
\end{eqnarray}
and set $\eta=m_2$ to obtain
\begin{eqnarray}
&&\frac{c^{3m}(\{q^{1/2},m_1^-\},\{m_2^-\},\{\},d^{-1/2})}
{c^{3m}(\{m_1^-\},\{m_2^-\},\{\},d^-)}\nn
&=&\frac{\spa q{m_2}}{\spa d{m_2}}-\frac{\spa{m_1}{m_2}}{\spa d{m_2}}
\frac{\H_{ABd}(q,m_2)}{\H_{ABd}(m_1,m_2)}\nn
&=&\frac{\spa {m_1}q}{\spa {m_1}d}\nn
&=&\frac{\H_{AAB}(m_1,q,m_2)}{\H_{ABd}(m_1,m_2)}.
\end{eqnarray}
The case chosen is, of course, deceptively easy since
$c^{3m}(\{q^{1/2},m_1^-\},\{m_2^{-1/2}\},\{\},d^-)=0$ as can be seen
by trying (and failing) to assign internal helicities in the quadruple
cut. Taking the more involved road, it can also be found by using the
SWI above with $\eta=m_1$. 

In most cases there are no large advantages of using the one method
(tracking the half unit of helicity through the vertices) over the
other (SWI) but in some cases the former should definitely be
avoided. Take as an example the gluon configuration $m_1\in A$,
$m_2,m_3\in B$ and move $-1/2$ helicity from $m_3$ to $q\in C$. When
analysing the possible internal helicity assignments we find that
there are in fact \emph{two} different ways it can be done,
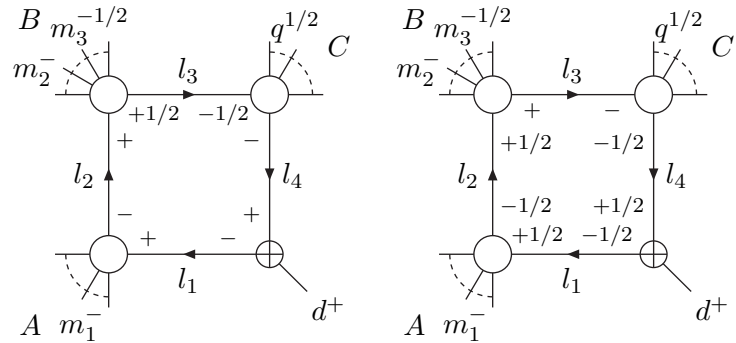
\begin{center}
\begin{picture}(140,140)(0,0)
\ArrowLine(40,40)(40,100)
\ArrowLine(40,100)(100,100)
\ArrowLine(100,100)(100,40)
\ArrowLine(100,40)(40,40)
\Text(70,35)[t]{$l_1$}
\Text(35,70)[r]{$l_2$}
\Text(70,105)[b]{$l_3$}
\Text(105,70)[l]{$l_4$}
\Text(85,43)[b]{\scriptsize $-$}
\Text(55,43)[b]{\scriptsize $+$}
\Text(43,55)[l]{\scriptsize $-$}
\Text(43,83)[l]{\scriptsize $+$}
\Text(57,97)[t]{\scriptsize $+1/2$}
\Text(83,97)[t]{\scriptsize $-1/2$}
\Text(97,83)[r]{\scriptsize $-$}
\Text(97,55)[r]{\scriptsize $+$}
\Line(40,40)(40,20)
\Line(40,40)(30,23)
\Line(40,40)(20,40)
\Text(15,17)[tr]{$A$}
\Text(30,20)[t]{$m_1^-$}
\DashCArc(40,40)(16,180,270){2}
\BCirc(40,40){7}
\Line(40,100)(20,100)
\Line(40,100)(23,110)
\Line(40,100)(30,117)
\Line(40,100)(40,120)
\Text(15,125)[br]{$B$}
\Text(21,110)[r]{$m_2^-$}
\Text(33,119)[b]{$m_3^{-1/2}$}
\DashCArc(40,100)(16,90,180){2}
\BCirc(40,100){7}
\Line(100,100)(100,120)
\Line(100,100)(110,117)
\Line(100,100)(120,100)
\Text(122,115)[bl]{$C$}
\Text(110,120)[b]{$q^{1/2}$}
\DashCArc(100,100)(16,0,90){2}
\BCirc(100,100){7}
\Line(100,40)(114,26)
\Text(116,24)[tl]{$d^+$}
\BCirc(100,40){5}
\Line(95,40)(105,40)
\Line(100,35)(100,45)
\end{picture}
\begin{picture}(140,140)(0,0)
\ArrowLine(40,40)(40,100)
\ArrowLine(40,100)(100,100)
\ArrowLine(100,100)(100,40)
\ArrowLine(100,40)(40,40)
\Text(70,35)[t]{$l_1$}
\Text(35,70)[r]{$l_2$}
\Text(70,105)[b]{$l_3$}
\Text(105,70)[l]{$l_4$}
\Text(83,43)[b]{\scriptsize $-1/2$}
\Text(57,43)[b]{\scriptsize $+1/2$}
\Text(43,58)[l]{\scriptsize $-1/2$}
\Text(43,82)[l]{\scriptsize $+1/2$}
\Text(55,97)[t]{\scriptsize $+$}
\Text(85,97)[t]{\scriptsize $-$}
\Text(97,82)[r]{\scriptsize $-1/2$}
\Text(97,58)[r]{\scriptsize $+1/2$}
\Line(40,40)(40,20)
\Line(40,40)(30,23)
\Line(40,40)(20,40)
\Text(15,17)[tr]{$A$}
\Text(30,20)[t]{$m_1^-$}
\DashCArc(40,40)(16,180,270){2}
\BCirc(40,40){7}
\Line(40,100)(20,100)
\Line(40,100)(23,110)
\Line(40,100)(30,117)
\Line(40,100)(40,120)
\Text(15,125)[br]{$B$}
\Text(21,110)[r]{$m_2^-$}
\Text(33,119)[b]{$m_3^{-1/2}$}
\DashCArc(40,100)(16,90,180){2}
\BCirc(40,100){7}
\Line(100,100)(100,120)
\Line(100,100)(110,117)
\Line(100,100)(120,100)
\Text(122,115)[bl]{$C$}
\Text(110,120)[b]{$q^{1/2}$}
\DashCArc(100,100)(16,0,90){2}
\BCirc(100,100){7}
\Line(100,40)(114,26)
\Text(116,24)[tl]{$d^+$}
\BCirc(100,40){5}
\Line(95,40)(105,40)
\Line(100,35)(100,45)
\end{picture}
\end{center}
Rather than calculating the result from this analysis, we can use that
(including a fermionic minus)\\
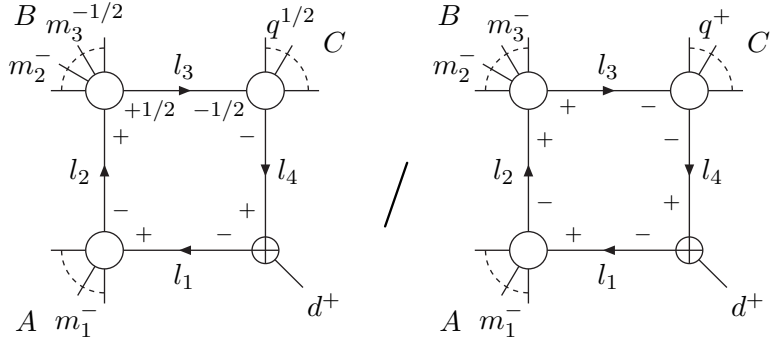
\begin{picture}(140,140)(0,0)
\ArrowLine(40,40)(40,100)
\ArrowLine(40,100)(100,100)
\ArrowLine(100,100)(100,40)
\ArrowLine(100,40)(40,40)
\Text(70,35)[t]{$l_1$}
\Text(35,70)[r]{$l_2$}
\Text(70,105)[b]{$l_3$}
\Text(105,70)[l]{$l_4$}
\Text(85,43)[b]{\scriptsize $-$}
\Text(55,43)[b]{\scriptsize $+$}
\Text(43,55)[l]{\scriptsize $-$}
\Text(43,83)[l]{\scriptsize $+$}
\Text(57,97)[t]{\scriptsize $+1/2$}
\Text(83,97)[t]{\scriptsize $-1/2$}
\Text(97,83)[r]{\scriptsize $-$}
\Text(97,55)[r]{\scriptsize $+$}
\Line(40,40)(40,20)
\Line(40,40)(30,23)
\Line(40,40)(20,40)
\Text(15,17)[tr]{$A$}
\Text(30,20)[t]{$m_1^-$}
\DashCArc(40,40)(16,180,270){2}
\BCirc(40,40){7}
\Line(40,100)(20,100)
\Line(40,100)(23,110)
\Line(40,100)(30,117)
\Line(40,100)(40,120)
\Text(15,125)[br]{$B$}
\Text(21,110)[r]{$m_2^-$}
\Text(33,119)[b]{$m_3^{-1/2}$}
\DashCArc(40,100)(16,90,180){2}
\BCirc(40,100){7}
\Line(100,100)(100,120)
\Line(100,100)(110,117)
\Line(100,100)(120,100)
\Text(122,115)[bl]{$C$}
\Text(110,120)[b]{$q^{1/2}$}
\DashCArc(100,100)(16,0,90){2}
\BCirc(100,100){7}
\Line(100,40)(114,26)
\Text(116,24)[tl]{$d^+$}
\BCirc(100,40){5}
\Line(95,40)(105,40)
\Line(100,35)(100,45)
\end{picture}
\raisebox{55pt}{\Huge /}
\begin{picture}(140,140)(0,0)
\ArrowLine(40,40)(40,100)
\ArrowLine(40,100)(100,100)
\ArrowLine(100,100)(100,40)
\ArrowLine(100,40)(40,40)
\Text(70,35)[t]{$l_1$}
\Text(35,70)[r]{$l_2$}
\Text(70,105)[b]{$l_3$}
\Text(105,70)[l]{$l_4$}
\Text(83,43)[b]{\scriptsize $-$}
\Text(57,43)[b]{\scriptsize $+$}
\Text(43,58)[l]{\scriptsize $-$}
\Text(43,82)[l]{\scriptsize $+$}
\Text(55,97)[t]{\scriptsize $+$}
\Text(85,97)[t]{\scriptsize $-$}
\Text(97,82)[r]{\scriptsize $-$}
\Text(97,58)[r]{\scriptsize $+$}
\Line(40,40)(40,20)
\Line(40,40)(30,23)
\Line(40,40)(20,40)
\Text(15,17)[tr]{$A$}
\Text(30,20)[t]{$m_1^-$}
\DashCArc(40,40)(16,180,270){2}
\BCirc(40,40){7}
\Line(40,100)(20,100)
\Line(40,100)(23,110)
\Line(40,100)(30,117)
\Line(40,100)(40,120)
\Text(15,125)[br]{$B$}
\Text(21,110)[r]{$m_2^-$}
\Text(33,119)[b]{$m_3^-$}
\DashCArc(40,100)(16,90,180){2}
\BCirc(40,100){7}
\Line(100,100)(100,120)
\Line(100,100)(110,117)
\Line(100,100)(120,100)
\Text(122,115)[bl]{$C$}
\Text(110,120)[b]{$q^+$}
\DashCArc(100,100)(16,0,90){2}
\BCirc(100,100){7}
\Line(100,40)(114,26)
\Text(116,24)[tl]{$d^+$}
\BCirc(100,40){5}
\Line(95,40)(105,40)
\Line(100,35)(100,45)
\end{picture}
\begin{eqnarray}
&=&\frac{c^{3m}(\{m_1^{-1/2}\},\{m_2^-,m_3^-\},\{q^{1/2}\},d^+)}
{c^{3m}(\{m_1^-\},\{m_2^-,m_3^-\},\{\},d^+)}\nn
&=&-\frac{\spa{l_2}{(-l_1)}}{\spa{l_2}{m_1}}
\frac{\spb d{l_1}}{\spb d{(-l_4)}}
\frac{\spa{(-l_3)}q}{\spa{(-l_3)}{l_4}}\nn
&=&-\frac{\spab{l_2}Ad\spa{l_3}q}{\spa{l_2}{m_1}\spba dC{l_3}}\nn
&=&-\frac{\spaa dABq}{\spaa dCB{m_1}}\nn
&=&-\frac{\H_{BBC}(m_2,m_3,q)}{\H_{ABB}(m_1,m_2,m_3)}
\end{eqnarray}
and use this in the corresponding SWI with $\eta=m_2$,
\begin{eqnarray}
&&\frac{c^{3m}(\{m_1^-\},\{m_2^-,m_3^{-1/2}\},\{q^{1/2}\},d^+)}
{c^{3m}(\{m_1^-\},\{m_2^-,m_3^-\},\{\},d^+)}\nn
&=&\frac{\spa q{m_2}}{\spa {m_3}{m_2}}
+\frac{\spa{m_1}{m_2}}{\spa{m_3}{m_2}}\frac{\H_{BBC}(m_2,m_3,q)}
{\H_{ABB}(m_1,m_2,m_3)}\nn
&=&\frac{\spa q{m_2}\spaa {m_1}BCd+\spa{m_1}{m_2}\spaa qBAd}
{\spa{m_3}{m_2}\spaa{m_1}BCd}\nn
&=&-\frac{\H_{ABC}(m_1,m_2,q)}{\H_{ABB}(m_1,m_2,m_3)}.
\end{eqnarray}
We see the continuation of the picture that moving $-1/2$ helicity
from one corner to another produces a ratio whose fourth power is the
ratio for moving $-2$ helicity between the same legs. This holds not
only when the external helicities constrain us to a single internal
state in the loop, but holds when more internal states open up. The
latter is a consequence of the $\N=4$ supersymmetry.

This becomes a helpful realization when we want to move helicity
around in the $ABC$ gluon configuration. No matter how we do it, we
will need to sum over at least eight states of the multiplet. This can
of course be done (and was indeed done in \cite{Risager05swi}) but it
is simpler to note \emph{e.g.}~that
\begin{eqnarray}
\spa q\eta\H_{ABC}(m_1,m_2,m_3)
&=&\spa {m_1}\eta\H_{ABC}(q,m_2,m_3)\nn
&&\quad-\spa{m_2}\eta\H_{AAC}(m_1,q,m_3)\nn
&&\quad-\spa{m_3}\eta\H_{AAB}(m_1,q,m_2),
\end{eqnarray}
and interpret this to be 'proportional to' the corresponding SWI
\begin{eqnarray}
&&\spa q\eta c^{3m}(\{m_1^-\},\{m_2^-\},\{m_3^-\},d^+)\nn
&=&\spa{m_1}\eta c^{3m}(\{m_1^{-1/2},q^{1/2}\},\{m_2^-\},\{m_3^-\}
,d^+)\nn
&&+\spa{m_2}\eta c^{3m}(\{m_1^-,q^{1/2}\},\{m_2^{-1/2}\},\{m_3^-\}
,d^+)\nn
&&+\spa{m_3}\eta c^{3m}(\{m_1^-,q^{1/2}\},\{m_2^-\},\{
m_3^{-1/2}\},d^+),
\end{eqnarray}
in the sense that
\begin{equation}
\frac{c^{3m}(\{m_1^{-1/2},q^{1/2}\},\{m_2^-\},\{m_3^-\},d^+)}
{c^{3m}(\{m_1^-\},\{m_2^-\},\{m_3^-\},d^+)}=
\frac{\H_{ABC}(q,m_2,m_3)}{\H_{ABC}(m_1,m_2,m_3)},
\end{equation}
\emph{etcetera}. Using the methods described in this section, one can
derive all box coefficients involving two gluinos. They are reported
in the tables below. The top line of each table states the gluon
configuration we start from and the associated $\H$ factor. Below, the
first column states where the $-1/2$ unit is moved to (the position of
$q$) and the second column states which of the negative helicity
gluons it came from. The third column then states which factor the
$\H$ in the top line is to be replaced by to obtain the correct box
coefficient. In the first three cases we have used identities such as
$-\H_{ABd}(m_1,m_2)=\H_{AAB}(q,m_1,m_2)$ to abbreviate results.
\begin{center}
%
%
\begin{tabular}{|c|c|r|}
\hline
\hline
\multicolumn{2}{|c|}{AAB}&$\H_{AAB}(m_1,m_2,m_3)$\\
\hline
\hline
$A$&$m_1$&$\H_{AAB}(q,m_2,m_3)$\\
$/d$&$m_2$&$\H_{AAB}(m_1,q,m_3)$\\
&$m_3$&$0$\\
\hline
$B$&$m_1$&$\H_{ABB}(m_2,q,m_3)$\\
&$m_2$&$\H_{ABB}(m_1,m_3,q)$\\
&$m_3$&$\H_{AAB}(m_1,m_2,q)$\\
\hline
$C$&$m_1$&$\H_{ABC}(m_2,m_3,q)$\\
&$m_2$&$-\H_{ABC}(m_1,m_3,q)$\\
&$m_3$&$-\H_{AAC}(m_1,m_2,q)$\\
\hline
\hline
\end{tabular}\qquad
%
%
\begin{tabular}{|c|c|r|}
\hline
\hline
\multicolumn{2}{|c|}{AAC}&$\H_{AAC}(m_1,m_2,m_3)$\\
\hline
\hline
$A$&$m_1$&$\H_{AAC}(q,m_2,m_3)$\\
$/d$&$m_2$&$\H_{AAC}(m_1,q,m_3)$\\
&$m_3$&$0$\\
\hline
$B$&$m_1$&$-\H_{ABC}(q,m_2,m_3)$\\
&$m_2$&$-\H_{ABC}(m_1,q,m_3)$\\
&$m_3$&$-\H_{AAC}(m_1,m_2,q)$\\
\hline
$C$&$m_1$&$\H_{ACC}(m_2,q,m_3)$\\
&$m_2$&$\H_{ACC}(m_1,m_3,q)$\\
&$m_3$&$\H_{AAC}(m_1,m_2,q)$\\
\hline
\hline
\end{tabular}\ \\ \ \\ \ \\
%
%
\begin{tabular}{|c|c|r|}
\hline
\hline
\multicolumn{2}{|c|}{ABB}&$\H_{ABB}(m_1,m_2,m_3)$\\
\hline
\hline
$A$&$m_1$&$\H_{ABB}(q,m_2,m_3)$\\
$/d$&$m_2$&$\H_{AAB}(q,m_1,m_3)$\\
&$m_3$&$\H_{AAB}(m_1,q,m_2)$\\
\hline
$B$&$m_1$&$0$\\
&$m_2$&$\H_{ABB}(m_1,q,m_3)$\\
&$m_3$&$\H_{ABB}(m_1,m_2,q)$\\
\hline
$C$&$m_1$&$\H_{BBC}(m_2,m_3,q)$\\
&$m_2$&$\H_{ABC}(m_1,m_3,q)$\\
&$m_3$&$-\H_{ABC}(m_1,m_2,q)$\\
\hline
\hline
\end{tabular}\qquad
%
%
\begin{tabular}{|c|c|r|}
\hline
\hline
\multicolumn{2}{|c|}{ABd}&$\H_{ABd}(m_1,m_2)$\\
\hline
\hline
$A$&$m_1$&$\H_{ABd}(q,m_2)$\\
&$m_2$&$0$\\
&$d$&$\H_{AAB}(m_1,q,m_2)$\\
\hline
$B$&$m_1$&$\H_{BBd}(m_2,q)$\\
&$m_2$&$\H_{ABd}(m_1,q)$\\
&$d$&$\H_{ABB}(m_1,m_2,q)$\\
\hline
$C$&$m_1$&$-\H_{BCd}(m_2,q)$\\
&$m_2$&$-\H_{ACd}(m_1,q)$\\
&$d$&$-\H_{ABC}(m_1,m_2,q)$\\
\hline
\hline
\end{tabular}\ \\ \ \\ \ \\ 
%
%
\begin{tabular}{|c|c|r|}
\hline
\hline
\multicolumn{2}{|c|}{ACd}&$\H_{ACd}(m_1,m_2)$\\
\hline
\hline
$A$&$m_1$&$\H_{ACd}(q,m_2)$\\
&$m_2$&$0$\\
&$d$&$\H_{ACd}(m_1,q)$\\
\hline
$B$&$m_1$&$-\H_{BCd}(q,m_2)$\\
&$m_2$&$-\H_{ABd}(m_1,q)$\\
&$d$&$-\H_{ABC}(m_1,q,m_2)$\\
\hline
$C$&$m_1$&$0$\\
&$m_2$&$\H_{ACd}(m_1,q)$\\
&$d$&$\H_{ACC}(m_1,m_2,q)$\\
\hline
\hline
\end{tabular}\qquad
%
%
\begin{tabular}{|c|c|r|}
\hline
\hline
\multicolumn{2}{|c|}{BBd}&$\H_{BBd}(m_1,m_2)$\\
\hline
\hline
$A$&$m_1$&$-\H_{ABd}(q,m_2)$\\
&$m_2$&$-\H_{ABd}(m_1,q)$\\
&$d$&$-\H_{ABB}(q,m_1,m_2)$\\
\hline
$B$&$m_1$&$\H_{BBd}(q,m_2)$\\
&$m_2$&$\H_{BBd}(m_1,q)$\\
&$d$&$0$\\
\hline
$C$&$m_1$&$-\H_{BCd}(m_2,q)$\\
&$m_2$&$-\H_{BCd}(m_1,q)$\\
&$d$&$-\H_{BBC}(m_1,m_2,q)$\\
\hline
\hline
\end{tabular}\ \\ \ \\ 
%
%
\begin{tabular}{|c|c|r|}
\hline
\hline
\multicolumn{2}{|c|}{ABC}&
$\H_{ABC}(m_1,m_2,m_3)$\\
\hline
\hline
$A$&$m_1$&$\H_{ABC}(q,m_2,m_3)$\\
&$m_2$&$\H_{AAC}(q,m_1,m_3)$\\
&$m_3$&$\H_{AAB}(q,m_1,m_2)$\\
\hline
$B$&$m_1$&$\H_{BBC}(q,m_2,m_3)$\\
&$m_2$&$\H_{ABC}(m_1,q,m_3)$\\
&$m_3$&$\H_{ABB}(m_1,q,m_2)$\\
\hline
$C$&$m_1$&$\H_{BCC}(m_2,q,m_3)$\\
&$m_2$&$\H_{ACC}(m_1,q,m_3)$\\
&$m_3$&$\H_{ABC}(m_1,m_2,q)$\\
\hline
$d$&$m_1$&$-\H_{BCd}(m_2,m_3)$\\
&$m_2$&$-\H_{ACd}(m_1,m_3)$\\
&$m_3$&$-\H_{ABd}(m_1,m_2)$\\
\hline
\hline
\end{tabular}
\end{center}

\subsection{Beyond Two Fermions}

Using the same methods as above and assisted by SWI's, the analysis
can be extended to other external particles than gluons and two
fermions. For instance, the ratio of coefficients between a
$(g^-,g^-,s_{ab},\tilde s_{ab})$ and a $(g^-,g^-,g^-,g^+)$ amplitude
is found by squaring the ratio between coefficients of
$(g^-,g^-,f^-,f^+)$ and a $(g^-,g^-,g^-,g^+)$, \emph{e.g.}
\begin{eqnarray}
&&c^{3m}(\{m_1^-,m_2^-\},\{m_3^{0-}\},\{ q^{0+}\},d^+)\nn
&=&\frac{-\H_{AAC}(m_1,m_2,q)}{\H_{AAB}(m_1,m_2,m_3)}
c^{3m}(\{m_1^-,m_2^-\},\{m_3^{-1/2}\},\{q^{1/2}\},d^+)\nn
&=&\bigg(\frac{-\H_{AAC}(m_1,m_2,q)}{\H_{AAB}(m_1,m_2,m_3)}\bigg)^2
c^{3m}(\{m_1^-,m_2^-\},\{m_3^-\},\{q^+\},d^+).
\end{eqnarray}
If we moved four units of $-1/2$ helicity from $m_3$ to $q$ we would
have to take the parenthesis to the power 4, which is of course
equivalent to exchanging $\H_{AAB}(m_1,m_2,m_3)^4$ with
$\H_{AAC}(m_1,m_2,q)^4$ in (\ref{eq:threemasscoef}) as it should be.

Amplitudes with both scalars and fermions can be constructed in
similar ways by viewing them as pure-glue amplitudes where units of
$-1/2$ helicity have been moved around. For each move, one of the
original $\H_k$ factors is removed and replaced by another which
captures where the helicity was moved to. In general this does not fix
the overall sign unambiguously, so we must resort to SWIs to be
completely sure. As an example we can try to compute the coefficient
$c^{3m}(\{m_1^-,m_2^{-1/2}\},\{m_3^{-1/2}\},\{q^{0+}\},d^+)$ by noting
that it comes from moving $-1/2$ helicity from $m_2$ to $q$ in
$c^{3m}(\{m_1^-,m_2^-\},\{m_3^{-1/2}\},\{q^{1/2}\},d^+)$. This tells
us that the relative factor must be
$\pm\H_{ABC}(m_1,m_3,q)/\H_{AAB}(m_1,m_2,m_3)$, and that we must use a
SWI of the type (\ref{eq:scalarnmhvswi}) which describes moving
helicity from somewhere to in an $AAB$ configuration to corner
$C$. Such an SWI is represented in the first table above in the last
three lines, namely
\begin{eqnarray}
0&=&\spa {m_1}\eta\H_{ABC}(m_2,m_3,q)-\spa
{m_2}\eta\H_{ABC}(m_1,m_3,q)\nn
&&-\spa {m_3}\eta\H_{AAC}(m_1,m_2,q)-\spa q\eta\H_{AAB}
(m_1,m_2,m_3).
\end{eqnarray}
This leads us to conclude that 
\begin{eqnarray}
&&c^{3m}(\{m_1^-,m_2^{-1/2}\},\{m_3^{-1/2}\},\{ q^{0+}\},d^+)\nn
&=&\frac{-\H_{ABC}(m_1,m_3,q)}{\H_{AAB}(m_1,m_2,m_3)}
c^{3m}(\{m_1^-,m_2^-\},\{m_3^{-1/2}\},\{q^{1/2}\},d^+)\nn
&=&\frac{\H_{ABC}(m_1,m_3,q)\H_{AAC}(m_1,m_2,q)}
{\H_{AAB}(m_1,m_2,m_3)^2}
c^{3m}(\{m_1^-,m_2^-\},\{m_3^-\},\{q^+\},d^+).\nn
\end{eqnarray}
With these methods, three-mass box function coefficients for any NMHV
external particle configuration can be calculated.

\subsection{Beyond Three-Mass Boxes}
\label{sec:beyondthreemass}

Up until now we completely focussed on three-mass boxes because they
were MHV constructible and have a well controlled
structure. Two-mass-hard boxes are also MHV constructible but they
have two contributions for each coefficient since either of the two
massless corners can be the googly MHV one. We can easily obtain those
coefficients, however, from the three-mass coefficients by taking
massive corners to have only one momentum on them, 
\begin{equation}
c^{2mh}(A,B,c,d)=c^{3m}(A,B,\{c\},d)+c^{3m}(\{d\},A,B,c).
\end{equation}
The two-mass-easy boxes are not MHV constructible because the opposite
massless corners are both googly MHV and the two massive corners must
be MHV and NMHV, respectively. Rather than using explicit expressions
for the the NMHV amplitudes, we can use the same procedure as
\cite{Bern04nmhv} where the known structure of IR divergences in
$\N=4$ super-Yang--Mills tells us that
\begin{equation}
\A^{\mathrm{1-loop}}\bigg|_{1/\epsilon}
=-\frac{\hat c_\Gamma}{\epsilon^2}\sum_{i=1}^n
\bigg(\frac{\mu^2}{-s_{i,i+1}}\bigg)^\epsilon\A^{\mathrm{tree}},
\end{equation}
which again translates to linear conditions on the box coefficients. These
can be inverted (at least for an odd number of external particles) to
produce expressions for the two-mass-easy coefficients in terms of the
three-mass coefficeints. Because we are considering $\N=4$
supersymmetry in both external and internal states, these identities
hold irrespective of the external states. Formally, the identity is
\begin{eqnarray}
c^{2me}(A,b,C,d)&=&\sum_{d\in\hat C}c^{3m}(\hat C,\ldots,\ldots,b)
+\sum_{b\in\hat A}c^{3m}(\hat A,\ldots,\ldots,d),
\end{eqnarray}
where the summations rum over all three-mass box coefficients that
fulfill the condition.

Just as two-mass-hard boxes were degenerate three-mass boxes, one-mass
boxes can be characterized as degenerate two-mass-easy and three-mass
boxes. This gives us the formula for the one-mass coefficients,
\begin{equation}
c^{1m}(A,b,c,d)=c^{2me}(A,b,\{c\},d)+c^{3m}(\{d\},A,\{b\},c).
\end{equation}
Since there are no further contributions to the one-loop amplitude,
this constitutes a complete characterization of the one-loop NMHV
amplitude in $\N=4$ super-Yang--Mills.

\chapter{One-Loop Amplitudes in $\N=8$ Supergravity}
\label{cha:sugra}

This chapter introduces the calculation of one-loop amplitudes in
$\N=8$ supergravity and the ``No Triangle Hypothesis''\footnote{Given
  the present weight of evidence it should be termed the ``No Triangle
  Conjecture'', but the original name seems to have stuck with
  it. Considering what it actually says, it ought to be called the
  ``Box Hypothesis'' or something similar.} for the
structure. It is based on the article \cite{Risager06notri}, which
extends the results of earlier work \cite{BjerrumBohr05notri}. 

\section{One-Loop Structure of Maximal Supergravity}

The results of sections \ref{sec:pv} and \ref{sec:integralbasis} apply
just as well to a gravity theory as to Yang--Mills. Thus, any one-loop
amplitude in four dimensions can be written as a sum of boxes,
triangles, bubbles and a rational function,
\begin{equation}
\label{eq:gravbasis}
\M^{\mathrm{1-loop}}=\sum_{i\in\mathcal C}c_iI_4^i
+\sum_{j\in\mathcal D}d_jI_3^j+\sum_{k\in\mathcal E}e_kI_2^k
+R.
\end{equation}
There are, however, no theorems that use supersymmetry to constrain
this further as with Yang--Mills, and generically, any theory of
gravity should be expected to include all types of terms.

For a low number of external gravitons, there are some results that do
constrain the form. If we go back to the integrand of the loop
integral, it will generally have $m\leq n$ propagator terms where $n$
is the number of external particles. Since all Feynman vertices in
gravity contain momentum squared, we should have $2m$ powers of loop
momentum in the numerator. By multiplying and dividing by all
propagators not in the expressions, this can be written as $2n$ powers
of loop momenum in the numerator and $n$ propagators,
\begin{equation}
\mathcal M_n\sim \int d^Dl\frac{[l^\mu]^{2n}}
{l_1^2 l_2^2\cdots l_n^2} 
\end{equation}
When we do Passarino--Veltman reduction on this, each propagator can
be exchanged for one or two fewer powers of loop momentum. For any $n$,
this is expected to result in a combination of all the terms appearing
in (\ref{eq:gravbasis}). However, we should also expect that there are
supersymmetry cancellations like in Yang--Mills that reduce the power
$2n$. This is indeed the case in the so-called string-based approach
to loop calculations \cite{Bern93stringrulesgrav,
Dunbar94stringrulessugra} (for the string-based approach in
Yang--Mills, consult \cite{Bern90stringrules1, Bern91stringrules2,
Bern92bosstringrules, Bern91stringrulesasgauge}) where cancellations
can be made explicit which reduce the numerator power of loop momentum
by 8 in $\N=8$,
\begin{equation}
\mathcal M_n^{\N=8}\sim \int d^Dl\frac{[l^\mu]^{2n-8}}
{l_1^2 l_2^2\cdots l_n^2}. 
\end{equation}
This explicitly tells us that four-point amplitudes (as explicitly
shown by a classic string calculation \cite{Green824gravitonloop}) can
only have a box integral, five-point can have boxes and triangles,
six-point can have boxes, triangles and bubbles, while seven-point and
above is again unconstrained.

Somewhat surprisingly, the situation seems to be even better. It has
been known since the end of the 90'es \cite{Bern98multileg} that
five-point and six-point MHV 1-loop amplitudes consisted only of boxes
and it was conjectured that this would hold for all
MHV amplitudes. When it was recently found that the box function part
of the NMHV six-graviton amplitude was sufficient to explain all IR,
soft and collinear properties of the amplitude, it was hypothesized
that all one-loop amplitudes in $\N=8$ supergravity consisted of boxes
only. In particular, since there this means no triangles, it was
called the No-Triangle Hypothesis. 

An equivalent way to express this at one-loop is that $\N=8$
supergravity has the same structure as $\N=4$ super Yang--Mills, in
particular that there are cancellations which remove additional powers
of loop momentum from the numerator of integrands to reduce it to 
\begin{equation}
\mathcal M_n^{\N=8}\sim \int d^Dl\frac{[l^\mu]^{n-4}}
{l_1^2 l_2^2\cdots l_n^2}.
\end{equation}
This is $n-4$ more cancellations than $\N=8$ supersymmetry is known to
give and the origin is unknown. As described in the introduction, this
and many other results have led to speculations that $\N=8$
supergravity is perturbatively UV finite to all orders and that this
may be caused by an unknown symmetry present in all gravitational
theories, independent of supersymmetry, which softens the UV
behaviour.

\section{Evidence for the No-Triangle Hypothesis}

The present strategy for supporting the No-Triangle Hypothesis is to
determine the part of all six and seven-point amplitudes depending on box
integrals and to argue that no further cut-constructible parts can be
present. This leaves rational parts which are known to be absent at
six-points; additional arguments are presented why these ought to
vanish at seven-point and higher.

\subsection{Box Coefficients and Soft Divergences}

Box coefficients can be determined by use of the quadruple cut method
of section \ref{sec:quadcuts}. There are two strategies for
determining these; the direct and the KLT-inspired. The latter works
by using the KLT relations on all corner amplitudes of the quadruple
cut to write the $\N=8$ coefficient as a product of kinematic
invariants and two $\N=4$ coefficients
\cite{Bern05oneloopsixpointgrav}. As an example, the coefficient of
$I_4(1^-,2^-,3^+,(4^+,5^+))$ given by the quadruple cut
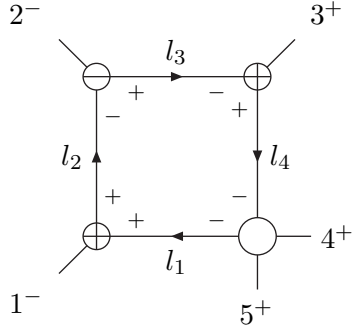
\begin{center}
\begin{picture}(140,140)(0,0)
\ArrowLine(40,40)(40,100)
\ArrowLine(40,100)(100,100)
\ArrowLine(100,100)(100,40)
\ArrowLine(100,40)(40,40)
\Text(70,35)[t]{$l_1$}
\Text(35,70)[r]{$l_2$}
\Text(70,105)[b]{$l_3$}
\Text(105,70)[l]{$l_4$}
\Text(85,43)[b]{\scriptsize $-$}
\Text(55,43)[b]{\scriptsize $+$}
\Text(43,55)[l]{\scriptsize $+$}
\Text(43,85)[l]{\scriptsize $-$}
\Text(55,97)[t]{\scriptsize $+$}
\Text(85,97)[t]{\scriptsize $-$}
\Text(97,88)[r]{\scriptsize $+$}
\Text(97,55)[r]{\scriptsize $-$}
\Line(40,40)(26,26)
\Text(20,20)[tr]{$1^-$}
\BCirc(40,40){5}
\Line(35,40)(45,40)
\Line(40,35)(40,45)
\Line(40,100)(26,114)
\Text(20,120)[br]{$2^-$}
\BCirc(40,100){5}
\Line(35,100)(45,100)
\Line(100,100)(114,114)
\Text(120,120)[bl]{$3^+$}
\BCirc(100,100){5}
\Line(95,100)(105,100)
\Line(100,95)(100,105)
\Line(100,40)(120,40)
\Line(100,40)(100,20)
\Text(124,40)[l]{$4^+$}
\Text(100,16)[t]{$5^+$}
\BCirc(100,40){7}
\end{picture}
\end{center}
which can be KLT'ed to
\begin{eqnarray}
c_{1^-,2^-,3^+,(4^+,5^+)}^{\N=8}&=&\frac12
\M(-l_1^+,1^-,l_2^+)\M(-l_2^-,2^-,l_3^+)\nn
&&\qquad\times\M(-l_3^-,3^+,l_4^+)
\M(-l_4^-,4^+,5^+,l_1^-)\nn
&=&\frac12\A(-l_1^+,1^-,l_2^+)^2\A(-l_2^-,2^-,l_3^+)^2
\A(-l_3^-,3^+,l_4^+)^2\nn
&&\qquad\times s_{45}\A(-l_4^-,4^+,5^+,l_1^-)\A(-l_4^-,5^+,4^+,l_1^-)\nn
&=&2s_{45}c_{1^-,2^-,3^+,(4^+,5^+)}^{\N=4}c_{1^-,2^-,3^+,(5^+,4^+)}^{\N=4}.
\end{eqnarray}
In the cases where there is a summation over the multiplet running in
the loop, the rewriting will still hold because the $\N=8$ multiplet
is two copies of the $\N=4$ multiplet, each being summed over in the
two $\N=4$ coefficients on the right hand side. Unfortunately, this
strategy becomes increasingly cumbersome at higher points. The KLT
relations become longer and involve permutations of legs which lead to
the $\N=4$ coefficients being the non-planar ones.

The more direct strategy is to use the expressions for $\N=8$
supergravity amplitudes. These can be derived from on-shell recursion
in more compact form than those given by the KLT relations, and
presumably give rise to the most compact forms of the coefficients. The
coefficients for general MHV amplitudes were given in
\cite{Bern98multileg} and the coefficients for the six-point NMHV
amplitude were given in \cite{Bern05oneloopsixpointgrav}. Thus, the
remaining seven-point amplitudes are the NMHV ones. A large portion of
these are MHV constructible (\emph{cf.}~section
\ref{sec:mhvconstructibility}) but some one-mass coefficients may
depend on the NMHV six-graviton tree amplitude. One even requires
summation over the full multiplet, and we thus need to extend the NMHV
six-point tree amplitude to the case where two external particles are
non-gravitons. This can be achieved by redoing the on-shell recursive
calculation of \cite{Britto05bcfw} and the result is given in
appendix C of \cite{Risager06notri}. The procedure like the one used
in section \ref{sec:beyondthreemass} cannot be used as it relies on the
previous knowledge that the amplitude contains only boxes, the fact
that we are trying to establish. All the seven-point NMHV box
coefficients are collected in appendix A of \cite{Risager06notri}.

The results for the box coefficients constrain the rest of the
amplitude because it must have the soft ($1/\epsilon$) divergence
\cite{Dunbar95softgravity},
\begin{equation}
\label{eq:gravsoftdiv}
\M^{\mathrm{1-loop}}\bigg|_{\frac{1}{\epsilon}}=\frac{i}{(4\pi)^2}
\bigg[\frac{\sum_{i<j}s_{ij}\log[-s_{ij}]}{2\epsilon}\bigg]
\M^{\mathrm{tree}}.
\end{equation}
If this holds for just the box integral part of the amplitude, the
triangles and bubbles must have their soft divergences add up to
zero. That this is indeed the case has been checked numerically for
the all amplitudes up to and including seven-point. In practise, this
is done by extracting the soft parts of all the box functions and
rephrasing (\ref{eq:gravsoftdiv}) as identities among the box coefficients.

The soft divergences of one and two-mass triangles are best
studied by rewriting them in a basis consisting of the functions
\begin{equation}
G(-K^2)=\frac{(-K^2)^{-\epsilon}}{\epsilon^2}
=\frac1{\epsilon^2}-\frac{\log(-K^2)}\epsilon+\ldots,
\end{equation}
where $K^2$ runs over all independent kinematic invariants. Since the
soft parts of this basis are independent, the vanishing of their sum
amounts to requiring the vanishing of every coefficient. Going back to
the basis of one and two-mass triangles this means that all
coefficents of these are also zero. The three-mass triangles have no
IR singularities and can thus not be ruled out in this way. The bubble
functions are proportional to
\begin{equation}
\frac{(-K^2)^{-\epsilon}}{\epsilon}
=\frac1{\epsilon}-\log(-K^2)+\ldots,
\end{equation}
so all we can say about them is that their coefficients add up to zero.

\subsection{Ruling Out Bubbles}

The safest way of ruling out bubble contributions is of course to
calculate them directly. For this, the article uses two methods which
are closely related. The first is the method of Britto, Buchbinder,
Cachazo and Feng \cite{Britto05sqcd1} where the unitarity
cuts of amplitudes are explicitly integrated and the results compared
with the cut of the expansion in terms of scalar integrals. By
inspecting all unitarity cuts of seven-point NMHV amplitudes in this
light, bubble integrals can indeed be ruled out.

Rather than going through this technical, albeit interesting,
calculation, we will use a method which ties together the behaviour of
tree amplitudes deformed by a complex parameter $z$ as in on-shell
recursion (\emph{cf.}~section \ref{sec:recursion}) and the UV behaviour of
one-loop amplitude. This choice is made to illustrate one of the
central points of this thesis, namely that properties of graviton tree
amplitudes (such as $z\to\infty$ behaviour) are tightly linked with
the whole perturbative expansion of $\N=8$ supergravity.

Using the expansion (\ref{eq:gravbasis}) we can write the unitarity
cut in a particular channel with momentum $P$ and lift the momentum
integration. This gives
\begin{equation}
\sum_{i\in\mathcal C'}\frac{c_i}{(l_1-K_1(i))^2(l_2-K_2(i))^2}
+\sum_{j\in\mathcal D'}\frac{d_j}{(l_1-K_3(j))^2}
+e_k+D(l_1,l_2),
\end{equation}
where $\mathcal C'$ and $\mathcal D'$ are the sets of box and triangle
functions which have a cut in the relevant channel and the
$l-K_{1,2,3}$ represent the remaining propagators. Only one bubble
function has the cut; this is represented by $e_k$. Finally, we have
added a total derivative term which is set to zero by the
integration. By the normal logic of unitarity, this integrand should
be equal to a product of two on-shell tree amplitudes, summed over
internal helicities,
\begin{equation}
\label{eq:prodofms}
\sum_{h_1,h_2=\pm}\M(\ldots,l_1^{h_1},-l_2^{-h_2})
\M(\ldots,l_2^{h_2},-l_1^{-h_1}).
\end{equation}
We can now deform this product in the same way as done in the BCFW
construction of on-shell recursion described in section
\ref{sec:recursion}, namely by making the shift
\begin{equation}
\label{eq:gravbubbleshift}
\spxa {\widehat l_1}=\spxa {l_1}+z\spxa {l_2},\qquad
\spxb {\widehat l_2}=\spxb {l_2}-z\spxb {l_1}.
\end{equation}
Under this shift, the integrand of boxes and triangles go as $z^{-2}$
or $z^{-1}$ as $z\to\infty$ and bubbles go as $z^0$. This immediately
suggests a condition for absence of a bubble: If (\ref{eq:prodofms})
tends to zero as $z\to\infty$ (and $D(l_1,l_2)$ does as well) the
bubble corresponding to the cut is absent. This condition can be
related directly to an equivalent condition for vanishing of bubbles
in the direct integration method mentioned above, so it is indeed an
appropriate check.

The behaviour of tree amplitudes under the under above shift
(\ref{eq:gravbubbleshift}) is not known in general, but two things are
known: For MHV amplitudes where the helicities of $(l_1,l_2)$ are
given as $(h_1,h_2)$, then the scaling can be derived from known
expressions for the amplitude
\cite{Bedford05gravbcfw,Berends88gravitonmhv} to be
\begin{eqnarray}
\label{eq:mhvzbehaviour}
&\M^{\mathrm{MHV}}\sim z^{-2}\quad
&\mathrm{when}\quad(h_1,h_2)=(+,+),(-,-),(+,-),\nn
&\M^{\mathrm{MHV}}\sim z^6\quad&\mathrm{when}\quad(h_1,h_2)=(-,+),\nn
&\M^{\mathrm{MHV}}\sim z^{2h+2}\quad&\mathrm{when}\quad
(h_1,h_2)=(-h,h).
\end{eqnarray}
It is also known generally \cite{Benincasa07gravrecur} that
\begin{equation}
\M\sim z^{-2}\quad\mathrm{when}\quad(h_1,h_2)=(+,-)
\end{equation}
for all graviton amplitudes. Furthermore it can be verified that the
two first lines of (\ref{eq:mhvzbehaviour}) holds for all six and
seven-point tree amplitudes needed here. Thus, it is not unreasonable
to assume that this is a general picture.

The known $z\to\infty$ behaviour of MHV amplitudes can be used
immediately to derive the behaviour of the cut integrand for one-loop
MHV amplitudes. A singlet cut integrand where there is only one state
running in the loop has the generic form
\begin{equation}
\M(1^-,2^-,3^+,\ldots,r^+,l_1^+,-l_2^+)
\M(-l_1^-,l_2^-,(r+1)^+,\ldots,n^+).
\end{equation}
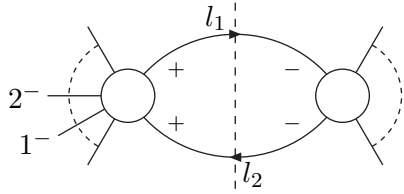
\begin{center}
\begin{picture}(180,100)(0,0)
\ArrowArcn(90,26.9)(46.2,150,30)
\ArrowArcn(90,73.1)(46.2,330,210)
\Text(65,63)[tl]{\scriptsize $+$}
\Text(115,63)[tr]{\scriptsize $-$}
\Text(115,37)[br]{\scriptsize $-$}
\Text(65,37)[bl]{\scriptsize $+$}
\Text(87,75)[br]{$l_1$}
\Text(93,25)[tl]{$l_2$}
\DashLine(90,15)(90,85){3}
\Line(50,50)(35,24)
\Line(50,50)(24,35)
\Line(50,50)(20,50)
\Line(50,50)(35,76)
\DashCArc(50,50)(22,120,240){2}
\BCirc(50,50){10}
\Text(22,37)[tr]{$1^-$}
\Text(18,50)[r]{$2^-$}
\Line(130,50)(145,76)
\Line(130,50)(145,24)
\DashCArc(130,50)(22,300,60){2}
\BCirc(130,50){10}
\end{picture}
\end{center}
Under the shift, both tree amplitudes go as $z^{-2}$ so the cut goes
as $z^{-4}$, and there are no bubble contributions. 

If the cut is non-singlet, the whole multiplet can run in
the loop,
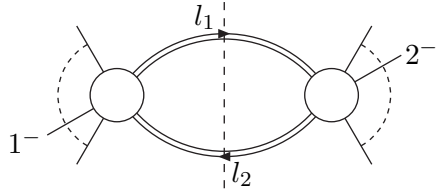
\begin{center}
\begin{picture}(180,100)(0,0)
\ArrowArcn(90,26.9)(46.2,150,30)
\ArrowArcn(90,73.1)(46.2,330,210)
\CArc(90,26.9)(44.2,30,150)
\CArc(90,73.1)(44.2,210,330)
\Text(87,75)[br]{$l_1$}
\Text(93,25)[tl]{$l_2$}
\DashLine(90,15)(90,85){3}
\Line(50,50)(35,24)
\Line(50,50)(24,35)
\Line(50,50)(35,76)
\DashCArc(50,50)(22,120,240){2}
\BCirc(50,50){10}
\Text(22,37)[tr]{$1^-$}
\Line(130,50)(145,76)
\Line(130,50)(156,65)
\Line(130,50)(145,24)
\DashCArc(130,50)(22,300,60){2}
\BCirc(130,50){10}
\Text(158,62)[bl]{$2^-$}
\end{picture}
\end{center}
the integrand becomes
\begin{eqnarray}
\label{eq:nonsingletmhvcut}
&&\sum_h(-1)^{2h}\bigg(\begin{array}{c}8\\4-2h\end{array}\bigg)\nn
&&\qquad\times\M(2^-,3^+,\ldots,r^+,l_1^h,-l_2^{-h})\M(-l_1^{-h},l_2^h,
(r+1)^+,\ldots,n^+)
\end{eqnarray}
 However, we know from SWIs that MHV amplitudes with different
 external particle content are related, \emph{e.g.},
\begin{eqnarray}
&&\M(2^-,3^+,\ldots,r^+,l_1^h,-l_2^{-h})=
\bigg(\frac{\spa 2{l_1}}{\spa 2{l_2}}\bigg)^{4-2h}
\M(2^-,3^+,\ldots,r^+,l_1^+,-l_2^-),\nn
\end{eqnarray}
so we can write (\ref{eq:nonsingletmhvcut}) as
\begin{equation}
\rho\times
\M(2^-,3^+,\ldots,r^+,l_1^+,-l_2^-)\M(-l_1^-,l_2^+,
(r+1)^+,\ldots,n^+)
\end{equation}
where
\begin{equation}
\rho=\sum_h(-1)^{2h}\bigg(\begin{array}{c}8\\4-2h\end{array}\bigg)
\bigg(\frac{\spa 2{l_1}\spa 1{l_2}}{\spa 2{l_2}\spa 1{l_1}}
\bigg)^{4-2h}=\bigg(\frac{\spa 12\spa {l_1}{l_2}}{\spa 2{l_2}\spa 
1{l_1}}\bigg)^8.
\end{equation}
Under the shift, the two tree amplitudes go as $z^{-2}$ and $z^6$,
respectively, while $\rho$, whose form is a consequence of $\N=8$
supersymmetry, goes as $z^{-8}$. In total, the integrand goes as
$z^{-4}$ which rules out bubbles in all one-loop MHV amplitudes.  

For singlet NMHV cuts, the situation is quite the same as for MHV; the
integrand is 
\begin{equation}
\M(l_1^-,\ldots,-l_2^-)\M(l_2^+,\ldots,-l_1^+),
\end{equation}
and since both tree amplitudes go as $z^{-2}$ under the shift, the
integrand goes as $z^{-4}$. This explicitly holds true for all cuts of
six and seven-graviton amplitudes, and continues to hold for any
singlet cut if the two first lines of (\ref{eq:mhvzbehaviour}) hold
beyond MHV.

The non-singlet NMHV cuts are more involved since they require the
summation over the $\N=8$ multiplet. For some integrands, such as
\begin{equation}
\sum_h(-1)^{2h}\bigg(\begin{array}c8\\4-2h\end{array}\bigg)
\M(1^-,2^-,4^+,l_1^{-h},-l_2^h)\M(-l_1^h,l_2^{-h},3^-,5^+,6^+),
\end{equation}
The calculation is essentially the same as for the MHV case because
the left NMHV amplitude is also a googly MHV amplitude where
\begin{equation}
\M(1^-,2^-,4^+,l_1^{-h},-l_2^h)
=\bigg(-\frac{\spb 4{l_1}}{\spb 4{l_2}}\bigg)^{4-2h}
\M(1^-,2^-,4^+,l_1^-,-l_2^+),
\end{equation}
such that the integrand becomes
\begin{equation}
\bigg(\frac{\spab 3{P_{12}}4}{\spa 3{l_2}\spb {l_2}4}\bigg)^8
\M(1^-,2^-,4^+,l_1^-,-l_2^+)\M(-l_1^+,l_2^-,3^-,5^+,6^+),
\end{equation}
and goes as $z^{-4}$. But in the seven-point calculation we will need
cuts like
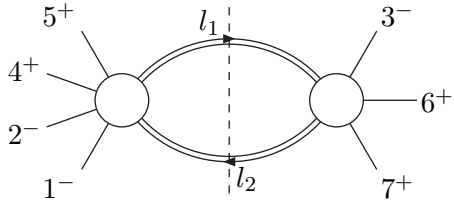
\begin{center}
\begin{picture}(180,100)(0,0)
\ArrowArcn(90,26.9)(46.2,150,30)
\ArrowArcn(90,73.1)(46.2,330,210)
\CArc(90,26.9)(44.2,30,150)
\CArc(90,73.1)(44.2,210,330)
\Text(87,75)[br]{$l_1$}
\Text(93,25)[tl]{$l_2$}
\DashLine(90,15)(90,85){3}
\Line(50,50)(35,24)
\Line(50,50)(22,40)
\Line(50,50)(22,60)
\Line(50,50)(35,76)
\BCirc(50,50){10}
\Text(33,22)[tr]{$1^-$}
\Text(20,38)[r]{$2^-$}
\Text(20,62)[r]{$4^+$}
\Text(33,78)[br]{$5^+$}
\Line(130,50)(145,76)
\Line(130,50)(160,50)
\Line(130,50)(145,24)
\BCirc(130,50){10}
\Text(147,78)[bl]{$3^-$}
\Text(162,50)[l]{$6^+$}
\Text(147,22)[tl]{$7^+$}
\end{picture}
\end{center}
with the integrand
\begin{eqnarray}
\label{eq:cutwithnmhv}
&&\sum_h(-1)^{2h}\bigg(\begin{array}c8\\4-2h\end{array}\bigg)
\M(1^-,2^-,4^+,5^+,l_1^{-h},-l_2^h)\M(-l_1^h,l_2^{-h},3^-,6^+,7^+),\nn
\end{eqnarray}
the expression depends on the six-point NMHV tree amplitude. As
mentioned above, this has been calculated in the paper and is found to
contain 14 terms, of which 12 have the structure $X\times Y^{2h-4}$
and the last two only contribute at $h=\pm2$. When this expression is
inserted into (\ref{eq:cutwithnmhv}) we find that it goes to zero as
$z\to\infty$ \emph{term by term} (after the $h$ summation) although it
does not go as well as $z^{-4}$. Again this highlights the connection
between recursion relations and UV structure: Using recursion
relations we can write down a form of the six-point amplitude which
rules out certain bubble contributions to the seven-point one-loop
amplitude without the need for further cancellations.

Since we have now calculated the $z\to\infty$ behaviour of every
independent cut at six and seven-points, and found that they approach
zero, we conclude that there are no bubble contributions to the
one-loop amplitudes, and that the absence can be tied to recursion
relations for gravity.

\subsection{Ruling Out Three-Mass Triangles}

As mentioned above, three-mass triangles have no IR singularities and
have to be ruled out explicitly. This can be done by inspecting the
triple cuts
\begin{eqnarray}
&&\sum_{h_1,h_2,h_3}\int d^Dl_1\delta(l_1^2)\delta(l_2^2)
\delta(l_3^2)\nn
&&\qquad\times
\M(\ldots,-l_1^{-h_1},l_2^{h_2})
\M(\ldots,-l_2^{-h_2},l_3^{h_3})
\M(\ldots,-l_3^{-h_3},l_1^{h_1}),
\end{eqnarray}
and arguing that they are all results of triple cuts of the boxes,
\begin{equation}
\int d^Dl_1\delta(l_1^2)\delta(l_2^2)\delta(l_3^2)
\sum_{i\in\mathcal C''}\frac{c_i}{(l_1-K(i))^2}
\end{equation}
where $l_1-K(i)$ is the momentum of the remaining uncut
propagator. The helpful fact here is that there are kinematic regimes
where the three-mass triple cut is real, and thus the remaining
integration can be performed numerically without appealing to complex
Minkowski space methods.

The box functions that have three-mass triple cuts are the 2-mass
hard, 3-mass and 4-mass boxes. Since we are only going up to
seven-point amplitudes we need not worry about the last of these. The
result of the cut integrations are
\begin{eqnarray}
I_4(k_1,k_2,K_3,K_4)\Big|_{\mathrm{cut}}
&=&\frac{\pi}{2(k_1+k_2)^2(k_2+K_3)^2}\nn
I_4(k_1,K_2,K_3,K_4)\Big|_{\mathrm{cut}}
&=&\frac{\pi}{2\Big((k_1+K_2)^2(K_2+K_3)^2-K_2^2K_4^2\Big)}\nn
I_3(K_1,K_2,K_3)\Big|_{\mathrm{cut}}
&=&\frac{\pi}{2\sqrt{K_1^4+K_2^4+K_3^4-2(K_1^2K_2^2+K_2^2K_3^2
+K_3^2K_1^2)}}\nn
\end{eqnarray}
With the knowledge of the box coefficients we have, it is fairly
straightforward to numerically integrate all the triple cuts and
compare them to the expectation from the boxes at some randomly chosen
kinematic points. This explicitly rules out three-mass triangles for
both six and seven-point amplitudes. For the six-point NMHV, the
independent triangle coefficients are
\begin{eqnarray}
&&d^{3m}(\{1^-,2^-\},\{3^-,4^+\},\{5^+,6^+\})\nn
&&d^{3m}(\{1^-,4^+\},\{2^-,5^+\},\{3^-,6^+\}).
\end{eqnarray}
The triple cut corresponding to the latter is a non-singlet cut which
requires a sum over the $\N=8$ multiplet. This means that it is
non-zero for $\N<8$. The first triangle coefficient can be expected to
be zero in any massless theory of gravity. For the seven-point NMHV, the
independent three-mass triangle coefficients are
\begin{eqnarray}
&&d^{3m}(\{1^-,2^-\},\{3^-,4^+\},\{5^+,6^+,7^+\})\nn
&&d^{3m}(\{1^-,2^-\},\{3^-,4^+,5^+\},\{6^+,7^+\})\nn
&&d^{3m}(\{1^-,2^-,4^+\},\{3^-,5^+\},\{6^+,7^+\})\nn
&&d^{3m}(\{1^-,4^+\},\{2^-,5^+\},\{3^-,6^+,7^+\}).
\end{eqnarray}
Again, the three first correspond to singlet cuts which are only
sensitive to the presence of the graviton and should be zero in any
massless gravity theory, while the last is non-zero for $\N<8$.

\subsection{Factorization and Rational Terms}

We have now ruled out contributions to the cut-constructible parts
from anything but boxes at six and seven-point. At six-point it is
already known that there are no rational parts, so this provides a
verification of the No-Triangle Hypothesis in that case. For
seven-point amplitudes, however, there is no general argument ruling
out rational contributions. 

Computing rational contributions is in general very hard in gravity
theories. A strategy that works in Yang--Mills theory is to do the
full calculation in $4-2\epsilon$ dimensions (\emph{cf.}~section
\ref{sec:ddimunitarity}) and extract the rational pieces by taking
$\epsilon\to0$. This works well because any massless field content can
be decomposed into an $\N=4$ multiplet, some number of $\N=1$ chiral
multiplets and some number of scalars, and because it can be proven
explicitly that the two supersymmetric theories have no rational
parts; only the scalar loop calculation needs to be done exactly in
$4-2\epsilon$ dimensions. For gravity, there is no such theorem saying
that supersymmetric theories have no rational contributions, so the
calculation needs to be done in $4-2\epsilon$ dimensions for all
states of the multiplet, thereby rendering most of the benefits of the
spinor-helicity formalism useless.

An alternative is to use the methods of section
\ref{sec:looprecursion} or, more generally, the known factorization
structure of one-loop gravity amplitudes. As in the gauge case, this
means in particular multiparticle, collinear and soft
factorization. The multiparticle factorization is, on the grounds of
general field theory arguments \cite{Bern95factorization}, the
same as for gauge theory (\ref{eq:ymmultiparticle}) with the obvious
corrections for the lack of colour ordering in gravity. The collinear
behaviour at one-loop is actually simpler than in Yang--Mills
(\ref{eq:ymcollinear}) as the splitting functions have no loop
corrections. This gives \cite{Bern98multileg}
\begin{eqnarray}
\label{eq:gravcollinear}
\M^{\mathrm{1-loop}}(a^{h_a},b^{h_b},\ldots)&\to&
\sum_{h=\pm2}\mathrm{Split}^{\mathrm{grav}}_{-h}(z,a^{h_a},b^{h_b})
\M^{\mathrm{1-loop}}(P^h,\ldots)
\end{eqnarray}
as $a\to zP$, $b\to (1-z)P$. The splitting functions for gravitons are
\begin{eqnarray}
\mathrm{Split}^{\mathrm{grav}}_+(z,a^+,b^+)
&=&0\nn
\mathrm{Split}^{\mathrm{grav}}_-(z,a^+,b^+)
&=&-\frac1{z(1-z)}\frac{\spb ab}{\spa ab}\nn
\mathrm{Split}^{\mathrm{grav}}_+(z,a^-,b^+)
&=&-\frac{z^3}{1-z}\frac{\spb ab}{\spa ab}.
\end{eqnarray}
There is also a universal soft behaviour when a graviton momentum
approaches zero, which also does not incur loop corrections,
\begin{equation}
\label{eq:gravsoft}
\M^{\mathrm{1-loop}}_n(s^\pm,\ldots)\to\mathcal S^{\mathrm{grav}}_
n(s^\pm)
\M^{\mathrm{1-loop}}_{n-1}(\ldots),
\end{equation}
as $k_s\to 0$. When the graviton going soft is number $n$ in some
arbitrary ordering, the soft factor is
\begin{equation}
\mathcal S^{\mathrm{grav}}_n(n^+)=-\frac1{\spa 1n\spa n{(n-1)}}
\sum_{i=2}^{n-2}\frac{\spa 1i\spa i{(n-1)}\spb in}{\spa in}.
\end{equation}

These limits put some quite strong constraints on the possible
rational parts which must have the same factorization as the
cut-constructible parts. We know that the rational parts of the
six-graviton amplitude are zero, so by (\ref{eq:gravcollinear}) and
(\ref{eq:gravsoft}) the rational part of the seven-graviton amplitude
can have no collinear or soft singularities. It is hard to imagine any
rational spinor expression which has none of these singularities while
being of mass dimension 4 and having the right holomorphic weight of
spinors as determined by their helicities. Indeed, it is hard to
imagine such an expression for \emph{any} number of external
gravitons, so by ``loose induction'' we should expect that all
rational parts are zero.

This argument could probably be formalized by using on-shell recursive
methods like those of section \ref{sec:looprecursion} for
gravity. Some progress has been made in this direction
\cite{Brandhuber07gravlooprecursion} but the approach is still too
undeveloped to be used for the problem at hand.

\section{Conclusions}

Using an array of different methods, we have explicitly calculated the
cut-containing parts of the NMHV six and seven graviton one-loop
amplitudes in $\N=8$ supergravity and found that only box integrals
contribute. At six points, this determines the amplitude completely
while at seven-point we have argued that an additional rational term
is highly unlikely. This proves the No-Triangle Hypothesis at six
points and renders it highly credible at seven. 

The methods used can be applied at any number of external particles,
and it does not appear that they would suddenly start failing. This
applies in particular to the ruling out of bubbles which relied on
very general scaling arguments for the amplitudes, scalings that are
expected to hold to any number of points. Together with the arguments
that the rational parts of amplitudes are likely to be zero, this
provides strong evidence that the one-loop amplitude in $\N=8$
supergravity consists exclusively of boxes just as $\N=4$ SYM.

As explained in the beginning of the chapter, this amounts to the
cancellation of $n-4$ powers of loop momentum in a $n$-point
amplitude, a cancellation which is unexplained. Similar cancellations
seem to occur at higher loop order also. They have been explicitly
found in the four-point amplitude at two \cite{Bern98twoloop} and
three \cite{Bern07threeloop} loops, where it was proven that the
divergences of maximal supergravity and maximal super-Yang--Mills are
the same in any dimension. This gives a strong hint that the
divergences of the two theories match in general, and that $\N=8$
supergravity in four dimensions is UV finite in particlular. It also
hints that there is some unknown symmetry or dynamical principle
present in $\N=8$ which is responsible, although there are no
candidates for such a symmetry. There is evidence that the
cancellations occur already in non-supersymmetric gravity
\cite{Bern07gravitycancellations}, and thus the symmetry should be
visible in some form without the presence of supersymmetry. The
identification of such a symmetry will be of immense interest.

\chapter{Amplitude with a Higgs}
\label{cha:higgs}

In this chapter we make the full determination of a particular
one-loop helicity amplitude,
$\A^1(\mathrm{Higgs},1^-,2^-,3^+,4^+)$. The chapter is based on
\cite{Badger07phimhv} but presents the calculation in a somewhat
different way. The article calculates the cut constructible parts,
using the one-loop MHV rules of section \ref{sec:bst}, of the
all-$n$ MHV amplitude $\A^1(\phi,1^-,2^-,3^+,\ldots,n^+)$ ($\phi$ to be
defined) and the rational part of $\A^1(\mathrm{Higgs},1^-,2^-,3^+,4^+)$
with on-shell recursion from section \ref{sec:looprecursion}. In this
chapter, we limit ourselves to the four-gluon case all the way through
and compute the cut-constructible parts using classical unitarity as
in section \ref{sec:unitarity}.

Even though the $pp\to Hjj$ process, where this amplitude is relevant,
is not included in table \ref{tab:wishlist} of wanted NLO
calculations, it is generally considered of the same importance. Some
semi-numerical results already exist \cite{ellis05higgs4parton}, but
clearer analytic results are still missing. The process forms a
background to Higgs production by vector boson fusion
\cite{Buttar06Leshouches}.

\section{Higgs in the Large Top Mass Limit}
\label{sec:Higgs}

In the present calculation we will make the assumption that the top
mass is large or, to be exact, that the kinematic scales involved in
the scattering are small compared to twice the top mass. The dominant
coupling of the Higgs to massless QCD particles comes through a top
quark triangle loop connecting to two gluons,
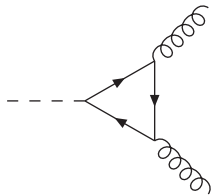
\begin{center}
\begin{picture}(75,70)(0,0)
\DashLine(0,35)(29,35)4
\ArrowLine(29,35)(55,50)
\ArrowLine(55,50)(55,20)
\ArrowLine(55,20)(29,35)
\Gluon(55,50)(75,70)34
\Gluon(55,20)(75,0)34
\end{picture}
\end{center}
In the large top mass limit, such an interaction is well approximated
by an effective term in the Lagrangian \cite{Shifman79higgs,
Wilczek77higgs}
\begin{equation}
\frac C2H\Tr F_{\mu\nu}F^{\mu\nu},
\end{equation}
where $H$ is the Higgs field and $C$ is an effective coupling
constant equal to \cite{Inami82higgscoupling}
\begin{equation}
\frac{\alpha_s}{6\pi v}\bigg(1+\frac{11\alpha_s}{4\pi}
+\mathcal O(\alpha_s^2)\bigg).
\end{equation}

In the addition to the Higgs, we can add a pseudoscalar Higgs $A$ such
that the total effective Lagrangian becomes
\begin{equation}
\frac C2\Big( H\Tr F_{\mu\nu}F^{\mu\nu}+iA\Tr F_{\mu\nu}{}^*F^{\mu\nu}
\Big)
\end{equation}
where 
\begin{equation}
{}^*F^{\mu\nu}=\frac i2\epsilon^{\mu\nu\kappa\lambda}F_{\kappa\lambda}.
\end{equation}
If we now change variables to
\begin{eqnarray}
&&\phi=\frac12(H+iA),\qquad \phi^\dagger=\frac12(H-iA),\nn
&&F_{SD}^{\mu\nu}=\frac12(F^{\mu\nu}+{}^*F^{\mu\nu}),\qquad
F_{ASD}^{\mu\nu}=\frac12(F^{\mu\nu}-{}^*F^{\mu\nu}),
\end{eqnarray}
the effective Lagrangian becomes
\begin{equation}
C\Big(\phi\Tr F_{SD\mu\nu}F_{SD}^{\mu\nu}
+\phi^\dagger\Tr F_{ASD\mu\nu}F_{ASD}^{\mu\nu}\Big),
\end{equation}
and Higgs amplitudes can be recovered by
\begin{equation}
\A(H,\ldots)=\A(\phi,\ldots)+\A(\phi^\dagger,\ldots).
\end{equation}
The Higgs is colourless and does not participate in the colour
ordering, but is conventionally written first.

The central observation is that tree amplitudes for $\phi$ and
$\phi^\dagger$ are separately simpler than amplitudes for $H$ or
$A$. In particular, they follow a pattern like that of the
Parke--Taylor amplitudes (\ref{eq:parketaylor}),
\begin{eqnarray}
\A^0(\phi,1^+,\ldots,n^+)&=&0\nn
\A^0(\phi,1^-,2^+,\ldots,n^+)&=&0\nn
\A^0(\phi,1^-,2^+,\ldots,i^-,\ldots,n^+)&=&\frac{\spa 1i^4}{\spa 12\spa
  23 \cdots\spa n1},
\end{eqnarray}
for $\phi$, and
\begin{eqnarray}
\A^0(\phi^\dagger,1^-,\ldots,n^-)&=&0\nn
\A^0(\phi^\dagger,1^+,2^-,\ldots,n^-)&=&0\nn
\A^0(\phi^\dagger,1^+,2^-,\ldots,i^+,\ldots,n^-)
&=&(-1)^n\frac{\spb 1i^4}{\spb 12\spb 23 \cdots\spb n1},
\end{eqnarray}
for $\phi^\dagger$. Even though the amplitudes look like the MHV and
googly-MHV, the condition between the gluon momenta for conservation
of momentum is different because of the presence of the $\phi$
momentum. Notice also that parity changes the sign of gluon helicities
\emph{and} daggers the $\phi$, so there are no easy
$(\phi,+,+,-,-,-,\ldots)$ amplitudes. In fact, there is a simple
$\phi$-all-minus amplitude (and $\phi^\dagger$-all-plus amplitude),
\begin{eqnarray}
\A^0(\phi,1^-,\ldots,n^-)&=&(-1)^n\frac{m_\phi^4}{\spb 12\spb 23\cdots
\spb n1},\nn
\A^0(\phi^\dagger,1^+,\ldots,n^+)&=&\frac{m_{\phi^\dagger}^4}
{\spa 12\spa 23\cdots\spa n1}.
\end{eqnarray}

The similarity of the $\phi$-MHV amplitude to the MHV amplitude have
led people to develop MHV rules for amplitudes with a single $\phi$,
where a $\phi$-MHV vertex must be used once but where the off-shell
prescription is unchanged
\cite{Dixon04phicsw,Badger04quarkphicsw}. The extension to one-loop
amplitudes was explored in \cite{Badger06allplus} and the article on
which this chapter is based \cite{Badger07phimhv}. The former deals
with the cut-constructible part of $\A^1(\phi,-,-,\ldots,-)$.

One-loop amplitudes involving a $\phi$ and at most one negative
helicity gluon are purely rational because the tree amplitude
vanishes. All-$n$ expressions for these amplitudes have been
constructed in \cite{Berger06phinite} where Schmidt's results for a
Higgs and three gluons \cite{Schmidt97H3g} is converted into the
$\phi/\phi^\dagger$ notation. Some of those we need are given at the
relevant point below.

\section{A One-Loop Higgs Amplitude}
\label{sec:phimmpp}

The methods presented in chapter \ref{cha:loop} can be applied just as
well to amplitudes containing a Higgs in the large $m_t$ limit. Here
we will apply them to the calculation of $A(H,1^-,2^-,3^+,4^+)$, that
is, the one-loop amplitude for a Higgs and four gluons with the
'adjacent MHV' helicity configuration. This configuration has a
non-zero tree amplitude and we can thus expect cut-containing pieces
at one-loop. In practice we will calculate $\A(\phi,1^-,2^-,3^+,4^+)$
and use that
\begin{eqnarray}
\A(H,1^-,2^-,3^+,4^+)&=&\A(\phi,1^-,2^-,3^+,4^+)+
\A(\phi^\dagger,1^-,2^-,3^+,4^+)\nn
&=&\A(\phi,1^-,2^-,3^+,4^+)+\A(\phi,3^-,4^-,1^+,2^+)^*.
\label{eq:higgsfromphi}
\end{eqnarray}
$\A(\phi,1^-,2^-,3^+,4^+)$ is split into its cut-containing pieces and
its rational pieces. For generality we calculate the cut-containing
pieces for $\A(\phi,1^-,2^-,3^+,\ldots,n^+)$. The whole calculation is
done in the four-dimensional helicity (FDH) renormalization scheme.

\subsection{Cut-Constructible Pieces}
\label{subsec:higgscut}

As described in section \ref{sec:susydecom}, the states running in the loop can
be written as a linear combination of an $\N=4$ multiplet, an $\N=1$
chiral multiplet, and a complex scalar referred to as $\N=0$. The
calculation of cut-containing pieces can be performed separately for
the three multiplets, and the result assembled from them.

Our plan of action is to first uncover any box functions by the use of
quadruple cuts. We then consider the (two-particle) unitarity cuts
where we isolate the terms coming from the (by now known) boxes and
relate the remainder to triangles and bubbles.

\subsubsection{Quadruple Cuts}

If the amplitude is cut in four places, the four corners must together
have six minuses (two external and four internal). These must
necessarily be distributed as two three-point googly corners
(opposite) and two MHV corners. Since the corner
with the $\phi$ cannot be three-point googly, it must be one of the
MHV corners. For $\N=4$, which contains the gluons, this gives us the
following six types of non-zero quadruple cuts:
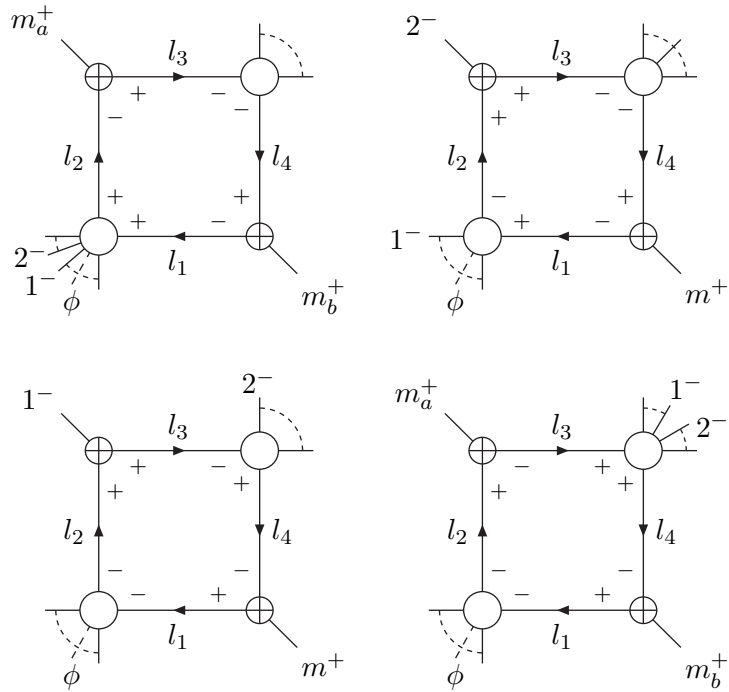
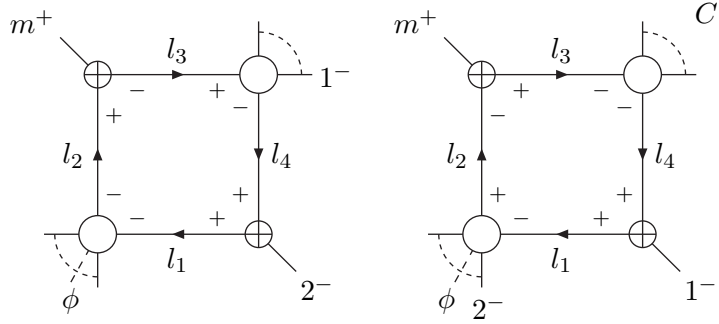
\begin{center}
\begin{picture}(140,140)(0,0)
\ArrowLine(40,40)(40,100)
\ArrowLine(40,100)(100,100)
\ArrowLine(100,100)(100,40)
\ArrowLine(100,40)(40,40)
\Text(70,35)[t]{$l_1$}
\Text(35,70)[r]{$l_2$}
\Text(70,105)[b]{$l_3$}
\Text(105,70)[l]{$l_4$}
\Text(85,43)[b]{\scriptsize $-$}
\Text(55,43)[b]{\scriptsize $+$}
\Text(43,55)[l]{\scriptsize $+$}
\Text(43,85)[l]{\scriptsize $-$}
\Text(55,97)[t]{\scriptsize $+$}
\Text(85,97)[t]{\scriptsize $-$}
\Text(97,88)[r]{\scriptsize $-$}
\Text(97,55)[r]{\scriptsize $+$}
\Line(40,40)(40,20)
\DashLine(40,40)(30,22){3}
\Line(40,40)(25,27)
\Line(40,40)(21,33)
\Line(40,40)(20,40)
\Text(30,20)[t]{$\phi$}
\Text(25,27)[tr]{$1^-$}
\Text(21,33)[r]{$2^-$}
\DashCArc(40,40)(16,180,200){2}
\DashCArc(40,40)(16,220,270){2}
\BCirc(40,40){7}
\Line(40,100)(26,114)
\Text(24,116)[br]{$m_a^+$}
\BCirc(40,100){5}
\Line(35,100)(45,100)
\Line(40,95)(40,105)
\Line(100,100)(100,120)
\Line(100,100)(120,100)
\DashCArc(100,100)(16,0,90){2}
\BCirc(100,100){7}
\Line(100,40)(114,26)
\Text(116,24)[tl]{$m_b^+$}
\BCirc(100,40){5}
\Line(95,40)(105,40)
\Line(100,35)(100,45)
\end{picture}
\begin{picture}(140,140)(0,0)
\ArrowLine(40,40)(40,100)
\ArrowLine(40,100)(100,100)
\ArrowLine(100,100)(100,40)
\ArrowLine(100,40)(40,40)
\Text(70,35)[t]{$l_1$}
\Text(35,70)[r]{$l_2$}
\Text(70,105)[b]{$l_3$}
\Text(105,70)[l]{$l_4$}
\Text(85,43)[b]{\scriptsize $-$}
\Text(55,43)[b]{\scriptsize $+$}
\Text(43,55)[l]{\scriptsize $-$}
\Text(43,85)[l]{\scriptsize $+$}
\Text(55,97)[t]{\scriptsize $+$}
\Text(85,97)[t]{\scriptsize $-$}
\Text(97,88)[r]{\scriptsize $-$}
\Text(97,55)[r]{\scriptsize $+$}
\Line(40,40)(40,20)
\DashLine(40,40)(30,22){3}
\Line(40,40)(20,40)
\Text(30,20)[t]{$\phi$}
\Text(18,40)[r]{$1^-$}
\DashCArc(40,40)(16,180,270){2}
\BCirc(40,40){7}
\Line(40,100)(26,114)
\Text(24,116)[br]{$2^-$}
\BCirc(40,100){5}
\Line(35,100)(45,100)
\Line(40,95)(40,105)
\Line(100,100)(100,120)
\Line(100,100)(114,114)
\Line(100,100)(120,100)
\DashCArc(100,100)(16,0,90){2}
\BCirc(100,100){7}
\Line(100,40)(114,26)
\Text(116,24)[tl]{$m^+$}
\BCirc(100,40){5}
\Line(95,40)(105,40)
\Line(100,35)(100,45)
\end{picture}\\
\begin{picture}(140,140)(0,0)
\ArrowLine(40,40)(40,100)
\ArrowLine(40,100)(100,100)
\ArrowLine(100,100)(100,40)
\ArrowLine(100,40)(40,40)
\Text(70,35)[t]{$l_1$}
\Text(35,70)[r]{$l_2$}
\Text(70,105)[b]{$l_3$}
\Text(105,70)[l]{$l_4$}
\Text(85,43)[b]{\scriptsize $+$}
\Text(55,43)[b]{\scriptsize $-$}
\Text(43,55)[l]{\scriptsize $-$}
\Text(43,85)[l]{\scriptsize $+$}
\Text(55,97)[t]{\scriptsize $+$}
\Text(85,97)[t]{\scriptsize $-$}
\Text(97,88)[r]{\scriptsize $+$}
\Text(97,55)[r]{\scriptsize $-$}
\Line(40,40)(40,20)
\DashLine(40,40)(30,22){3}
\Line(40,40)(20,40)
\Text(30,20)[t]{$\phi$}
\DashCArc(40,40)(16,180,270){2}
\BCirc(40,40){7}
\Line(40,100)(26,114)
\Text(24,116)[br]{$1^-$}
\BCirc(40,100){5}
\Line(35,100)(45,100)
\Line(40,95)(40,105)
\Line(100,100)(100,120)
\Line(100,100)(120,100)
\Text(100,122)[bb]{$2^-$}
\DashCArc(100,100)(16,0,90){2}
\BCirc(100,100){7}
\Line(100,40)(114,26)
\Text(116,24)[tl]{$m^+$}
\BCirc(100,40){5}
\Line(95,40)(105,40)
\Line(100,35)(100,45)
\end{picture}
\begin{picture}(140,140)(0,0)
\ArrowLine(40,40)(40,100)
\ArrowLine(40,100)(100,100)
\ArrowLine(100,100)(100,40)
\ArrowLine(100,40)(40,40)
\Text(70,35)[t]{$l_1$}
\Text(35,70)[r]{$l_2$}
\Text(70,105)[b]{$l_3$}
\Text(105,70)[l]{$l_4$}
\Text(85,43)[b]{\scriptsize $+$}
\Text(55,43)[b]{\scriptsize $-$}
\Text(43,55)[l]{\scriptsize $-$}
\Text(43,85)[l]{\scriptsize $+$}
\Text(55,97)[t]{\scriptsize $-$}
\Text(85,97)[t]{\scriptsize $+$}
\Text(97,88)[r]{\scriptsize $+$}
\Text(97,55)[r]{\scriptsize $-$}
\Line(40,40)(40,20)
\DashLine(40,40)(30,22){3}
\Line(40,40)(20,40)
\Text(30,20)[t]{$\phi$}
\DashCArc(40,40)(16,180,270){2}
\BCirc(40,40){7}
\Line(40,100)(26,114)
\Text(24,116)[br]{$m_a^+$}
\BCirc(40,100){5}
\Line(35,100)(45,100)
\Line(40,95)(40,105)
\Line(100,100)(100,120)
\Line(100,100)(110,117)
\Line(100,100)(117,110)
\Line(100,100)(120,100)
\Text(110,120)[bl]{$1^-$}
\Text(120,110)[l]{$2^-$}
\DashCArc(100,100)(16,0,30){2}
\DashCArc(100,100)(16,60,90){2}
\BCirc(100,100){7}
\Line(100,40)(114,26)
\Text(116,24)[tl]{$m_b^+$}
\BCirc(100,40){5}
\Line(95,40)(105,40)
\Line(100,35)(100,45)
\end{picture}\\
\begin{picture}(140,140)(0,0)
\ArrowLine(40,40)(40,100)
\ArrowLine(40,100)(100,100)
\ArrowLine(100,100)(100,40)
\ArrowLine(100,40)(40,40)
\Text(70,35)[t]{$l_1$}
\Text(35,70)[r]{$l_2$}
\Text(70,105)[b]{$l_3$}
\Text(105,70)[l]{$l_4$}
\Text(85,43)[b]{\scriptsize $+$}
\Text(55,43)[b]{\scriptsize $-$}
\Text(43,55)[l]{\scriptsize $-$}
\Text(43,85)[l]{\scriptsize $+$}
\Text(55,97)[t]{\scriptsize $-$}
\Text(85,97)[t]{\scriptsize $+$}
\Text(97,88)[r]{\scriptsize $-$}
\Text(97,55)[r]{\scriptsize $+$}
\Line(40,40)(40,20)
\DashLine(40,40)(30,22){3}
\Line(40,40)(20,40)
\Text(30,20)[t]{$\phi$}
\DashCArc(40,40)(16,180,270){2}
\BCirc(40,40){7}
\Line(40,100)(26,114)
\Text(24,116)[br]{$m^+$}
\BCirc(40,100){5}
\Line(35,100)(45,100)
\Line(40,95)(40,105)
\Line(100,100)(100,120)
\Line(100,100)(120,100)
\Text(123,100)[l]{$1^-$}
\DashCArc(100,100)(16,0,90){2}
\BCirc(100,100){7}
\Line(100,40)(114,26)
\Text(116,24)[tl]{$2^-$}
\BCirc(100,40){5}
\Line(95,40)(105,40)
\Line(100,35)(100,45)
\end{picture}
\begin{picture}(140,140)(0,0)
\ArrowLine(40,40)(40,100)
\ArrowLine(40,100)(100,100)
\ArrowLine(100,100)(100,40)
\ArrowLine(100,40)(40,40)
\Text(70,35)[t]{$l_1$}
\Text(35,70)[r]{$l_2$}
\Text(70,105)[b]{$l_3$}
\Text(105,70)[l]{$l_4$}
\Text(85,43)[b]{\scriptsize $+$}
\Text(55,43)[b]{\scriptsize $-$}
\Text(43,55)[l]{\scriptsize $+$}
\Text(43,85)[l]{\scriptsize $-$}
\Text(55,97)[t]{\scriptsize $+$}
\Text(85,97)[t]{\scriptsize $-$}
\Text(97,88)[r]{\scriptsize $-$}
\Text(97,55)[r]{\scriptsize $+$}
\Line(40,40)(40,20)
\DashLine(40,40)(30,22){3}
\Line(40,40)(20,40)
\Text(27,20)[t]{$\phi$}
\Text(43,17)[t]{$2^-$}
\DashCArc(40,40)(16,180,270){2}
\BCirc(40,40){7}
\Line(40,100)(26,114)
\Text(24,116)[br]{$m^+$}
\BCirc(40,100){5}
\Line(35,100)(45,100)
\Line(40,95)(40,105)
\Line(100,100)(100,120)
\Line(100,100)(120,100)
\Text(120,120)[bl]{$C$}
\DashCArc(100,100)(16,0,90){2}
\BCirc(100,100){7}
\Line(100,40)(114,26)
\Text(116,24)[tl]{$1^-$}
\BCirc(100,40){5}
\Line(95,40)(105,40)
\Line(100,35)(100,45)
\end{picture}
\end{center}
Notice that all permitted cuts necessarily have gluons running in the
loop. This immediately shows us that $\N=1$ and $\N=0$ cannot
contribute since they do not have gluons running in loops.

The calculation of the prefactors of the box functions is almost
identical to the one without the $\phi$. Thus, we will only present
one example calculation, namely that of the coefficient $c$ of the box
corresponding to the second cut above. From (\ref{eq:2meplussoln}) in
appendix \ref{app:quad} we derive that
\begin{eqnarray}
&&l_1=\frac{\spxa m\spaxx 2{P_{3,m}}}{\spa m2},\qquad
l_2=\frac{\spxa 2\spaxx m{P_{2,m-1}}}{\spa m2},\nn
&&l_3=\frac{\spxa 2\spaxx m{P_{3,m-1}}}{\spa m2},\qquad
l_4=\frac{\spxa m\spaxx 2{P_{3,m-1}}}{\spa m2},
\end{eqnarray}
such that
\begin{eqnarray}
c&=&\frac12\frac{\spa 1{l_2}^3}{\spa {l_2}{(-l_1)}\spa {(-l_1)}{(m+1)}
\langle (m+1)\cdots 1\rangle}
\frac{\spb {l_3}{(-l_2)}^3}{\spb{(-l_2)}2\spb 2{l_3}}\nn
&&\qquad\times
\frac{\spa {l_4}{(-l_3)}^3}{\spa{(-l_3)}3\langle 3\cdots (m-1)
\rangle\spa {(m-1)}{l_4}}
\frac{\spb {(-l_4)}m^3}{\spb m{l_1}\spb {l_1}{(-l_4)}}\nn
&=&\frac12\frac{\spa 12\spa 23\spa {(m-1)}m\spa m{(m+1)}
\spa 1{l_2}^3\spb {l_2}{l_3}^3\spa {l_3}{l_4}^3
\spb {l_4}m^3}{\hcyc\spa {l_2}{l_1}
\spa {l_1}{(m+1)}\spb {l_2}2\spb 2{l_3}
\spa {l_3}3\spa {(m-1)}{l_4}\spb m{l_1}\spb {l_1}{l_4}}\nn
&=&\frac12\frac{\spa 12\spa 23\spa {(m-1)}m\spa m{(m+1)}
\spa 12^3(-\spaa m{P_{2,m-1}}{P_{3,m-1}}m)^3}{\hcyc
\spa m2^4\spa 2m\spa m{(m+1)}\spab m{P_{2,m-1}}2(-\spba 2{P_{3,m-1}}m
)\spa 23}\nn
&&\qquad\times\frac{\spa 2m^3\spab 2{P_{3,m-1}}m^3}{\spa {(m-1)}m
(-\spba m{P_{3,m}}2)(-\spaa 2{P_{3,m}}{P_{3,m-1}}2)}\nn
&=&\frac{\spa 12^4}{\hcyc}\bigg(-\frac{\spab 2{P_{3,m-1}}m
\spab m{P_{3,m-1}}2}2\bigg).
\end{eqnarray}
As usual for MHV amplitudes, this is just the Gram determinant times
the tree amplitude. This holds for all coefficients, so the boxy part
of the amplitude is given by
\begin{equation}
\frac{\spa 12^4}{\hcyc}\sum_{m_1=1}^n \sum_{m_2=m_1+2}^{m_1-1}
F_4(m_1,P_{m_1+1,m_2-1},m_2),
\end{equation}
which is a combination of 1-mass and 2-mass-easy box functions. The
difference in comparison to the result without the $\phi$ is primarily
that there are twice as many 2-mass boxes because the $\phi$ makes the
two massive corners non-identical.

\subsubsection{Two Particle Cuts, $\N=4$ Case}

By counting minuses, we can again see that the two tree amplitudes
separated by the cut must be MHV and $\phi$-MHV, respectively:
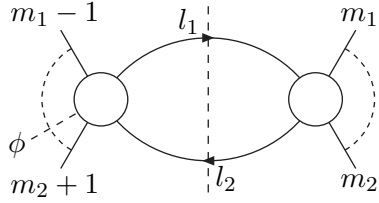
\begin{center}
\begin{picture}(180,100)(0,0)
\ArrowArcn(90,26.9)(46.2,150,30)
\ArrowArcn(90,73.1)(46.2,330,210)
\Text(87,75)[br]{$l_1$}
\Text(93,25)[tl]{$l_2$}
\DashLine(90,15)(90,85){3}
\Line(50,50)(35,24)
\DashLine(50,50)(24,35)3
\Line(50,50)(35,76)
\DashCArc(50,50)(22,120,240){2}
\BCirc(50,50){10}
\Text(22,33)[r]{$\phi$}
\Text(33,22)[t]{$m_2+1$}
\Text(33,78)[b]{$m_1-1$}
\Line(130,50)(145,76)
\Line(130,50)(145,24)
\DashCArc(130,50)(22,300,60){2}
\BCirc(130,50){10}
\Text(147,78)[b]{$m_1$}
\Text(147,22)[t]{$m_2$}
\end{picture}
\end{center}
Regardless of the position of 1 and 2, the cut integrand becomes
\begin{equation}
\frac{\spa 12^4}{\hcyc}\frac{\spa {(m_1-1)}{m_1}\spa {l_1}{l_2}
\spa {m_2}{(m_2+1)}\spa {l_2}{l_1}}{\spa {(m_1-1)}{l_1}\spa {l_1}{m_1}
\spa {m_2}{l_2}\spa {l_2}{(m_2+1)}}.
\end{equation}
For generic values of $m_1$ and $m_2$ it can be shown, analogously to
the pure gluon case, that this is equivalent to the same cut of
\begin{eqnarray}
&&\frac{\spa 12^4}{\hcyc}\Big(F_4(m_1-1,P_{m_1,m_2-1},m_2)
+F_4(m_1-1,P_{m_1,m_2},m_2+1)\nn
&&\qquad +F_4(m_1,P_{m_1+1,m_2-1},m_2)
+F_4(m_1,P_{m_1+1,m_2},m_2+1)\Big).
\end{eqnarray}
The pure glue requirement that $m_1-1\neq m_2+1$ is, however, not
present when there is a $\phi$ on the left tree amplitude where the
cut
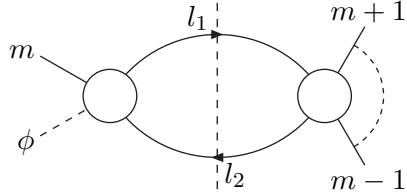
\begin{center}
\begin{picture}(180,100)(0,0)
\ArrowArcn(90,26.9)(46.2,150,30)
\ArrowArcn(90,73.1)(46.2,330,210)
\Text(87,75)[br]{$l_1$}
\Text(93,25)[tl]{$l_2$}
\DashLine(90,15)(90,85){3}
\DashLine(50,50)(24,35)3
\Line(50,50)(24,65)
\BCirc(50,50){10}
\Text(22,33)[r]{$\phi$}
\Text(22,67)[r]{$m$}
\Line(130,50)(145,76)
\Line(130,50)(145,24)
\DashCArc(130,50)(22,300,60){2}
\BCirc(130,50){10}
\Text(147,78)[b]{$m+1$}
\Text(147,22)[t]{$m-1$}
\end{picture}
\end{center}
must be taken into account. The decomposition is the same as above,
but the box function
$F_4(m_1-1,P_{m_1,m_2},m_2+1)=F_4(m,P_{m+1,m-1},m)$ is physically
nonsensical as it stands. Luckily, its cut can be rewritten,
\begin{eqnarray}
&&-\frac12\frac{\langle (m_1-1)P_{m_1,m_2}(m_2+1)P_{m_1,m_2}
(m_1-1)]}{(l_1-(m_1-1))^2(l_2+(m_2+1))^2}\nn
&=&\frac12\frac{\spab m{P_{m+1,m-1}}m^2}{\spab m{l_1}m\spab
 m{l_2}m}\nn
&=&\frac{\spab m{P_{m+1,m-1}}m}{2(l_1-m)^2}+\frac{\spab 
m{P_{m+1,m-1}}m}{2(l_2+m)^2},
\end{eqnarray}
as a sum of the cuts of the two triangle functions. But since a
triangle functions is independent of the ordering of its momenta,
these two triangles are actually the same, leaving us exactly with the
cut of $F_3(m,P_{m+1,m-1})$. There are no contributions from bubble
functions in the $\N=4$ case.

The conclusion is, that the amplitude should have the following
dependence on triangles
\begin{equation}
\frac{\spa 12^4}{\hcyc}\sum_{m=1}^{n}F_3(m,P_{m+1,m-1}).
\end{equation}

\subsubsection{Two Particle Cuts, $\N=1$ and $\N=0$ Cases}

With the two particle cuts, we have a possibility to let a full
multiplet run in the loop if we have $1^-$ and $2^-$ sitting on
different sides of the cut, that is, in the cases
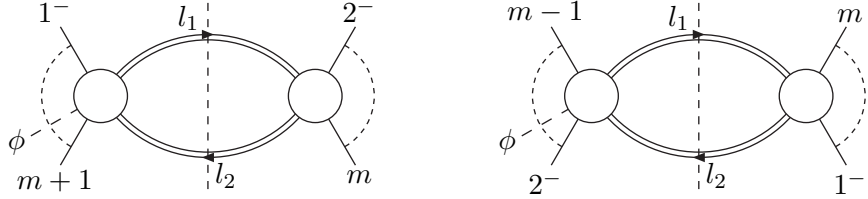
\begin{center}
\begin{picture}(180,100)(0,0)
\ArrowArcn(90,26.9)(46.2,150,30)
\ArrowArcn(90,73.1)(46.2,330,210)
\CArc(90,26.9)(44.2,30,150)
\CArc(90,73.1)(44.2,210,330)
\Text(87,75)[br]{$l_1$}
\Text(93,25)[tl]{$l_2$}
\DashLine(90,15)(90,85){3}
\Line(50,50)(35,24)
\DashLine(50,50)(24,35)4
\Line(50,50)(35,76)
\DashCArc(50,50)(22,120,240){2}
\BCirc(50,50){10}
\Text(22,33)[r]{$\phi$}
\Text(33,22)[t]{$m+1$}
\Text(33,78)[b]{$1^-$}
\Line(130,50)(145,76)
\Line(130,50)(145,24)
\DashCArc(130,50)(22,300,60){2}
\BCirc(130,50){10}
\Text(147,78)[b]{$2^-$}
\Text(147,22)[t]{$m$}
\end{picture}
\begin{picture}(180,100)(0,0)
\ArrowArcn(90,26.9)(46.2,150,30)
\ArrowArcn(90,73.1)(46.2,330,210)
\CArc(90,26.9)(44.2,30,150)
\CArc(90,73.1)(44.2,210,330)
\Text(87,75)[br]{$l_1$}
\Text(93,25)[tl]{$l_2$}
\DashLine(90,15)(90,85){3}
\Line(50,50)(35,24)
\DashLine(50,50)(24,35)4
\Line(50,50)(35,76)
\DashCArc(50,50)(22,120,240){2}
\BCirc(50,50){10}
\Text(22,33)[r]{$\phi$}
\Text(33,22)[t]{$2^-$}
\Text(33,78)[b]{$m-1$}
\Line(130,50)(145,76)
\Line(130,50)(145,24)
\DashCArc(130,50)(22,300,60){2}
\BCirc(130,50){10}
\Text(147,78)[b]{$m$}
\Text(147,22)[t]{$1^-$}
\end{picture}
\end{center}
where $m$ runs from 3 to $n$ in both. For simplicity, we only treat
the first of these.

For $\N=1$, the cut is given by
\begin{eqnarray}
&&\frac{\spa 1{l_1}\spa 1{l_2}\spa 2{l_1}\spa 2{l_2}
\Big(-\Big(\spa 1{l_1}\spa 2{l_2}-\spa 1{l_2}\spa 2{l_1}\Big)^2\Big)}
{\spa {l_1}{l_2}\spa {l_2}{(m+1)}\langle (m+1)\cdots 1\rangle
\spa 1{l_1}\spa {l_2}{l_1}\spa {l_1}2\langle 2\cdots m\rangle
\spa m{l_2}}\nn
&=&\frac{\spa 12^3\spa m{(m+1)}}{\hcyc}\frac{\spa 1{l_2}\spa{l_2}2}
{\spa m{l_2}\spa {l_2}{(m+1)}}\nn
&=&\frac{\spa 12^3}{\hcyc}\bigg(\frac{\spa m1\spa 2{l_2}}{\spa m{l_2}}
+\frac{\spa 1{(m+1)}\spa 2{l_2}}{\spa {(m+1)}{l_2}}\bigg)\nn
&=&\frac{\spa 12^3}\hcyc\bigg(\frac{\spab 2{l_2}m\spa m1}
{\spab m{l_2}m}-\frac{\spab 2{l_2}{(m+1)}\spa {(m+1)}1}{\spab
{(m+1)}{l_2}{(m+1)}}\bigg).
\end{eqnarray}
We have now written this as cuts of vector triangles. To put this into
the standard basis of scalar integrals we need to perform a
Passarino--Veltman reduction \cite{Passarino78pv}. The details of this
procedure is given in appendix \ref{app:pv}, and the result is (using
$P\equiv -P_{2,m}$):
\begin{equation}
\frac{\spa 12^3}\hcyc\bigg(\frac{\spaa 2{P_{2,m}}m1}
{\spab m{P_{2,m}}m}-\frac{\spaa 2{P_{2,m}}{(m+1)}1}{\spab
{(m+1)}{P_{2,m}}{(m+1)}}\bigg).
\end{equation}
Here we have to remember that we have suppressed the integration, so
what this really tells us is that the cuts we have calculated are the
cuts of 
\begin{equation}
c_\Gamma \sum_{m=3}^n\frac{\spa 12^3}\hcyc\bigg(\frac{\spaa 2{P_{2,m}}m1}
{\spab m{P_{2,m}}m}-\frac{\spaa 2{P_{2,m}}{(m+1)}1}{\spab
{(m+1)}{P_{2,m}}{(m+1)}}\bigg)F_2(P_{2,m}).
\end{equation}
Using that $F_2(P_{2,m})-F_2(P_{2,m-1})=-\log(P_{2,m}^2/P_{2,m-1}^2)$,
we can rewrite this as
\begin{equation}
-c_\Gamma\frac{\spa 12^3}\hcyc\sum_{m=4}^n\frac{\spaa 2{P_{2,m}}m1}
{\spab m{P_{2,m}}m}\log\bigg(\frac{P_{2,m}^2}{P_{2,m-1}^2}\bigg).
\end{equation}
The contribution from the second class of diagrams is
\begin{equation}
c_\Gamma\sum_{m=3}^n\frac{\spa 12^3}\hcyc\bigg(\frac{\spaa 1{P_{m,1}}{(m-1)}2}
{\spab {(m-1)}{P_{m,1}}{(m-1)}}-\frac{\spaa 1{P_{m,1}}m2}{\spab
m{P_{m,1}}m}\bigg)F_2(P_{m,1}),
\end{equation}
or
\begin{equation}
-c_\Gamma\frac{\spa 12^3}\hcyc\sum_{m=4}^n\frac{\spaa 1{P_{m,1}}{(m-1)}2}
{\spab {(m-1)}{P_{m,1}}{(m-1)}}\log\bigg(\frac{P_{m,1}^2}{P_{m-1,1}^2}
\bigg).
\end{equation}

The $\N=0$ case is complicated by the need for integral
reduction. However, if we plunge right into it, the cut corresponding
to the first diagram above is
\begin{eqnarray}
&&2\frac{\spa 1{l_1}^2\spa 1{l_2}^2\spa 2{l_1}^2\spa 2{l_2}^2}
{\spa {l_1}{l_2}\spa {l_2}{(m+1)}\langle (m+1)\cdots 1\rangle
\spa 1{l_1}\spa {l_2}{l_1}\spa {l_1}2\langle 2\cdots m\rangle
\spa m{l_2}}\nn
&=&2\frac{\spa 12\spa m{(m+1)}}\hcyc\frac{\spa 1{l_1}
\spa 1{l_2}^2\spa 2{l_1}\spa 2{l_2}^2}{\spa {l_1}{l_2}^2
\spa m{l_2}\spa {l_2}{(m+1)}}\nn
&=&2\frac{\spa 12}\hcyc\frac{\spa 1{l_1}\spa 1{l_2}\spa 2{l_1}
\spa 2{l_2}^2}{\spa{l_1}{l_2}^2}\bigg(\frac{\spa 1m}{\spa m{l_2}}
+\frac{\spa 1{(m+1)}}{\spa {l_2}{(m+1)}}\bigg)\nn
&=&2\frac{\spa 12}\hcyc\frac{\spaa 1{l_1}{l_2}1\spaa 2{l_1}{l_2}2}
{P_{2,m}^4}\bigg(\frac{\spaa 2{l_2}{(m+1)}1}{\spab {(m+1)}{l_2}
{(m+1)}}-\frac{\spaa 2{l_2}m1}{\spab m{l_2}m}\bigg)\nn
&=&2\frac{\spa 12}\hcyc\frac{\spaa 1{P_{2,m}}{l_2}1\spaa 2{P_{2,m}}{l_2}2}
{P_{2,m}^4}\bigg(\frac{\spaa 2{l_2}{(m+1)}1}{\spab {(m+1)}{l_2}
{(m+1)}}-\frac{\spaa 2{l_2}m1}{\spab m{l_2}m}\bigg).
\end{eqnarray}
These are cuts of triangles with three-index tensors in the
numerators. Again, we use the Passarino--Veltman reduction from
appendix \ref{app:pv} (using $P\equiv -P_{2,m}$) to reduce
\begin{eqnarray}
&&\frac{\spaa 1{P_{2,m}}{l_2}1\spaa 2{P_{2,m}}{l_2}2}
{P_{2,m}^4}\frac{\spaa 2{l_2}m1}{\spab m{l_2}m}
\end{eqnarray}
to
\begin{eqnarray}
&&\frac{\spaa 1{P_{2,m}}m1\spaa 2{P_{2,m}}m2\spaa 2{P_{2,m}}m1}
{3\spab m{P_{2,m}}m^3}-\frac{\spaa 1{P_{2,m}}{\gamma^\mu}1
\spaa 2{P_{2,m}}m2\spaa 2{\gamma_\mu}m1}{12\spab m{P_{2,m}}m^2}\nn
&&\qquad -\frac{\spaa 1{P_{2,m}}{\gamma^\mu}1\spaa 2{P_{2,m}}
{\gamma_\mu}2\spaa 2{P_{2,m}}m1}{12P_{2,m}^2\spab m{P_{2,m}}m}\nn
&=&\frac{\spaa 1{P_{2,m}}m1\spaa 2{P_{2,m}}m2\spaa 2{P_{2,m}}m1}
{3\spab m{P_{2,m}}m^3}+\frac{\spaa 2{P_{2,m}}m2\spaa 1{P_{2,m}}m1
\spa 12}{6\spab m{P_{2,m}}m^2}\nn
&&\qquad -\frac{\spa 12^2\spaa 2{P_{2,m}}m1}{6\spab m{P_{2,m}}m}\nn
&=&\frac{\spaa 1{P_{2,m}}m1\spaa 2{P_{2,m}}m2\langle 
2[{P_{2,m}},m]1\rangle}
{6\spab m{P_{2,m}}m^3}-\frac{\spa 12^2\spaa 2{P_{2,m}}m1}
{6\spab m{P_{2,m}}m}.
\end{eqnarray}
Thus, the bubble functions coming from the first diagram are
\begin{eqnarray}
&&c_\Gamma \frac{\spa 12}{3\hcyc}\bigg(\frac{\spaa 1{P_{2,m}}{(m+1)}1
\spaa 2{P_{2,m}}{(m+1)}2\langle 2[{P_{2,m}},(m+1)]1\rangle}
{\spab {(m+1)}{P_{2,m}}{(m+1)}^3}\nn
&&\qquad -\frac{\spa 12^2\spaa 2{P_{2,m}}
{(m+1)}1}{\spab {(m+1)}{P_{2,m}}{(m+1)}}
-\frac{\spaa 1{P_{2,m}}m1\spaa 2{P_{2,m}}m2\langle 
2[{P_{2,m}},m]1\rangle}
{\spab m{P_{2,m}}m^3}\nn
&&\qquad +\frac{\spa 12^2\spaa 2{P_{2,m}}m1}
{\spab m{P_{2,m}}m}\bigg)F_2(P_{2,m}),
\end{eqnarray}
while those coming from the second are
\begin{eqnarray}
&&c_\Gamma\frac{\spa 12}{3\hcyc}\bigg(-\frac{\spaa 1{P_{m,1}}{(m-1)}1
\spaa 2{P_{m,1}}{(m-1)}2\langle 2[{P_{m,1}},(m-1)]1\rangle}
{\spab {(m-1)}{P_{m,1}}{(m-1)}^3}\nn
&&\qquad +\frac{\spa 12^2\spaa 1{P_{m,1}}
{(m-1)}2}{\spab {(m-1)}{P_{m,1}}{(m-1)}}
+\frac{\spaa 1{P_{m,1}}m1\spaa 2{P_{m,1}}m2\langle 
2[{P_{m,1}},m]1\rangle}
{\spab m{P_{m,1}}m^3}\nn
&&\qquad -\frac{\spa 12^2\spaa 1{P_{m,1}}m2}
{\spab m{P_{m,1}}m}\bigg)F_2(P_{m,1}).
\end{eqnarray}
Again, these can be rewritten as
\begin{eqnarray}
\lefteqn{c_\Gamma\frac{\spa 12}{3\hcyc}\sum_{m=4}^n\bigg[\bigg(
\frac{\spaa 1{P_{2,m}}m1\spaa 2{P_{2,m}}m2\langle 
2[{P_{2,m}},m]1\rangle}{\spab m{P_{2,m}}m^3}}\nn
&&\qquad\qquad-
\frac{\spa 12^2\spaa 2{P_{2,m}}m1}{\spab m{P_{2,m}}m}\bigg)
\log\bigg(\frac{P_{2,m}^2}{P_{2,m-1}^2}\bigg)\nn
&&\qquad+\bigg(\frac{\spaa 1{P_{m,1}}{(m-1)}1
\spaa 2{P_{m,1}}{(m-1)}2\langle 2[{P_{m,1}},(m-1)]1\rangle}
{\spab {(m-1)}{P_{m,1}}{(m-1)}^3}\nn
&&\qquad\qquad-\frac{\spa 12^2\spaa 1{P_{m,1}}
{(m-1)}2}{\spab {(m-1)}{P_{m,1}}{(m-1)}}\bigg)
\log\bigg(\frac{P_{m,1}^2}{P_{m-1,1}^2}\bigg)\bigg].
\end{eqnarray}

\subsubsection{Summing Up}

We can now sum up the cut containing pieces for arbitrary fermion and
scalar content. In place of the logs we use
\begin{equation}
L_1(s,t)=\frac{\log(s/t)}{s-t},\qquad
L_3(s,t)=\frac{\log(s/t)}{(s-t)^3}.
\end{equation}
The result is
\begin{eqnarray}
\lefteqn{c_\Gamma \frac{\spa 12^4}\hcyc\Bigg\{\sum_{m_1=1}^n 
\bigg(F_3(m_1,P_{m_1+1,m_1-1})+\sum_{m_2=m_1+2}^{m_1-1}
F_4(m_1,P_{m_1+1,m_2-1},m_2)\bigg)}\nn
&&\qquad +\sum_{m=4}^{n}\Bigg[\bigg(1-\frac{n_f}N+
\frac{n_s}N\bigg)\bigg(\nn
&&\qquad\frac{\spaa 1{P_{2,m}}m1\spaa 2{P_{2,m}}m2\langle 
2[{P_{2,m}},m]1\rangle}{3\spa 12^3}L_3(P_{2,m}^2
,P_{2,m-1}^2)\nn
&&\qquad -\frac{\spaa 1{P_{m,1}}{(m-1)}1
\spaa 2{P_{m,1}}{(m-1)}2\langle 2[{P_{m,1}},(m-1)]1\rangle}
{3\spa 12^3}\nn
&&\qquad\times L_3(P_{m,1}^2,P_{m-1,1}^2)\bigg)
+\bigg(\frac{11}3-\frac{2n_f}{3N}-\frac{n_s}{3N}\bigg)
\bigg(\nn
&&\qquad\frac{\spaa 2{P_{2,m}}m1}{\spa 12}
L_1(P_{2,m}^2,P_{2,m-1}^2)\nn
&&\qquad-\frac{\spaa 1{P_{m,1}}{(m-1)}2}{\spa 12}
L_1(P_{m,1}^2,P_{m-1,1}^2)\bigg)\Bigg]\Bigg\}.
\end{eqnarray}
We can now specialize to $n=4$, since this is the amplitude we
ultimately want to calculate. There, the result is
\begin{eqnarray}
\label{eq:phiccresult}
&&c_\Gamma\frac{\spa 12^4}\hcyc\sum_{m=1}^4\bigg(F^{1m}_4(
m,m+1,m+2)+F^{2me}_4(m,P_{m+1,m+2},m+3)\nn
&&\qquad +F^{2m}_3(m,P_{m+1,m+3})\bigg)
+c_\Gamma\frac{\spb 34}{3\spa 34}
\bigg(1-\frac{n_f}N+\frac{n_s}N\bigg)\Big(\\
&&\qquad \spaa 1{P_{23}}42\langle 2[P_{23},4]1\rangle
L_3(s_{234},s_{23})+ \spaa 2{P_{41}}31
\langle 2[P_{41},3]1\rangle L_3(s_{41},s_{341})\Big)\nn
&&\qquad +c_\Gamma\frac{\spb 34\spa 12^2}{\spa 34}
\bigg(\frac{11}3-\frac{2n_f}{3N}-\frac{n_s}{3N}\bigg)
\Big(L_1(s_{234},s_{23})+L_1(s_{41},s_{341})\Big)\nonumber.
\label{eq:phimmppcc}
\end{eqnarray}

\subsection{Rational Pieces}
\label{subsec:higgsrational}

To compute the remaining parts of the amplitude we use the methods
described in section \ref{sec:looprecursion}. The total one-loop
amplitude is
\begin{equation}
\A^1=C+R,
\end{equation}
but $C$ and $R$ contain unphysical poles which cancel against each
other, but prevent us from finding $R$ directly from
recursion. Instead, we add and subtract a term $CR$ with the same
unphysical singularities as $C$,
\begin{equation}
\A^1=(C+CR)+(R-CR),
\end{equation}
such that only physical poles are left in each of the parentheses. Our
first observation about (\ref{eq:phimmppcc}) will be that the
unphysical poles only reside in the $L_3$ functions, and by rewriting
it as
\begin{eqnarray}
L_3(s,t)&=&\frac12\frac{\log\Big[1+(s-t)/t\Big]}{(s-t)^3}
+\frac12\frac{\log\Big[1+(t-s)/s\Big]}{(t-s)^3}\nn
&=&\frac12\frac1{t(s-t)^2}-\frac14\frac1{t^2(s-t)}
+\frac12\frac1{s(t-s)^2}-\frac14\frac1{s^2(t-s)}
+\mathcal O((s-t)^0)\nn
&=&\frac12\bigg(\frac1t+\frac1s\bigg)\frac1{(s-t)^2}
-\frac14\frac{s^2-t^2}{s^2t^2(s-t)}+\mathcal O((s-t)^0)\nn
&=&\frac12\bigg(\frac1t+\frac1s\bigg)\frac1{(s-t)^2}
+\mathcal O((s-t)^0),
\end{eqnarray}
we see that $CR$ must be constructed by replacing $L_3(s,t)$ by
\begin{equation}
-\frac12\bigg(\frac1t+\frac1s\bigg)\frac1{(s-t)^2}.
\end{equation}
Thus,
\begin{eqnarray}
CR&=&-\frac{c_\Gamma\spb 34}{6\spa 34}
\bigg(1-\frac{n_f}N+\frac{n_s}N\bigg)
\bigg[\frac{\spaa 1{P_{23}}42\langle 2[P_{23},4]1\rangle}
{\spab 4{P_{23}}4^2}\bigg(\frac1{s_{234}}+\frac1{s_{23}}\bigg)\nn
&&\qquad+\frac{\spaa 2{P_{41}}31\langle 2[P_{41},3]1\rangle}
{\spab 3{P_{41}}3^2}\bigg(\frac1{s_{341}}+\frac1{s_{41}}\bigg)
\bigg].
\end{eqnarray}

Knowing now that $R-CR$ has only physical poles, we can attack it with
recursion relations. For reasons which will become clear later, we
choose the recursive shift
\begin{equation}
\spxa{\wh 1}=\spxa 1+z\spxa 2,\qquad \spxb{\wh 2}=\spxb2 -z\spxb 1.
\label{eq:higgsshift}
\end{equation}
Rather than proving that this shift has the proper $z\to\infty$
behaviour, we will note that it does indeed work without the $\phi$
\cite{Kosower05mhvfinite}, and that the tests to be explained in the next
section show no signs of bad $z\to\infty$ behaviour. The results of
doing recursion on $CR$ are the overlap terms $O_{23}$, $O_{234}$,
$O_{41}$, and $O_{341}$, coming from the singularities in $s_{23}$,
\emph{etc}. These are found by leaving the kinematic invariant,
shifting its coefficient according to (\ref{eq:higgsshift}), and
inserting the $z$ which puts us exactly at the pole. Taking $O_{23}$ as
an example, we note that $CR$ has a pole where
\begin{equation}
0=\wh s_{23}=\spa 23\spb 3{\wh 2}=\spa 23(\spb 32-z\spb 31)
\quad \Rightarrow \quad z=\frac{\spb 32}{\spb 31},
\end{equation}
such that
\begin{equation}
\spxa {\wh 1}=\frac{\spxxb {P_{123}}3}{\spb 13},\qquad
\spxb {\wh 2}=\spxb 3\frac{\spb 12}{\spb 13},\qquad
\wh P_{23}=\frac{\spxxb {P_{123}}1\spbx 3}{\spb 31}, 
\label{eq:overlap23hats}
\end{equation}
and find that
\begin{eqnarray}
O_{23}&=&-\frac{c_\Gamma}6\bigg(1-\frac{n_f}N+\frac{n_s}N\bigg)
\frac{\spb 34\spaa {\wh 1}{P_{123}}42\langle 2(34-4P_{123})\wh 1\rangle}
{\spa 34\spab 4{\wh P_{23}}4^2 s_{23}}\nn
&=&-\frac{c_\Gamma}6\bigg(1-\frac{n_f}N+\frac{n_s}N\bigg)
\frac{\spb 34s_{123}\spb 34\spa 42\Big(s_{123}\spa 24\spb 43-\spa 23\spb 34
\spab 4{P_{12}}3\Big)}{\spa 34\spab 4{P_{23}}1^2\spb 34^2 s_{23}}\nn
&=&-\frac{c_\Gamma}6\bigg(1-\frac{n_f}N+\frac{n_s}N\bigg)
\frac{\spb 34s_{123}\spa 42\Big(s_{123}\spa 42
+\spab 4{P_{12}}3\spa 32\Big)}{\spa 34\spab 4{P_{23}}1^2 s_{23}}.
\end{eqnarray}
The other overlaps are
\begin{eqnarray}
O_{234}&=&-\frac{c_\Gamma}6\bigg(1-\frac{n_f}N+\frac{n_s}N\bigg)
\frac{\spb 34s_{1234}\spa 42\Big(s_{1234}\spa 42+\spaa 4{P_{123}}{P_{34}}2
\Big)}{\spa 34\spab 4{P_{23}}1^2s_{234}},\\
O_{41}&=&-\frac{c_\Gamma}6\bigg(1-\frac{n_f}N+\frac{n_s}N\bigg)
\frac{\spb 34\spa 12^2}{\spa 34s_{41}},\\
O_{341}&=&-\frac{c_\Gamma}6\bigg(1-\frac{n_f}N+\frac{n_s}N\bigg)
\frac{\spb 34\spab 2{P_{13}}4\Big(\spab 2{P_{13}}4+\spa 23\spb 34\Big)}
{\spa 34\spb 41^2 s_{341}}.
\end{eqnarray}
The hattings used in these and later calculations are given in the
table below:
\begin{displaymath}
\begin{array}{|c|c|c|c|}
\hline
i & \spxa{\wh 1} & \spxb{\wh 2} & \wh P_i\\
\hline
23 
& \displaystyle\frac{\spxxb{P_{123}}3}{\spb 13} 
& \displaystyle\spxb 3\frac{\spb 12}{\spb 13} 
& \displaystyle\frac{\spxxb{P_{123}}1\spbx 3}{\spb 31}\\
234 
& \displaystyle \frac{\spxxxa {P_{1234}}{P_{34}}2}{\spab 2{P_{34}}1}
& \displaystyle -\frac{\spxxxb {P_{34}}{P_{1234}}1}{\spab 2{P_{34}}1}
& \displaystyle\frac{\spxxb{P_{1234}}1\spaxx 2{P_{34}}}{\spab 2{P_{34}}1}\\
41 
& \displaystyle \spxa 4\frac{\spa 21}{\spa 24}
& \displaystyle \frac{\spxxa {P_{412}}4}{\spa 24}
& \displaystyle\frac{\spxa 4\spaxx 2{P_{412}}}{\spa 24}\\
341 
& \displaystyle -\frac{\spxxxa {P_{34}}{P_{1234}}2}{\spab 2{P_{34}}1}
& \displaystyle \frac{\spxxxb {P_{1234}}{P_{34}}1}{\spab 2{P_{34}}1}
& \displaystyle \frac{\spxxb{P_{34}}1\spaxx 2{P_{1234}}}{\spab 2{P_{34}}1}\\
\hline 
\end{array}
\end{displaymath}

We can now turn to the results of doing recursion on $R$. This does
not give us the $R$ since we are summing over a limited set of poles,
but rather gives us the 'direct recursive' terms $DR_{23}$,
\emph{etc}, composed of (a sum of) an unshifted propagator times a
shifted tree amplitude times a shifted rational part of a one-loop
amplitude. One complication is that the three-point one-loop vertices
cannot really be interpreted as the corresponding amplitudes, rather
they have to be deduced from the one-loop splitting functions. In the
case at hand, however, we are considering a scalar running in the loop
and we have chosen our shift such that the two external gluons in a
three-point vertex have opposite signs. Fortunately, the rational
parts of the corresponding splitting functions are all zero
(\emph{cf.}~(\ref{eq:rationalsplit})) so there are no contributions in
our case from three-point one-loop vertices. The non-zero diagrams are
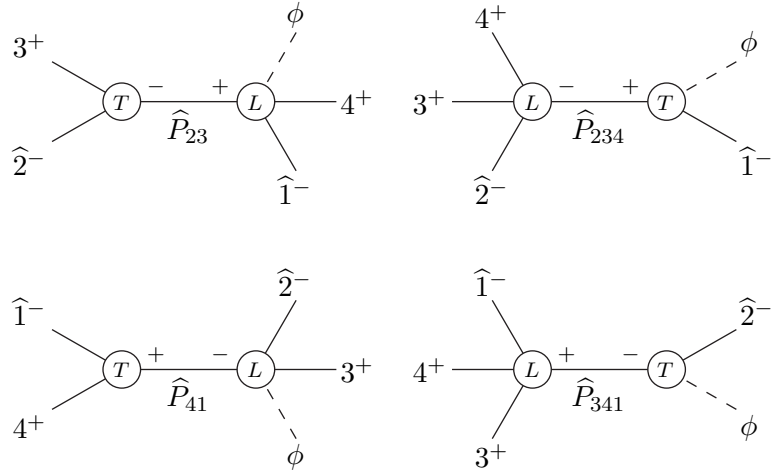
\begin{center}
\begin{picture}(150,100)(0,0)
\Line(50,50)(100,50)
\Text(63,53)[b]{\scriptsize $-$}
\Text(87,53)[b]{\scriptsize $+$}
\Text(75,47)[t]{$\widehat P_{23}$}
\Line(50,50)(24,35)
\Line(50,50)(24,65)
\Text(22,33)[rt]{$\widehat 2^-$}
\Text(22,67)[rb]{$3^+$}
\BCirc(50,50){7}
\Text(50,50)[]{\scriptsize $T$}
\DashLine(100,50)(115,76)4
\Line(100,50)(130,50)
\Line(100,50)(115,24)
\Text(115,78)[b]{$\phi$}
\Text(132,50)[l]{$4^+$}
\Text(115,22)[t]{$\widehat 1^-$}
\BCirc(100,50){7}
\Text(100,50)[]{\scriptsize $L$}
\end{picture}
\begin{picture}(150,100)(0,0)
\Line(50,50)(100,50)
\Text(63,53)[b]{\scriptsize $-$}
\Text(87,53)[b]{\scriptsize $+$}
\Text(75,47)[t]{$\widehat P_{234}$}
\Line(50,50)(35,24)
\Line(50,50)(20,50)
\Line(50,50)(35,76)
\Text(35,22)[t]{$\widehat 2^-$}
\Text(18,50)[r]{$3^+$}
\Text(35,78)[b]{$4^+$}
\BCirc(50,50){7}
\Text(50,50)[]{\scriptsize $L$}
\DashLine(100,50)(126,65)4
\Line(100,50)(126,35)
\Text(128,67)[lb]{$\phi$}
\Text(128,33)[lt]{$\widehat 1^-$}
\BCirc(100,50){7}
\Text(100,50)[]{\scriptsize $T$}
\end{picture}\\
\begin{picture}(150,100)(0,0)
\Line(50,50)(100,50)
\Text(63,53)[b]{\scriptsize $+$}
\Text(87,53)[b]{\scriptsize $-$}
\Text(75,47)[t]{$\widehat P_{41}$}
\Line(50,50)(24,35)
\Line(50,50)(24,65)
\Text(22,33)[rt]{$4^+$}
\Text(22,67)[rb]{$\widehat 1^-$}
\BCirc(50,50){7}
\Text(50,50)[]{\scriptsize $T$}
\Line(100,50)(115,76)
\Line(100,50)(130,50)
\DashLine(100,50)(115,24)4
\Text(115,78)[b]{$\widehat 2^-$}
\Text(132,50)[l]{$3^+$}
\Text(115,22)[t]{$\phi$}
\BCirc(100,50){7}
\Text(100,50)[]{\scriptsize $L$}
\end{picture}
\begin{picture}(150,100)(0,0)
\Line(50,50)(100,50)
\Text(63,53)[b]{\scriptsize $+$}
\Text(87,53)[b]{\scriptsize $-$}
\Text(75,47)[t]{$\widehat P_{341}$}
\Line(50,50)(35,24)
\Line(50,50)(20,50)
\Line(50,50)(35,76)
\Text(35,22)[t]{$3^+$}
\Text(18,50)[r]{$4^+$}
\Text(35,78)[b]{$\widehat 1^-$}
\BCirc(50,50){7}
\Text(50,50)[]{\scriptsize $L$}
\Line(100,50)(126,65)
\DashLine(100,50)(126,35)4
\Text(128,67)[lb]{$\widehat 2^-$}
\Text(128,33)[lt]{$\phi$}
\BCirc(100,50){7}
\Text(100,50)[]{\scriptsize $T$}
\end{picture}
\end{center}
The relevant lower-point amplitudes with a $\phi$ can be found in
\cite{Berger06phinite}
\begin{eqnarray}
A^1(\phi,1^-,2^-)&=&\frac1{8\pi^2}A^0(\phi,1^-,2^-)\nn
R(\phi,1^-,2^-,3^+)&=&\frac1{8\pi^2}A^0(\phi,1^-,2^-,3^+)\nn
A^1(\phi,1^-,2^+,3^+)&=&\frac1{48\pi^2}\bigg(1-\frac{n_f}N
+\frac{n_s}N\bigg)\frac{\spa 12\spa 31\spb 23}{\spa 23^2}\nn
&&\qquad -\frac1{8\pi^2}A^0(\phi^\dagger,1^-,2^+,3^+)\nn
\end{eqnarray}
while the last was calculated in \cite{Bern91stringrules2}
\begin{eqnarray}
A^1(1^-,2^+,3^+,4^+)&=&\frac1{48\pi^2}\bigg(1-\frac{n_f}N
+\frac{n_s}N\bigg)\frac{\spa 24\spb 24^3}{\spb 12\spa 23\spa 34
\spb 41}.
\end{eqnarray}
This gives the direct recursive terms
\begin{eqnarray}
DR_{23}&=&
\A_3(\wh 2^-,3^+,-\wh P_{23}^-)\frac1{P_{23}^2}R_4(\phi,4^+,\wh 1^-,\wh
 P_{23}^+)\nn
&=&\frac{c_\Gamma}3\bigg(1-\frac{n_f}N+\frac{n_s}N\bigg)
\frac{\spb 34 s_{123}\spb 13\spab 4{P_{12}}3}{\spb 12\spb 23
\spab 4{P_{23}}1^2}\nn
&&\qquad -2c_\Gamma \A^0(\phi^\dagger,1^-,2^-,3^+,4^+)
\end{eqnarray}
\begin{eqnarray}
DR_{234}&=&
R_4(\wh 2^-,3^+,4^+,-\wh P_{234}^+)\frac1{P_{234}^2}\A_3(\phi,\wh 1^-,\wh
 P_{234}^-)\nn
&=&\frac{c_\Gamma}3\bigg(1-\frac{n_f}N+\frac{n_s}N\bigg)
\frac{\spb 34 s_{1234}^2\spa 24^3\spab 3{P_{24}}1}
{\spa 34^2s_{234}\spab 2{P_{34}}1\spab 4{P_{23}}1^2}
\end{eqnarray}
\begin{eqnarray}
DR_{41}&=&
\A_3(4^+,\wh 1^-,-\wh P_{41}^+)\frac1{P_{41}^2}R_4(\phi,\wh 2^-,3^+,\wh
 P_{41}^-)\nn
&=&2c_\Gamma \A^0(\phi,1^-,2^-,3^+,4^+)
\end{eqnarray}
\begin{eqnarray}
DR_{341}&=&
R_4(3^+,4^+,\wh 1^-,-\wh P_{341}^+)\frac1{P_{341}^2}\A_3(\phi,\wh 2^-,\wh
 P_{341}^-)\nn
&=&\frac{c_\Gamma}3\bigg(1-\frac{n_f}N+\frac{n_s}N\bigg)
\frac{\spb 31\spab 2{P_{13}}4^3}{\spa 34 s_{341}\spb 41^2
\spab 2{P_{34}}1}.
\end{eqnarray}
To obtain $R$, we now have to add the $DR$ terms and $CR$ and subtract
the $O$ terms. Some lengthy but trivial manipulations give
\begin{eqnarray}
\label{eq:phimhvrational}
R&=&
2c_\Gamma i\A^0(A,1^-,2^-,3^+,4^+)+CR+\frac{c_\Gamma\spb 34}{3\spa 34}
\bigg(1-\frac{n_f}N+\frac{n_s}N\bigg)\bigg[\nn
&&\qquad \frac{\spa 23\spab 1{P_{24}}3^2}{\spa 34\spb 43
\spb 32 s_{234}}-\frac{\spa 41\spab 3{P_{12}}3}{\spa 34\spb 12
\spb 32}+\frac{\spa 14\spab 2{P_{13}}4^2}{\spa 34\spb 43\spb 41 
s_{341}}\nn
&&\qquad -\frac{\spa 32\spab 4{P_{12}}4}{\spa 34\spb 12\spb 41}
+\frac{\spa 12\spab 2{P_{13}}4}{2\spb 41s_{341}}
-\frac{\spa 12\spab 1{P_{24}}3}{2\spb 32s_{234}}\nn
&&\qquad -\spa 12^2\bigg(\frac1{s_{34}}+\frac1{s_{12}}+\frac1{2s_{23}}
+\frac1{2s_{41}}\bigg)\bigg].
\end{eqnarray}
It is of special interest here that the first term is present
regardless of supersymmetry, where we might have expected that an
amplitude with supersymmetric particle content running in the loop
would be fully cut-constructible. The catch is, of course, that
cut-constructibility requires also the external particles to be
supersymmetric, something which is not the case with $\phi$.
With those arguments it may seem slightly surprising (although
welcome) that it drops out of $\A^1(H,1^-,2^-,3^+,4^+)$ when we use
(\ref{eq:higgsfromphi}). In $\A^1(A,1^-,2^-,3^+,4^+)$, of course, the
additional rational piece stays.

\subsection{Tests}
\label{subsec:higgstests}

There are a series of tests which we can perform to check that the
result is correct. The first regards only the cut-containing part, in
particular the IR divergences in $\epsilon$. These are given by
(\emph{cf.}~section \ref{sec:looplimit})
\begin{equation}
\A^1=-\frac{c_\Gamma}{\epsilon^2}\A^0\sum_{i=1}^4\bigg(\frac{
\mu^2}{-s_{i,i+1}}\bigg)^\epsilon+\mathcal O(\epsilon^0).
\end{equation}
This can be checked readily by taking the IR divergent parts of the
box and triangle functions from appendix \ref{app:integrals} and
putting them into (\ref{eq:phiccresult}). There are no contributions
from the bubble-like and rational parts.

\subsubsection{Collinear Limits}

The second test is that of collinear limits. The full checking of all
collinear limits is a rather tedious exercise and will not be repeated
here. It is contained in the original article in detail. Since the
cut-containing parts have been fixed by unitarity, we expect the right
collinear limits automatically; for the rational parts, the use of
recursion amounts exactly to enforcing the correct collinear limits
for gluons 4 and 1, and for gluons 2 and 3. Checking these limits
provide primarily a check on the algebra. 

From the perspective of on-shell recursion, the important collinear
limits are the rational ones which do not participate in the direct
recursive terms. If these turn out wrong, it is a sign that the use of
recursion was not justified because $R(z)-CR(z)$ did not vanish as
$z\to\infty$. In the case at hand, the important limits in this sense
is when 1 and 2 become collinear and when 3 and 4 become collinear.

For these checks, we can use the rational parts of the splitting
functions for a scalar running in the loop (\ref{eq:rationalsplit})
along with the tree splitting functions (\ref{eq:treesplit}). These
predict that
\begin{eqnarray}
R(\phi,1^-,2^-,3^+,4^+)&\to&
-\frac1{\sqrt{z(1-z)}\spb 12}R(\phi,P^-,3^+,4^+)
\end{eqnarray}
as 1 and 2 go collinear, and
\begin{eqnarray}
\label{eq:col34}
R(\phi,1^-,2^-,3^+,4^+)&\to&
\frac1{\sqrt{z(1-z)}\spa 34}R(\phi,1^-,2^-,P^+)\nn
&&+\frac{c_\Gamma}3\bigg(1-\frac{n_f}N+\frac{n_s}N\bigg)\sqrt{z(1-z)}\nn
&&\qquad\times\bigg(-\frac1{\spa 34}
\A^0(\phi,1^-,2^-,P^+)\nn
&&\qquad\qquad+\frac{\spb 34}{\spa 34^2}
\A^0(\phi,1^-,2^-,P^-)\bigg)
\end{eqnarray}
as 3 and 4 go collinear. We can prove the first by first extracting
the part of (\ref{eq:phimhvrational}) that has the correct
singularity, 
\begin{eqnarray}
R&\to&2c_\Gamma\frac{-\spb 34^3}{\spb 12\spb 23\spb 41}
-\frac{c_\Gamma\spb 34}{3\spa 34^2}
\bigg(1-\frac{n_f}N+\frac{n_s}N\bigg)
\bigg[\frac{\spa 41\spab 3{P_{12}}3}{\spb 12\spb 32}
+\frac{\spa 32\spab 4{P_{12}}4}{\spb 12\spb 41}\bigg]\nn
&=&\frac{-1}{\sqrt{z(1-z)}\spb 12}\bigg[
-2c_\Gamma \A^0(\phi^\dagger,P^-,3^+,4^+)\nn
&&\qquad+\frac{c_\Gamma\spb 34}
{3\spa 34^2}\bigg(1-\frac{n_f}N+\frac{n_s}N\bigg)\spa 4P\spa P3
\bigg]\nn
&\to&\frac{-1}{\sqrt{z(1-z)}\spb 12}R(\phi,P^-,3^+,4^+),
\end{eqnarray}
as wanted. For the collinear limit between 3 and 4 things are a bit
more messy. Because $R(\phi,1^-,2^-,P^+)\sim \A^0(\phi,1^-,2^-,P^+)$,
the first line of (\ref{eq:col34}) comes from the first term of
(\ref{eq:phimhvrational}) in the same way as for 1 and 2. The
remaining terms with the collinear singularity are
\begin{eqnarray}
&&\frac{c_\Gamma\spb 34}{3\spa 34}
\bigg(1-\frac{n_f}N+\frac{n_s}N\bigg)\bigg[
\frac{\spa 23\spab 1{P_{24}}3^2}{\spa 34\spb 43\spb 32s_{234}}
+\frac{\spa 14\spab 2{P_{31}}4^2}{\spa 34\spb 43\spb 41s_{341}}\nn
&&\qquad -\frac{\spa 41\spab 3{P_{12}}3}{\spa 34\spb 12\spb 32}
-\frac{\spa 32\spab 4{P_{12}}4}{\spa 34\spb 12\spb 41}
-\frac{\spa 12^2}{\spa 34\spb 43}\bigg]\nn
&\to&\frac{c_\Gamma\spb 34}{3\spa 34}
\bigg(1-\frac{n_f}N+\frac{n_s}N\bigg)\bigg[
\frac{\spa 23\spa 12\spb23\Big(\spa 12\spb 23+2\spa 14\spb 43\Big)}
{\spa 34\spb 43\spb 32s_{234}}\nn
&&\qquad +\frac{\spa 14\spa 21\spb 14\Big(\spa 21\spb 14
+2\spa 23\spb 34\Big)}{\spa 34\spb 43\spb 41s_{341}}
-\frac{\spa 12^2}{\spa 34\spb 43}\nn
&&\qquad -\frac{\sqrt{z(1-z)}}{\spa 34}\bigg(\frac{\spa P1
\spab P{P_{12}}P}{\spb 12\spb P2}+\frac{\spa P2\spab P{P_{12}}P}
{\spb 12\spb P1}\bigg)\bigg]\nn
&=&\frac{c_\Gamma\spb 34}{3\spa 34}
\bigg(1-\frac{n_f}N+\frac{n_s}N\bigg)\bigg[
\frac{\spa 12^2}{s_{34}}\Big(\frac{s_{23}}{s_{234}}+\frac{s_{41}}
{s_{341}}-1\Big)+2\frac{\spa 12\spa 23\spa 41}{\spa 34}\nn
&&\qquad\times\Big(\frac1{s_{234}}+\frac1{s_{341}}\Big)
-\frac{\sqrt{z(1-z)}}{\spa 34}\frac{\spab P{P_{12}}P^2}
{\spb 12\spb 2P\spb P1}\bigg]\nn
&\to&\frac{c_\Gamma\spb 34}{3\spa 34}
\bigg(1-\frac{n_f}N+\frac{n_s}N\bigg)\bigg[
\frac{\spa 12^3\spb 24\spb 13}{\spb 34s_{234}s_{341}}+
\frac{\spa 12^2\spb 21\spa 41\spa 23}{\spa 34s_{234}s_{341}}\nn
&&\qquad +2\frac{\spa 12\spa 23\spa 41(\spab 3{P_{12}}3
+\spab 4{P_{12}}4)}{\spa 34s_{234}s_{341}}
-\frac{\sqrt{z(1-z)}}{\spa 34}\frac{\spab P{P_{12}}P^2}
{\spb 12\spb 2P\spb P1}\bigg]\nn
&\to&\frac{c_\Gamma\sqrt{z(1-z)}}3
\bigg(1-\frac{n_f}N+\frac{n_s}N\bigg)\bigg[
-\frac{\spa 12^3}{\spa 34\spa 2P\spa P1}
-\frac{\spb 34m_\phi^4}{\spa 34^2
\spb 12\spb 2P\spb P1}\bigg]\nn
&=&\frac{c_\Gamma\sqrt{z(1-z)}}3
\bigg(1-\frac{n_f}N+\frac{n_s}N\bigg)\nn
&&\qquad\times\bigg[-
\frac{\A^0(\phi,1^-,2^-,P^+)}{\spa 34}+\frac{\spb 34
\A^0(\phi,1^-,2^-,P^-)}{\spa 34^2}\bigg],
\end{eqnarray}
as wanted. Notice how we had to rewrite a contribution which
superficially went as $s_{34}^{-1}$ to one with either a $\spa 34^{-1}$
or a $\spb 34^{-1}$ divergence. As noted above, the correctness of
these two collinear limits are an indicative necessary condition for
the recursion relation to hold.

\subsubsection{Soft Higgs Limit}

The very last test we will make is the limit as the Higgs momentum
goes soft. In fact, it turns out to fail, but only because the naive
expectation for the soft limit is wrong. The naive expectation comes
from noting that as the Higgs momentum goes to zero, the $H$ field in
the effective correction $CHF_{\mu\nu}F^{\mu\nu}$ becomes a constant,
and the term becomes proportional to the Higgsless gluon term. This
tells us that the first order term in $C$ should be obtainable in this
limit as
\begin{eqnarray}
\A^l(k_H\to 0,n\mathrm{~gluons})&=&Cg\frac{\partial}{\partial g}
\A^l(n\mathrm{~gluons})\nn
&=&C(n-2+2l)\A^l(n\mathrm{~gluons}),
\end{eqnarray}
where $g$ is the gauge coupling and $l$ is the number of loops. From
looking at the form of the $\phi$-MHV amplitude and the
$\phi^\dagger$-googly-MHV amplitude, it is reasonable to guess that
the corresponding limits for $\phi$ and $\phi^\dagger$ are
\begin{eqnarray}
\A^l(k_\phi\to 0,n_-g^-,n_+g^+)
&=&(n_--1+l) \A^l(n_-g^-,n_+g^+),\nn
\A^l(k_{\phi^\dagger}\to 0,n_-g^-,n_+g^+)
&=&(n_+-1+l) \A^l(n_-g^-,n_+g^+).
\end{eqnarray}
In other words, we should be trying to prove that
\begin{equation}
\A^1(\phi,1^-,2^-,3^+,4^+)\to 2\A^1(1^-,2^-,3^+,4^+)
\end{equation}
as $k_\phi\to 0$. $\A^1(1^-,2^-,3^+,4^+)$ in the FDH renormalization
scheme has the cut-containing and rational parts \cite{Bern91stringrules2} 
\begin{eqnarray}
C(1^-,2^-,3^+,4^+)&=&c_\Gamma
\A^0(1^-,2^-,3^+,4^+)\bigg[2F^{0m}_4(1,2,3,4)\nn
&&\qquad-\bigg(\frac{11}3
-\frac{2n_f}{3N}-\frac{n_s}{3N}\bigg)F_2(s_{23})\bigg]\nn
R(1^-,2^-,3^+,4^+)&=&\frac1{72\pi^2}\bigg(1-\frac{n_f}N+
\frac{n_s}N\bigg)\A^0(1^-,2^-,3^+,4^+).
\end{eqnarray}

The cut-containing part can be checked quite easily by taking the
expression (\ref{eq:phiccresult}) and converting the $L_k$ functions
back into $F_2$ bubble functions and then treating each term in
turn. In short, only the one-mass box (which becomes zero-mass) and
$F_2(s_{23})$ survive the limit, and for the bubble functions it is
only the last line of (\ref{eq:phiccresult}) that survives.

The limit of the rational part is in principle done in the same
way. The term proportional to $\A^0(A,1^-,2^-,3^+,4^+)$ vanishes in the
limit, and there are several terms which must be considered together
if the limit has to be taken consistently. In the end we find that
\begin{equation}
R(\phi,1^-,2^-,3^+,4^+)\to
-\frac1{48\pi^2}\bigg(1-\frac{n_f}N+\frac{n_s}N\bigg) 
\A^0(1^-,2^-,3^+,4^+),
\end{equation}
as $k_\phi\to 0$. The corresponding $\phi^\dagger$ rational part has
the soft limit
\begin{eqnarray}
R(\phi^\dagger,1^-,2^-,3^+,4^+)&=&R(\phi,1^+,2^+,3^-,4^-)^*\nn
&\to&-\frac1{48\pi^2}\Bigg(1-\frac{n_f}N+\frac{n_s}N\bigg)
\A^0(1^+,2^+,3^-,4^-)^*\nn
&=&-\frac1{48\pi^2}\Bigg(1-\frac{n_f}N+\frac{n_s}N\bigg)
\A^0(1^-,2^-,3^+,4^+),
\end{eqnarray}
such that the Higgs soft limit is
\begin{equation}
R(H,1^-,2^-,3^+,4^+)\to
-\frac1{24\pi^2}\bigg(1-\frac{n_f}N+\frac{n_s}N
\bigg)\A^0(1^-,2^-,3^+,4^+),
\end{equation}
which is a factor $-1/3$ away from the expectation. This sort of
violation was anticipated already in \cite{Berger06phinite} where it
was pointed out that the Higgs momentum can act as a kind of effective
infrared regulator and that this would cause an exchange-of-limits
problem with the dimensional regularization. Thus, we conclude that
our results are indeed correct.

\appendix

\chapter{Integral Functions}
\label{app:integrals}

The box, triangle and bubble integrals used in this thesis can be
obtained from \cite{Bern95oneloopn1mhv}, though with slightly
different conventions. The boxes are given by
\begin{eqnarray}
&&I_4(K_1,K_2,K_3,K_4)\nn
&=&-i\mu^{2\epsilon}
\int\frac{d^{4-2\epsilon}L}{(2\pi)^{4-2\epsilon}}
\frac1{L^2(L+K_1)^2(L+K_1+K_2)^2(L-K_4)^2}
\end{eqnarray}
where the usual $i\epsilon$ prescribtion is understood. The integrals
are classified according to which of the momenta are massive and which
are massless. 1-mass integrals have one massive momentum, 4-mass
integrals have four. There are two different 2-mass configurations; if
the massive momenta are opposite it is called the 2-mass-easy and if
they are adjacent it is called the 2-mass-hard.

We give the integrals in terms of the box functions $F_4(\ldots)$
which are dimensionless because they are multiplied by the Gram
determinant $\mathcal G$ of that integral. We also take out a factor
$r_\Gamma$ coming from dimensional reduction,
\begin{equation}
I_4(\ldots)=\frac{c_\Gamma}{\mathcal G} F_4(\ldots)
\end{equation}
where
\begin{equation}
\label{eq:cGamma}
c_\Gamma=\frac1{(4\pi)^{2-\epsilon}}\frac{\Gamma(1+\epsilon)\Gamma^2
(1-\epsilon)}{\Gamma(1-2\epsilon)}=\frac1{16\pi^2}+\mathcal O(\epsilon).
\end{equation}
We also define the kinematic invariants
\begin{equation}
s=(K_1+K_2)^2,\qquad t=(K_2+K_3)^2.
\end{equation}
The box functions through 3-mass are,
\begin{eqnarray}
&&F_4^{0m}(k_1,k_2,k_3,k_4)\nn
&=&-\frac1{\epsilon^2}\bigg[\bigg(\frac{\mu^2}{-s}\bigg)^\epsilon
+\bigg(\frac{\mu^2}{-t}\bigg)^\epsilon\bigg]
+\log^2\bigg(\frac{-s}{-t}\bigg)+\pi^2\\
&&F_4^{1m}(k_1,k_2,k_3,K_4)\nn
&=&-\frac1{\epsilon^2}\bigg[\bigg(\frac{\mu^2}{-s}\bigg)^\epsilon
+\bigg(\frac{\mu^2}{-t}\bigg)^\epsilon
-\bigg(\frac{\mu^2}{-K_4^2}\bigg)^\epsilon\bigg]\nn
&&\qquad +\Li\bigg(1-\frac{K_4^2}s\bigg)
+\Li\bigg(1-\frac{K_4^2}t\bigg)
+\frac12\log^2\bigg(\frac{-s}{-t}\bigg)
+\frac{\pi^2}6\\
&&F_4^{2me}(k_1,K_2,k_3,K_4)\nn
&=&-\frac1{\epsilon^2}\bigg[\bigg(\frac{\mu^2}{-s}\bigg)^\epsilon
-\bigg(\frac{\mu^2}{-K_4^2}\bigg)^\epsilon
+\bigg(\frac{\mu^2}{-t}\bigg)^\epsilon
-\bigg(\frac{\mu^2}{-K_2^2}\bigg)^\epsilon
-\bigg(\frac{\mu^2}{-K_4^2}\bigg)^\epsilon\bigg]\nn &&\qquad
+\Li\bigg(1-\frac{K_2^2}{s}\bigg)+\Li\bigg(1-\frac{K_2^2}
{t}\bigg)+\Li\bigg(1-\frac{K_4^2}{s}\bigg)\nn
&&\qquad+\Li\bigg(1-\frac{K_4^2}{t}\bigg)-\Li\bigg(1-\frac{K_2^2K_4^2}
{st}\bigg)+\frac12\log^2\bigg(\frac st\bigg)\\
&&F_4^{2mh}(k_1,k_2,K_3,K_4)\nn
&=&-\frac 1{\epsilon^2}\bigg[\bigg(\frac{\mu^2}{-s}\bigg)^\epsilon
+\bigg(\frac{\mu^2}{-t}\bigg)^\epsilon-\bigg(\frac{\mu^2}{-K_3^2}
\bigg)^\epsilon-\bigg(\frac{\mu^2}{-K_4^2}\bigg)^\epsilon
+\frac12\bigg(-\frac{\mu^2s}{K_3^2K_4^2}\bigg)^\epsilon\bigg]\nn
&&\qquad +\Li\bigg(1-\frac{K_3^2}t\bigg)+\Li\bigg(1-\frac{K_4^2}t
\bigg)+\frac12\log^2\bigg(\frac st\bigg)\\
&&F_4^{3m}(k_1,K_2,K_3,K_4)\nn
&=&-\frac1{\epsilon^2}\bigg[\bigg(\frac{\mu^2}{-s}\bigg)^\epsilon
+\bigg(\frac{\mu^2}{-t}\bigg)^\epsilon-\bigg(\frac{\mu^2}
{-K_2^2}\bigg)^\epsilon-\bigg(\frac{\mu^2}{-K_3^2}
\bigg)^\epsilon-\bigg(\frac{\mu^2}{-K_4^2}\bigg)^\epsilon\nn
&&\qquad +\frac12\bigg(-\frac{\mu^2t}{K_2^2K_3^2}\bigg)^\epsilon
+\frac12\bigg(-\frac{\mu^2s}{K_3^2K_4^2}\bigg)^\epsilon\bigg]
+\Li\bigg(1-\frac{K_2^2}{s}\bigg)\nn
&&\qquad +\Li\bigg(1-\frac{K_4^2}{t}\bigg)-\Li\bigg(1-\frac{K_2^2
K_4^2}{st}\bigg)+\frac12\log^2\bigg(\frac st\bigg)
\end{eqnarray}
The Gram determinants are given by
\begin{equation}
\mathcal G^{0m}=
\mathcal G^{1m}=
\mathcal G^{2mh}=-\frac12st,
\end{equation}
\begin{equation}
\mathcal G^{2me}=
\mathcal G^{3m}=-\frac12\Big(st-K_2^2K_4^2\Big).
\end{equation}
Triangles are defined by the integral
\begin{equation}
I_3(K_1,K_2,K_3)=i\mu^2\int\frac{d^{4-2\epsilon}L}{(2\pi)^{4-2\epsilon}}
\frac1{L^2(L+K_1)^2(L-K_3)^2},
\end{equation}
again with $i\epsilon$ understood. They have the same classification
in terms of massive corners as the boxes. As in the box case, we can
take factors out of the integral and define triangle functions
\begin{equation}
I_3=\frac{c_\Gamma}{\mathcal G}F_3.
\end{equation}
The triangle functions are (omitting the 3-mass) 
\begin{eqnarray}
F_3^{1m}(K_1,k_2,k_3)
&=&\frac1{\epsilon^2}\bigg(\frac{\mu^2}{-K_1^2}
\bigg)^\epsilon\\
F_3^{2m}(K_1,K_2,k_3)
&=&\frac1{\epsilon^2}\bigg[\bigg(\frac{\mu^2}{-K_1^2}\bigg)^\epsilon
-\bigg(\frac{\mu^2}{-K_2^2}\bigg)^\epsilon\bigg]
\end{eqnarray}
where
\begin{eqnarray}
\mathcal G^{1m}&=&-K_1^2\nn
\mathcal G^{2m}&=&-K_1^2+K_2^2.
\end{eqnarray}
The bubble integral is given by
\begin{eqnarray}
I_2(K)&=&-i\mu^{2\epsilon}\int\frac{d^{4-2\epsilon}L}{(2\pi)^{4-2\epsilon}}
\frac1{L^2(L+K)^2}\nn
&=&\frac{c_\Gamma}{\epsilon(1-2\epsilon)}\bigg(\frac{\mu^2}{-K^2}
\bigg)^\epsilon\nn
&=&c_\Gamma\bigg[\frac1\epsilon+\log\bigg(\frac{\mu^2}{-K^2}\bigg)+2
\bigg]+\mathcal O(\epsilon)\nn
&\equiv&c_\Gamma F_2(K^2).
\end{eqnarray}

\chapter{Solutions of Quadruple Cut Constraints}
\label{app:quad}

This appendix lists the solutions to the cut constraints in a
quadruple cut of a one-loop amplitude. For each number of massive
corners there are two solutions which are distinguished by which
massless corners are MHV and which are googly-MHV. The two solutions
will always be related by flipping
($\spxa\cdot\leftrightarrow\spxb\cdot$) but they are all included for
clarity. The four-mass cut is not used in this thesis and the reader
is referred to the original article for those solutions
\cite{Britto04quad}. It should also be noted that all the solutions
here follow from the three-mass cases (which follow from four-mass) by
taking external momenta lightlike, but again, all cases are included
to benefit maximally from the masslessness of some corners.

Our conventions are to denote the internal on-shell momenta $l_1$
through $l_4$ starting from the bottom of the diagram and moving
clockwise around the loop. These momenta point in the clockwise
direction. The corners are denoted by $A$ through $D$ starting in the
lower left corner. The outgoing momenta at those corners are denoted
by the corresponding letter in uppercase when it is massive and in
lowercase when it is massless. Massless corners are denoted by
\begin{picture}(10,10)(0,0)
\BCirc(5,5){5}
\Line(0,5)(10,5)
\end{picture}
 when they are MHV and by 
\begin{picture}(10,10)(0,0)
\BCirc(5,5){5}
\Line(0,5)(10,5)
\Line(5,0)(5,10)
\end{picture}
when they are googly-MHV.\\
\parbox{55mm}
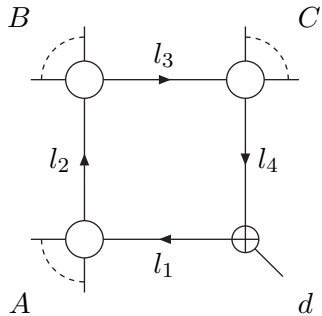
{
\begin{picture}(140,140)(0,0)
\ArrowLine(40,40)(40,100)
\ArrowLine(40,100)(100,100)
\ArrowLine(100,100)(100,40)
\ArrowLine(100,40)(40,40)
\Text(70,35)[t]{$l_1$}
\Text(35,70)[r]{$l_2$}
\Text(70,105)[b]{$l_3$}
\Text(105,70)[l]{$l_4$}
\Line(40,40)(40,20)
\Line(40,40)(20,40)
\Text(20,20)[tr]{$A$}
\DashCArc(40,40)(16,180,270){2}
\BCirc(40,40){7}
\Line(40,100)(20,100)
\Line(40,100)(40,120)
\Text(20,120)[br]{$B$}
\DashCArc(40,100)(16,90,180){2}
\BCirc(40,100){7}
\Line(100,100)(100,120)
\Line(100,100)(120,100)
\Text(120,120)[bl]{$C$}
\DashCArc(100,100)(16,0,90){2}
\BCirc(100,100){7}
\Line(100,40)(114,26)
\Text(120,20)[tl]{$d$}
\BCirc(100,40){5}
\Line(95,40)(105,40)
\Line(100,35)(100,45)
\end{picture}}
\parbox{80mm}{
\begin{eqnarray}
&&l_1=\frac{\spxa d\langle dCBA|}{\spaa dACd}\nn
&&l_2=\frac{|BCd\rangle\spaxx dA}{\spaa dACd}\nn
&&l_3=\frac{|BAd\rangle\spaxx dC}{\spaa dACd}\nn
&&l_4=\frac{\spxa d\langle dABC|}{\spaa dACd}
\end{eqnarray}}
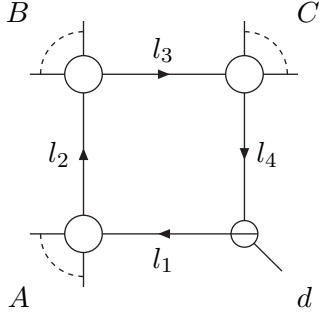
\parbox{55mm}{
\begin{picture}(140,140)(0,0)
\ArrowLine(40,40)(40,100)
\ArrowLine(40,100)(100,100)
\ArrowLine(100,100)(100,40)
\ArrowLine(100,40)(40,40)
\Text(70,35)[t]{$l_1$}
\Text(35,70)[r]{$l_2$}
\Text(70,105)[b]{$l_3$}
\Text(105,70)[l]{$l_4$}
\Line(40,40)(40,20)
\Line(40,40)(20,40)
\Text(20,20)[tr]{$A$}
\DashCArc(40,40)(16,180,270){2}
\BCirc(40,40){7}
\Line(40,100)(20,100)
\Line(40,100)(40,120)
\Text(20,120)[br]{$B$}
\DashCArc(40,100)(16,90,180){2}
\BCirc(40,100){7}
\Line(100,100)(100,120)
\Line(100,100)(120,100)
\Text(120,120)[bl]{$C$}
\DashCArc(100,100)(16,0,90){2}
\BCirc(100,100){7}
\Line(100,40)(114,26)
\Text(120,20)[tl]{$d$}
\BCirc(100,40){5}
\Line(95,40)(105,40)
\end{picture}}
\parbox{80mm}{
\begin{eqnarray}
&&l_1=\frac{|ABCd]\spbx d}{\spbb dCAd}\nn
&&l_2=\frac{\spxxb Ad\spbxxx dCB}{\spbb dCAd}\nn
&&l_3=\frac{\spxxb Cd\spbxxx dAB}{\spbb dCAd}\nn
&&l_4=\frac{|CBAd]\spbx d}{\spbb dCAd}
\end{eqnarray}}
\parbox{55mm}
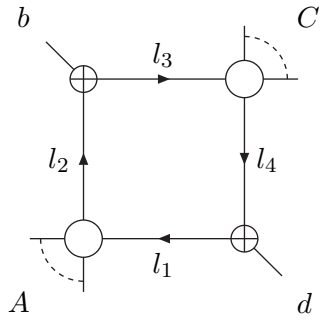
{
\begin{picture}(140,140)(0,0)
\ArrowLine(40,40)(40,100)
\ArrowLine(40,100)(100,100)
\ArrowLine(100,100)(100,40)
\ArrowLine(100,40)(40,40)
\Text(70,35)[t]{$l_1$}
\Text(35,70)[r]{$l_2$}
\Text(70,105)[b]{$l_3$}
\Text(105,70)[l]{$l_4$}
\Line(40,40)(40,20)
\Line(40,40)(20,40)
\Text(20,20)[tr]{$A$}
\DashCArc(40,40)(16,180,270){2}
\BCirc(40,40){7}
\Line(40,100)(26,114)
\Text(20,120)[br]{$b$}
\BCirc(40,100){5}
\Line(35,100)(45,100)
\Line(40,95)(40,105)
\Line(100,100)(100,120)
\Line(100,100)(120,100)
\Text(120,120)[bl]{$C$}
\DashCArc(100,100)(16,0,90){2}
\BCirc(100,100){7}
\Line(100,40)(114,26)
\Text(120,20)[tl]{$d$}
\BCirc(100,40){5}
\Line(95,40)(105,40)
\Line(100,35)(100,45)
\end{picture}}
\parbox{80mm}{
\begin{eqnarray}
\label{eq:2meplussoln}
&&l_1=\frac{\spxa d\spaxx bA}{\spa bd}\nn
&&l_2=\frac{\spxa b\spaxx dA}{\spa bd}\nn
&&l_3=\frac{\spxa b\spaxx dC}{\spa db}\nn
&&l_4=\frac{\spxa d\spaxx bC}{\spa db}
\end{eqnarray}}
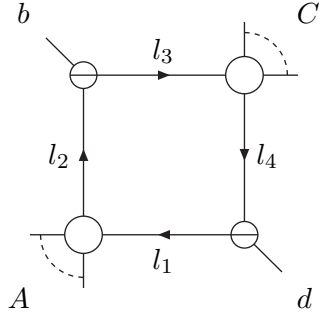
\parbox{55mm}{
\begin{picture}(140,140)(0,0)
\ArrowLine(40,40)(40,100)
\ArrowLine(40,100)(100,100)
\ArrowLine(100,100)(100,40)
\ArrowLine(100,40)(40,40)
\Text(70,35)[t]{$l_1$}
\Text(35,70)[r]{$l_2$}
\Text(70,105)[b]{$l_3$}
\Text(105,70)[l]{$l_4$}
\Line(40,40)(40,20)
\Line(40,40)(20,40)
\Text(20,20)[tr]{$A$}
\DashCArc(40,40)(16,180,270){2}
\BCirc(40,40){7}
\Line(40,100)(26,114)
\Text(20,120)[br]{$b$}
\BCirc(40,100){5}
\Line(35,100)(45,100)
\Line(100,100)(100,120)
\Line(100,100)(120,100)
\Text(120,120)[bl]{$C$}
\DashCArc(100,100)(16,0,90){2}
\BCirc(100,100){7}
\Line(100,40)(114,26)
\Text(120,20)[tl]{$d$}
\BCirc(100,40){5}
\Line(95,40)(105,40)
\end{picture}}
\parbox{80mm}{
\begin{eqnarray}
&&l_1=\frac{\spxxb Ab\spbx d}{\spb db}\nn
&&l_2=\frac{\spxxb Ad\spbx b}{\spb db}\nn
&&l_3=\frac{\spxxb Cd\spbx b}{\spb bd}\nn
&&l_4=\frac{\spxxb Cb\spbx d}{\spb bd}
\end{eqnarray}}
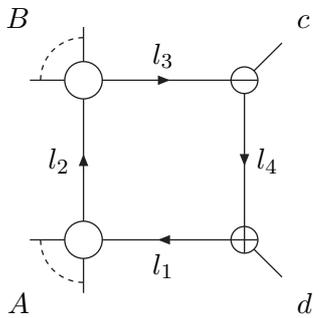
\parbox{55mm}{
\begin{picture}(140,140)(0,0)
\ArrowLine(40,40)(40,100)
\ArrowLine(40,100)(100,100)
\ArrowLine(100,100)(100,40)
\ArrowLine(100,40)(40,40)
\Text(70,35)[t]{$l_1$}
\Text(35,70)[r]{$l_2$}
\Text(70,105)[b]{$l_3$}
\Text(105,70)[l]{$l_4$}
\Line(40,40)(40,20)
\Line(40,40)(20,40)
\Text(20,20)[tr]{$A$}
\DashCArc(40,40)(16,180,270){2}
\BCirc(40,40){7}
\Line(40,100)(20,100)
\Line(40,100)(40,120)
\Text(20,120)[br]{$B$}
\DashCArc(40,100)(16,90,180){2}
\BCirc(40,100){7}
\Line(100,100)(114,114)
\Text(120,120)[bl]{$c$}
\BCirc(100,100){5}
\Line(95,100)(105,100)
\Line(100,40)(114,26)
\Text(120,20)[tl]{$d$}
\BCirc(100,40){5}
\Line(95,40)(105,40)
\Line(100,35)(100,45)
\end{picture}}
\parbox{80mm}{
\begin{eqnarray}
&&l_1=\frac{\spxa d\spbxxx cBA}{\spab dBc}\nn
&&l_2=\frac{\spxxb Bc\spaxx dA}{\spab dAc}\nn
&&l_3=\frac{\spxxxa BAd\spbx c}{\spab dBc}\nn
&&l_4=\spxa d\frac{(B+c)^2}{\spab dAc}\spbx c
\end{eqnarray}}
\parbox{55mm}
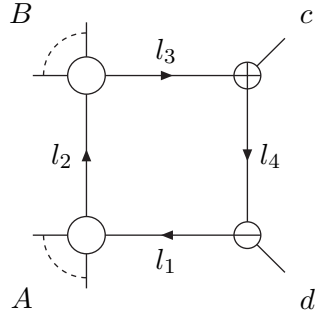
{
\begin{picture}(140,140)(0,0)
\ArrowLine(40,40)(40,100)
\ArrowLine(40,100)(100,100)
\ArrowLine(100,100)(100,40)
\ArrowLine(100,40)(40,40)
\Text(70,35)[t]{$l_1$}
\Text(35,70)[r]{$l_2$}
\Text(70,105)[b]{$l_3$}
\Text(105,70)[l]{$l_4$}
\Line(40,40)(40,20)
\Line(40,40)(20,40)
\Text(20,20)[tr]{$A$}
\DashCArc(40,40)(16,180,270){2}
\BCirc(40,40){7}
\Line(40,100)(20,100)
\Line(40,100)(40,120)
\Text(20,120)[br]{$B$}
\DashCArc(40,100)(16,90,180){2}
\BCirc(40,100){7}
\Line(100,100)(114,114)
\Text(120,120)[bl]{$c$}
\BCirc(100,100){5}
\Line(95,100)(105,100)
\Line(100,95)(100,105)
\Line(100,40)(114,26)
\Text(120,20)[tl]{$d$}
\BCirc(100,40){5}
\Line(95,40)(105,40)
\end{picture}}
\parbox{80mm}{
\begin{eqnarray}
&&l_1=\frac{\spxxxa ABc\spbx d}{\spab cBd}\nn
&&l_2=\frac{\spxxb Ad\spaxx cB}{\spab cAd}\nn
&&l_3=\frac{\spxa c\spbxxx dAB}{\spab cBd}\nn
&&l_4=\spxa c\frac{(B+c)^2}{\spab cAd}\spbx d
\end{eqnarray}}
\parbox{55mm}{
\begin{picture}(140,140)(0,0)
\ArrowLine(40,40)(40,100)
\ArrowLine(40,100)(100,100)
\ArrowLine(100,100)(100,40)
\ArrowLine(100,40)(40,40)
\Text(70,35)[t]{$l_1$}
\Text(35,70)[r]{$l_2$}
\Text(70,105)[b]{$l_3$}
\Text(105,70)[l]{$l_4$}
\Line(40,40)(40,20)
\Line(40,40)(20,40)
\Text(20,20)[tr]{$A$}
\DashCArc(40,40)(16,180,270){2}
\BCirc(40,40){7}
\Line(40,100)(26,114)
\Text(20,120)[br]{$b$}
\BCirc(40,100){5}
\Line(35,100)(45,100)
\Line(40,95)(40,105)
\Line(100,100)(114,114)
\Text(120,120)[bl]{$c$}
\BCirc(100,100){5}
\Line(95,100)(105,100)
\Line(100,40)(114,26)
\Text(120,20)[tl]{$d$}
\BCirc(100,40){5}
\Line(95,40)(105,40)
\Line(100,35)(100,45)
\end{picture}}
\parbox{80mm}{
\begin{eqnarray}
&&l_1=\frac{\spxa d\spaxx bA}{\spa bd}\nn
&&l_2=\frac{\spxa b\spaxx dA}{\spa bd}\nn
&&l_3=\spxa b\frac{\spa dc}{\spa db}\spbx c\nn
&&l_4=\spxa d\frac{\spa bc}{\spa db}\spbx c
\end{eqnarray}}
\parbox{55mm}{
\begin{picture}(140,140)(0,0)
\ArrowLine(40,40)(40,100)
\ArrowLine(40,100)(100,100)
\ArrowLine(100,100)(100,40)
\ArrowLine(100,40)(40,40)
\Text(70,35)[t]{$l_1$}
\Text(35,70)[r]{$l_2$}
\Text(70,105)[b]{$l_3$}
\Text(105,70)[l]{$l_4$}
\Line(40,40)(40,20)
\Line(40,40)(20,40)
\Text(20,20)[tr]{$A$}
\DashCArc(40,40)(16,180,270){2}
\BCirc(40,40){7}
\Line(40,100)(26,114)
\Text(20,120)[br]{$b$}
\BCirc(40,100){5}
\Line(35,100)(45,100)
\Line(100,100)(114,114)
\Text(120,120)[bl]{$c$}
\BCirc(100,100){5}
\Line(95,100)(105,100)
\Line(100,95)(100,105)
\Line(100,40)(114,26)
\Text(120,20)[tl]{$d$}
\BCirc(100,40){5}
\Line(95,40)(105,40)
\end{picture}}
\parbox{80mm}{
\begin{eqnarray}
&&l_1=\frac{\spxxb Ab\spbx d}{\spb db}\nn
&&l_2=\frac{\spxxb Ad\spbx b}{\spb db}\nn
&&l_3=\spxa c\frac{\spb cd}{\spb bd}\spbx b\nn
&&l_4=\spxa c\frac{\spb cb}{\spb bd}\spbx d
\end{eqnarray}}
\parbox{55mm}{
\begin{picture}(140,140)(0,0)
\ArrowLine(40,40)(40,100)
\ArrowLine(40,100)(100,100)
\ArrowLine(100,100)(100,40)
\ArrowLine(100,40)(40,40)
\Text(70,35)[t]{$l_1$}
\Text(35,70)[r]{$l_2$}
\Text(70,105)[b]{$l_3$}
\Text(105,70)[l]{$l_4$}
\Line(40,40)(26,26)
\Text(20,20)[tr]{$a$}
\BCirc(40,40){5}
\Line(35,40)(45,40)
\Line(40,100)(26,114)
\Text(20,120)[br]{$b$}
\BCirc(40,100){5}
\Line(35,100)(45,100)
\Line(40,95)(40,105)
\Line(100,100)(114,114)
\Text(120,120)[bl]{$c$}
\BCirc(100,100){5}
\Line(95,100)(105,100)
\Line(100,40)(114,26)
\Text(120,20)[tl]{$d$}
\BCirc(100,40){5}
\Line(95,40)(105,40)
\Line(100,35)(100,45)
\end{picture}}
\parbox{80mm}{
\begin{eqnarray}
&&l_1=\spxa d\frac{\spa ba}{\spa bd}\spbx a\nn
&&l_2=\spxa b\frac{\spa da}{\spa bd}\spbx a\nn
&&l_3=\spxa b\frac{\spa dc}{\spa db}\spbx c\nn
&&l_4=\spxa d\frac{\spa bc}{\spa db}\spbx c
\end{eqnarray}}
\parbox{55mm}{
\begin{picture}(140,140)(0,0)
\ArrowLine(40,40)(40,100)
\ArrowLine(40,100)(100,100)
\ArrowLine(100,100)(100,40)
\ArrowLine(100,40)(40,40)
\Text(70,35)[t]{$l_1$}
\Text(35,70)[r]{$l_2$}
\Text(70,105)[b]{$l_3$}
\Text(105,70)[l]{$l_4$}
\Line(40,40)(26,26)
\Text(20,20)[tr]{$a$}
\BCirc(40,40){5}
\Line(35,40)(45,40)
\Line(40,35)(40,45)
\Line(40,100)(26,114)
\Text(20,120)[br]{$b$}
\BCirc(40,100){5}
\Line(35,100)(45,100)
\Line(100,100)(114,114)
\Text(120,120)[bl]{$c$}
\BCirc(100,100){5}
\Line(95,100)(105,100)
\Line(100,95)(100,105)
\Line(100,40)(114,26)
\Text(120,20)[tl]{$d$}
\BCirc(100,40){5}
\Line(95,40)(105,40)
\end{picture}}
\parbox{80mm}{
\begin{eqnarray}
&&l_1=\spxa a\frac{\spb ab}{\spa db}\spbx d\nn
&&l_2=\spxa a\frac{\spb ad}{\spa db}\spbx b\nn
&&l_3=\spxa c\frac{\spb cd}{\spb bd}\spbx b\nn
&&l_4=\spxa c\frac{\spb cb}{\spb bd}\spbx d
\end{eqnarray}}

\chapter{Passarino--Veltman Reduction of Cuts}
\label{app:pv}

This appendix deals with the reduction of cut integrals which have loop
momentum dependence in their numerators. Those needed for this thesis
are:
\begin{equation}
I_0=\int d\mu,\qquad I_1^\mu=\int l^\mu d\mu,\qquad
I_2^{\mu\nu}=\int l^\mu l^\nu d\mu,
\end{equation}
\begin{equation}
J_0=\int\frac{d\mu}{\spab mlm},\qquad J_1^\mu=\int\frac{l^\mu}{
\spab mlm}d\mu,\qquad J_2^{\mu\nu}=\int\frac{l^\mu l^\nu}{\spab mlm}
d\mu,
\end{equation}
\begin{equation}
J_3^{\mu\nu\kappa}=\int\frac{l^\mu l^\nu l^\kappa}
{\spab mlm}d\mu,
\end{equation}
where
\begin{equation}
d\mu=\delta(l^2)\delta((l-P)^2)d^4l.
\end{equation}
The goal is to express all of these integrals in terms of $m^\mu$,
$P^\mu$, $g^{\mu\nu}$, $I_0$, and $J_0$. The method can be illustrated
for $J_1^\mu$ by noting Lorentz invariance forces it to take on the
structure
\begin{equation}
J_1^\mu=P^\mu I'+m^\mu I'',
\end{equation}
where $I'$ and $I''$ are scalar integrals. These can be determined
from the equations
\begin{equation}
\spab m{J_1}m=\int\frac{\spab mlm}{\spab mlm}d\mu
=I_0=\spab mPm I'
\end{equation}
and
\begin{equation}
2P\cdot J_1=\int \frac{2P\cdot l}{\spab mlm}d\mu
=P^2J_0=2P^2 I'+\spab mPm I''.
\end{equation}
Similarly, $I_2^{\mu\nu}$ can be computed by noting that it must take
on the form
\begin{equation}
I_2^{\mu\nu}=g^{\mu\nu}I'+P^\mu P^\nu I''
\end{equation}
and using the contractions
\begin{equation}
I_{2\mu}^{\mu}=\int l^2 d\mu=0=4I'+P^2I'',
\end{equation}
and
\begin{equation}
2P^\nu I_2^{\mu\nu}=\int (2P\cdot l)l^\mu d\mu
=P^2I_1^\mu=2P^\mu I'+2P^2 P^\mu I'',
\end{equation}
which relate $I'$ and $I''$ to $I_1^\mu$. This procedure will allow us
to construct the required integrals recursively. Rather than deriving
the results in detail, we will just note that they are
\begin{eqnarray}
I_1^\mu&=&\frac12 P^\mu I_0\\
I_2^{\mu\nu}&=&-\frac{P^2}{12}I_0+\frac13 P^\mu P^\nu I_0\\
J_1^\mu&=&\frac{P^\mu}{\spab mPm}I_0+m^\mu\bigg(-\frac{2P^2}
{\spab mPm^2}I_0+\frac{P^2}{\spab mPm}J_0\bigg)\\
J_2^{\mu\nu}&=&-\frac{g^{\mu\nu}P^2}{4\spab mPm}I_0+
\frac{P^\mu P^\nu}{2\spab mPm}I_0+\frac{(m^\mu P^\nu+P^\mu m^\nu)
P^2}{2\spab mPm^2}I_0\nn
&&\qquad+m^\mu m^\nu\bigg(-\frac{3P^4}{\spab mPm^3}I_0+
\frac{P^4}{\spab mPm^2}J_0\bigg),\\
J_3^{\mu\nu\kappa}&=&-\frac{(g^{\mu\nu}P^\kappa+g^{\nu\kappa}P^\mu
+g^{\kappa\mu}P^\nu)P^2}{12\spab mPm}I_0
+\frac{P^\mu P^\nu P^\kappa}{3\spab mPm}I_0\\
&&\qquad +\frac{(P^\mu P^\nu m^\kappa+m^\mu P^\nu P^\kappa+
P^\mu m^\kappa P^\nu)P^2}{6\spab mPm^2}I_0\\
&&\qquad +\frac{(m^\mu m^\nu P^\kappa+P^\mu m^\nu m^\kappa+
m^\mu P^\nu m^\kappa)P^4}{3\spab mPm^3}I_0\\
&&\qquad-\frac{(g^{\mu\nu}m^\kappa+g^{\nu\kappa}m^\mu
+g^{\kappa\mu}m^\nu)P^4}{12\spab mPm^2}I_0\\
&&\qquad +m^\mu m^\nu m^\kappa\bigg(-\frac{11}{9}\frac{P^6}
{\spab mPm^4}I_0+\frac{P^6}{3\spab mPm^3}J_0\bigg).
\end{eqnarray}

\end{document}